\documentclass[useAMS,usenatbib,a4paper]{mn2e}
\usepackage{aastex_hack}
\usepackage{epsfig}
\usepackage{times}

\newcommand{\myemail}{lam@astro.caltech.edu}
\newcommand{\arcmpt}{\hbox to 1pt{}\rlap{\arcmin}.\hbox to 2pt{}}
\newcommand{\arcsecpt}{\hbox to 1pt{}\rlap{\arcsec}.\hbox to 3pt{}}
\newcommand{\littless}{\ifmmode{\mbox{\scriptsize s} }
     \else{\hbox{\scriptsize s}}\fi}
\newcommand{\littlemm}{\ifmmode{\mbox{\scriptsize m} }
     \else{\hbox{\scriptsize m}}\fi}
\newcommand{\littlecirc}{\ifmmode{\mbox{\scriptsize \circ} }
     \else{\hbox{\scriptsize \circ }}\fi}
\newcommand{\littlehour}{\ifmmode{\mbox{\scriptsize h} }
     \else{\hbox{\scriptsize h}}\fi}
\newcommand{\toph}{\raise .9ex \hbox{\littlehour}}
\newcommand{\hourpt}{\hbox to 2pt{}\rlap{\hskip -.5ex \toph}.\hbox to 2pt{}}
\newcommand{\topemm}{\raise .9ex \hbox{\littlemm}}
\newcommand{\magpt}{\hbox to 2pt{}\rlap{\hskip -.5ex \topemm}.\hbox to 2pt{}}
\newcommand{\arcss}{\raise .9ex \hbox{\littless}}
\newcommand{\arcspt}{\hbox to 1pt{}\rlap{\arcss}.\hbox to 2pt{}}

\newcommand{\Neff}{\ifmmode N_{eff} \else $N_{eff}$\,\fi}
\newcommand{\Meff}{\ifmmode M_{eff} \else $M_{eff}$\,\fi}
\newcommand{\lt}{\ifmmode\,<\,\else \,$<$\,\fi}
\newcommand{\kms}{\ifmmode\,{\rm km}\,{\rm s}^{-1}\else km$\,$s$^{-1}$\fi}
\newcommand{\sersic}{S\'{e}rsic}
\newcommand{\chisqr}{$\chi^2$}
\newcommand{\magarc}{\ifmmode {{{{\rm mag}~{\rm arcsec}}^{-2}}}
             \else {{{mag}$~${arcsec}$^{-2}$}}
             \fi}
\newcommand{\taueff}{$\hat{\tau}$}
\newcommand{\avgFe}{$\langle$Fe$\rangle$}
\newcommand{\avgA}{$\langle$A$\rangle$}
\newcommand{\avgAl}{\ifmmode {\langle{\rm A}\rangle_{l}} \else {$\langle$A$\rangle_{l}$} \fi}
\newcommand{\avgAm}{\ifmmode {\langle{\rm A}\rangle_{m}} \else {$\langle$A$\rangle_{m}$} \fi}
\newcommand{\avgZ}{$\langle${\it Z}$\rangle$}
\newcommand{\avgZl}{\ifmmode {\langle{\rm Z}\rangle_{l}} \else {$\langle$Z$\rangle_{l}$} \fi}
\newcommand{\avgZm}{\ifmmode {\langle{\rm Z}\rangle_{m}} \else {$\langle$Z$\rangle_{m}$} \fi}
\newcommand{\etal}{et~al.\@}
\newcommand{\eg}{e.g.\@}
\newcommand{\ie}{i.e.\@}
\newcommand{\hii}{\ion{H}{2}}
\newcommand{\oi}{[\ion{O}{1}]}

\newcommand{\oiii}{[\ion{O}{3}]}
\newcommand{\oiiil}{[\ion{O}{3}] $\lambda$5007}
\newcommand{\oiiill}{[\ion{O}{3}] $\lambda \lambda$5007,4959}
\newcommand{\nii}{[\ion{N}{2}]}

\newcommand{\sii}{[\ion{S}{2}]}
\newcommand{\caii}{\ion{Ca}{2}}
\def\halpha   {H$\alpha$}

\def\hbeta    {H$\beta$}
\def\hgamma   {H$\gamma$}
\def\hdelta   {H$\delta$}

\def\hgammaA   {H$\gamma_{A}$}

\def\hdeltaA   {H$\delta_{A}$}
\def\sm {$\sim\,$}

\title[Population Synthesis of Spiral Galaxies]{Stellar Population and
Kinematic Profiles In Spiral Bulges \& Disks: Population Synthesis of
Integrated Spectra}

\author [L.A. MacArthur \etal]
{Lauren A. MacArthur$^{1}$\thanks{E-mail: \myemail},
J. Jes{\'u}s Gonz{\'a}lez$^{2}$\thanks{E-mail: jesus@astroscu.unam.mx}, and
St{\' e}phane Courteau$^{3}$\thanks{E-mail: courteau@astro.queensu.ca}\\
$^{1}$Department of Astrophysics, California Institute of 
                Technology, MS 105-24, Pasadena, CA 91125\\
$^{2}$Instituto de Astronomia, Universidad Nacional Aut{\'o}noma 
                de M{\'e}xico, Apdo Postal 70-264, Cd. Universitaria, 
                04510 M{\'e}xico\\
$^{3}$Department of Physics, Engineering Physics \& Astronomy, 
                 Queen's University, Kingston, ON K7L 3N6, Canada
}

\date{Accepted Jan 16, 2009}

\pagerange{\pageref{firstpage}--\pageref{lastpage}} \pubyear{2008}

\begin{document}

\label{firstpage}

\maketitle

\begin{abstract}

We present a detailed study of the stellar populations (SPs) and
kinematics of the bulge and inner disk regions of eight nearby spiral
galaxies (Sa--Sd) based on deep Gemini/GMOS data. The
long-slit spectra extend to 1--2 disk scale lengths with
S/N/\AA\,$\geq\,$50.  Several different model fitting techniques 
involving absorption-line indices and full spectrum fitting are
explored and found to weigh age, metallicity, and abundance ratios
differently. We find that SPs of spiral galaxies are not well
matched by single episodes of star formation; more
representative SPs must involve average SP values integrated over the
star formation history (SFH) of the galaxy.  Our ``full
population synthesis'' method is an optimised linear combination
of model templates to the full spectrum with masking of regions 
poorly represented by the models.  Realistic determinations of
the SP parameters and kinematics (rotation and velocity dispersion)
also rely on careful attention to data/model matching
(resolution and flux calibration).  The population fits reveal a wide
range of age and metallicity gradients (from negative to
positive) in the bulge, allowing for diverse formation
mechanisms. The observed positive age gradients within the
effective radius of some late-type bulges helps reconcile the
long-standing conundrum of the co-existence of secular-like
kinematics, light profile shape, and stellar bar with the
``classical''-like old and $\alpha$-enhanced SPs in the Milky
Way bulge.  The disks, on the other hand, almost always show mildly
decreasing to flat profiles in both age and metallicity, consistent
with inside-out formation.  Our spiral bulges follow the same
correlations of increasing light-weighted age and metallicity with
central velocity dispersion as those of elliptical galaxies and
early-type bulges found in other studies, but when SFHs more complex
and realistic than a single burst are invoked, the trend with
age is shallower and the scatter much reduced.  In a
mass-weighted context, however, all bulges are predominantly composed
of old and metal-rich SPs.  While secular contributions to the
evolution of many of our bulges is clearly evident, with young
(0.001--1\,Gyr) SPs contributing as much as 90\% of the optical
($V$-band) light, the bulge mass fraction from young stars is
small ($\la$\,25\%).  This implies a bulge formation dominated
by early processes that are common to all spheroids, whether they
currently reside in disks or not.  While monolithic collapse cannot be
ruled out in some cases, merging must be invoked to explain the SP
gradients in most bulges.  Further bulge growth via secular processes
or ``rejuvenated'' star formation generally contributes minimally to
the stellar mass budget, with the relative secular weight
increasing with decreasing central velocity dispersion.

\end{abstract}


\section{Introduction}\label{sec:intro}
In the context of the currently favored cosmological $\Lambda$-Cold
Dark Matter model ($\Lambda$CDM) of our Universe, the formation and
evolution of galaxies remains a major unsolved problem.  In this
model, structures form in a bottom-up, hierarchical, manner by which
smaller fragments merge together to form more massive systems.  While
this model has been very successful at reproducing observations on
large scales, a number of issues remain to be resolved at the
galaxy-scale regime (Moore \etal\ 1999; Navarro \& Steinmetz 2000;
Primack 2007, and references therein).  In particular, the formation of
disk galaxies is not well represented in current implementations of
simulations based on $\Lambda$CDM (\eg\ Bell \etal\ 2003; Kaufmann \etal\
2007; Dutton \etal\ 2007).  The discrepancies do not necessarily
indicate a failure in the $\Lambda$CDM model, but rather point to
physical regimes and processes that are either not well understood or
difficult to implement in large simulations.  Given the theoretical
difficulties and limitations faced by current models, a true
understanding of disk galaxies must include information from an
observational perspective.

All aspects of baryonic physics, from hydrodynamical processes
involving gas, through star formation and feedback from supernovae and
AGN, can play a role in the formation and regulation processes that
shape galaxies, particularly where dynamically violent processes such
as major mergers and rapid collapse are not dominant.  For spiral
galaxies, the very presence of a disk component implies that no such
major event has occurred since the formation of the dynamically
fragile disk.  Additionally, most disk galaxies harbor central bulge
components whose prominence in terms of total mass or light weight
spans a wide range from truly bulgeless systems to the bulge-dominated
early-type S0/Sa's.  A question that naturally arises is if the final
step to the disk-less, \ie\ pure elliptical, galaxies is a natural
extension of bulge-to-total ratios in systems along the Hubble
sequence.  Such an extension would imply that the dominant bulges of
early-type spirals have followed a similar evolutionary path to those
of pure elliptical galaxies.  However, coming from the opposite end, a
connection between the tiny bulges of late-type spirals, through
early-types, to pure ellipticals may not seem as obvious, and
allowance must still be made for the pure disk galaxies in any
scenario.  The emerging observational picture of bulge formation
currently implies similar evolutionary paths for early-type spirals
and pure ellipticals, while the later-type bulges exhibit much more
diversity in their observed properties, consistent with having formed
through a secular redistribution of material from the disk (\eg\
Kormendy \& Kennicutt 2004).  The nomenclature predominant in the
literature to distinguish between formation scenarios for bulges names
those appearing very similar to pure elliptical systems 
(\ie\ rapid and/or violent formation which includes both the 
monolithic collapse and major merging scenarios) as ``classical
bulges'' and those formed secularly from the disk as ``pseudobulges''.
However, uncertainties in making a clear-cut distinction between these
cases still remain, particularly in regards to the stellar populations
(SPs) of spiral bulges.

A very useful probe in discerning between formation scenarios is a
detailed breakdown of the age, metallicity ($Z$), and kinematic
properties of the SPs comprising bulges of all types.  Information
about both light- and mass-weighted quantities is needed to form a
comprehensive picture of the star formation history (SFH) of a given
system.  However, beyond our Local Group, we are limited to
observations of the integrated light along a given line of sight.  Of
relevance is whether the integrated light at any location in a galaxy
can be deconvolved into the relative fractions of stars of a given
population that contribute to the total light.  This question is
especially acute for spiral galaxies that are known to harbor a
mixture of young and old stars and can suffer from the extinction
effects of dust.

Observations of colour gradients and absorption line-indices in
galaxies can both be interpreted as a single-burst, single-$Z$ stellar
population (SSP), from which light-weighted SSP-equivalent ages and
metallicities can be drawn with the help of spectral synthesis models
(\eg\ Vazdekis 1999; Bruzual \& Charlot 2003, hereafter BC03; Thomas,
Maraston, \& Bender 2003; Le Borgne \etal\ 2004; Maraston 2005;
Schiavon 2007).  Elliptical galaxies (\eg\ Trager, Faber, \& Dressler
2008) and globular clusters (\eg\ Cohen, Blakeslee \& Rhyzhov 1998;
Puzia \etal\ 2005; Hernandez \& Valls-Gabaud 2008), are often assumed
to result from the evolution of a single burst of star formation and
thus can be represented by a SSP of a given age and $Z$.  This
assumption, however, breaks down for spiral galaxies that are believed
to have been converting gas into stars at a relatively constant rate
over most of their lifetime (\eg\ Kennicutt 1983; James, Prescott, \&
Baldry 2008).  Since the presence of a young SP will dominate the
optical light even when its contribution to the mass is minimal, the
light-weighted ages will not be representative of the entire
population.  Even pure ellipticals show signs of ``frostings'' of
young SPs and may not be well represented by a single SSP (Serra \&
Trager 2007).

Greater potential for disentangling the SFH of spiral galaxies exists
if they can be assumed to have proceeded smoothly with time. This
approach was followed by MacArthur \etal\ (2004) using broad-band
optical colours for 172 low-inclination disk galaxies.  To the extent
that the nearby spiral galaxies in this sample share a similar
underlying SFH and that no burst involving more than $\sim$\,10\% of
the galaxy mass has occurred within the past 1--2\,Gyr, the method
provides reliable relative results.  Among others, it was found that
the SFH of spiral galaxies depends strongly on the galaxy potential
and halo spin parameter.

Conclusions based on colour gradients are, however, plagued by a
degeneracy between the effects of age, $Z$, and dust, all leading to
redder colours.  Spectroscopic techniques, such as those based on
absorption-line equivalent widths that are largely impervious to dust
effects (MacArthur 2005), or fitting of the full spectral energy
distribution (SED) which can incorporate a dust component, offer a
more detailed and discriminating view, especially in light of the
latest implementations of SP synthesis models (\eg\ Vazdekis 1999;
BC03; Le Borgne \etal\ 2004).

Early-type spirals with their large bulge-to-disk ratio offer a
significant observational advantage over late-types since their
bulges, which rise significantly above and below the disk, can be
studied spectroscopically in the edge-on perspective free from disk
contamination and extinction from dust.  Late-type bulges, on the
other hand, are small and often no thicker than the galaxy disk
itself, thus requiring a face-on projection for their study.  While
dust and contamination from the inner disk will still thwart any
pristine observation of the bulge, the face-on orientation minimizes
line-of-sight integrations and keeps the light distribution of the
bulge and disk free of inclination effects.

Spectroscopic studies of bulges spanning the full range of Hubble
types are few, and the results are often conflicting.  The
absorption-line studies of the central regions of Sa\,--\,Sc spirals
of Trager \etal\ (1999) and Proctor \& Sansom (2002, hereafter PS02)
both find that late-type bulges cannot be reproduced using primordial
collapse models and invoke extended gas infall onto the central bulge
to explain the observations.  In a similar study, Goudfrooij \etal\
(1999) conclude the opposite, their findings being more compatible
with predictions of the `dissipative collapse' model than with those
of the `secular evolution' model for bulge formation.  Their sample
is, however, dominated by early-type spirals and these conclusions may
not apply to the full range of spiral types.  Finally, the analysis of
Moorthy \& Holtzman (2006) of line strengths in the bulges and inner
disks of 38 spirals also favours a scenario whereby
merging is the dominant mechanism for bulge formation.

Thomas \& Davies (2006) reanalyzed the PS02 data and found no
difference between the SPs of spiral bulges and Es at a given central
velocity dispersion, $\sigma_0$, concluding that processes involving
disk material cannot be responsible for the recent star formation
implied by the young ages.  In an analysis of stellar absorption line
gradients in the bulges of 32 nearby edge-on spirals (S0\,--\,Sc),
Jablonka \etal\ (2007) also find that, when compared at a at a given
velocity dispersion, galaxy bulges resemble elliptical galaxies.
Recently, MacArthur \etal\ (2008) have extended these results to
bulges at intermediate redshifts (0.1\,$<$\,$z$\,$<$\,1) finding that
the mass assembly history of bulges, inferred via the Fundamental
Plane, is indistinguishable from that of pure spheroidal galaxies
(E/S0s) at a given mass.  

A significant outlier from this trend, however, is the bulge of our
own Milky Way (MW).  Detailed photometric (Zoccali \etal\ 2003) and
spectroscopic (Rich \& Origlia 2005; Zoccali \etal\ 2006) studies of
abundance ratios of individual MW bulge stars reveal old and
$\alpha$-enhanced SPs that must have formed long ago and on short
timescales, and no evidence is found for the presence of a younger SP.
This observation is difficult to reconcile with the MW's small
$\sigma_0$ and the presence of a stellar bar (\eg\
L{\'o}pez-Corredoira \etal\ 2007).  As Thomas \& Davies point out, the
MW studies sample a larger physical radius than those of external
galaxies so this conundrum could be resolved if there is a positive
age gradient in the bulge.

Adding further complexity to the picture, in a study of absorption-line
maps of 24 early-type spirals, Peletier \etal\ (2007) find that while
inclined samples, which sample the disk-free outer bulge regions,
reveal uniformly old SPs, randomly oriented samples,
spanning the same range in Hubble type, show a wide range in SP
parameters.  This discrepancy, along with their observation of central
dips in the velocity dispersion profiles of half of their bulges,
indicative of the presence of a colder central component, can be
reconciled in a picture whereby the centers of most early-type spirals
contain multiple kinematic components: an old and slowly rotating
elliptical-like component, and one or more disk-like, rotationally
supported, components which are typically young, but can be old.  The
sum of all central components make up the photometrically defined
bulge, and the different components contribute differently to the
integrated light depending on the viewing angle.  Their conclusion of
a rejuvenation of the central region through dynamical disk-like
processes, however, is difficult to understand in the context of the
above results in that the ``rejuvenated'' low-$\sigma$ ellipticals do
not have disk components.

At the heart of these conflicts is the fact that extant studies of
local bulges are limited in one way or another (sample size, range in
Hubble type, data of insufficient depth and/or spatial resolution for
studies of gradients and assessment of disk contamination, etc.).  Our
aim is to remedy this situation with a systematic and homogeneous
study of the SPs and kinematics of bulges for the full range of Hubble
types.  A first, and important, step is to establish a reliable
procedure for extracting a proper representation of the SP content
that does not suffer from the degeneracies and limitations discussed
above.  In this paper, we present a pilot sample of long-slit optical
spectroscopy of eight nearby, low-inclination, spiral galaxies
observed with Gemini/GMOS with which we develop such a technique and
attempt to address the formation of spiral bulges through a study of
their radial profiles in SPs and kinematics.

The organization of the paper is as follows.  In \S\ref{sec:data} we
detail the sample selection, observational strategy, and reduction
procedures.  \S\ref{sec:SPs} outlines our technique for
extracting SP parameters from integrated spectra including details of
the fitting algorithm, providing specifics about the weighting of
individual pixels and our so-called ``$\sigma$-clipping'' procedure
for masking deviant pixels, as well as a description of the model SP
templates used in the fits. Our results are presented in
\S\ref{sec:results} which includes the derived SP parameters and their
radial profiles, kinematic profiles (rotation velocity and velocity
dispersion), an investigation of correlations of central parameters,
and a comparison with previous studies.  We discuss in
\S\ref{sec:discuss} results for each individual galaxy in the context
of bulge formation scenarios, and \S\ref{sec:summary} summarizes the
salient points from the entire analysis.  Finally, we provide several
appendical sections: \S\ref{sec:spectra} presents the observed
radially resolved spectra for our galaxies.  \S\ref{sec:AZfits}
discusses derivations of SP parameters for our spiral galaxies based
on the Lick-index system and of full spectrum fitting of single SSPs.
All techniques are compared and contrasted in \S\ref{sec:comparefits}.

\section{Data}\label{sec:data}
\begin{table*}
\centering
\begin{minipage}{0.76\textwidth}
\caption{Galaxy Sample: catalog information.}\label{tab:galaxies}
\begin{tabular}{rrrc@{\,\,\,$\times$}cccccc}
\hline
\multicolumn{2}{c}{Names} &
\multicolumn{1}{c}{Hubble} &
\multicolumn{2}{c}{Diam} &
\multicolumn{1}{c}{RA} &
\multicolumn{1}{c}{DEC} &
\multicolumn{1}{c}{V$_{helio}$} &
\multicolumn{1}{c}{M$_B$} &
\multicolumn{1}{c}{A$_B$}\\
\multicolumn{1}{c}{\scriptsize NGC/IC} &
\multicolumn{1}{c}{\scriptsize UGC} &
\multicolumn{1}{c}{Type} &
\multicolumn{2}{c}{a(\arcmin)\,\,$\times$\,\,b(\arcmin)} &
\multicolumn{2}{c}{(J2000)} &
\multicolumn{1}{c}{(km s$^{-1}$)} &
\multicolumn{1}{c}{\scriptsize{(mag)}} &
\multicolumn{1}{c}{\scriptsize{(mag)}} \\
\multicolumn{1}{c}{(1)} & \multicolumn{1}{c}{(2)} & \multicolumn{1}{c}{(3)} & 
\multicolumn{1}{c}{(4)} &
\multicolumn{1}{c}{(5)} & \multicolumn{1}{c}{(6)} & \multicolumn{1}{c}{(7)} & 
\multicolumn{1}{c}{(8)} &
\multicolumn{1}{c}{(9)} & \multicolumn{1}{c}{(10)}\\
\hline
{\bf N0173} & U369   & SA(rs)c   &  \phn3.2 & 2.6 & 00\toph37\topemm12\arcspt47 &   +01\degr56\arcmin32\farcs1 & 4366 & 13.70 & 0.110 \\
{\bf N0628} & U1149  & SA(s)c    & 10.5 & 9.5 & 01\toph36\topemm41\arcspt77 &   +15\degr47\arcmin00\farcs5 &  \phn657 & \phn9.95 & 0.301 \\
N1015 & {\bf U2124}  & SB(r)a    &  \phn2.6 & 2.6 & 02\toph38\topemm11\arcspt56 & $-$01\degr19\arcmin07\farcs3 & 2629 & 12.98 & 0.139 \\
{\bf N7490} & U12379 & Sbc       &  \phn2.8 & 2.6 & 23\toph07\topemm25\arcspt17 &   +32\degr22\arcmin30\farcs2 & 6213 & 13.05 & 0.362 \\
{\bf N7495} & U12391 & SAB(s)c   &  \phn1.8 & 1.7 & 23\toph08\topemm57\arcspt18 &   +12\degr02\arcmin52\farcs9 & 4887 & 13.73 & 0.371 \\
{\bf N7610} & U12511 & Scd       &  \phn2.5 & 1.9 & 23\toph19\topemm41\arcspt37 &   +10\degr11\arcmin06\farcs0 & 3554 & 13.44 & 0.171 \\
{\bf N7741} & U12754 & SB(s)cd   &  \phn4.4 & 3.0 & 23\toph43\topemm54\arcspt37 &   +26\degr04\arcmin32\farcs2 &  \phn751 & 11.84 & 0.323 \\
{\bf I0239} & U2080  & SAB(rs)cd &  \phn4.6 & 4.2 & 02\toph36\topemm27\arcspt88 & +38\degr58\arcmin11\farcs7 &  \phn903 & 11.80 & 0.307 \\
\hline
\end{tabular}
{\scriptsize {\it Notes} ---
Cols.\@ (1) \& (2): NGC/IC, and UGC Galaxy IDs.  Names highlighted in bold 
are those used subsequently to refer to the particular galaxy. 
Col.\@ (3): Hubble Type (from NED).
Cols.\@ (4 \& 5): Major ($a$) and minor ($b$) axis diameters (in arcmin).
Cols.\@ (6 \& 7): Galaxy RA and DEC (J2000).
Col.\@ (8): Heliocentric radial velocity.
Col.\@ (9): Total $B$-band magnitude (Vega) from NED.
Col.\@ (10): $B$-band galactic extinction from Schlegel, Finkbeiner, \& 
            Davis (1998).
}
\end{minipage}
\end{table*}

The long-slit spectroscopic data for this study were collected using
the Gemini Multi-Object Spectrograph (GMOS; Hook \etal\ 2004) on the
8-m Gemini North telescope at Mauna Kea in Hawaii.  The GMOS detector
consists of three 2048$\times$4608 CCDs with 13.5\,$\mu$m pixels
providing a spatial resolution of 0.072\arcsec/pix and a dispersion of
0.45\,\AA/pix with the B600\_G5303 grating.  The internal stability of
the slit mask requires two equally spaced bridges between the slit
edges which lie along the spatial direction.  The spacings between the
3 CCD detectors manifest as small ($\sim$17\,\AA) gaps in wavelength.

The high sensitivity of the B600\_G5303 grating in the blue matches
well our simultaneous spectral coverage of
$\sim$\,4050--6750\,\AA.  This range includes most of the major
atomic and molecular features to disentangle age and metallicity
effects in integrated galaxy spectra (Worthey 1994).  
The slit field of view (FOV) was 5\arcmin\
(length) $\times$ 2\arcsec\ (width) for all of our observations.  The
choice of slit-width was prescribed by the need to maximize
signal-to-noise (S/N) in the outer disk while maintaining adequate
spectral resolution throughout.  The 2\arcsec\ slit and B600 grating
give a top-hat 10.81\,\AA\ full width at half maximum (FWHM)
resolution and a Gaussian instrumental resolution of
0.8\,$\pm$\,0.02\,\AA\ (as measured from the width of the narrowest sky
emission lines).

\subsection{Galaxy Sample}\label{sec:Gemini_sample}

Ideally, we desire a large and homogeneous sample of nearby spiral
galaxies spanning the full Hubble sequence to study spectroscopically
the SPs of spiral bulges and disks.  The sample should also include
barred and non-barred spirals in order to test for the expected mixing
effects by a bar.  The size of the current sample was a compromise
between a reasonable telescope time request for a pilot study and the
need to measure systematic variations within that sample.  We have
thus narrowed in on the following criteria:\\
$\bullet$ Hubble type Sa\,--\,Sd (with emphasis on later types)\\
$\bullet$ Mix of barred/unbarred systems and SB profile types\\
$\bullet$ Face-on (inclination $\la$\,35\degr)\\
$\bullet$ Blue Galactic extinction 
A$_{B}$\,=\,4\,$\times$\,$E(B-V)$\,$\le$\,0.5\,mag (Schlegel \etal\ 1998)\\
$\bullet$ Blue major axis $\la$\,2\,$\times$\,slit length
($\la$\,10\arcmpt5), but large enough to resolve bulge and inner
disk ($\ga$\,2\arcsec).\\

The galaxy sample and catalog information from the NASA Extragalactic
Database (NED)\footnote{http://nedwww.ipac.caltech.edu/} are shown in
Table~\ref{tab:galaxies}.  Optical imaging for all eight galaxies was
collected using the Large Format Camera on the 200-inch Hale telescope
at the Palomar Observatory in December 2006.  Observations were
obtained in the BVRI Bessel filters for all galaxies except N0173, for
which only B\&V were obtained (due to bad weather).  Surface
brightness profiles were extracted from these images and bulge and
disk photometric parameters were derived from bulge-to-disk (B/D)
decompositions into an exponential disk plus a generalized power-law
(\sersic) profile for the bulge, as listed in
Table~\ref{tab:gal_pars}.  We refer the reader to MacArthur, Courteau,
\& Holtzman (2003, hereafter Mac03) for details about profile
extraction and B/D decompositions.  Note that here we refer to the
disk scale length as $r_d$ and the bulge effective (half-light) radius
as $r_e$ for a profile characterized by the \sersic\ $n$ shape
parameter.

As is typical for late-type bulges (\eg\ de~Jong 1996; Graham 2001;
Mac03), most of our decompositions reveal bulge profile shapes close
to exponential, which translates to a \sersic\ index $n$\,=\,1.  Most
notably, the bulge of N7495 was best fit with $n$\,=\,2.3.  Bulges
with $n$\,$\ga$\,2 are typically found in earlier-type bulges thought
to be more akin to elliptical galaxies (\eg\ Fisher \& Drory 2008).
This suggests perhaps that the Hubble type of this galaxy is earlier
than Sc.  In fact, as noted in NED, the original UGC catalogue does
refer to this as an ``early spiral'' (Nilson 1973).  On the other
hand, our only Sa bulge, U2124, is best fit with $n$\,=\,$1.5$ which is low
for an early-type bulge.  However, the strong bar in this galaxy
renders the decomposition less certain.
\begin{table}
\centering
\begin{minipage}{0.26\textwidth}
\caption{Bulge and Disk Photometric Scale Parameters \label{tab:gal_pars}}
\begin{tabular}{rccc}
\hline
\multicolumn{1}{c}{Name} &
\multicolumn{1}{c}{$r_{e}$} &
\multicolumn{1}{c}{S\'{e}rsic} &
\multicolumn{1}{c}{$r_{d}$} \\
\multicolumn{1}{c}{\scriptsize NGC} &
\multicolumn{1}{c}{(\arcsec)} &
\multicolumn{1}{c}{$n$} &
\multicolumn{1}{c}{(\arcsec)} \\
\multicolumn{1}{c}{(1)} & \multicolumn{1}{c}{(2)} & \multicolumn{1}{c}{(3)} & \multicolumn{1}{c}{(4)} \\
\hline
N0173 &  3.8 & 1.5 & 21.9 \\
N0628 & 10.5 & 1.1 & 68.2 \\
U2124 &  5.7 & 1.5 & 12.2 \\
N7490 &  4.5 & 1.6 & 21.4 \\ 
N7495 &  3.6 & 2.3 & 17.2 \\
N7610 &  1.7 & 0.8 & 17.6 \\
N7741 &  6.7 & 1.0 & 42.3 \\
I0239 &  8.2 & 0.9 & 43.8 \\
\hline
\end{tabular}
\scriptsize{{\it Notes} --- Photometric bulge and disk scale
parameters based on optical imaging (either $R$ or $I$-band, but $V$
for N0173) from the Palomar 200-inch telescope.  Errors on the disk
scale lengths are $\la$\,10\% and $\la$\,20\% for bulge
parameters.  See Mac03 for details of our B/D decomposition scheme.
Col.\@ (1): Galaxy ID.  
Cols.\@ (2--4): B/D decomposition values for bulge effective radius 
($r_e$), \sersic\ $n$, and disk scale length ($r_d$).}
\end{minipage}
\end{table}

\subsection{Standard Star Sample}\label{sec:Gemini_stars}

We observed seven Lick/IDS standard stars (Worthey \etal\ 1994) with
the same observational set-up as the main targets for flux calibration
of the galaxy spectra.  These included dwarfs and giants with a range
of physical parameters relevant for star-forming galaxies (spectral
type: A1--F5; effective temperature: $T_{eff}$\,$\sim$\,4700--9000\,K,
surface gravity: log($g$)\,=\,2.4--4.3; and metallicity:
[Fe/H]\,=\,$-$0.45--0.07).  Flux calibration was achieved by
comparison of our standard star observations with flux calibrated
spectra of stars in common with the ELODIE archive (Moultaka \etal\
2004; see http://atlas.obs-hp.fr/elodie).

\subsection{Integration Times and Observing Strategy}\label{sec:Gemini_int}

\begin{figure*}
  \vskip 2.5in
 {\includegraphics{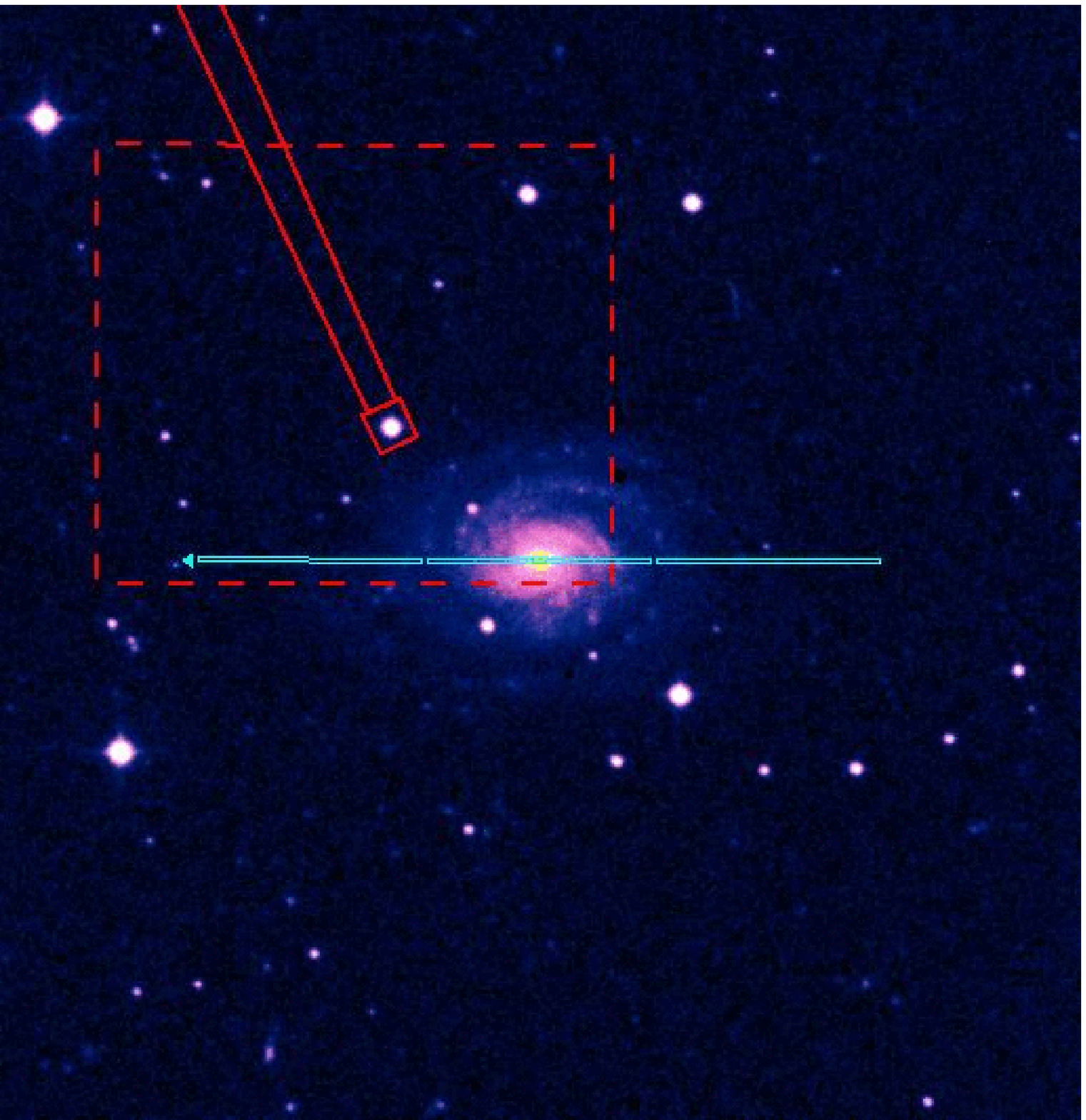}}
 {\includegraphics{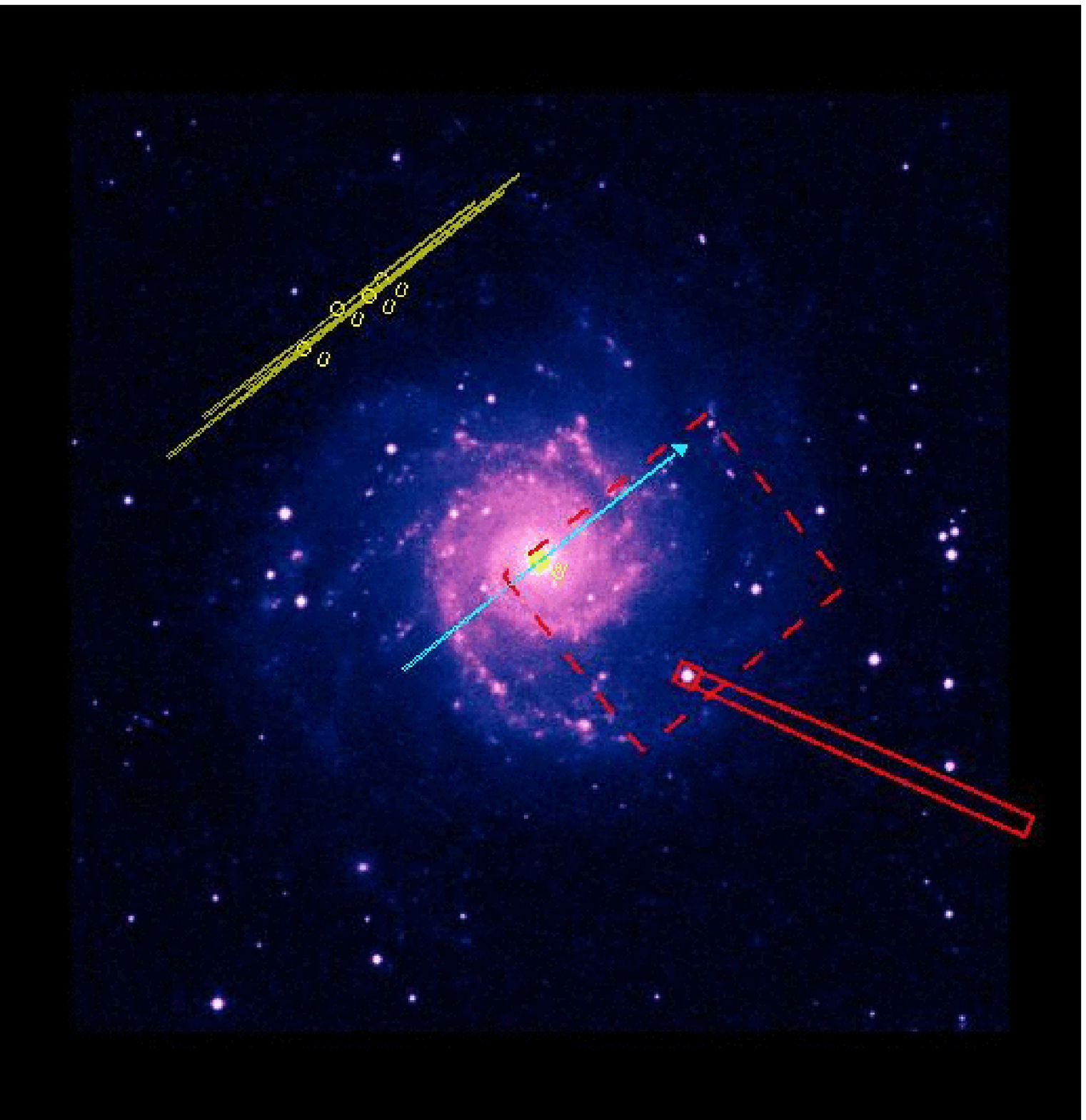}}
 {\includegraphics{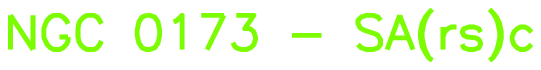}}
 {\includegraphics{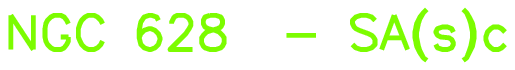}}

 {\includegraphics{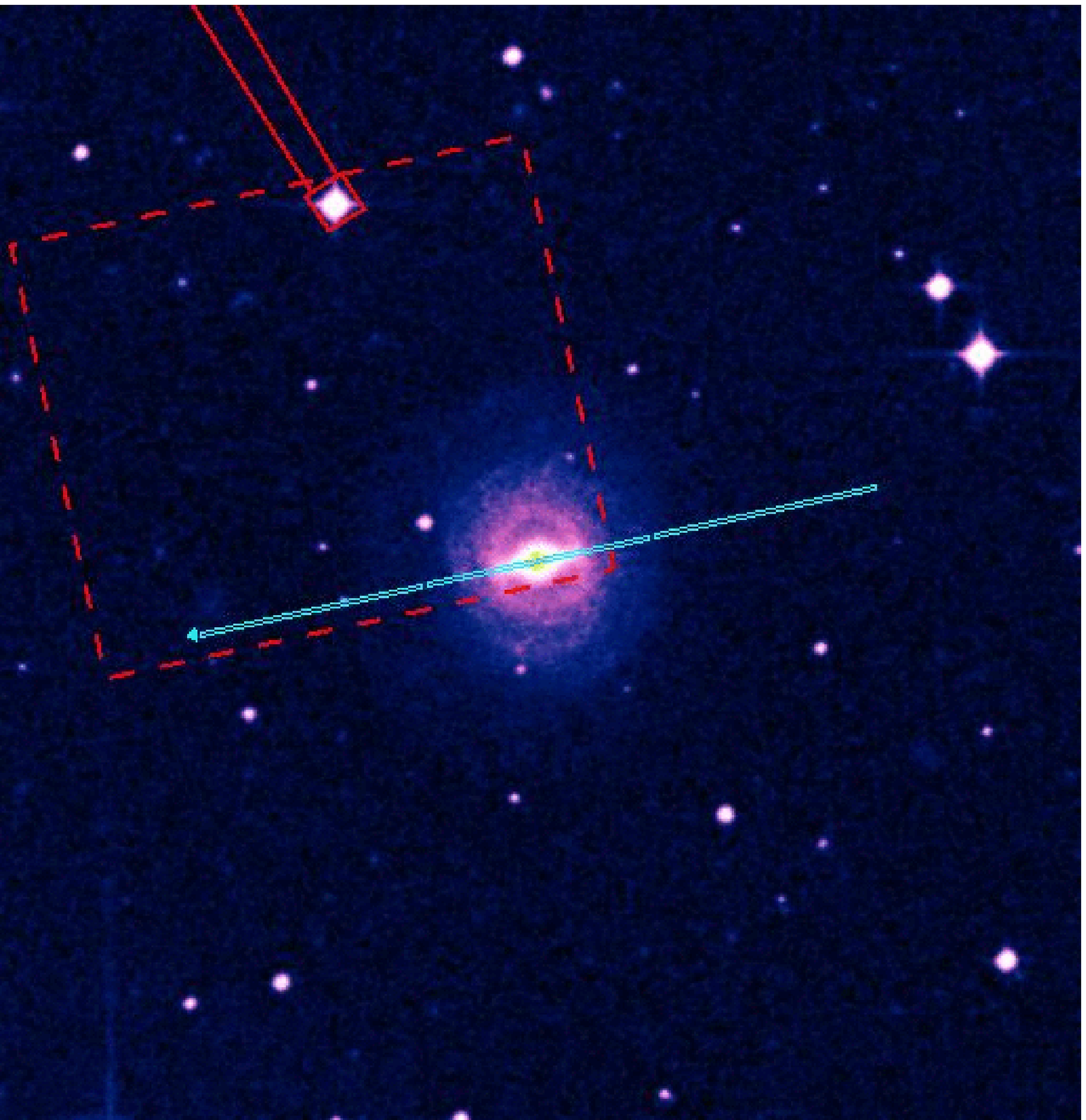}}
 {\includegraphics{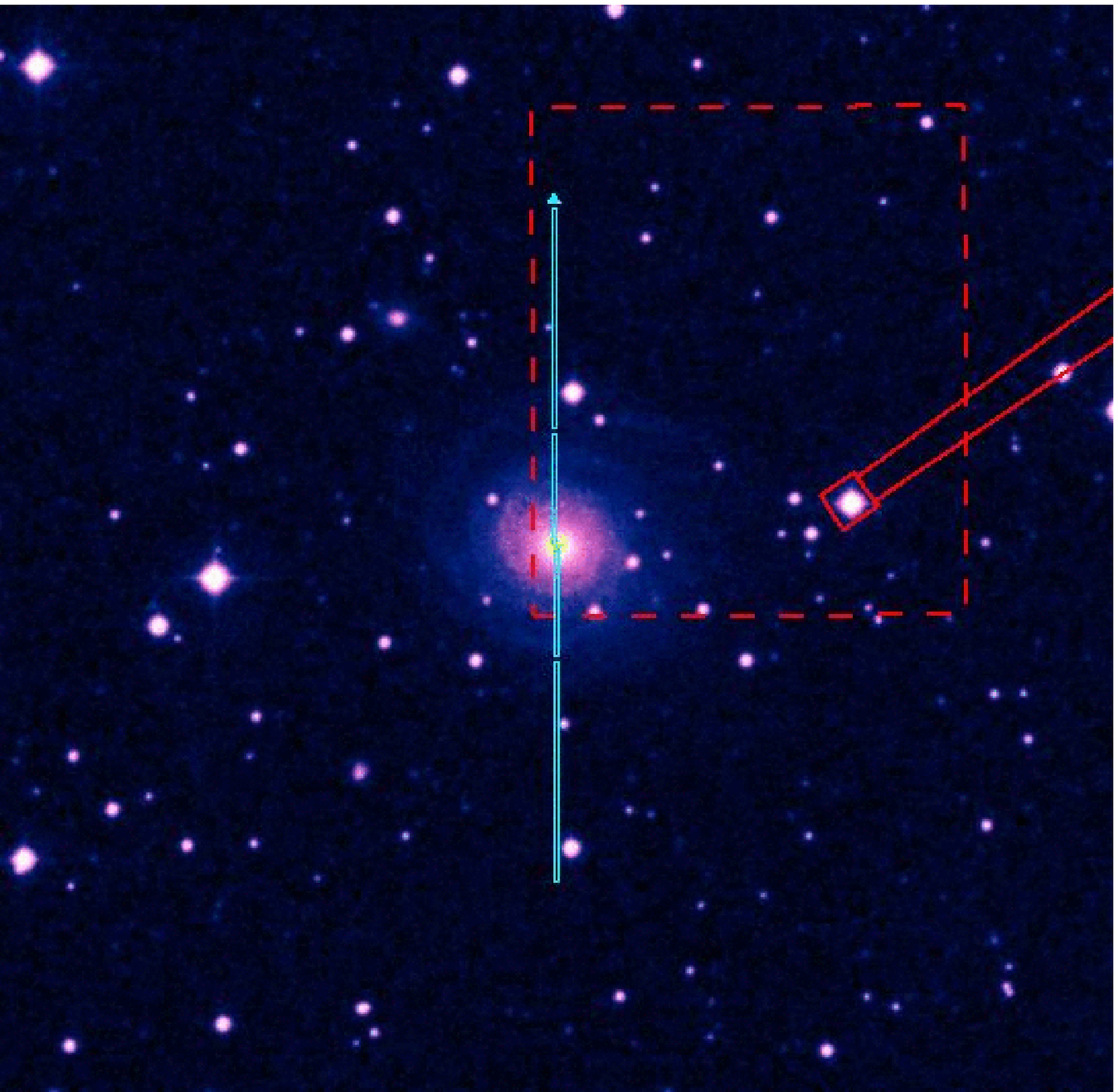}}
 {\includegraphics{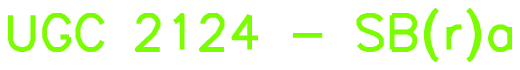}}
 {\includegraphics{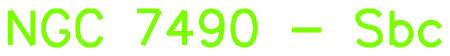}}

 {\includegraphics{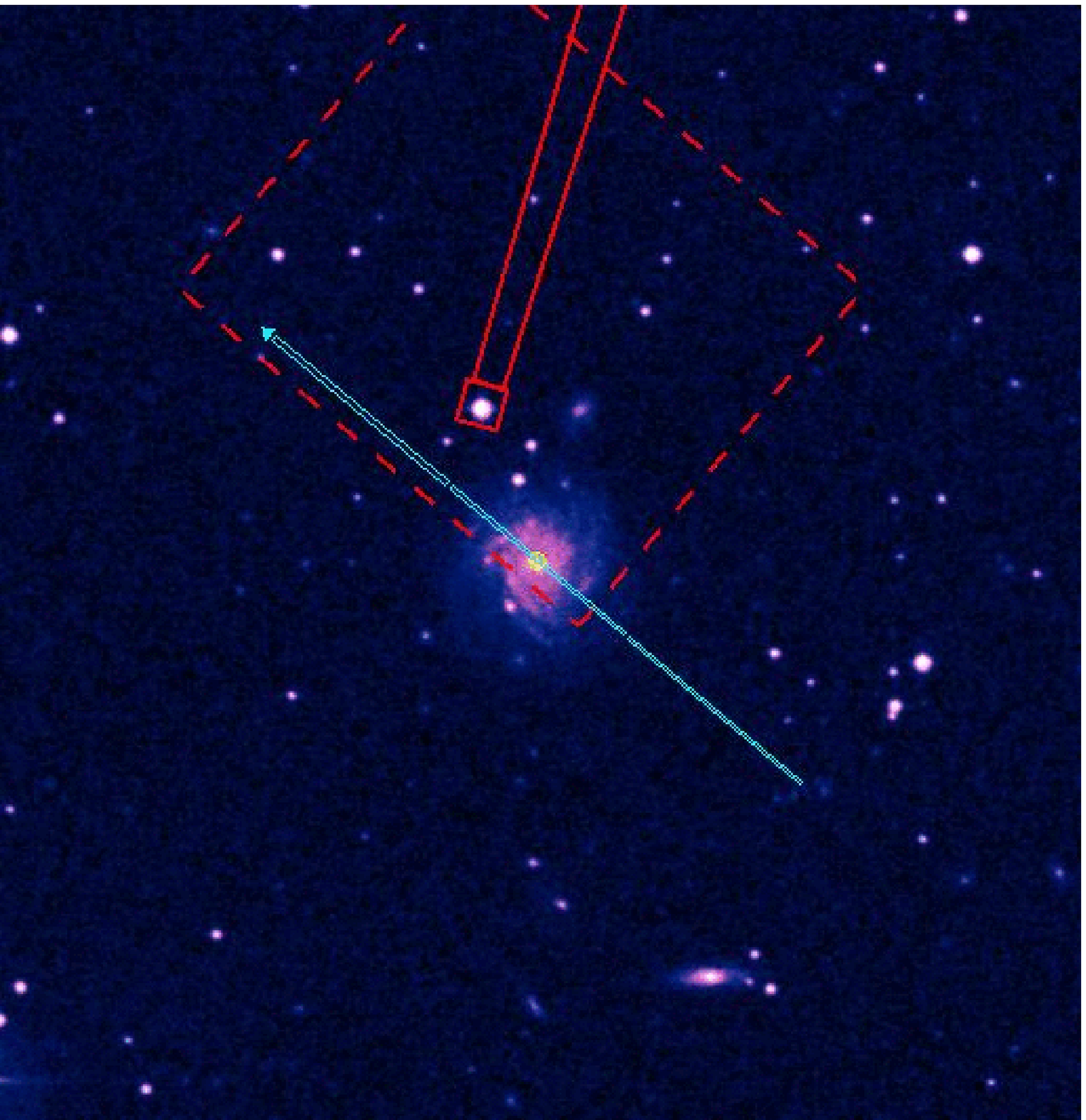}}
 {\includegraphics{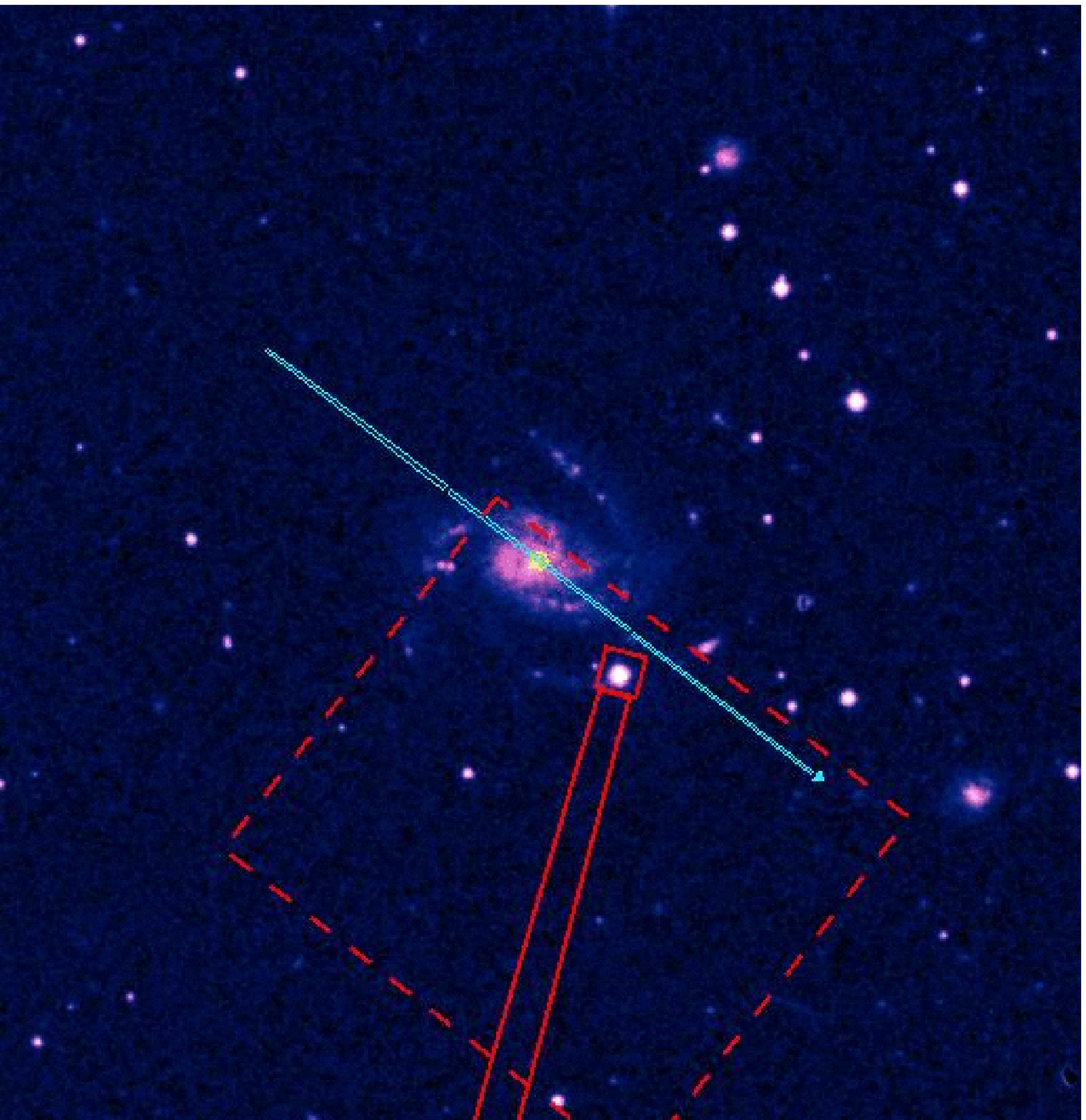}}
 {\includegraphics{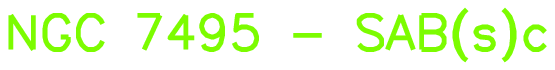}}
 {\includegraphics{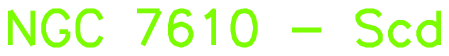}}

 {\includegraphics{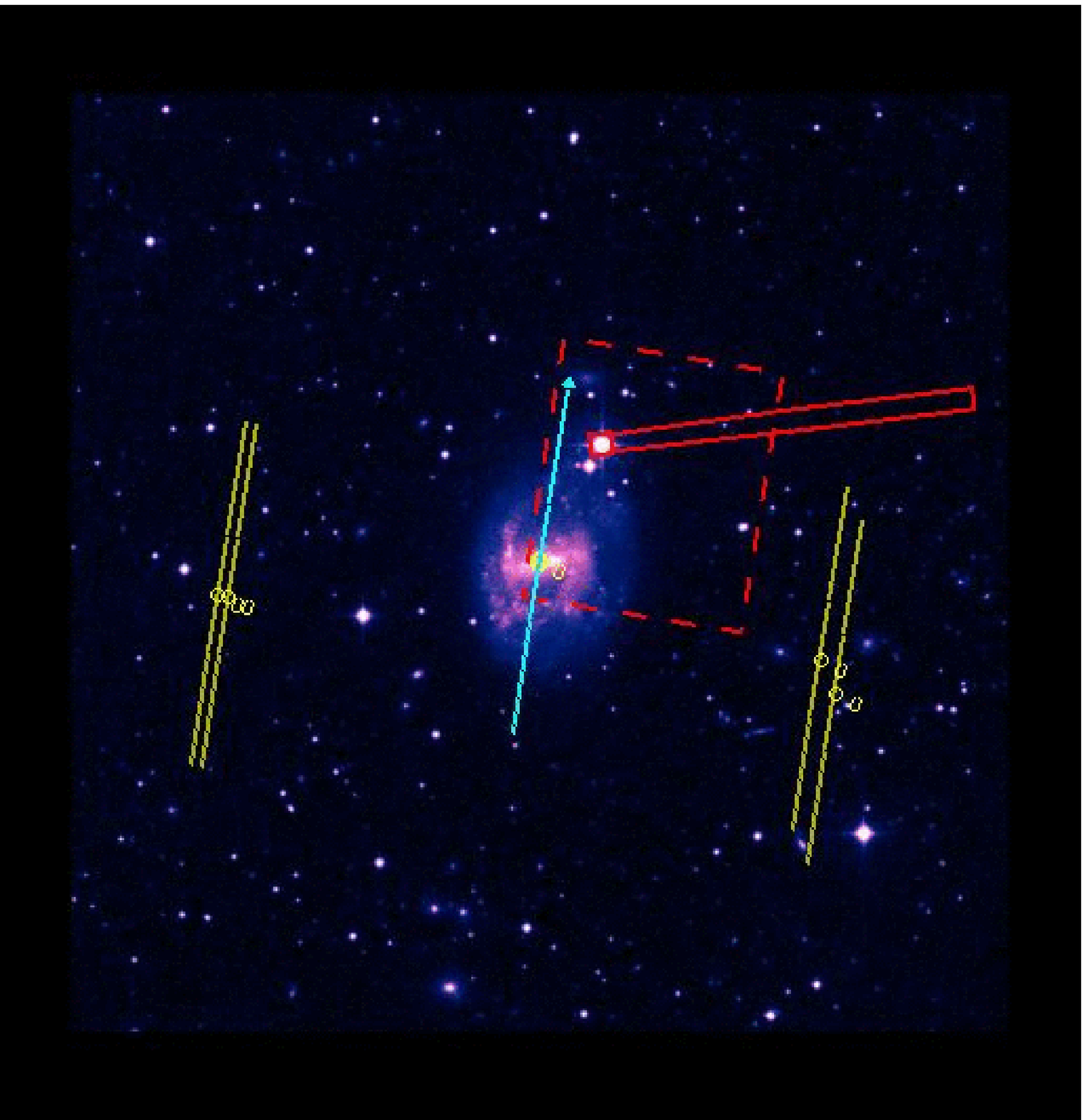}}
 {\includegraphics{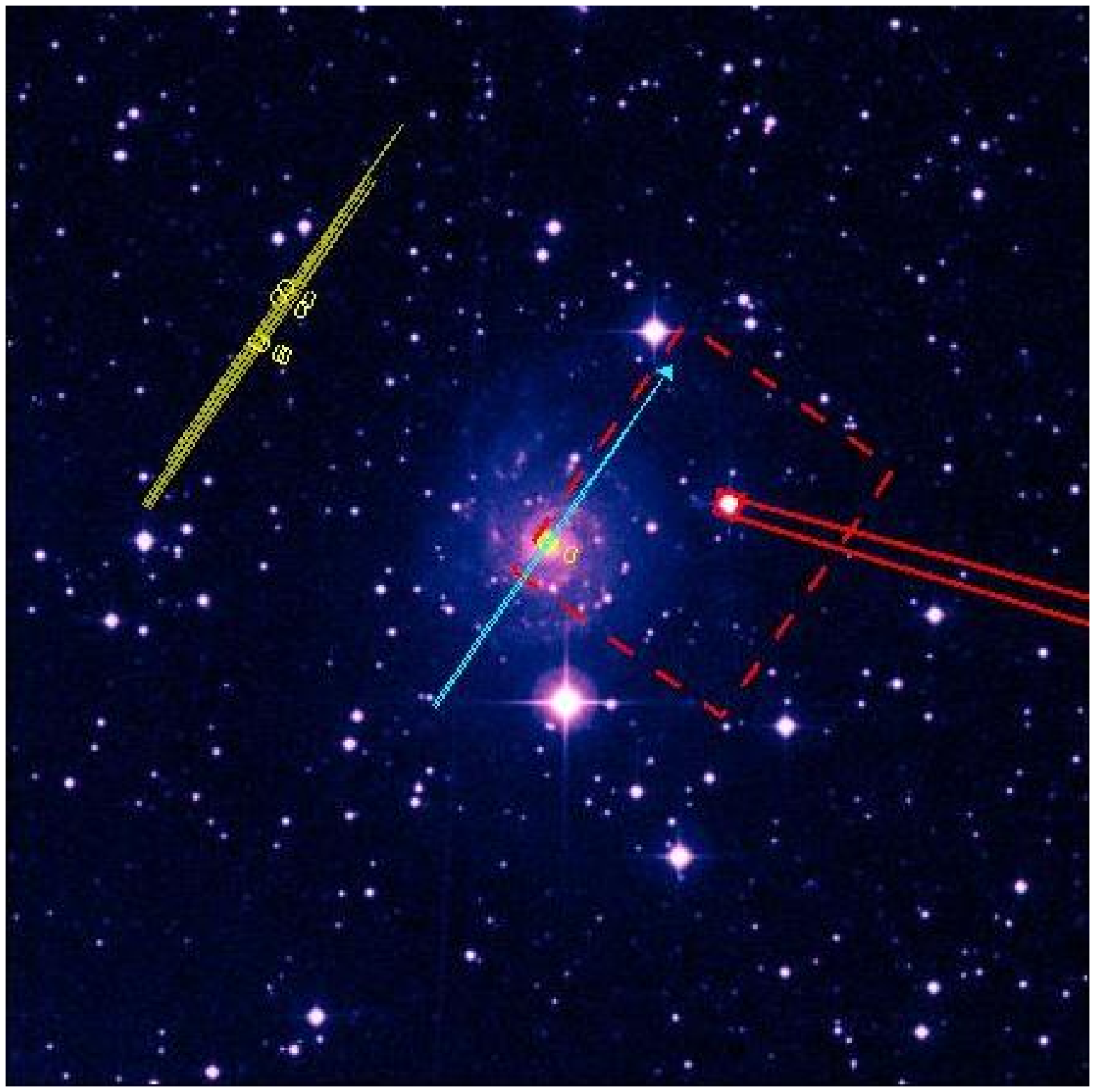}}
 {\includegraphics{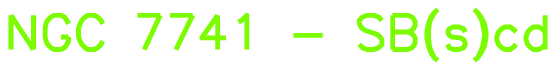}}
 {\includegraphics{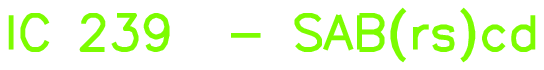}}
  \vskip 1.4in
    \caption{Observational set-up for the eight galaxies in our sample.
            The background images are from the Canadian Astronomy Data 
            Centre's Digitized Sky Server (CADC; http://cadcwww.dao.nrc.ca/). 
            The blue line represents the slit, the red (dashed) box and 
            long arm represent the FOV of the GMOS wavefront sensor camera,
            with the box at the end of the arm centered on the guide star.  
            The panels for large galaxies (NGC 628, NGC 7741, and IC 239) 
	    also show, as yellow lines, the sky offset positions.  The FOV for 
            the CADC pictures differ in all the panels but the slit length 
            is everywhere the same (5\arcmin).}
	    \label{fig:setup} 
\end{figure*}

The on-target integrations were divided into separate exposures to
filter out cosmic rays and confirm weak features.  On-target
Integration times were 3\,$\times$\,15 min for $M_B$\,$<$\,10.5,
3\,$\times$\,18 min for 10.5\,$<$\,$M_B$\,$<$\,12, and 4\,$\times$\,18
min for $M_B$\,$>$\,12, for total integrations of 45--72 min per
galaxy.  An estimate of the sky background was measured from the slit
edges for galaxies with major axis diameters $<$\,4\arcmin.  For
galaxies wider than the slit, we divided our exposures into
sky/galaxy/sky sequences with integration times for sky (off-target)
pointings of $\sim$1/5 of the on-target integrations\footnote{In
hindsight, sky integration times of 1/3--1/2 of the on-target times
would have been preferred.  The short sky exposures prevented a deeper
exploration of the low-SB disk regions.}.  See Fig.\@~\ref{fig:setup}
for the observational set-up for each galaxy.  Sky positions, when
necessary, are shown by the yellow slits.  Sky exposures were binned
to achieve the same S/N as the galaxy.  

\subsection{Data Reduction}
The relevant measurements for this study do not require absolute
spectrophotometric calibration of the data.  However, they do require
extreme care in removing instrumental effects in order to achieve
accurate relative calibration along both the dispersion and spatial
directions.  Proper subtraction of the sky and instrumental and
far-field scattered light is also crucial.  Finally, accurate flux
calibration is required for the galaxy observations, particularly for
long-baseline measurements such as those of molecular indices and full
spectrum fitting.  We went to great lengths to assure the highest
possible quality in our final data products.  In the interest of
space, the detailed data reduction procedures will be described in a
forthcoming paper. 

\section{Extracting Stellar Population Parameters From Spectra}\label{sec:SPs}

Many different techniques can be used to characterize the stellar
population parameters of an integrated (optical) spectrum.  We have
explored several of these in detail.  Appendix~\S\ref{sec:AZfits}
presents our analyses based on the Lick system of absorption-line
indices and that of full spectrum single SSP fits.  Both methods
produce SP parameters based on SSPs and are thus extremely sensitive
to the most recent episode of SF, even when its contribution to the SP
in terms of mass is insignificant.  The line-index technique is also
very sensitive to any ongoing SF as the emission-lines from \hii\
regions fill in the underlying absorption features, thus thwarting
reliable age estimates.  We conclude in
Appendix~\S\ref{sec:comparefits} that the SPs of our spiral galaxies
are not well represented by SSPs and an accurate representation must
represent true average values of the SP parameters, \ie\ integrated
over the SFH of the galaxy.  In the following sections we describe our
devised method for achieving this goal, which refer to as ``full
population synthesis''.

\subsection{Full Population Synthesis}\label{sec:synthesis}

Population synthesis is a technique applied to the integrated light of
composite systems to determine the fraction of light (or mass)
contributed by each of the main stellar groups comprising the system.
Using a linear combination of stellar or model templates that cover
the optical spectral range, the temperature mix of a galaxian
(composite) system can be determined and used to explore its
evolutionary history and to predict the flux of the system at other
wavelengths as a test of the synthesis (Jacoby, Hunter, \& Christian
1984; Pickels 1985; Heavens \etal\ 2000; Cid~Fernandes 2007; Tojeiro
\etal\ 2007).

The ``full population synthesis'' technique developed here consists of
deconvolving an observed galaxy spectrum into basic components
representing the relative contributions of a basis set of template
spectra.  It is important for the method of decomposition to allow for
constraints to be placed on the computed parameters for which we turn
to the method of {\sl bound constrained optimization} (Byrd \etal\
1995; Zhu, Byrd, \& Nocedal 1994, 1997).  The problem can be
summarized as solving for
\begin{equation}
\begin{array}{c}
{\rm min}\ \it f(x) \\
{\rm subject\ to}\ l \le x \le u
\end{array}
\ ,
\end{equation}
where $f(x)$ is the merit function which depends non-linearly, in
general, on the fit parameters $x_j$, and $l_j$ and $u_j$ are lower
and upper bounds respectively on each parameter $x_j$.  Additionally,
the gradient $g$ of the merit function $f$ with respect to the fit
parameters must be available.  The algorithm performs the minimization
iteratively where, at each iteration, the current fit parameters are
varied by amounts determined by the current partial derivatives of the
merit function.  To start the iteration an initial configuration must
be supplied.

Ideally, the locus of the merit function's minimum will be a single
point in the vector space defined by the basis library of template
spectra.  The necessary conditions for this ideal case are: (1) the
basis set is complete and non-redundant and (2) there is no noise in
the data.  In this situation the solution will be independent of the
initial configuration.  However, in any real synthesis we are limited
by observational uncertainties, resolution, physical differences
between the observed spectrum and model templates (\eg\ abundance
ratios, dust, emission lines from \hii\ regions, any non-stellar
contribution from AGN, etc.)  and possible degeneracies in the library
itself.  Hence, the locus of the minimum merit function will be a
finite volume in the basis space.  The ultimate success of a
population synthesis is therefore limited by the observational
accuracy of the spectra and the match between data and template
library.

\subsection{The Program}\label{sec:algorithm}
The problem of bound constrained optimization arises in numerous
scientific and technical applications and has been an active area of
research in the field of numerical analysis.  As a result, there are
now a number of robust optimization routines available; XRQP
(Bartholomew-Biggs 1979) and LANCELOT (Conn, Gould, \& Toint 1992), to
name a few.  The program of choice for the present problem is that of
Zhu \etal\ (1994; 1997) called L-BFGS-B\footnote{Can be downloaded from
http://www.ece.northwestern.edu/\\$\sim$nocedal/lbfgsb.html as of
writing.}.  This is a FORTRAN 77 code that solves large non-linear
optimization problems with simple bounds.  It has been well tested and
is ideal for the current problem.  A full description of the L-BFGS-B
algorithm is beyond the scope of this paper, but the interested
reader is referred to Byrd \etal\ (1995) and Zhu \etal\ (1994; 1997).

A driver program {\tt popsynth} was written which calls the L-BFGS-B 
routine.  {\tt popsynth} performs the following steps:
\begin{enumerate}
\item Set up input values for call to {\tt setulb}, the calling statement of 
      L-BFGS-B
\item Set up initial values for fit parameters $x_j$
\item Read in the galaxy spectra and associated variance vectors,
      and the template library
\item Compute the merit function and its gradient for the given set of 
      fit parameters
\item Call the {\tt setulb} subroutine with the new merit function and 
      gradient values
\item Test for convergence, if not repeat (iv)--(vi)
\item Output stellar fractions to data files.
\end{enumerate}
Zhu \etal\ (1997) provide a complete description of the input
parameters, but a brief introduction is relevant here in order to
understand the convergence criteria of the algorithm.

The input variable \tt factr \rm is a double precision number which defines 
a tolerance level in a termination test for the algorithm.  The iteration 
will stop when
\[\frac{f_{k}-f_{k+1}}{max(|f_{k}|,\ |f_{k+1}|,\ 1)}\le\tt{factr}*\tt{epsmch},\]
\rm where {\tt epsmch} is the machine precision which is automatically 
generated by the L-BFGS-B code (equal to $2.22\times 10^{-16}$ on the
Linux machine used for this project.)  A value of {\tt factr}\,=\,$10^{7}$ 
was chosen for ``moderate'' accuracy.  Similarly, {\tt pgtol} is a 
tolerance level for termination of an iteration if
\[max(|{\rm proj}\ g_{i}|,\  i = 1..n) \le \tt{pgtol}\]
where proj $g_{i}$ is the $i^{th}$ component of the projected gradient.  
For the current application {\tt pgtol} was set to $10^{-5}$.  The 
iteration terminated according to the latter criterion for all tests and 
runs of the program.


The merit function to be minimized is defined as
\begin{equation}\label{eq:merit}
{\rm Merit} =  \left(\frac{1}{N-M}\sum_{i=1}^{N}{R^2_i}\right)^{\frac{1}{2}}, 
\end{equation}
where 
\begin{equation}\label{eq:merit2}
R_i = w_i (G_{i} - {S_i}).
\end{equation}
In the above expressions, $N$ is the number of data points included in
the fit, $M$ is the number of templates in the library,
$G_i$ is the observed galaxy flux at wavelength $\lambda_i$,
$w_i$ is the weight of the $i^{th}$ pixel (see \S\ref{sec:weights}), and
$S_i$ is the modeled galaxy flux at $\lambda_i$ given by
\begin{equation}\label{eq:model}
S_i = \sum_{j=1}^{M}{x_j}{F_{ji}},
\end{equation}
where $F_{ji}$ denotes the flux at the $i^{th}$ wavelength of the
$j^{th}$ template, and $x_{j}$ is the relative contribution of the
$j^{th}$ template to the synthesized spectrum.

In order to prevent negative stellar contributions, as a non-physical artifact 
of the decomposition, the spectral coefficients are subject to a lower bound 
on the fit parameters such that
\[x_j \ge 0.\]

\subsection{Weighting}\label{sec:weights}

The nature of our fitting algorithm is particularly useful for masking
out undesirable regions in the spectrum.  The following pathologies,
which differ for each galaxy due, in most part, to velocity
differences, are examples of regions that one may want to omit: (i)
gaps in wavelength coverage due to the spacing between the three GMOS
CCDs, (ii) strong and variable night sky-line regions, and (iii)
emission line regions (due to \hii\ regions and/or planetary nebulae).
As a first guess, all of (i), (ii), \& (iii) are masked for the initial fit of 
each spectrum.  Subsequently, we allow for all points to enter back into the 
fit (subject to the constraints defined below) except for the gap regions, in 
which no real data exist, thus are always fixed to zero weight.

\subsubsection{$\sigma$ Clipping}\label{sec:sigmaclip}
We define a mask value, $m_i$, for each datum such that
$$ m_i = 
\cases{1, & {\rm include in current fit} \cr
       0, & {\rm remove from current fit}} $$
and the weight, $w_i$, as,
\begin{equation}\label{eq:weight}
w_i = m_i\, \frac{1}{\sigma_i^2},
\end{equation}
where $\sigma_i$ is the error on $G_i$.  

In a general weighted fit, very dissimilar weights can be distributed
among the data points.  As a result, the number of datum, $N$, does
not always relate directly to the actual information feeding the fit.
Instead, an effective number of points, \Neff, can be estimated by the
sum of the weights divided by the mean weight (Gonz{\' a}lez 1993)
(note that in the following, all sums run from $i$\,=\,$1..N$, where
$N$ is the total number of data points):
\begin{equation}
N_{eff} = \frac{\sum w_i}{\langle w_i \rangle }
        = \frac{\sum w_i}{(\sum w_i*w_i) / \sum w_i}
        = \frac{(\sum w_i)^2}{ \sum w_i^2 }.
\end{equation}

Next we define the effective number of templates entering each fit, 
\Meff.  This is required as it is possible for a significant fraction
of the total number of templates considered to end up with zero contribution
to the fit.  \Meff\ is defined as follows (here sums run from $j$\,=\,$1..M$, 
where $M$ is the total number of templates considered):
\begin{equation}
M_{eff} = \frac{\sum F_j}{\sum F_j^2/\sum F_j}
        = \frac{(\sum F_j)^2}{\sum F_j^2}
        = \frac{1}{\sum F_j^2},
\end{equation}
where $F_j$ is the fractional contribution of SSP model $M_j$ (thus 
$\sum F_j$\,=\,1 by definition).

The expected variance (of the mean) based on the measurement errors is 
\begin{equation}
\sigma_{exp}^2 = \frac{N_{eff}}{\sum w_i},
\end{equation}
and the typical variance from the fit is,
\begin{equation}\label{eq:sigfit}
\sigma_{fit}^2 =
= \frac{N_{eff}}{N_{eff}-M_{eff}}\left [\frac{\sum (w_i\,\Delta_i^{2})}{\sum w_i}\right ],
\end{equation}
where $\Delta_i \equiv |G_i - S_i|$.  

Finally, the fit \chisqr\ is the ratio of the mean typical and
expected variances,
\begin{equation}
\chi^2 = \sigma_{fit}^2/\sigma_{exp}^2.
\end{equation}

At each iteration, the quantities in
Eqs.\@~\ref{eq:model}--\ref{eq:sigfit} are computed and the
individual masks, $m_i$, are assigned according to the criterion,
$$ m_i = 
\cases{1, & {\rm if } $\Delta_i < N_{\sigma}\,\sigma_{fit}$ \cr
       0, & {\rm otherwise}} $$
On the first iteration we set $N_{\sigma}$\,=\,5, the second has 
$N_{\sigma}$\,=\,3.2, and two final iterations have $N_{\sigma}$\,=\,2.8.
We refer to this iterative masking scheme as ``$\sigma$-clipping''.

For our data, the typical number of data points for each spectrum is
in the range $N$\,=\,1280--1340; after the ``$\sigma$-clipping''
iterations described above, the typical effective number of points in
the final fit spans the range $\Neff$\,=\,850--1250.  In this analysis
we use $M$\,=\,70 model templates (see \S\ref{sec:BC03SSPs}) and
typical values for \Meff are in the range 2--6.

\subsection{Stellar Population Templates}\label{sec:BC03SSPs}
The stellar population templates used throughout this analysis are
from the models of BC03 which provide spectra of simple stellar
populations (SSPs), single-$Z$/single-bursts of star formation. These
cover the wavelength range 3200--9500\,\AA, a large range of ages
(0.0001--20\,Gyr), and metallicities ($Z$\,=\,0.0001--0.05) at a
reported resolution of $\sim 3$\,\AA\,FWHM.  We use here the models
with the Chabrier (2003) initial mass function (IMF).

\subsubsection{Template Characterization}\label{sec:templates}

Given the importance of knowing the precise resolution of the
templates for comparison with galaxy data and for measuring the small
velocity dispersions of spiral galaxies (see \S\ref{sec:kinem}), we
undertook a deep investigation into the true resolution of the BC03
models, which use the spectral library STELIB of Le Borgne \etal\
(2003), for a range of model SSPs and allowing for a wavelength
dependent (velocity-like, \ie\ constant with $\Delta\lambda/\lambda$)
term in addition to a term that is constant with wavelength.  Our
analysis is carried out using an upgraded version of the {\tt Movel}
algorithm described in Gonz{\' a}lez (1993).  In brief, {\tt Movel}
derives velocity shift, velocity dispersion, and $\gamma$ (a measure
of the relative line strengths, with $\gamma$\,=\,1 indicating a
perfect match) through a Fourier transform procedure.  In velocity
space, the observed galaxy spectrum can be characterized by the
spectrum of its population convolved with the instrumental response
function and a broadening function which characterizes the
distribution of stellar radial velocities along the line of sight
(LOSVD).  The ``population'' is represented by a template (observed or
model) of higher resolution than the galaxy spectrum.  {\tt Movel}
derives the broadening function that best matches the template to the
galaxy spectrum.  To then infer the absolute rotation and velocity
dispersion parameters of the observed spectrum, one must know very
accurately the characterization of the template used.
 
For the purpose of characterizing the BC03 resolution (both globally
and as a function of wavelength), we use as a template spectrum the
very high-resolution (effectively infinite for this application) solar
spectrum of Wallace, Hinkle, \& Livingston (1998) along with the BC03
model SSPs, which are each effectively different coadditions (in
relative weight and stars used) of stars in STELIB. The solar template
and model SSPs are processed through {\tt Movel} in 200\,\AA\
wavelength intervals centered around 26 spectral features, from
3575\,\AA\ up to 7400\,\AA\ defined by 21 Lick-like indices plus the
H$\alpha$ and \caii\ H\&K lines, and 3 more somewhat ad hoc regions
shortward of \caii\ H\&K and longward of H$\alpha$ to cover the full
optical range.  Spectral features in any given SSP that are not well
characterized by the solar template are weighted out accordingly.  We
then fit the trend of resolution vs.\@ $\lambda$ for a large range of
SSP templates using three different constraints; (i) fitting for
constant terms with both $\lambda$ and velocity with both terms as
free parameters, (ii) fixing FWHM\,=\,3\,\AA\ and fitting for a
velocity term, and (iii) forcing only a linear FWHM component.  This
procedure revealed that {\it the resolution trend of the BC03 models
is much better represented by a FWHM constant with $\lambda$ plus a
velocity term (constant with $\Delta\lambda/\lambda$).}

To investigate the cause of this discrepancy, we downloaded the STELIB
library\footnote{http://www.ast.obs-mip.fr/users/leborgne/stelib/index.html}
and examined the individual spectra.  First we noted that the starting
$\lambda$ and dispersion are not all 3200 \& 1\,\AA\ as stated in the
documentation on their website.  The reason is that the 2003 v3.2
library stars had their radial velocities removed, but the spectra
were not rebinned to the $\lambda_0$\,=\,3200 \& disp\,=\,1\,\AA.  A
look into the headers revealed a spread in the starting wavelength of
RMS$_{\lambda_0}$\,=\,1.09\,\AA\ and in dispersion of
RMS$_{disp}$\,=\,0.00033\,\AA.  As these two differences in the
relative wavelength calibration are correlated and, if fixed values
are assumed when coadding the individual stars, the combined effect
will degrade the effective resolution of the model SSPs.  In measuring
the resolution as a function of wavelength for 12 individual
solar-type stars in the STELIB library we found, not surprisingly,
that the individual stars are consistent with a pure constant
resolution (no velocity-like broadening at all) and we measured the
best constant FWHM to be 3.08\,\AA ($\pm$\,0.03).  We thus concluded
that the above-mentioned $\lambda_0$ \& dispersion mismatch was not
taken into account when the stars were coadded to create the BC03
model SSPs, thus resulting in a velocity-like broadening in addition
to the constant resolution term.  We also found a spread in the
velocities of the 12 stars (whose measured relative velocities were
removed in STELIB) with an RMS\,=\,11.4\,\kms, which will also
contribute to the velocity-like broadening of the SSPs.

The above procedure leads to the following conclusions about the true
effective resolution of the BC03 SSPs: (i) The actual spectral
resolution of the models is worse than 3.0\,\AA\ (significantly larger
than for individual stars in STELIB); (ii) The best constant (no
velocity broadening) FWHM is 3.40\,\AA ($\pm$\,0.04); (iii) If the
3.08\,\AA\ resolution derived for individual stars is assumed, the
SSP models (co-added STELIB stars) need a significant additional
velocity broadening (35.8\,$\pm$\,3.6\,\kms), but the fit is not as
good as when assuming a constant $\sim$\,3.4\,\AA\ resolution.  After
exploring the fit results under all the above assumptions, the adopted
(best) values for the FWHM resolution and velocity broadening for the
BC03 models are found to be:
\[{\rm FWHM}_{{\rm BC03}} = 3.375\,{\rm \AA} \,\,\,\& \,\,\,
\sigma_{{\rm BC03}} = 11.40\,\kms. \]
These results can be understood as the combination (convolution) of:
(i) the 3.08\,\AA\ FWHM resolution of each individual star, plus (ii)
the 11.40\,\kms\ radial velocity RMS error in STELIB, and (iii) an
additional 1.38\,\AA\ residual FWHM degradation in the models due to
the errors in the assumed initial wavelength and dispersions in STELIB
in the coaddition of BC03.

Finally, we used the same solar template to map the wavelength-scale
of STELIB.  The systematic deviations are no larger than
$\Delta\lambda$\,=\,$-$0.5 to +0.5\,\AA\ across the observed
wavelength range, yet are relevant enough for a fair comparison with
high-quality modern data.  These true systematic trends in resolution
and wavelength scale of the BC03/STELIB spectra are in excellent
agreement with those found in the similar analysis by Koleva \etal\
(2008).

In our full population synthesis fitting, all of the above is
accounted for.  In particular, we first distort our observed galaxy
wavelength scale to match that of STELIB.  We then account for
resolution effects by convolving all model and galaxy spectra (at all
radii) to the same resolution: top-hat\,=\,10.81\,\AA,
$\sigma$\,=\,168\,\kms.  These account for the slit-width and
approximately highest velocity dispersion in our galaxy sample
(measured on the first iteration and limited to the central, high S/N
regions), respectively.  This dispersion is sufficiently greater than
that of the native BC03 models such that any residual variations, \eg\
due to different combinations of stars in different SSPs, will be
smoothed out.  We discuss in \S\ref{sec:kinem} our accounting for
the BC03 effective resolution for kinematic measurements, which
are iteratively derived from the full spatial and spectral resolution
data.

\subsubsection{SSP Template Library}

\begin{figure*}
\epsfig{file=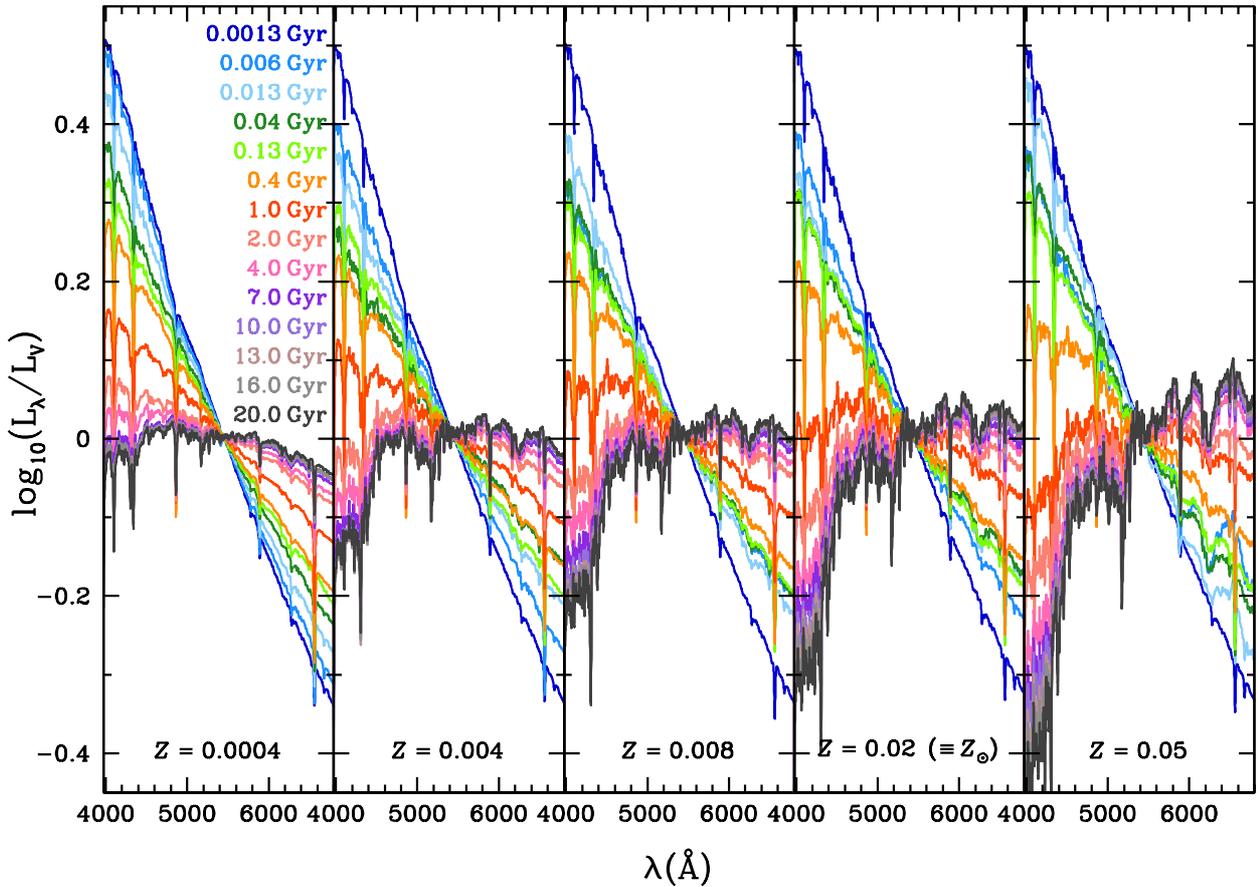,width=0.95\textwidth,bb=18 144 592 548}
   \caption{Spectra of the 70 SSP templates used in the population synthesis
            fits.  Metallicity increases from left to right.  Different
            ages have different colours and are labeled in leftmost panel.
            All spectra are normalized to their $V$-band flux.}
   \label{fig:templates}
\end{figure*}

We are limited here to a template library that comprises a discrete
sampling of SSP ages and metallicities.  Ultimately, we would like our
library to be a complete and non-redundant basis set, but this would
require an exact match between models and data that is not realized in
practice.  Through a series of tests and careful examination of the
individual SSP spectra, we converged on a selection of SSPs at 14
ages, roughly logarithmically spaced\footnote{The spacing in not
exactly logarithmic, particularly at ages of about 0.004\,Gyr at the
lowest $Z$s, where we noted anomalous SSP spectra in the BC03 models.
The spectra around this age were inspected individually to find the
``best behaved'' age (\ie\ bluer for younger ages).  The 0.006\,Gyr
SSPs had the most reasonable spectra.}, and 5 metallicities (excluding
only the lowest metallicity provided in the BC03 models which is the
least well modeled and exhibits anomalous SSP spectra).  This provides
a total basis set of 70 SSP spectra, all shown in
Fig.\@~\ref{fig:templates}\footnote{Note that we are not concerned here
with model age predictions that are older than the age of the
Universe.  Model ages are not precisely calibrated and, as such, we
are concerned primarily with relative trends.}.

We are interested in both the light- and mass-weighted contributions
of each SSP to the integrated galaxian light.  All SSPs in our library
have been normalized to their $V$-band flux, so the $V$-band
light-weight for SSP$_j$ is computed as (here sums run from
$j$\,=\,$1..M$, where $M$\,=\,70 is the total number of SSP templates
considered): 
\begin{equation}
l_{j} = x_j/\sum_{j} x_j,
\end{equation}
and the mass-weight as
\begin{equation}
m_{j} = x_j*(M/L_{V})_{j}/\sum_{j}(x_{j}*(M/L_{V})_{j}), 
\end{equation}
where $(M/L_{V})_{j}$ is the $V$-band mass-to-light ratio of the $j^{th}$
SSP.  Thus the average light-weighted quantities are computed as:
\begin{equation}
\langle {\rm A} \rangle_{l} = \sum_{j} l_j*{\rm A}_j 
\,\,\,\,\,\,\,\,\,\,\, {\rm and} 
\,\,\,\,\,\,\,\,\,\,\, \langle Z \rangle_{l} = \sum_{j} l_j*Z_j \ ,
\end{equation}
and the mass-weighted as: 
\begin{equation}
\langle {\rm A} \rangle_{m} = \sum_{j} m_j*{\rm A}_j 
\,\,\,\,\,\, {\rm and} 
\,\,\,\,\,\, \langle Z \rangle_{m} = \sum_{j} m_j*Z_j \ .
\end{equation}
Finally, we also compute the stellar 
$V$-band $M/L$ ratio of the composite system as
$M/L_{V}$\,=\,$\sum_{j}(l_{j}*(M/L_{V})_{j})$.

A number of caveats exist when applying model SSPs to the integrated
light of a star forming galaxy.  First, the models are based on
empirical libraries of stars which are necessarily limited to the
range of $T_{eff}$, [Fe/H], log($g$), and [$\alpha$/Fe] of stars
within the solar neighborhood.  Stellar populations in other regions
in our Galaxy and in external galaxies will not be confined to this
parameter space and thus not be faithfully represented.  However,
extrapolations in the ($T_{eff}$, [Fe/H], log($g$))-space within the
models do provide extensive coverage, with the extremes (low
$T_{eff}$, small [Fe/H]) being the least well modeled.  
Additionally, the BC03 models do not include the effects of 
non-solar abundance ratios and a full assessment of potential
biases awaits the upcoming implementations of stellar population
models which include abundance pattern variations.  Second, the models
do not account for emission from planetary nebulae and \hii\ regions
which can be dominant in actively star forming galaxies such as the
late-type spirals in our sample.  Indeed, many of our galaxy spectra
show strong emission lines (\eg\ H$\beta$, \oiiill, H$\alpha$, \nii,
\sii, etc.; see Fig.\@~\ref{fig:spec_rad}).  We
successfully circumvent this issue with our iterative masking
procedure described in \S\ref{sec:sigmaclip}.  Third, and again
particularly important in star forming systems, the optical stellar
light will likely suffer from the extinction and reddening
effects of dust.  While there is no straightforward way to account for
dust effects on integrated spectra, we attempt to determine if a
particular spectrum is affected by extinction by comparing dust-free
fits with those incorporating a model for dust extinction.  We adopt
the two-component dust model of Charlot \& Fall (2000, hereafter
CF00).  The CF00 model accounts for the finite lifetime
(${\sim}10^7$\,yr) of stellar birth clouds through the parameter
$\mu$, which represents the fraction of total dust absorption
contributed by diffuse interstellar medium (cirrus) dust.  The
wavelength dependence of the effective absorption curve is
$\propto\lambda^{-0.7}$, so the ``effective'' extinction in a given
band, $\hat{\tau}_\lambda$, is given by,
\begin{equation}
\hat{\tau}_{\lambda} = \left\{ \begin{array}{rll}
\hat{\tau}_{V}({\lambda}/5500 \,\,{\rm \AA})^{-0.7} & {\rm for} & t \le 10^{7}\,\,{\rm yr,} \\
\mu\,\hat{\tau}_{V}({\lambda}/5500 \,\,{\rm \AA})^{-0.7} &  {\rm for} & t > 10^{7}\,\,{\rm yr,}
\end{array}
\right.
\label{eq:dust}
\end{equation}
where $t$ is the age of any single stellar generation. 

For each SSP spectrum in our library, we create extincted versions
with a fixed $\mu$\,=\,0.3 (the average value for star forming
galaxies) and varying $\hat{\tau}_{V}$ in steps of 0.5.  For
reference, in this model, $\hat{\tau}_{V}$\,=\,2 is equivalent to a
colour excess E($B-V$)\,$\sim$\,0.1 for a solar metallicity template
(see Fig.\@~1 in MacArthur 2005).  We perform the full synthesis first
on the dust-free SSPs.  The results are recorded and the Merit
function (Eq.\@~\ref{eq:merit}) tabulated.  The fits are then
performed on the extincted SSPs, starting with
$\hat{\tau}_{V}$\,=\,0.5.  If the Merit function has decreased from
the previous fit, the next extinction level is fit, until a minimum in
the Merit function is reached.  While the reddening effects of dust
add an extra degeneracy to the fits, we always reach a convergence
(minimum Merit in the Age/{\it Z}/dust-space).

\section{Results}\label{sec:results}
\subsection{Full Population Synthesis Fits}\label{sec:fits}
\begin{figure*}
\begin{center}
\includegraphics[width=0.47\textwidth,bb=18 144 592 518]{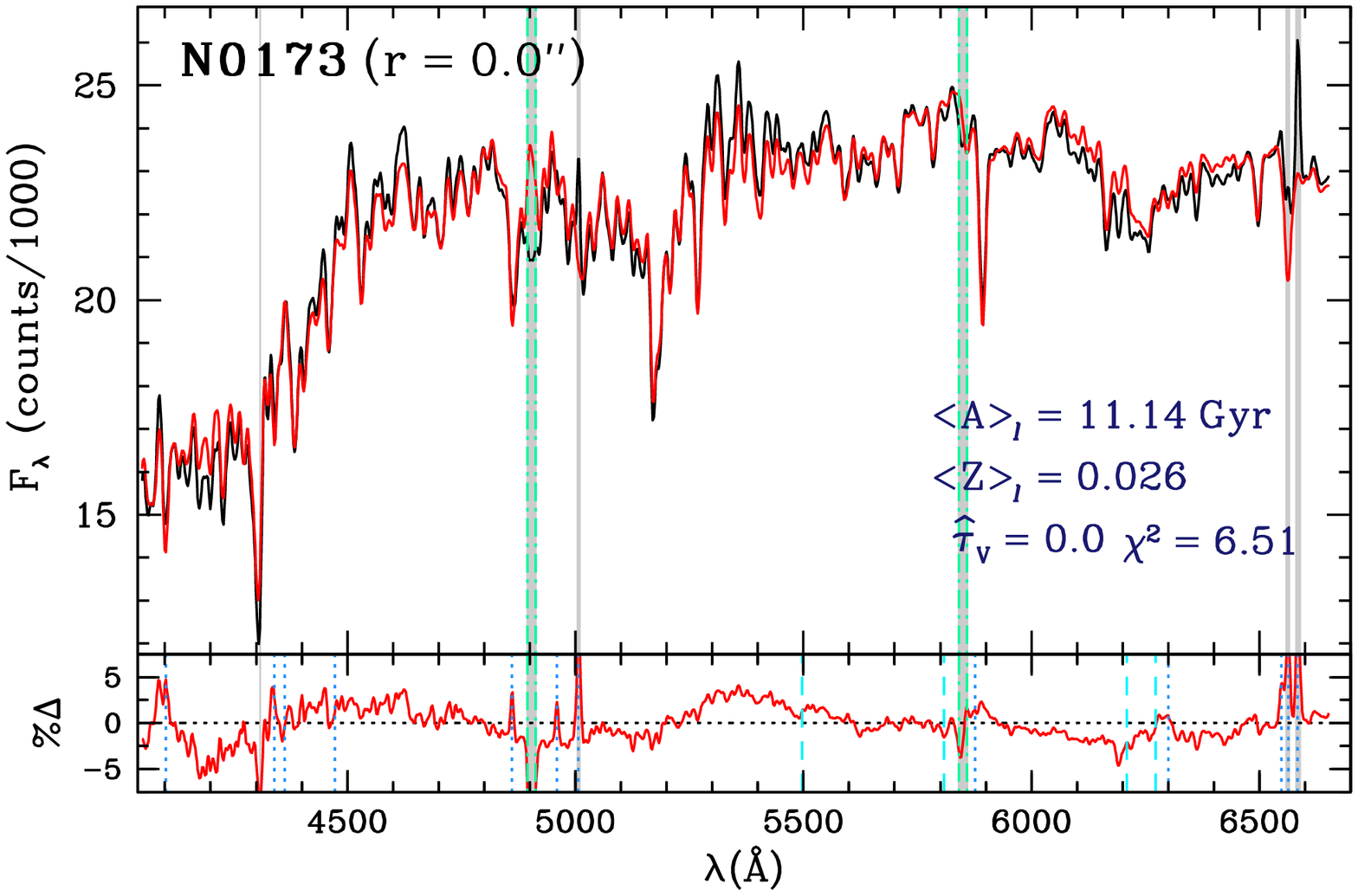}
\includegraphics[width=0.47\textwidth,bb=18 144 592 518]{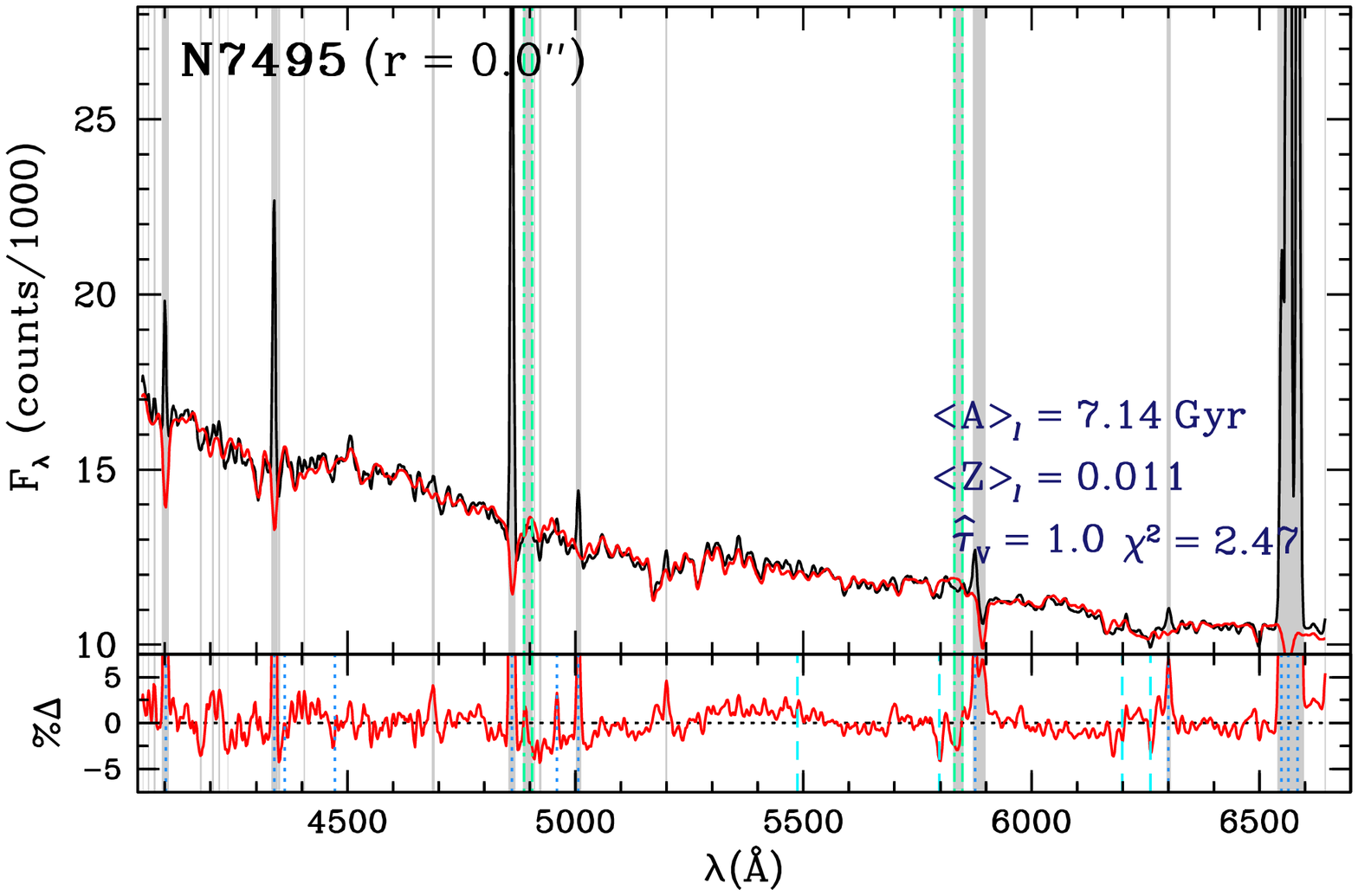} \\
\includegraphics[width=0.47\textwidth,bb=18 144 592 518]{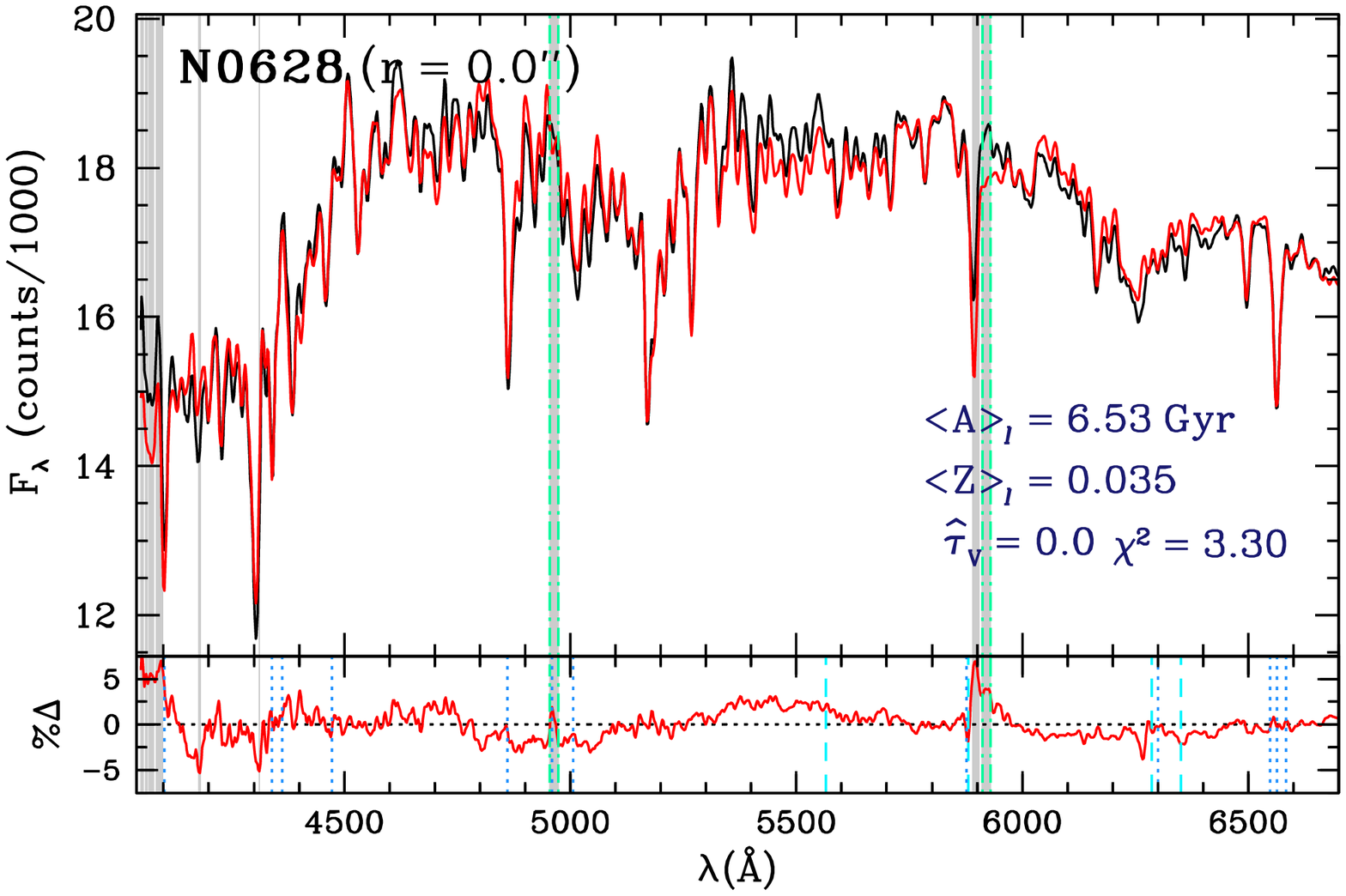} 
\includegraphics[width=0.47\textwidth,bb=18 144 592 518]{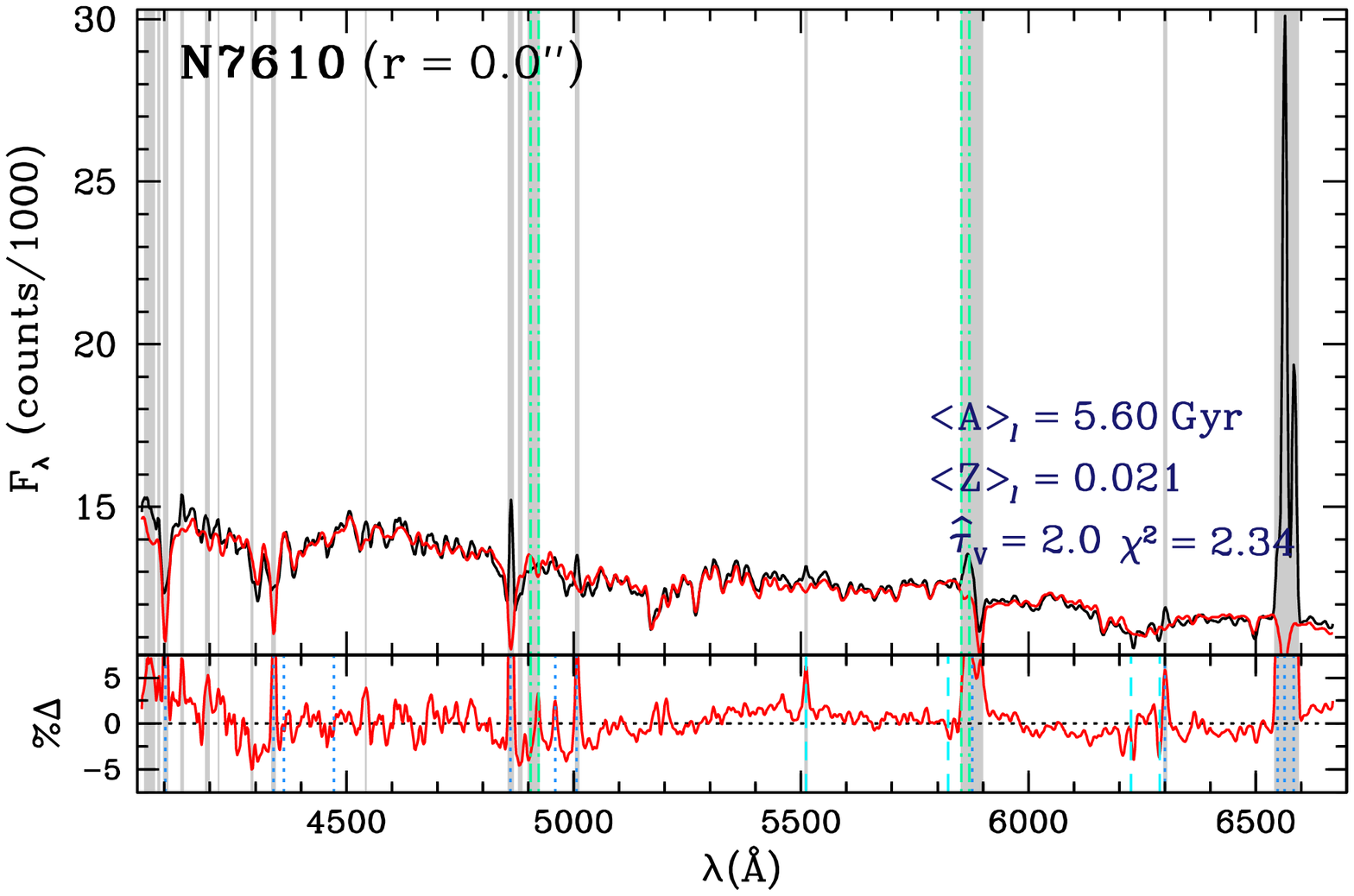} \\
\includegraphics[width=0.47\textwidth,bb=18 144 592 518]{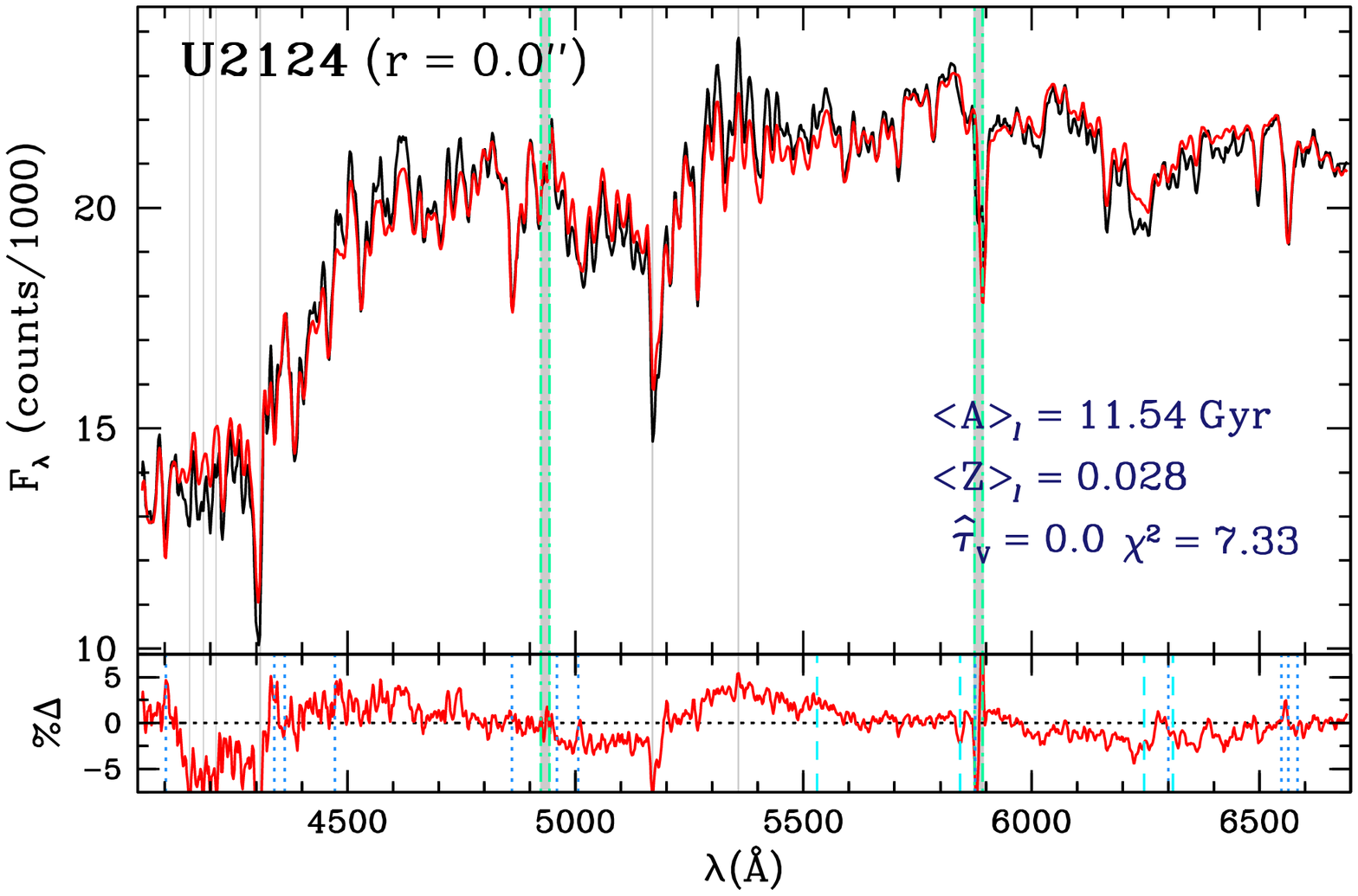}
\includegraphics[width=0.47\textwidth,bb=18 144 592 518]{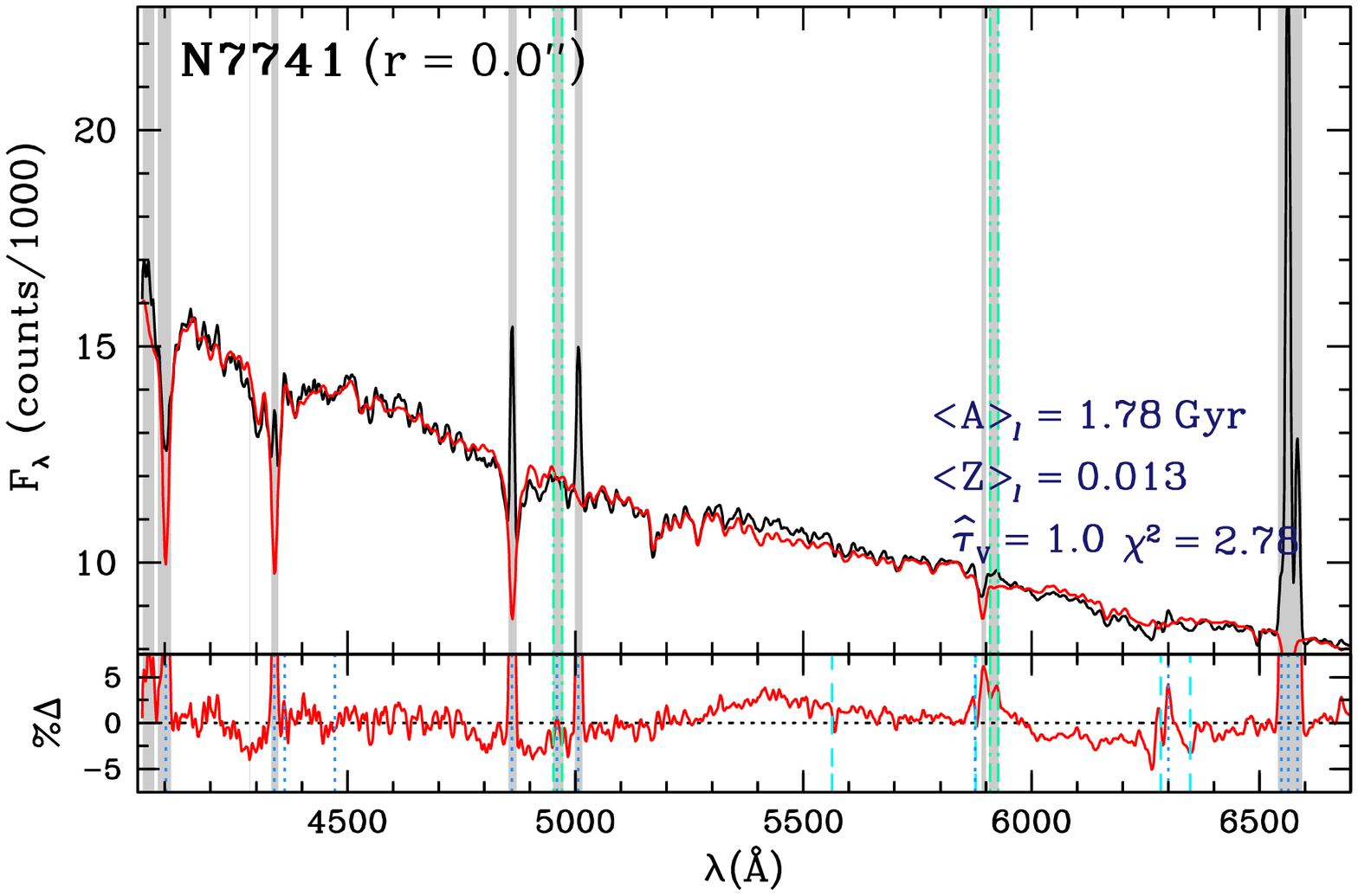} \\
\includegraphics[width=0.47\textwidth,bb=18 144 592 518]{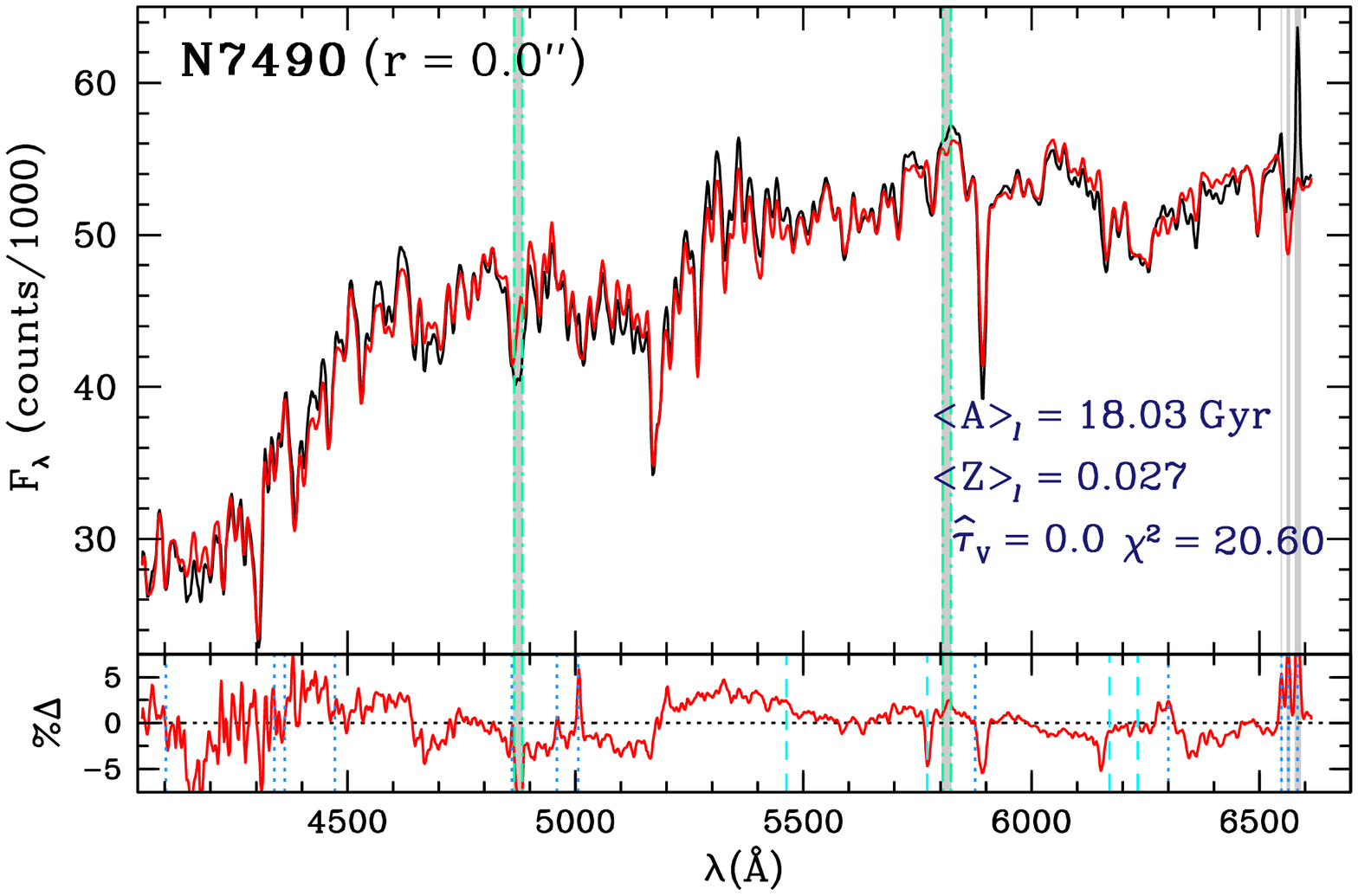}
\includegraphics[width=0.47\textwidth,bb=18 144 592 518]{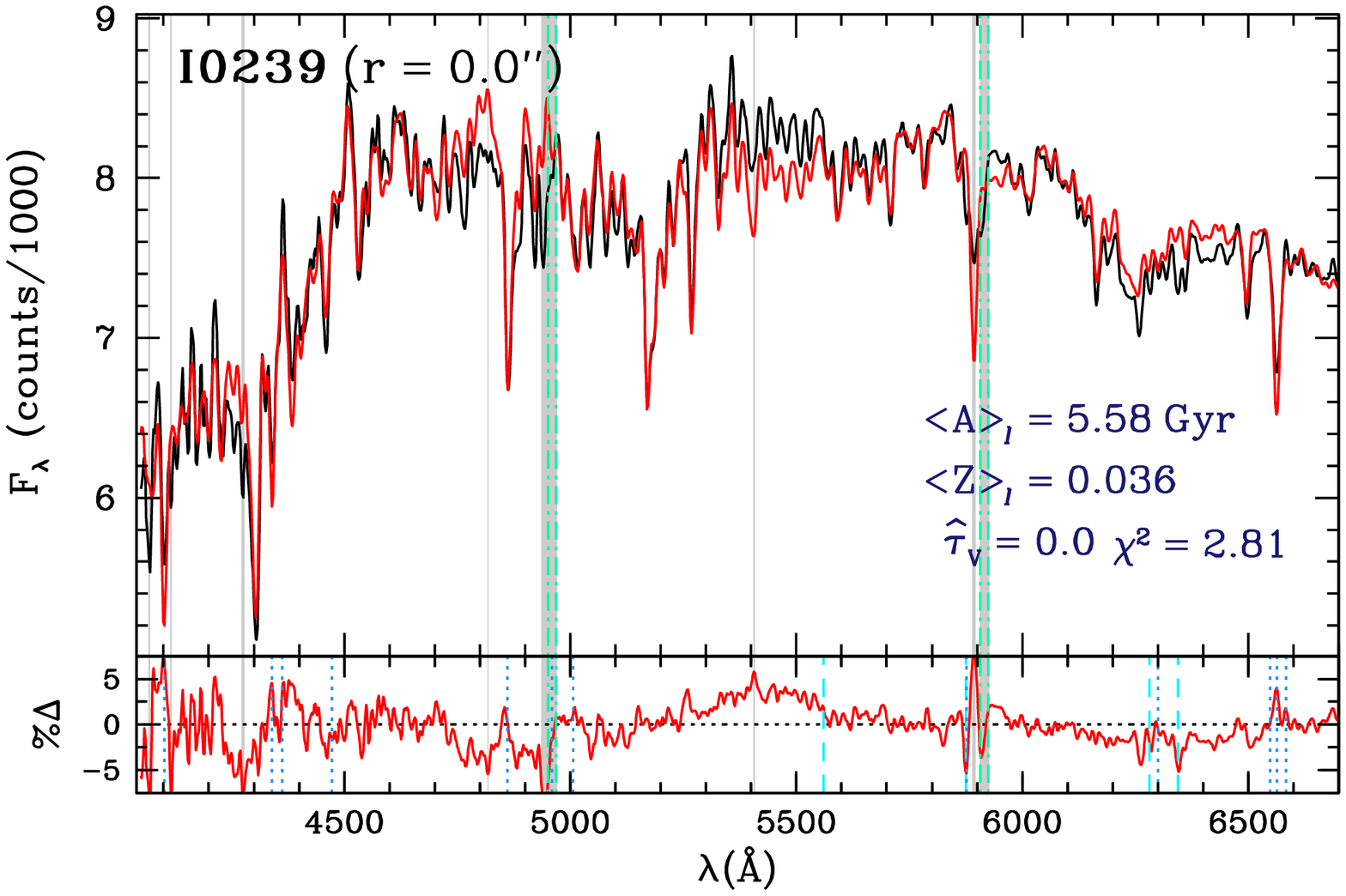}\hfill
   \caption{Central observed spectra (black) and full population synthesis
            fit (red) for all eight galaxies.  Gray shading indicates
            regions masked in the fit as determined by our iterative
            ``$\sigma$-clipping'' procedure (see
            \S\ref{sec:sigmaclip}) as well as the CCD gap regions
            (green vertical dash-dotted lines) which are always
            masked.  Shown at lower right on each panel are the
            average light-weighted age, \avgAl, and metallicity,
            \avgZl, effective \taueff$_V$, and \chisqr\ of the fit.
            The bottom panels show the percent data$-$model
            residuals. In the bottom panels, dashed vertical lines
            indicate variable sky-lines (which can be difficult to
            account for during sky-subtraction), and dotted vertical
            lines indicate emission lines prevalent in star forming
            (\hii) regions. }
   \label{fig:fits}
\end{center}
\end{figure*}

We here present example fits from the code and procedure described in
\S~\ref{sec:synthesis}.  Fig.\@~\ref{fig:fits} shows the central,
$r$\,=\,0\arcsec, spectrum fit for all eight galaxies.  In each plot,
black and red lines are respectively the data and synthesis fits.  The
gray shading indicates regions masked in the fits as determined by our
``$\sigma$-clipping'' procedure described in \S~\ref{sec:sigmaclip}.
Shown at lower right on each panel are the average light-weighted age,
\avgAl, and metallicity, \avgZl, effective \taueff$_V$, and \chisqr\
of the fit.  The bottom panels show the percent data$-$model
residuals.  In all cases, the full synthesis fits represent the galaxy
spectra remarkably well.  Our $\sigma$-clipping procedure clearly does
a good job at masking out regions where emission lines are present.

Closer examination of the data$-$model residuals for all of the
spectra that show little or no evidence for current SF reveal the same
overall rolling shape, with the most significant feature occurring
around $\lambda$\,$\sim$\,5150--5450, a region rich with Fe and Mg
absorption lines.  While this may be due to a flux calibration
problem, it could also be an indication of non-solar abundance ratios
that are not taken into account in the BC03 models.  A similar pattern
was observed in the fit residuals of SDSS galaxy spectra in the
studies of Panter \etal\ (2007) and Cid~Fernandes (2007).  The fact
that these independent data sets show the same features effectively
rules out a flux calibration problem.  Panter \etal\ (2007) suggest
that these features may be due to an error in the balance of K--M
giants in the BC03 models, whereas Cid~Fernandes (2007) attributes
them to non-solar $\alpha$-element abundance ratios.  We may be able
to test this in future work with new SSP models that include non-solar
$\alpha$/Fe.

\begin{figure}
\includegraphics[width=0.49\textwidth,bb=18 144 492 718]{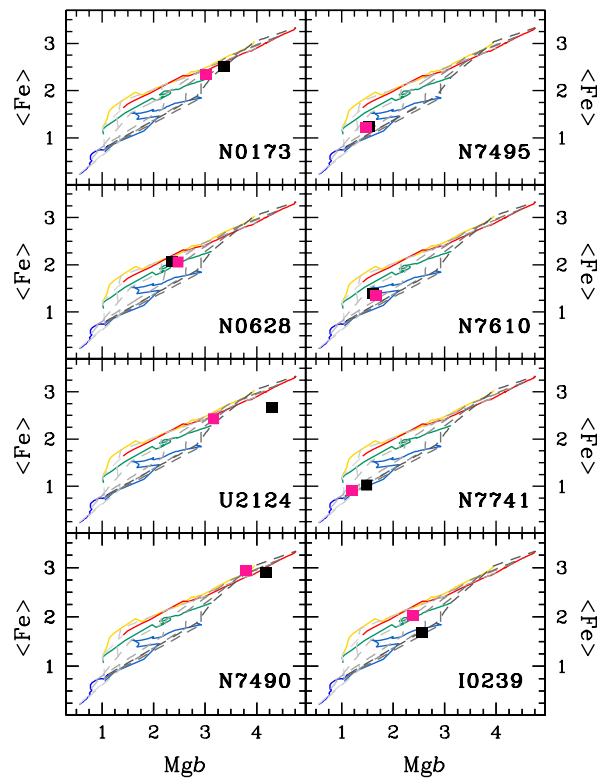}
\caption{\avgFe\ vs.\@ Mg{\it b} as an indication of non-solar
         abundance ratios.  Solid lines are iso-$Z$ tracks with colour
         scheme: $Z$\,=\,0.05 (red), 0.02 ($Z_{\odot}$) (yellow),
         0.008 (green), 0.004 (light blue), 0.0004 (blue), \& 0.0001
         (purple).  Gray-shaded dashed lines are iso-age tracks at
         ages of 1, 2, 5, 10, 13, \& 20\,Gyr (from light to dark
         shades; roughly left to right).  The black squares are
         measured from the data.  The pink squares are measured on the
         model fit spectra.}
   \label{fig:avgFeMgb}
\end{figure}

For now, however, the best diagnostic we have of $\alpha$/Fe abundance
comes from the Lick index measurements.  In particular, in a plot of
an index primarily sensitive to Fe versus one sensitive to an
$\alpha$-element, an enhancement in the latter can be identified if
the data extend beyond the solar $\alpha$/Fe model grids.  We examine
possible $\alpha$-enhancement in our data in Fig.\@~\ref{fig:avgFeMgb}
where we plot \avgFe\ vs.\@ Mg{\it b}.  For clarity, only the central,
$r=0$\arcsec, data point is plotted for each galaxy (for direct
comparison with central spectra shown in Fig.\@~\ref{fig:fits}).  The
black squares are measured directly from the galaxy spectra, whereas
the pink squares are measured from the full synthesis model fits (red
spectra in Fig.\@~\ref{fig:fits}).  By definition, the pink squares
must lie within the model grids, thus a difference between the black
and pink squares indicates that the models failed to reproduce the
data simultaneously in this combination of narrow index regions.
Indeed a few galaxies show signs of super-solar $\alpha$/Fe ratios in
their centers, with the most significant enhancement in U2124, which
also shows the most prominent features in its full synthesis fit
residuals (Fig.\@~\ref{fig:fits}).  This lends credence to the
non-solar abundance ratio interpretation for the cause of the most
prominent residuals that are not attributable to non-stellar radiation
sources.

\begin{figure}
\begin{center}
\includegraphics[width=0.49\textwidth]{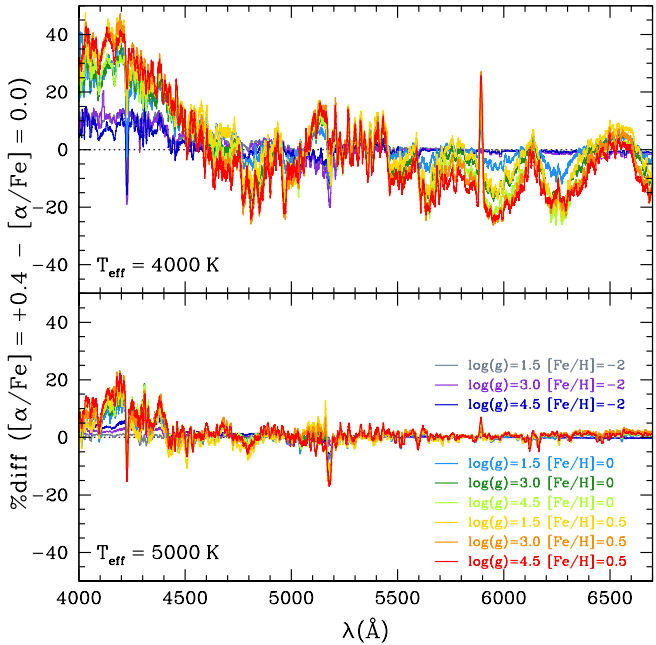}
\caption{Comparison of solar vs.\@ [$\alpha$/Fe]\,=\,+0.4 from models of
         Coelho \etal\ (2005), plotted as the \%-difference of enhanced
         minus solar-scaled models, for effective temperatures of 
         4000\,K (top) and 5000\,K (bottom), and for a range of
         surface gravity and metallicity (as labeled in bottom panel).}
   \label{fig:diffalphaFe}
\end{center}
\end{figure}

Another possibility to examine the conjecture that the dominant
features of the fit residuals in Fig.\@~\ref{fig:fits} are due to
non-solar [$\alpha$/Fe] uses the synthetic spectral library of Coelho
\etal\ (2005) who present models for a range of $T_{eff}$, log($g$),
and metallicity, here expressed as [Fe/H], with varying abundance
ratios.  The effects of super-solar $\alpha$/Fe are shown in
Fig.\@~\ref{fig:diffalphaFe} as the difference between the solar and
enhanced $\alpha$/Fe models of Coelho \etal\ (2005) for
$T_{eff}$\,=\,4000\,K (top) and 5000\,K (bottom), for a range of
log($g$) and [Fe/H].  The effects depend strongly on $T_{eff}$ \&
log($g$), but some of the features and wiggles are similar to those
seen in our galaxy--model fit residuals, thus strengthening the
conclusion that the wiggles are due, at least in part, to abundance
ratio effects.  Individual features such as those seen in
Fig.\@~\ref{fig:diffalphaFe} at $\lambda$\,$\sim$\,4150--4400, 4600,
5150--5450 (with a prominent feature at $\sim$\,5180\,\AA), and
6200--6300 \AA, are most evident in the residuals for the central spectrum
of U2124 in Fig.\@~\ref{fig:fits}.  The issue of abundance ratio
variations will be examined in more detail in a future work.

Cid~Fernandes (2007) also noted another striking feature in the data$-$model
residuals of their mean spectrum for star-forming galaxies in the
SDSS, selected on the basis of emission line diagnostic-diagrams.
This feature takes the form of a shallow but broad ``absorption'' band
around \hbeta, that appears whenever stars of $\sim$\,100\,Myr contribute
significantly to the spectrum.  While Cid~Fernandes points out that no
straightforward physical explanation nor element identification for
this trough exists, he also speculates that it is a side-effect of flux
calibrations issues in the STELIB spectral library.  We do not see
such absorption troughs around \hbeta\ for our star-forming galaxies
(see, \eg, N7495 in Fig.\@~\ref{fig:fits}).  We thus speculate that
the trough seen by Cid~Fernandes could be a residual trend from the
resolution issues of the BC03 models described in
\S\ref{sec:BC03SSPs}.

\subsubsection{Relative Contribution of Individual Templates}\label{sec:SSPcont}
It is interesting to investigate which of the 70 SSP templates are
most represented in the fits.  We have no reason {\it a priori} to
think that the SSPs at any given age and $Z$ will be more likely to
represent the SPs of our spiral galaxy data, but it is possible that
certain model SSPs suffer some systematic anomaly rendering them poor
representations of real SPs (we certainly observed some
such pathologies when selecting our 70 template SSPs, see
\S\ref{sec:BC03SSPs}).  Table~\ref{tab:fitstats} shows the cumulative
statistics of the contribution of each individual SSP template to the
fits.  Clearly, certain SSPs are favored, with many never entering any
fit.  This is partly the reason for the stability in the fits.  With
70 templates to choose from, one might expect the fits to be very
sensitive to initial conditions, for example.  We have tested the fits
using various initial conditions and the results are extremely
consistent.  Given that typically fewer than 10 templates have
non-zero weight in the fit, the stability is not as surprising.
Interestingly, the solar $Z$\,=\,0.02 SSP rarely makes it into any of
the fits and only at the oldest three age bins, whereas all other $Z$s
have some contribution at a range of ages.  While it is difficult
to identify the exact reason for this (\eg\ perhaps the real parameter
space is not fully represented in the models, or some systematic in
the data forces it to certain areas of the model parameter space), an
investigation based on fits to other SPS models may provide some
insight, and will be the subject of a future investigation.

\begin{table*}
\begin{minipage}{\textwidth}
\centering
\caption{Statistics of Individual SSP Templates Included in Full Population 
Synthesis Fits of Integrated Galaxy Spectra}\label{tab:fitstats}
\begin{tabular}{@{}l@{}rrrrrrrrrrrrrr@{}}
\hline
\multicolumn{1}{@{}l@{}}{Age(Gyr)} & \multicolumn{1}{r}{0.001} &  
\multicolumn{1}{r}{0.006} &  
\multicolumn{1}{r}{0.013} &  \multicolumn{1}{r}{0.04} &  
\multicolumn{1}{r}{0.128} &  \multicolumn{1}{r}{0.404} &  
\multicolumn{1}{r}{1.02} &  \multicolumn{1}{r}{2.0} &  
\multicolumn{1}{r}{4.0} &  \multicolumn{1}{r}{7.0} & 
\multicolumn{1}{r}{10.0} & \multicolumn{1}{r}{13.0} & 
\multicolumn{1}{r}{16.0} & \multicolumn{1}{r@{}}{20.0} \\
\hline
\multicolumn{1}{@{}c@{}}{$Z$} &  
\multicolumn{14}{c@{}}{Number of times each SSP template appears in any fit}\\
\hline
  0.0004 & 282 &  20 &   1 &   0 &   0 &  10 &   6 &  27 & 108 &   2 &  26 &   0 &   0 &   2 \\
  0.004 &    0 &  11 &   0 &   7 &   5 & 194 &   1 &   0 &  26 &   0 &  10 & 129 &  99 &  66 \\
  0.008 &   70 &   0 &   0 &   1 &   0 &   6 &   6 &   0 &   0 &   9 &  33 &   3 &   0 &   0 \\
  0.02  &    0 &   0 &   0 &   0 &   0 &   0 &   0 &   0 &   0 &   0 &   0 &   2 &   6 & 109 \\
  0.05  &    0 &  65 &   5 &   0 &   1 &  17 & 233 &  28 &  37 &  48 &  10 &  23 & 187 &  97 \\
\hline
\multicolumn{1}{@{}c@{}}{$Z$} &  
\multicolumn{14}{c@{}}{Cumulative {\it light} fraction for each SSP in all fits}\\
\hline
  0.0004 &2979.9 & 260.0 &   4.7 &   0.0 &   0.0 &  56.2 &  62.4 & 426.5 &1898.3 &   9.4 & 601.5 &   0.0 &   0.0 &  15.7 \\
  0.004 &    0.0 & 110.4 &   0.0 &  99.4 &  81.4 &2745.8 &  40.1 &   0.0 & 476.1 &   0.0 & 240.5 &4593.8 &1759.1 &1834.2 \\
  0.008 &  911.3 &   0.0 &   0.0 &  12.5 &   0.0 &  31.0 & 123.3 &   0.0 &   0.0 &  77.6 & 662.2 &  56.1 &   0.0 &   0.0 \\
  0.02  &    0.0 &   0.0 &   0.0 &   0.0 &   0.0 &   0.0 &   0.0 &   0.0 &   0.0 &   0.0 &   0.0 &  22.0 &  65.9 &2255.3 \\
  0.05  &    0.0 & 847.6 &  39.9 &   0.0 &   2.4 &  47.5 &8483.2 & 257.7 & 357.9 & 532.0 & 116.4 & 242.0 &2100.8 &1159.9 \\
\hline
\multicolumn{1}{@{}c@{}}{$Z$} &  
\multicolumn{14}{c@{}}{Cumulative {\it mass} fraction for each SSP in all fits}\\
\hline
  0.0004 & 353.2 &   2.9 &   0.1 &   0.0 &   0.0 &   8.3 &  11.0 &  95.0 & 927.1 &  14.8 & 471.6 &   0.0 &   0.0 &  30.5 \\
  0.004 &    0.0 &   3.9 &   0.0 &   4.1 &  12.0 & 414.4 &  33.1 &   0.0 & 487.8 &   0.0 & 290.9 &5053.0 &2820.2 &2928.9 \\
  0.008 &   41.3 &   0.0 &   0.0 &   0.1 &   0.0 &   4.1 & 126.7 &   0.0 &   0.0 &  82.3 & 869.8 & 128.4 &   0.0 &   0.0 \\
  0.02  &    0.0 &   0.0 &   0.0 &   0.0 &   0.0 &   0.0 &   0.0 &   0.0 &   0.0 &   0.0 &   0.0 &  35.7 & 128.0 &3983.9 \\
  0.05  &    0.0 &  37.3 &   1.3 &   0.0 &   0.3 &   8.0 &2910.5 & 136.9 & 735.7 &1271.2 & 229.1 & 669.2 &6979.5 &4357.8\\
\hline
\end{tabular}
\end{minipage}
\end{table*}

\subsubsection{Reliability of the Fits}\label{sec:errors}

\begin{figure}
\begin{center}
\includegraphics[width=0.48\textwidth]{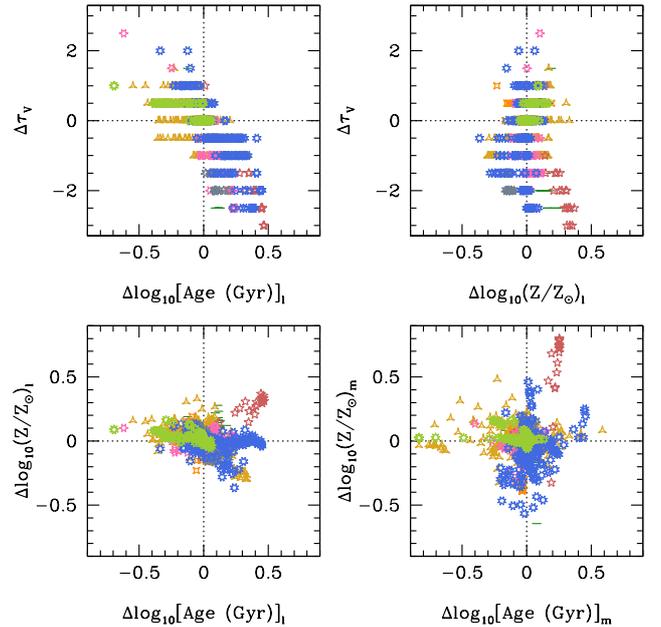}
\caption{Results from Monte Carlo simulations.  Plotted are the
         logarithmic differences between the parameters measured
         in each MC realization and the best-fit value.  Point types
         and colours distinguish the different galaxies (see legend
         of Fig.\@~\ref{fig:FSvsSSPvsInd}).}
\label{fig:MC}
\end{center}
\end{figure}

To test the robustness of our method and results, and to
associate reliable error estimates to the derived individual age, $Z$,
and \taueff$_V$ estimates, we perform a Monte Carlo (MC) analysis on a
subset of radial bins for each galaxy.  A full-blown MC analysis for
every spectrum in our sample is computationally prohibitive, so the
subset analyzed was selected to be representative of the entire data
set spanning the full range of S/N/\AA, SED shape, emission strength,
and radial coverage of each galaxy.  Twenty realizations for each
selected spectrum, after adding random noise, were then processed
through our fitting algorithm in an identical fashion as the original
galaxy data.  Characteristic errors for the derived parameters are
then taken as half the interval containing 68\% of the MC realization
results (\ie, the 1-$\sigma$ confidence interval).  Fig.~\ref{fig:MC}
summarizes the results from this exercise, showing the difference
between the parameters derived from each MC realization relative to
the best-fit value.  There is evidence for a mild age/$Z$ error
correlation, and where not present, it turns up as an error degeneracy
with \taueff$_V$.
The errors are typically $<<$\,10\% and, as expected, tend to
increase with decreasing age, $Z$, and S/N/\AA.  Error bars from these
simulations are plotted in all subsequent figures.  In particular,
Fig.\@~\ref{fig:grads} reveals the radial sampling of the MC analysis.

Beyond the small errors of the mean SP parameters, the MC analysis
also shows that the relative contribution of the individual SSPs, as
well as the inferred SFHs (see \S\ref{sec:SFH}), are preserved
providing further confidence on the robustness of our results and
method.

\subsection{Inferred Star Formation Histories}\label{sec:SFH}
Using full spectral synthesis, we are determining the best-fit
combination of SSPs at different ages and $Z$s that make up a given
spectrum.  As such, the SSPs included and their relative contributions
to the spectrum can be regarded as a stochastically sampled SFH.  We
can thus examine the radial variation in SFH for each galaxy and look
for signatures of star formation episodes that have occurred at
specific locations within the galaxy.  Fig.\@~\ref{fig:SFH} displays a
graphical representation of the radial variation in SFHs for all eight
galaxies in our sample.  In each panel, the vertical axis represents
galaxy radius with the dotted line at $r$\,=\,0\arcsec, and the
horizontal axis represents SSP age.  The point size is proportional to
the relative weight in the fit and the colours and point types code
the SSP metallicity.  Left panels show the light-weighted
contributions and right panels show their associated mass-weighted
contributions.
\begin{figure*}
\begin{center}
\includegraphics[width=0.45\textwidth]{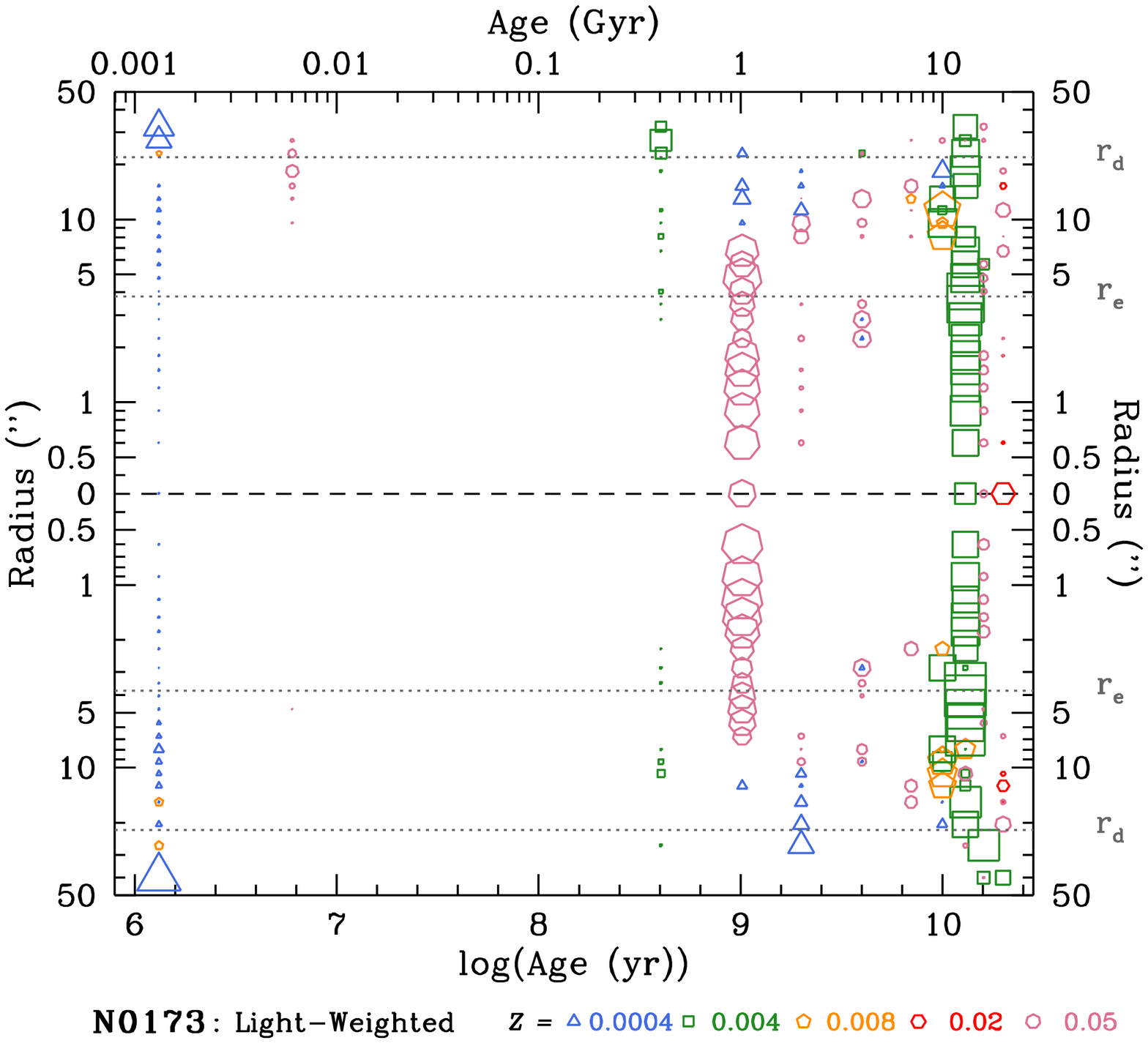}
\includegraphics[width=0.45\textwidth]{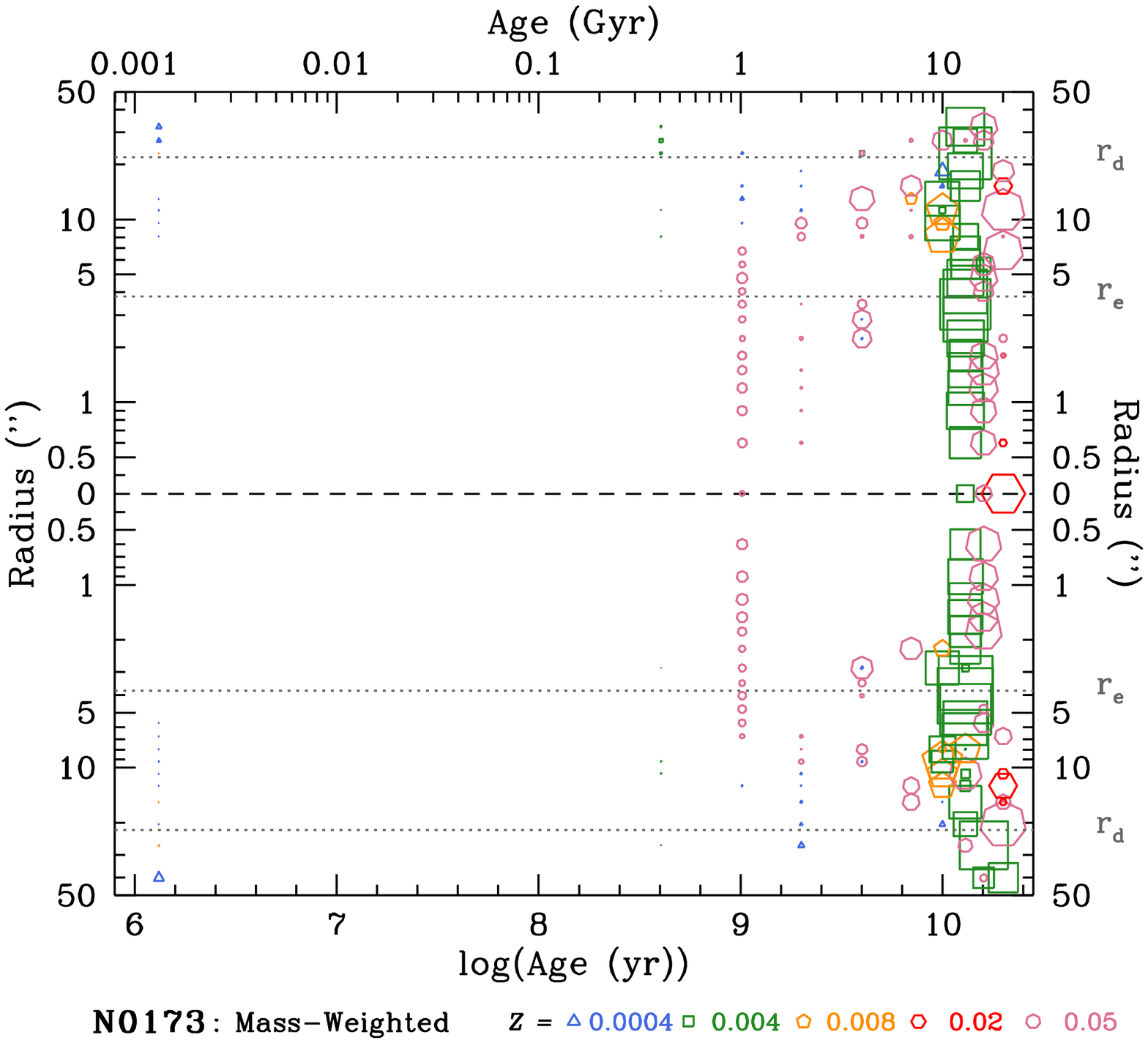}\\
\includegraphics[width=0.45\textwidth]{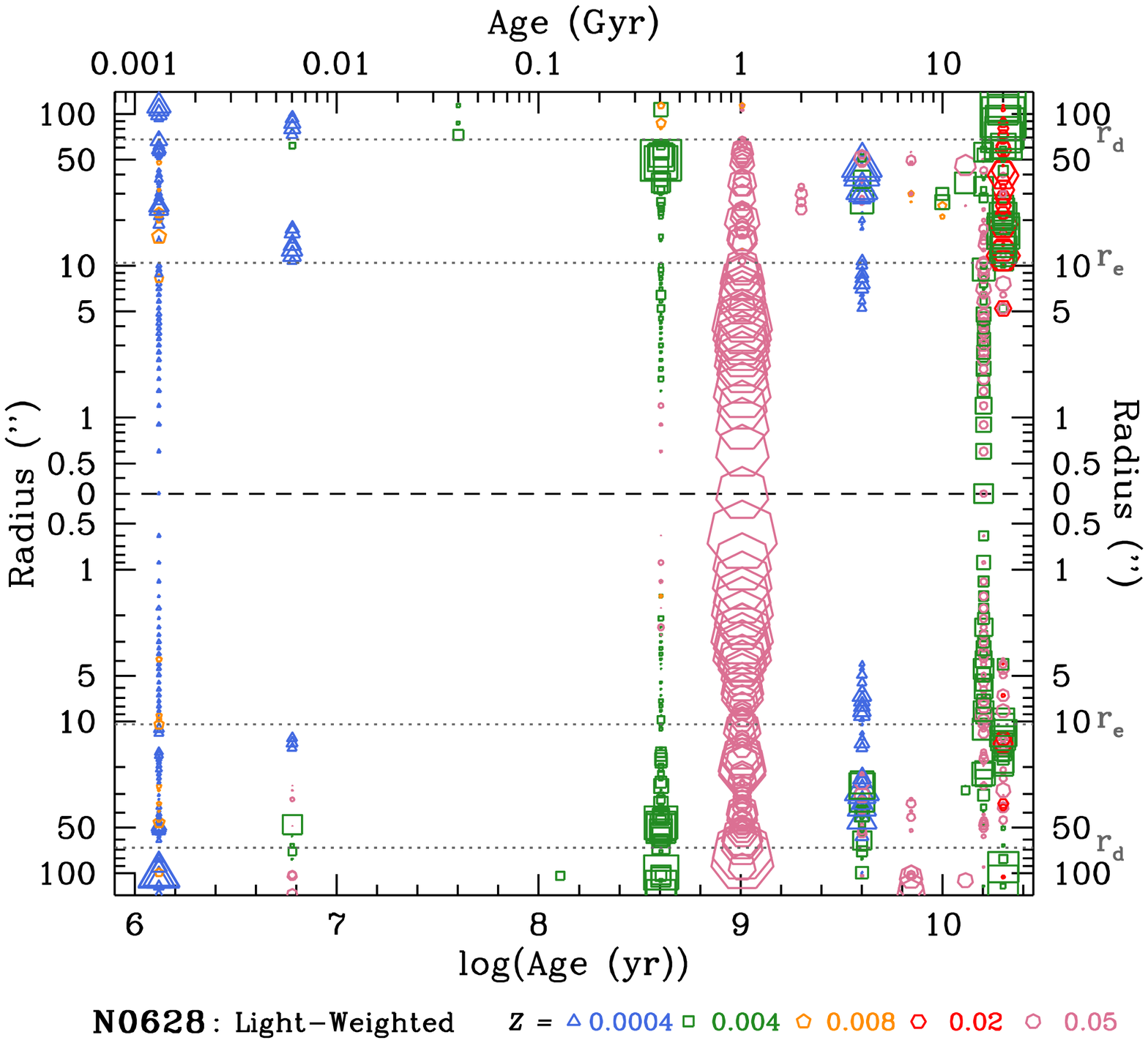}
\includegraphics[width=0.45\textwidth]{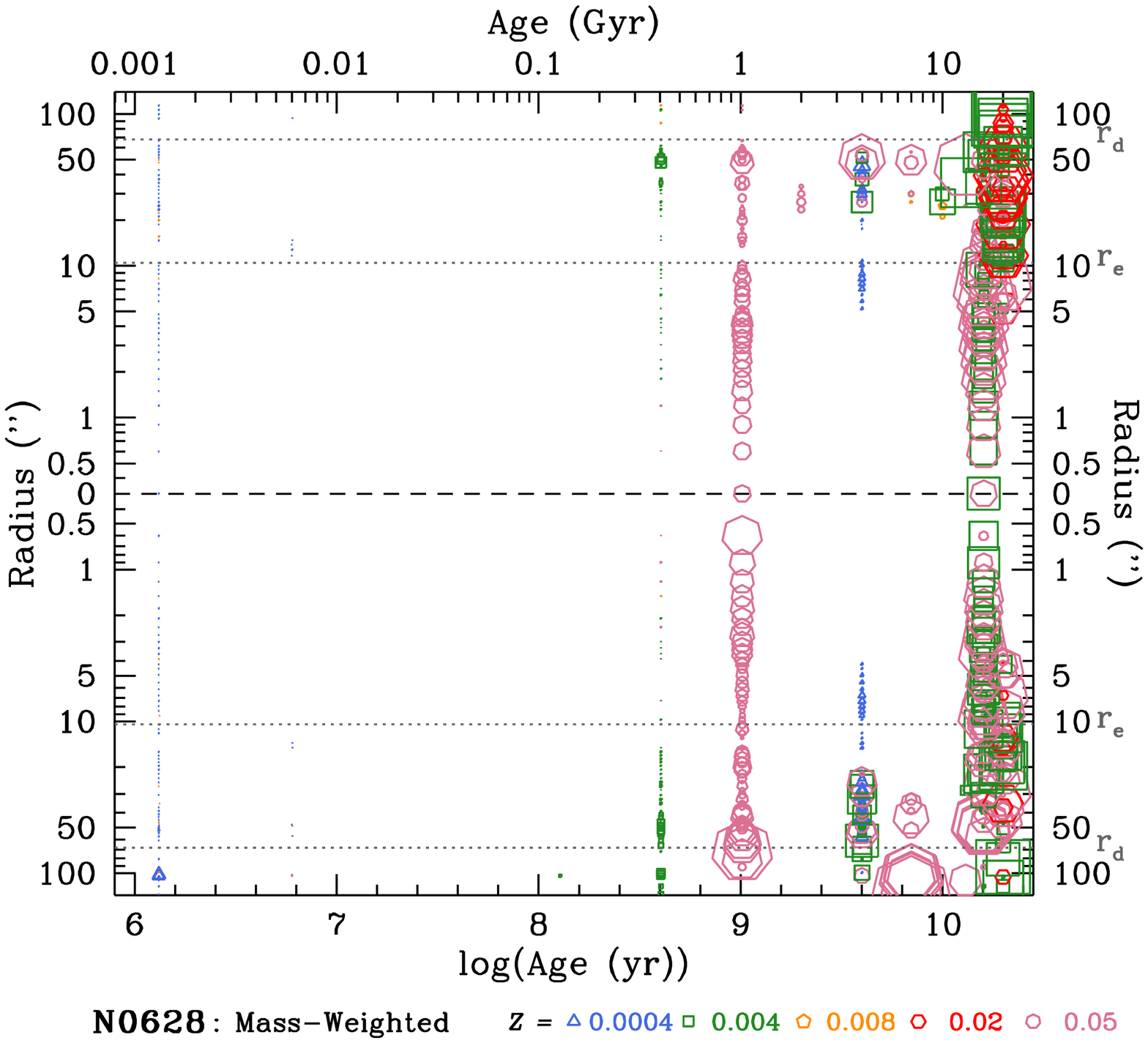}\hfill
\caption{Light (left) and mass (right) weighted SFHs for all eight
         galaxies.  In each panel, the vertical axis represents galaxy
         radius with the dotted line at $r$\,=\,0\arcsec, and the
         horizontal axis represents SSP age.  Point size is
         proportional to the relative weight in the fit and the
         colours and point types code the SSP metallicity.  Left
         panels show the light-weighted contributions and right panels
         show their associated mass-weighted contributions.  In terms
         of mass, the integrated spectrum at most radii is generally
         dominated by old and metal-rich population.  The contribution
         of very recent star formation, while it can dominate the
         light (cf.\@ N7495), is typically only a minor contribution
         to the stellar mass.}
\label{fig:SFH}
\end{center}
\end{figure*}
\begin{figure*}
\begin{center}
\includegraphics[width=0.45\textwidth]{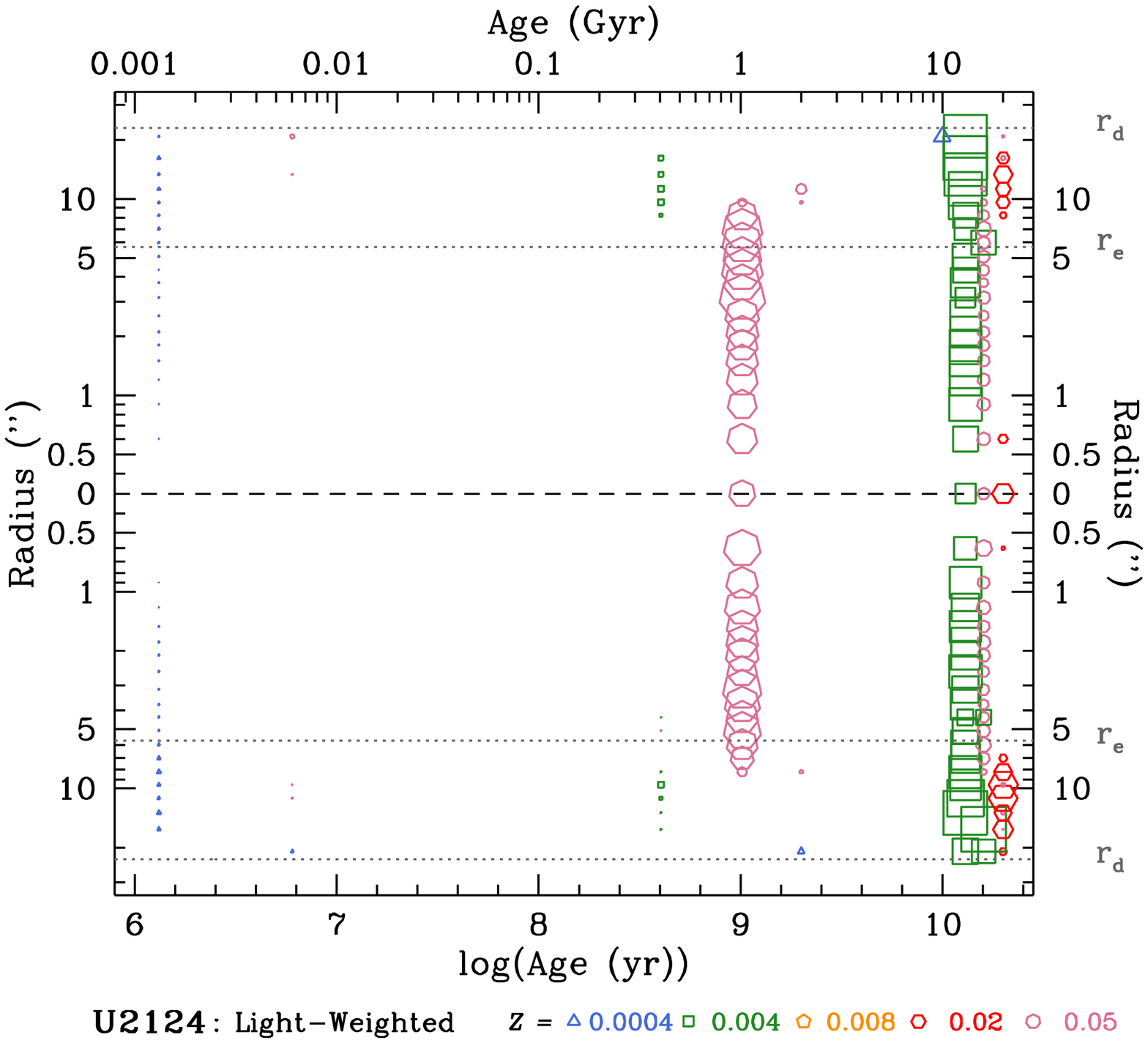}
\includegraphics[width=0.45\textwidth]{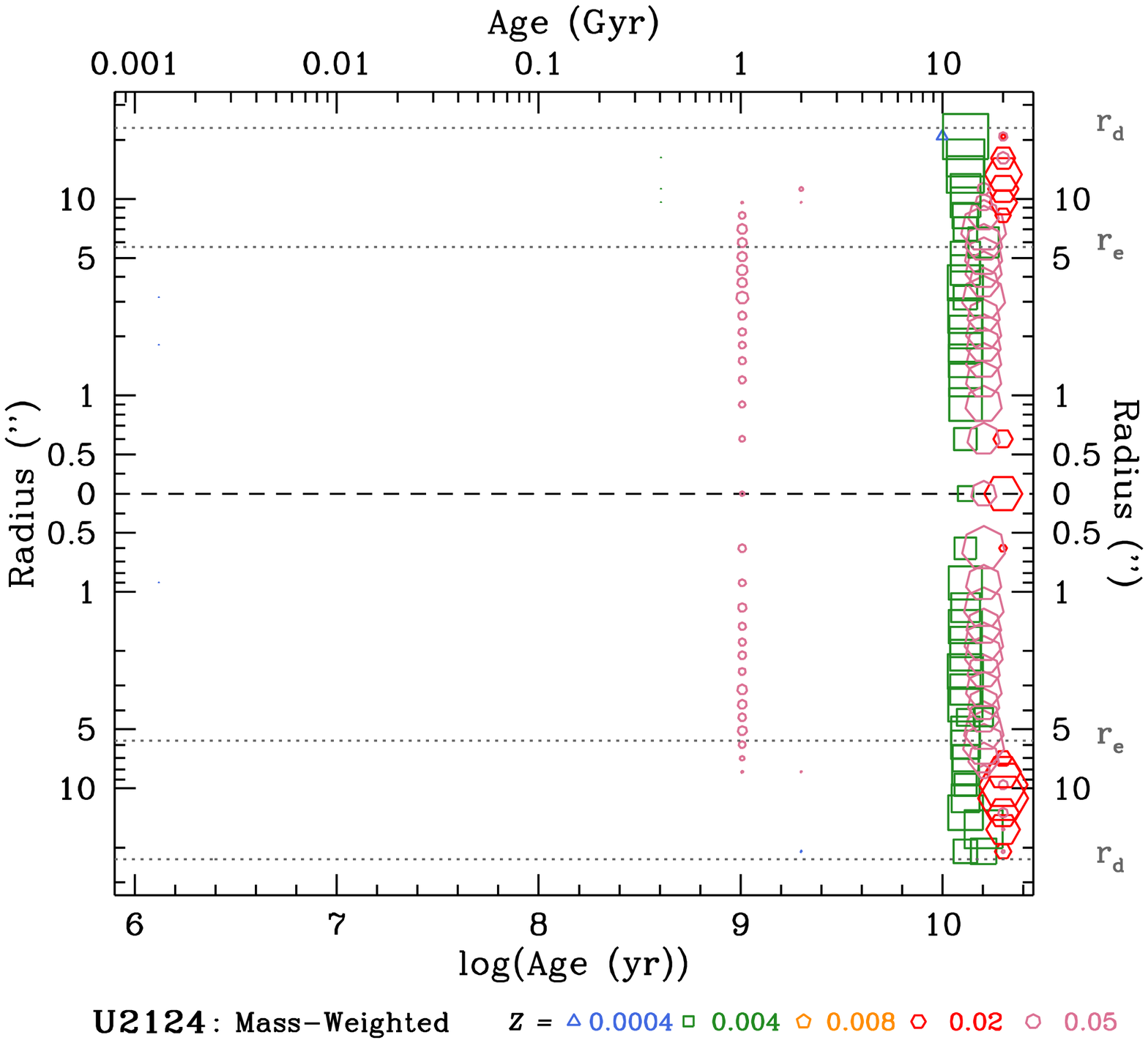}\\
\includegraphics[width=0.45\textwidth]{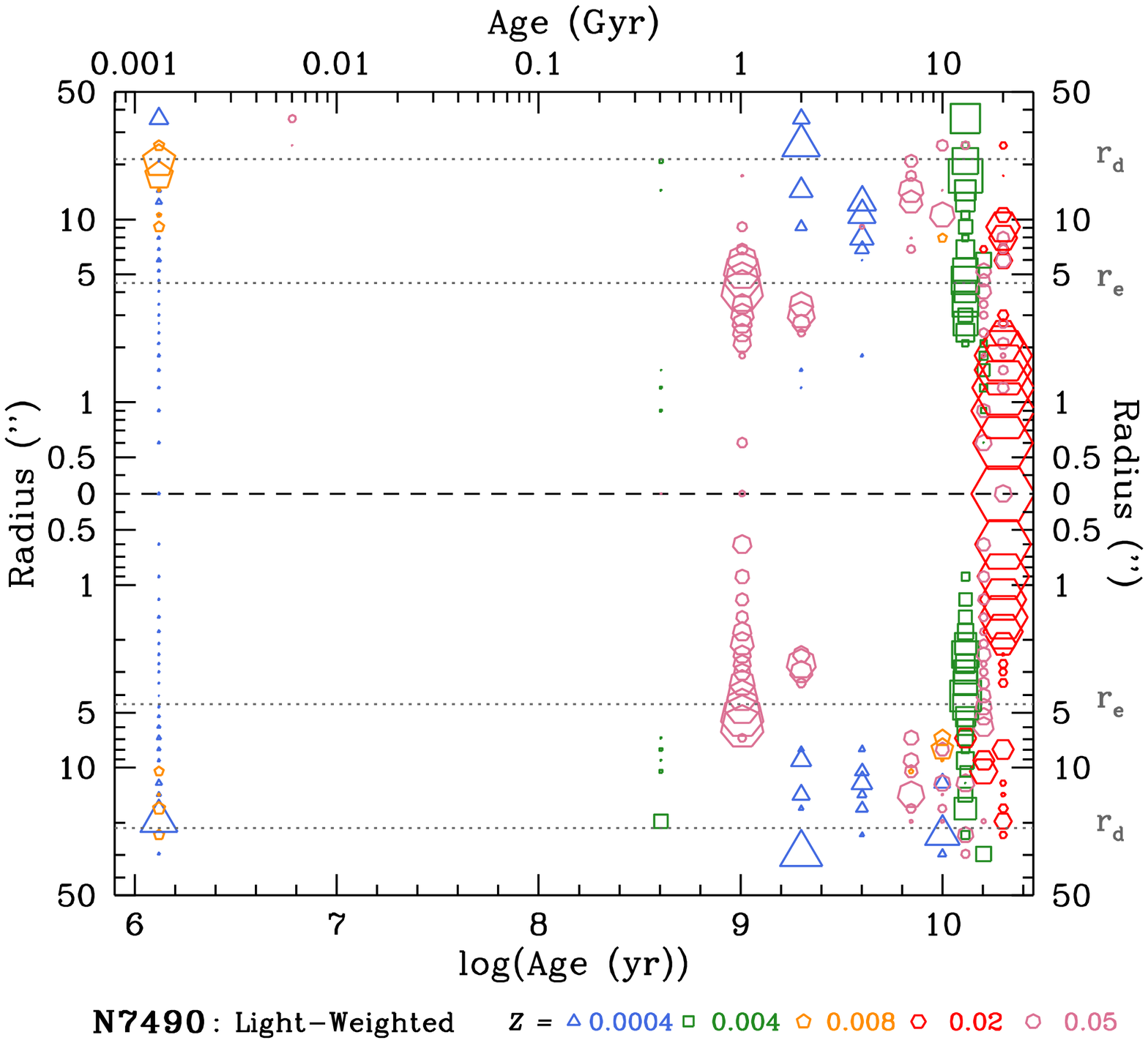}
\includegraphics[width=0.45\textwidth]{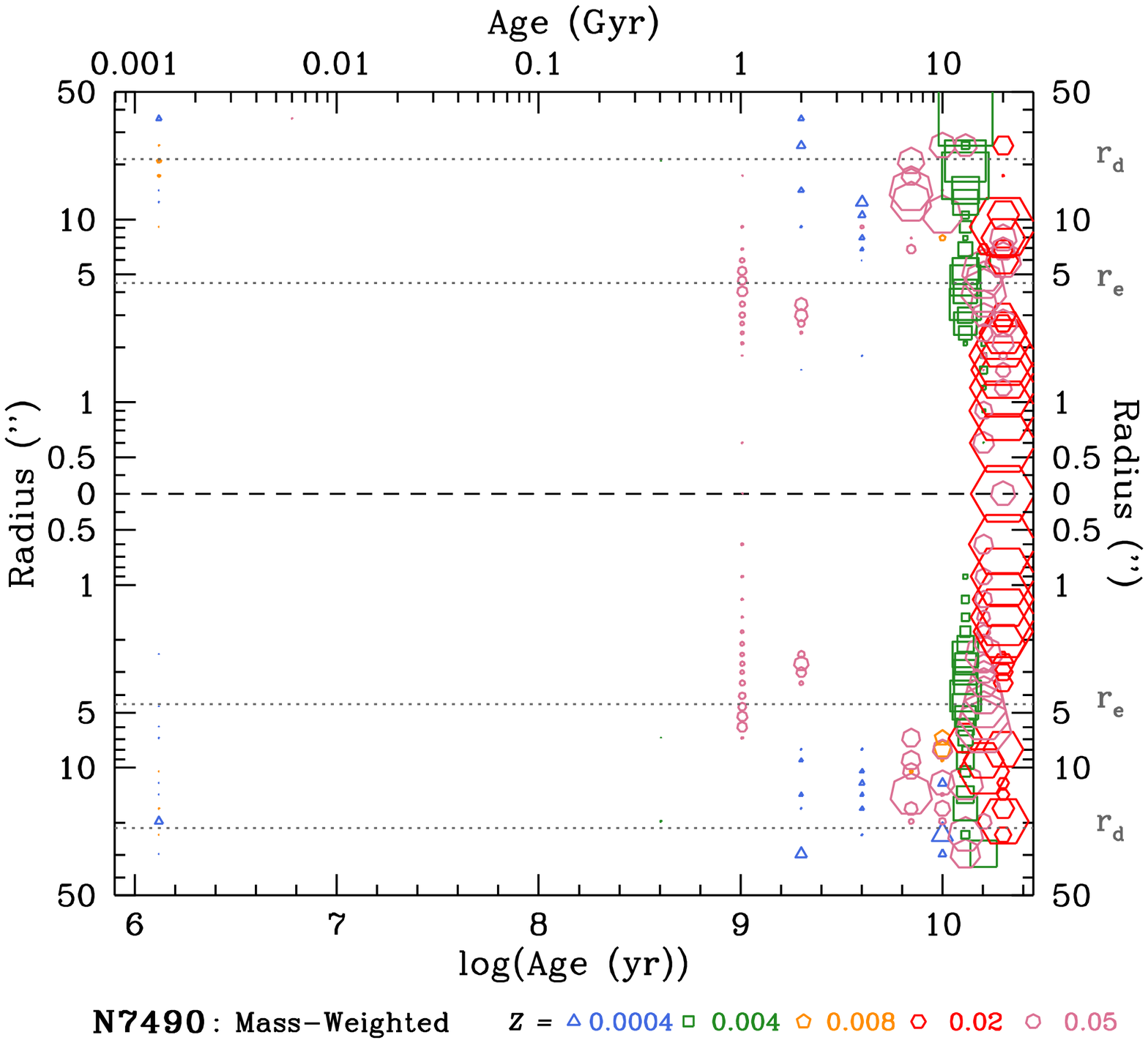}\\
    \contcaption{}
\end{center}
\end{figure*}
\begin{figure*}
\begin{center}
\includegraphics[width=0.45\textwidth]{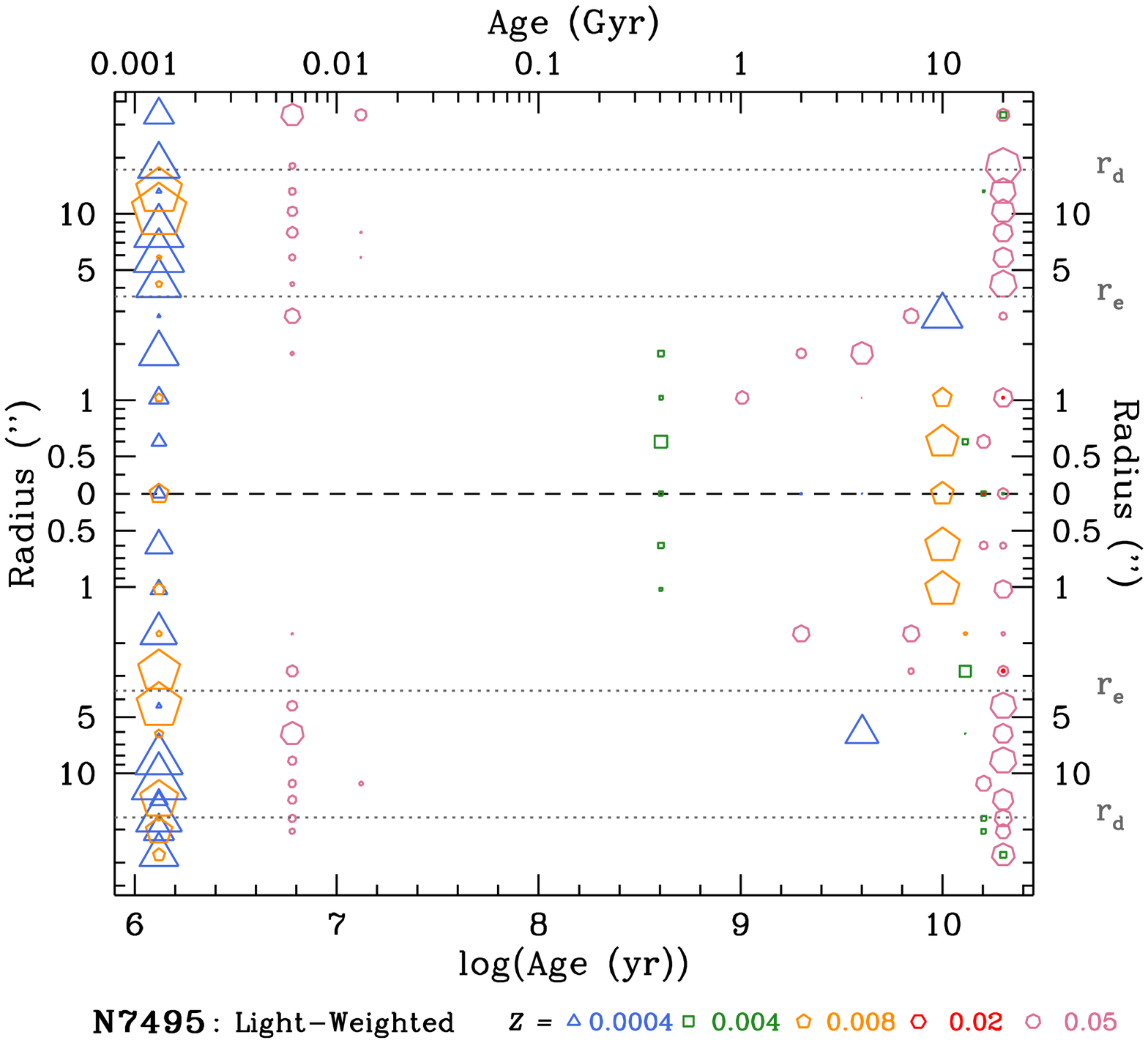}
\includegraphics[width=0.45\textwidth]{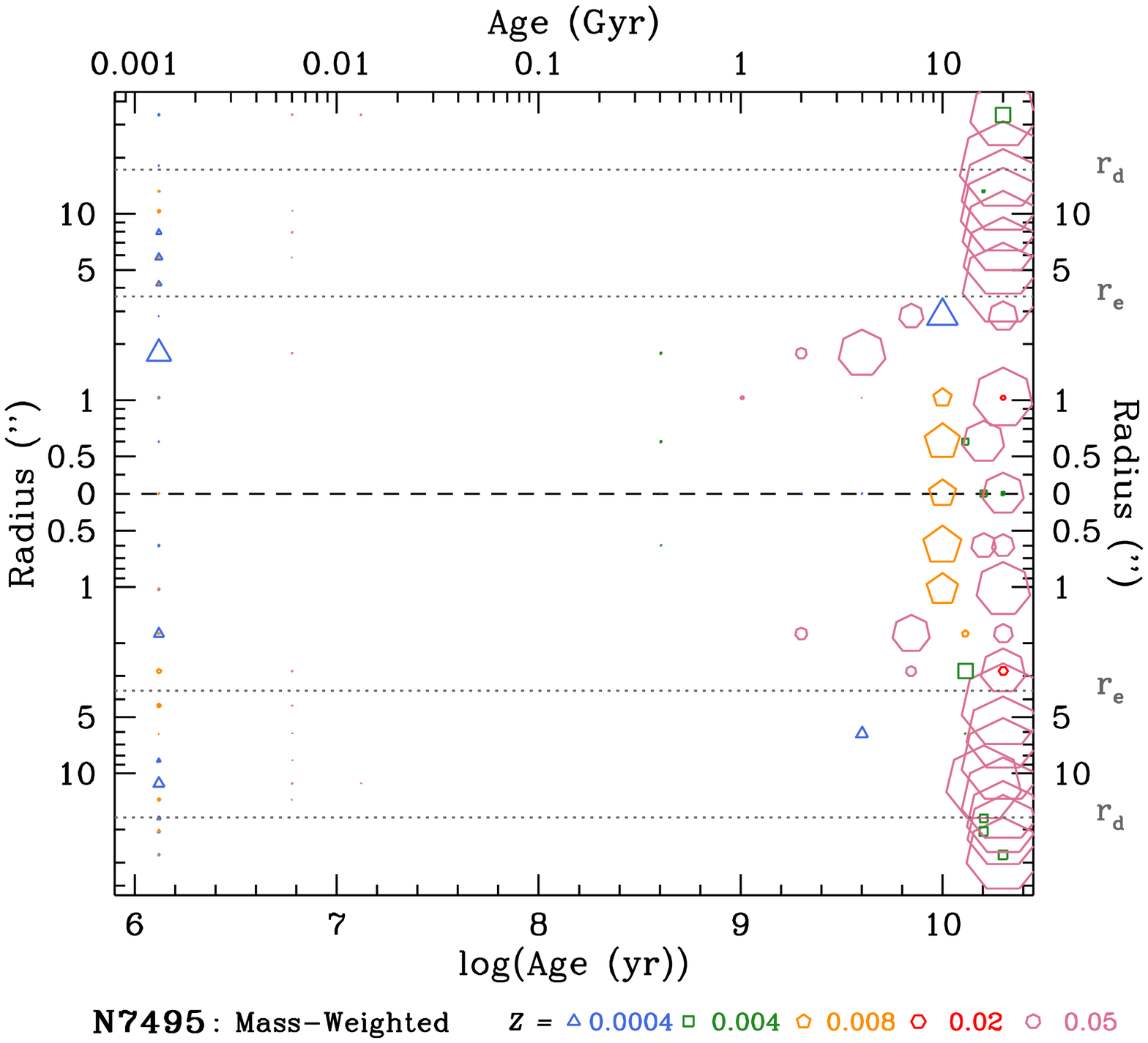}\\
\includegraphics[width=0.45\textwidth]{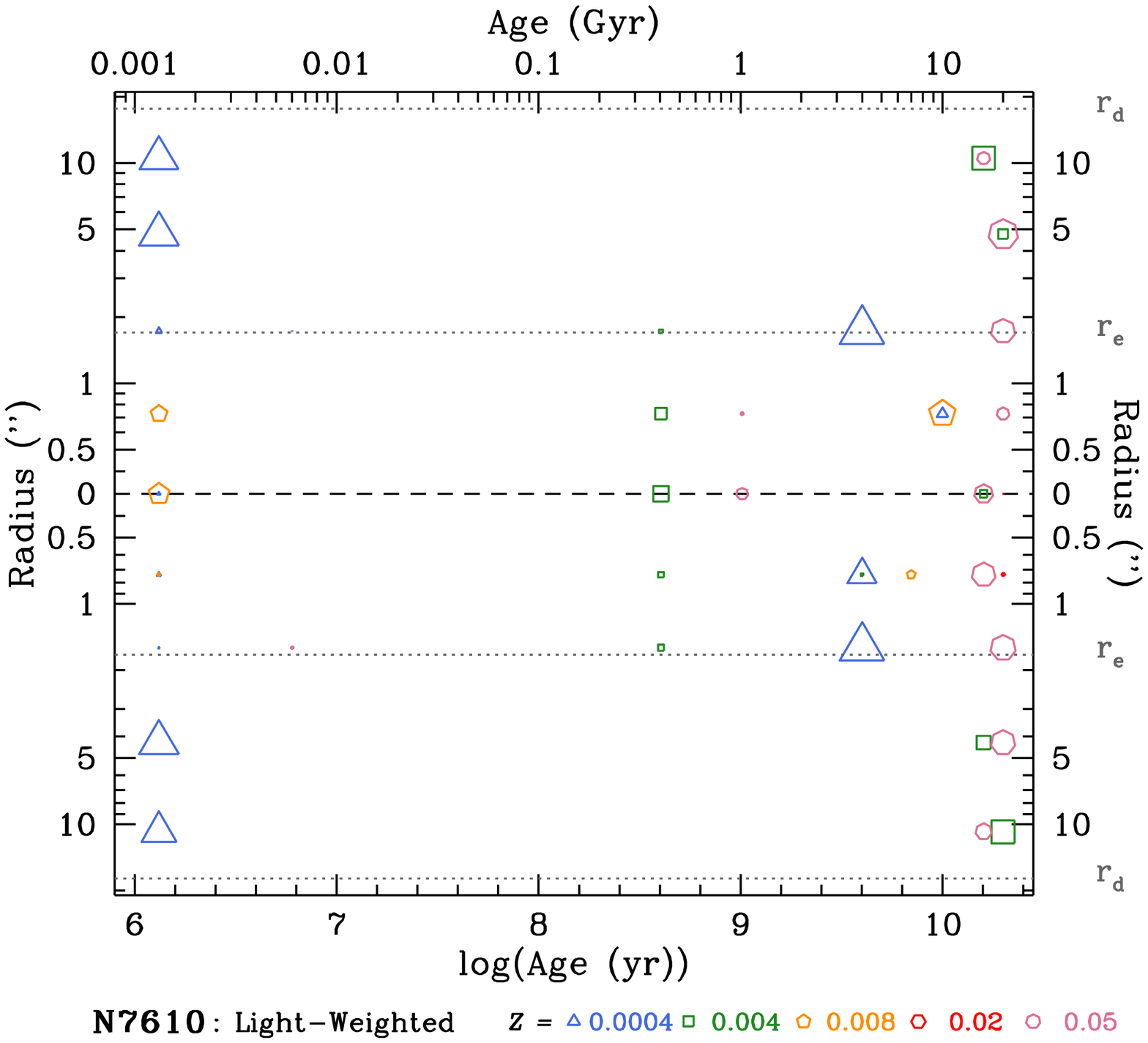}
\includegraphics[width=0.45\textwidth]{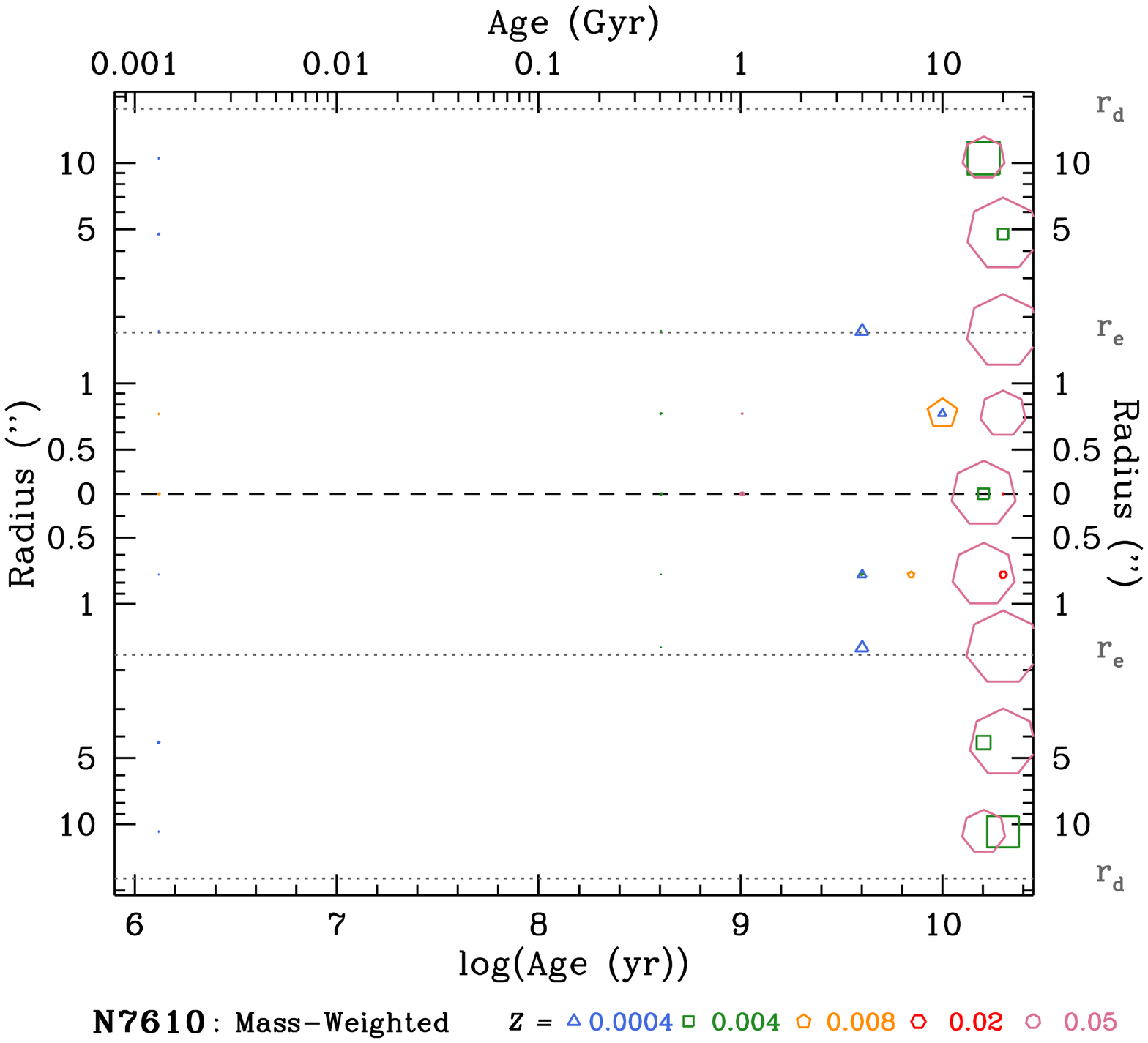}\\ 
    \contcaption{}
\end{center}
\end{figure*}
\begin{figure*}
\begin{center}
\includegraphics[width=0.45\textwidth]{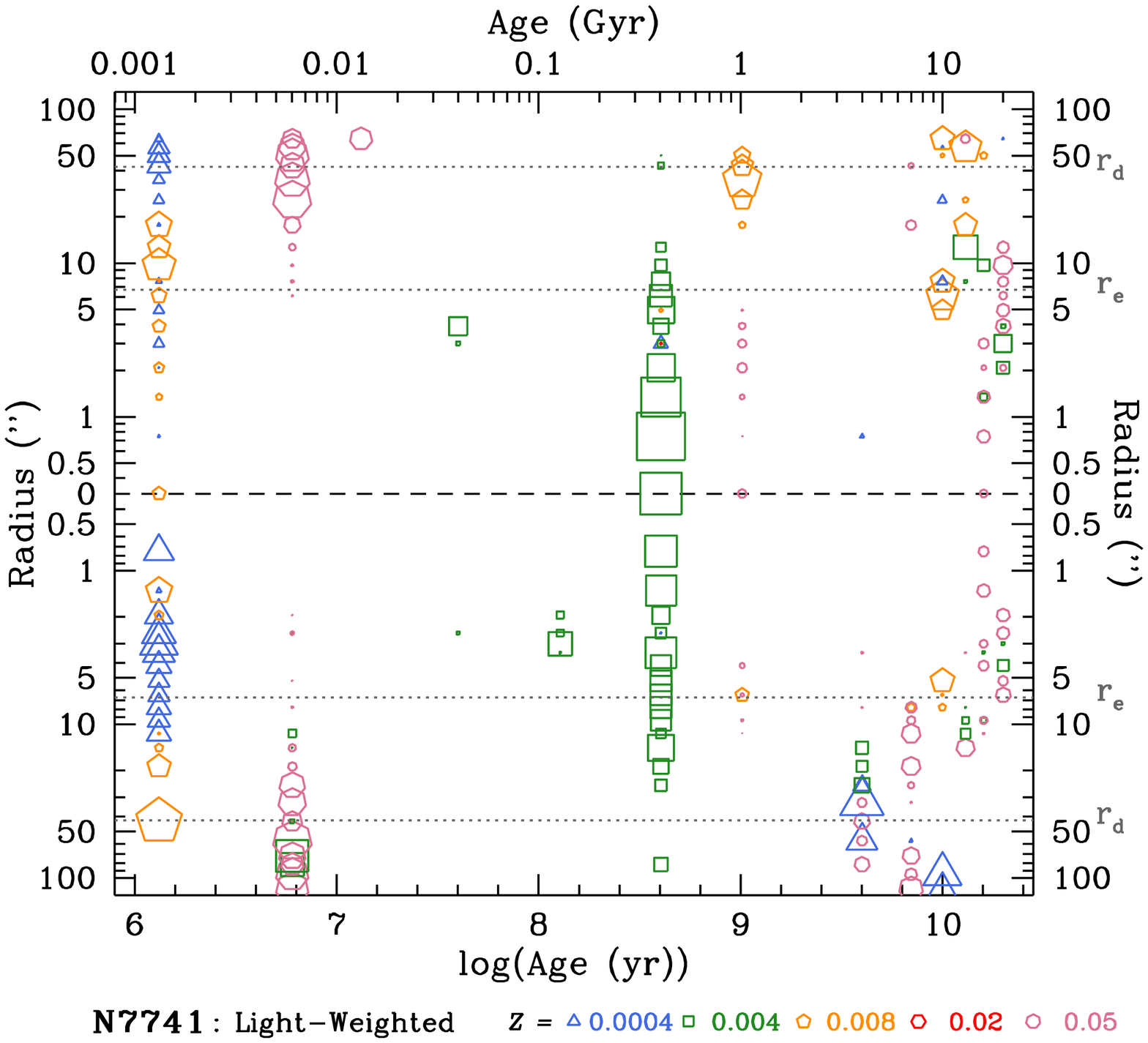}
\includegraphics[width=0.45\textwidth]{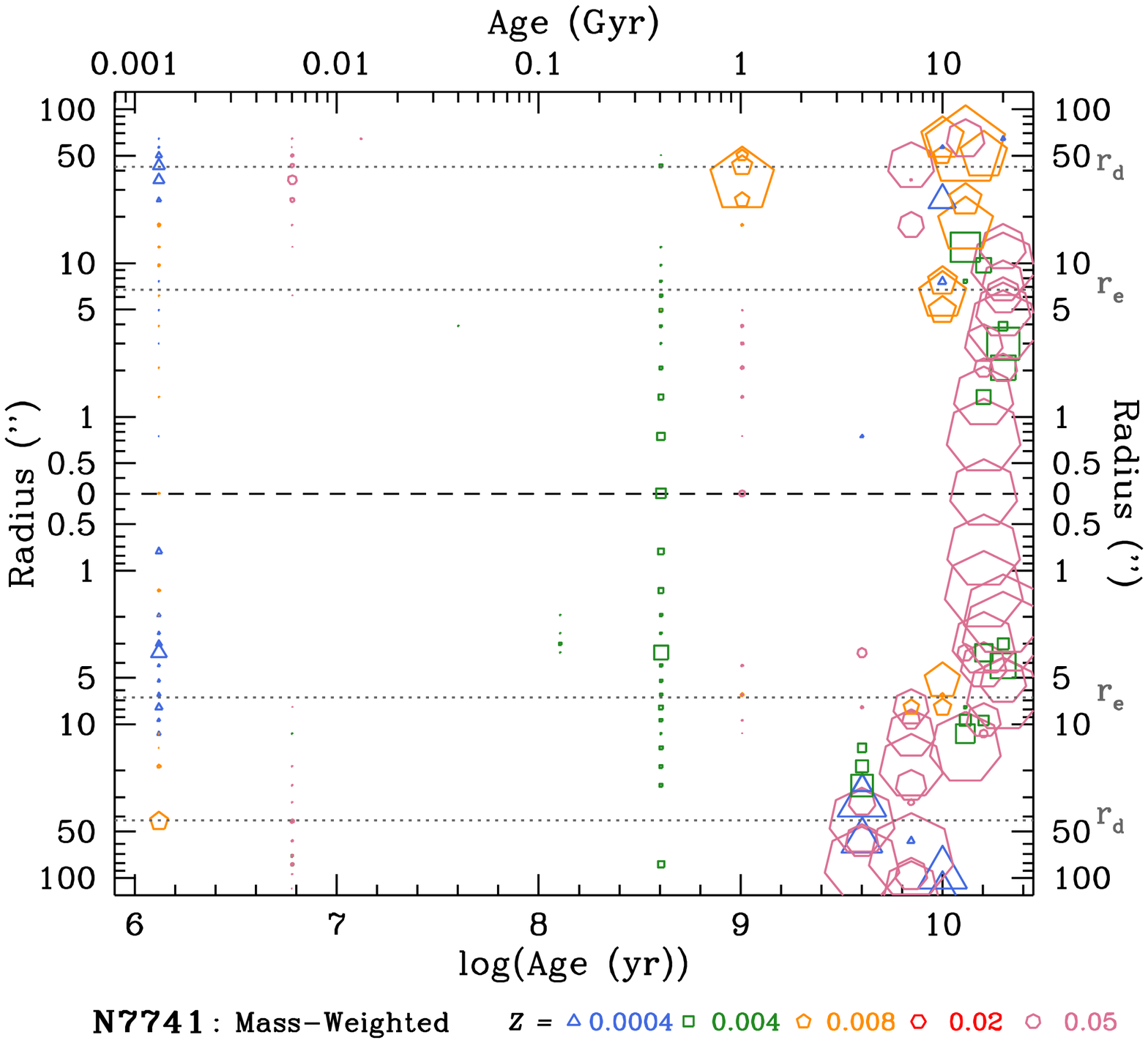}\\
\includegraphics[width=0.45\textwidth]{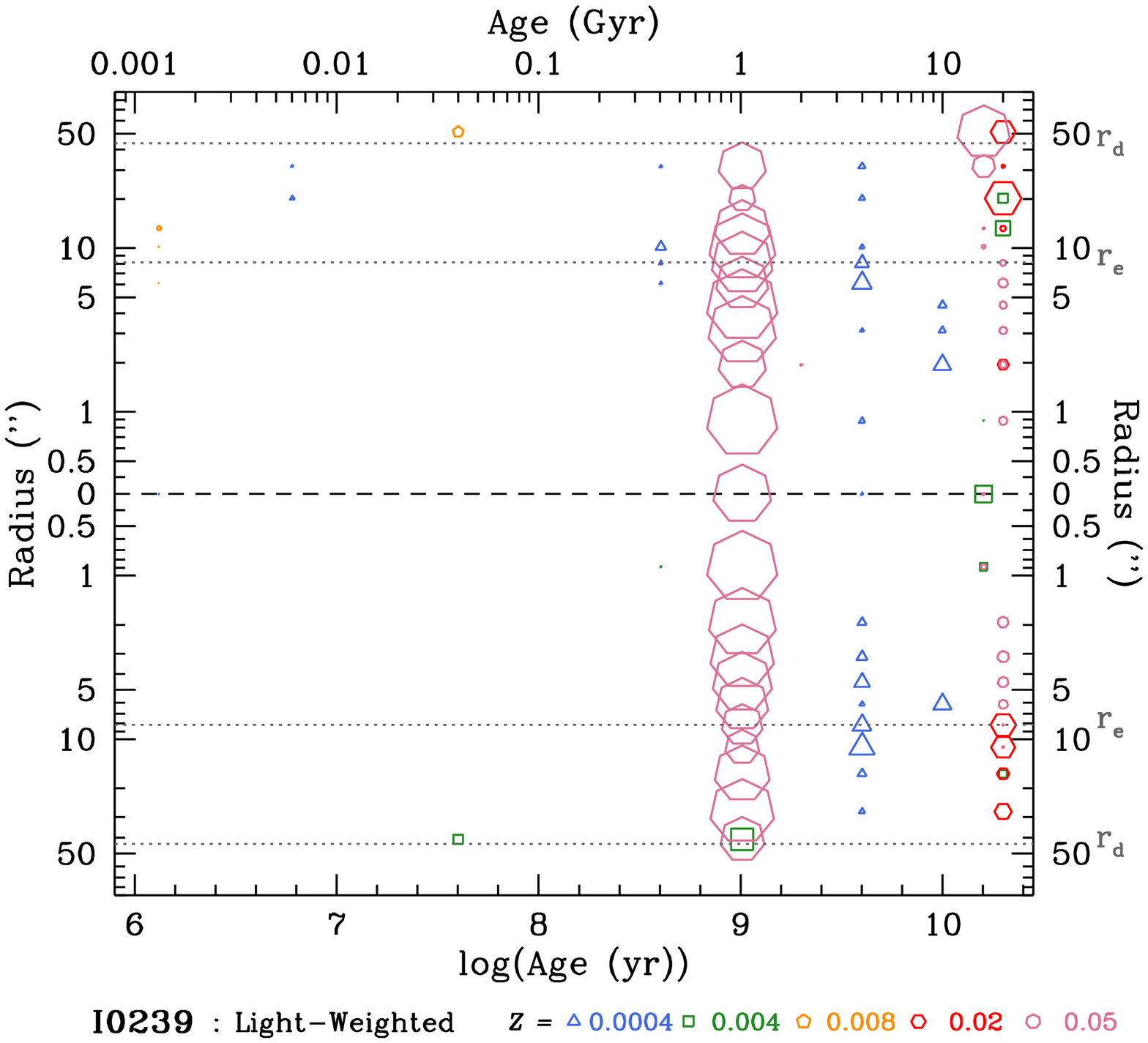}
\includegraphics[width=0.45\textwidth]{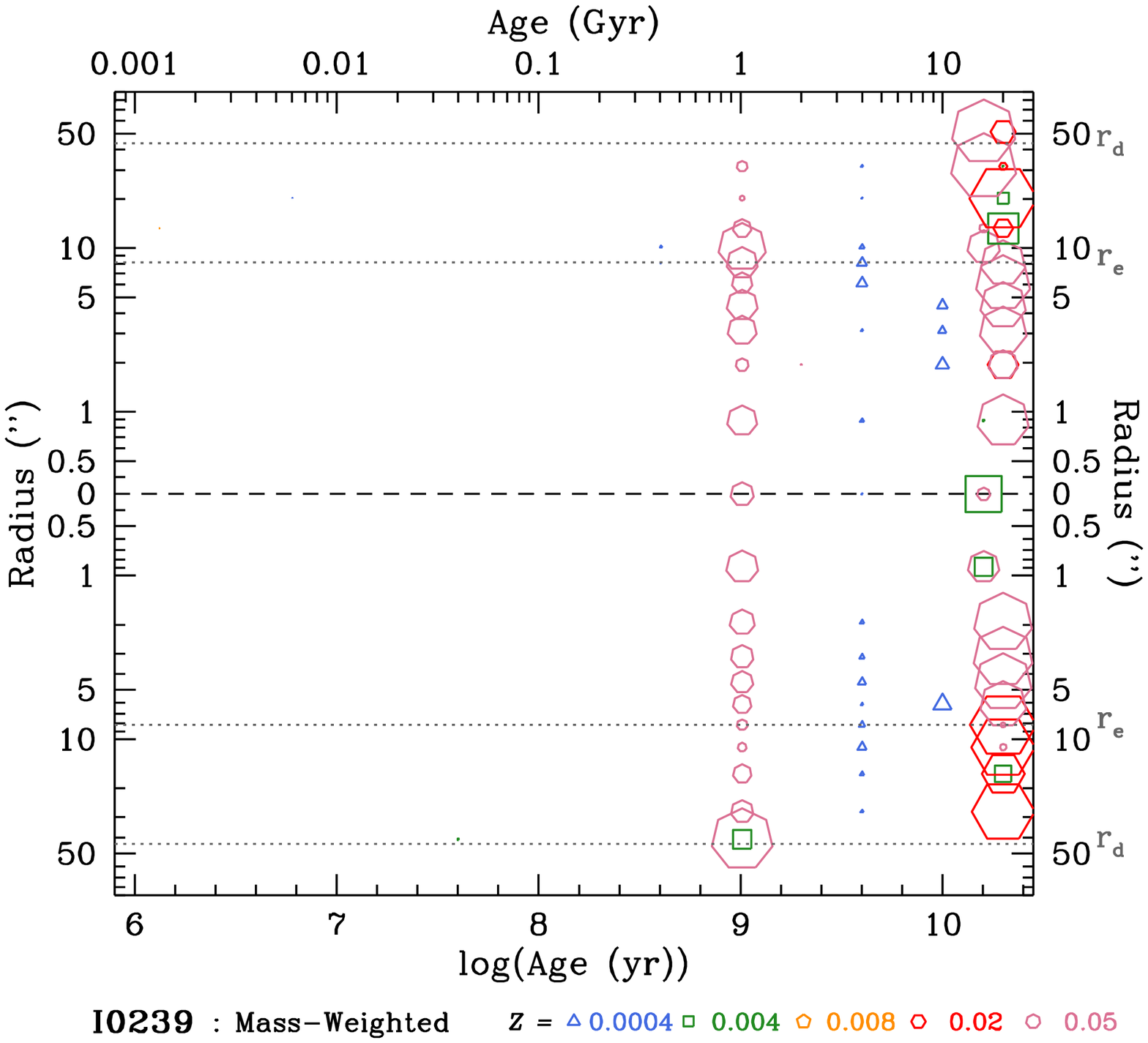}\\
    \contcaption{}
\end{center}
\end{figure*}

Looking at a specific example, examination of the SFH for N7490
reveals that the bulge region is dominated in light and mass by an old
population with solar metallicity.  Further out into the disk, the
light and mass become more dominated by a somewhat younger and more
metal-poor population.  It appears that there was an episode of high
metallicity SF 1\,Gyr ago whose biggest contribution was at about the
bulge/disk transition region ($r_e$\,=\,5.3\arcsec\ for N7490), \ie\
in a ring at that radius.  While the contribution of this recent SF
episode to the light is quite significant at some radii, it does not
contribute significantly in terms of mass.  Alternatively, looking at
N7495, which has significant emission in its spectra at most radial
bins (see Fig.\@~\ref{fig:spec_rad}) , we see that the
light is dominated by a very young ($\sim$\,1\,Myr), metal-poor
population, likely due to a recent accretion of pristine gas, but
the mass is still dominated at all radii by an old and metal-rich
population.  

In Table~\ref{tab:percent} we list for each galaxy the \%-light and
mass contributions of ``young'' (0.001--1\,Gyr), ``intermediate''
(2--7\,Gyr), and ``old'' (10--20\,Gyr) age SSPs to the fits at three
representative radii; $r$\,=\,0, $r_e$, and $r_e$.  With only a few
exceptions ($r_d$ for N0628 \& N7741 and $r_e$ for I0239), the
contribution to the stellar mass by the ``old'' population is
$\ga$\,65\% at all radii, while its corresponding contribution to the
light ranges from $\sim$\,10--90\%.  Again with those exceptions,
the contribution to the mass by the ``young'' SPs is $\la$\,25\%, while
its light contribution is in the range $\sim$\,10--90\%.
\begin{table}
\begin{minipage}{0.48\textwidth}
\centering
\caption{Percent Contributions of all SSPs in Given Age Ranges 
(Young\,=\,0.001--1\,Gyr; Intermediate\,=\,2--7\,Gyr; Old\,=\,10--20\,Gyr)
to Fit for Spectra at $r$\,=\,0, $r_e$, and $r_d$ (when reached) Weighted by
Light ($V$-band normalized) and Mass.} \label{tab:percent}
\begin{tabular}{@{}r@{}@{}c@{}rrrrrr@{}}
\hline
\multicolumn{1}{@{}c@{}}{} &
\multicolumn{1}{@{}c@{}}{} &
\multicolumn{3}{c}{Light-weight} &
\multicolumn{3}{c@{}}{Mass-weight} \\
\cline{3-8}
\cline{3-8}\\
\multicolumn{1}{@{}l@{}}{Name} &
\multicolumn{1}{@{}r@{}}{Rad} &
\multicolumn{1}{c}{0.001--1} &
\multicolumn{1}{c}{2--7} &
\multicolumn{1}{c}{10--20} &
\multicolumn{1}{c}{0.001--1} &
\multicolumn{1}{c}{2--7} &
\multicolumn{1}{c@{}}{10--20} \\
\multicolumn{1}{@{}c@{}}{ } &
\multicolumn{1}{@{}c@{}}{ } &
\multicolumn{1}{c}{(Gyr)} &
\multicolumn{1}{c}{(Gyr)} &
\multicolumn{1}{c}{(Gyr)} &
\multicolumn{1}{c}{(Gyr)} &
\multicolumn{1}{c}{(Gyr)} &
\multicolumn{1}{c@{}}{(Gyr)} \\
\multicolumn{1}{@{}c@{}}{(1)} & \multicolumn{1}{@{}r@{}}{(2)} & 
\multicolumn{1}{c}{(3)} & 
\multicolumn{1}{c}{(4)} &
\multicolumn{1}{c}{(5)} & \multicolumn{1}{c}{(6)} & \multicolumn{1}{c}{(7)} &
\multicolumn{1}{c@{}}{(8)}\\
\hline
 N0173 &   0   & 32.75 &  0.00 & 67.24 &  5.66 &  0.00 & 94.34 \\
       & $r_e$ & 27.98 &  7.85 & 64.18 &  7.10 &  8.34 & 84.56 \\
       & $r_d$ &  7.82 & 21.44 & 70.73 &  0.14 &  3.54 & 96.32 \\
\hline
 N0628 &   0   & 62.98 &  0.00 & 37.02 & 19.41 &  0.00 & 80.59 \\
       & $r_e$ & 45.43 &  6.78 & 47.80 &  7.54 &  2.26 & 90.20 \\
       & $r_d$ & 79.53 & 16.56 &  3.90 & 50.20 & 27.44 & 22.35 \\
\hline
 U2124 &   0   & 30.00 &  0.00 & 70.00 &  5.03 &  0.00 & 94.97 \\
       & $r_e$ & 45.32 &  0.00 & 54.68 & 10.94 &  0.00 & 89.06 \\
       & $r_d$ &       &       &       &       &       &       \\
\hline
 N7490 &   0   & 10.20 &  0.00 & 89.80 &  0.73 &  0.00 & 99.27 \\
       & $r_e$ & 42.29 &  0.00 & 57.71 &  9.73 &  0.00 & 90.27 \\
       & $r_d$ & 45.31 & 14.27 & 40.41 &  5.56 & 29.82 & 64.62 \\
\hline
 N7495 &   0   & 47.02 &  2.61 & 50.37 &  2.17 &  0.64 & 97.19 \\
       & $r_e$ & 61.18 &  6.46 & 32.36 &  5.40 & 11.27 & 83.33 \\
       & $r_d$ & 68.45 &  0.00 & 31.55 &  1.81 &  0.00 & 98.19 \\
\hline
 N7610 &   0   & 66.59 &  0.00 & 33.41 &  8.08 &  0.00 & 91.92 \\
       & $r_e$ & 14.48 & 56.96 & 28.55 &  0.70 & 16.12 & 83.18 \\
       & $r_d$ &       &       &       &       &       &       \\
\hline
 N7741 &   0   & 91.16 &  0.00 &  8.84 & 23.51 &  0.00 & 76.49 \\
       & $r_e$ & 53.22 &  0.00 & 46.78 &  4.23 &  0.00 & 95.77 \\
       & $r_d$ & 99.88 &  0.12 &  0.00 & 98.20 &  1.80 &  0.00 \\
\hline
 I0239 &   0   & 66.75 &  3.42 & 29.83 & 26.50 &  1.92 & 71.58 \\
       & $r_e$ & 74.87 & 18.27 &  6.86 & 36.37 & 13.09 & 50.55 \\
       & $r_d$ & 60.10 &  8.83 & 31.07 & 12.57 &  2.83 & 84.59 \\
\hline
\hline
\end{tabular}
\end{minipage}
\end{table}

Such observations are important for galaxy evolution studies in terms
of the quantities that most strongly affect the SP content and
predictions of a particular model.  If one is more interested in the
dominant component in terms of mass, the current SED can be very
misleading, especially when interpreted in terms of SSP values.  As we
demonstrate in Appendix~\ref{sec:SSPfits}, many of these integrated
late-type galaxy spectra are simply not well represented by single
SSPs, particularly when there is current or recent SF (see
Fig.\@~\ref{fig:SSPfits}, right panel).  While the concept of an
integrated spectrum of a galaxy being dominated in mass by an old
populations, but in light by a minor ``frosting'' of recent SF is not
new (\eg\ Trager \etal\ 2000), the practice of using
``SSP-equivalent'' ages is still in widespread use and could lead to
misleading interpretations in their comparison with different types of
galaxy evolution models.

Fig.\@~\ref{fig:SFH} also demonstrates the relatively few SSP
templates that enter each fit (see Table~\ref{tab:fitstats}), as
noted above.  We return to a discussion of these SFHs in
\S\ref{sec:indiv}.

\subsection{Stellar Population Gradients}\label{sec:gradients}

\begin{figure*}
\begin{center}
\includegraphics[width=0.43\textwidth,bb=0 290 610 718]{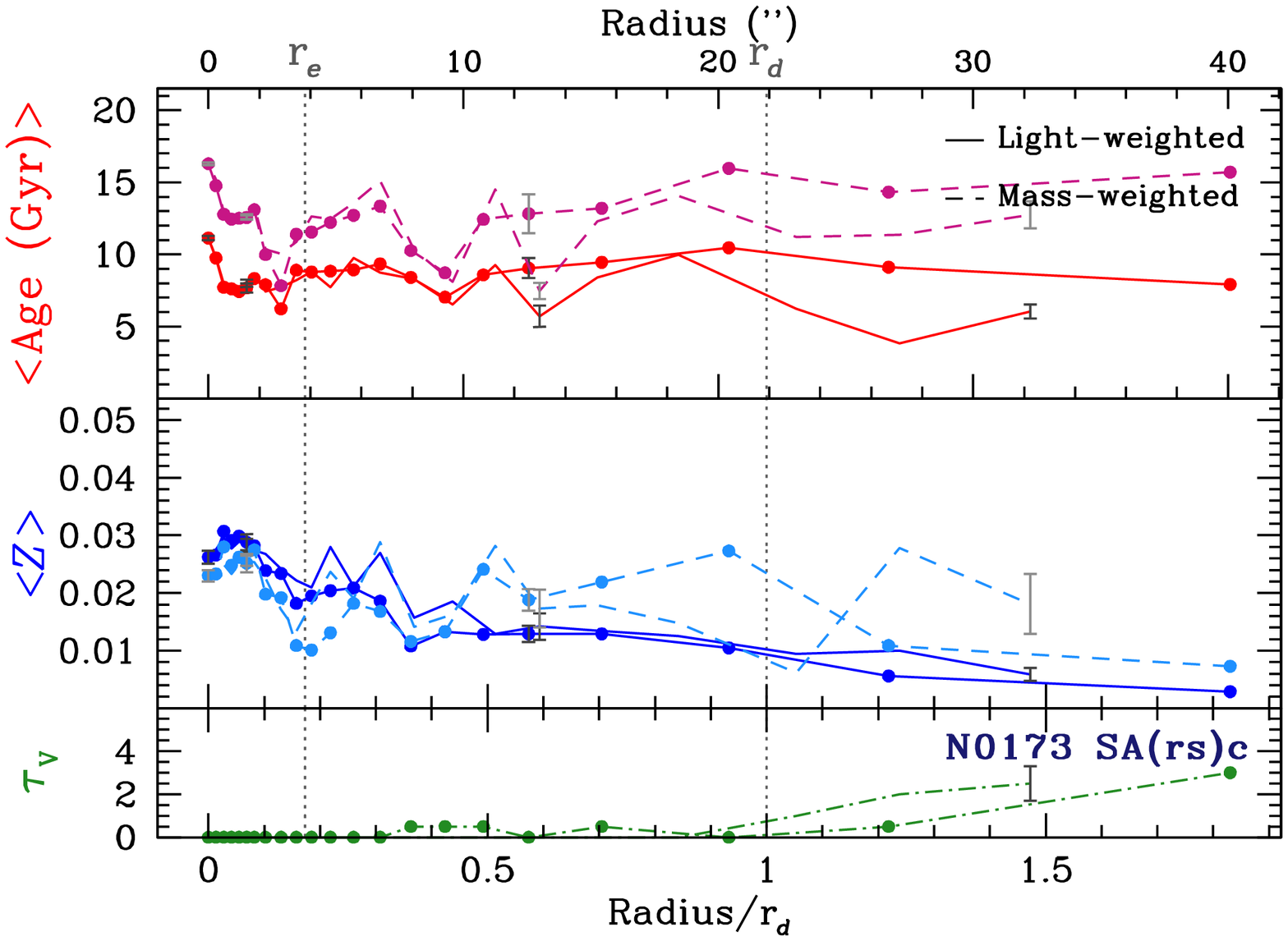}
\includegraphics[width=0.43\textwidth,bb=0 290 610 718]{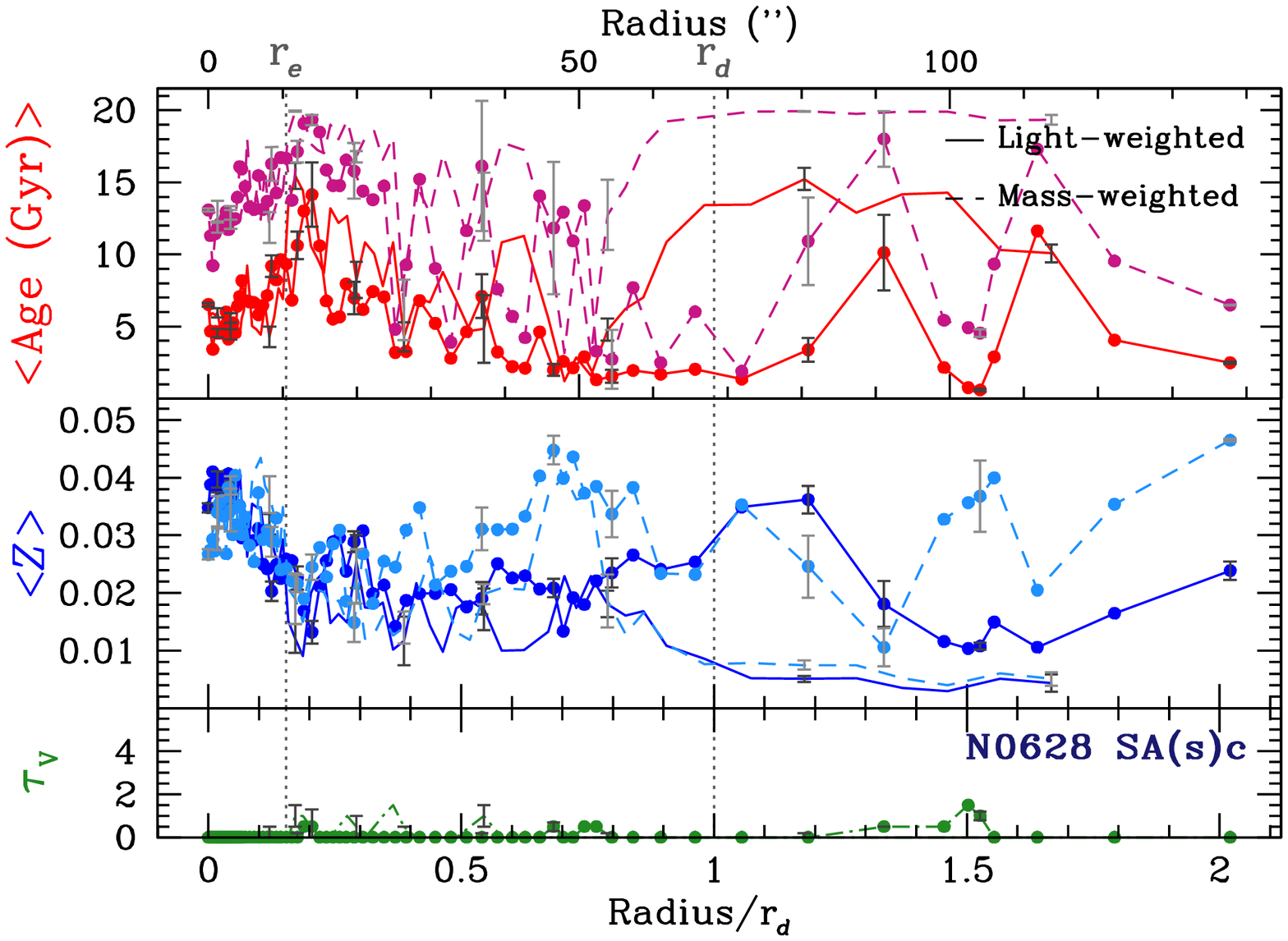}\hfill
\includegraphics[width=0.43\textwidth,bb=0 290 610 718]{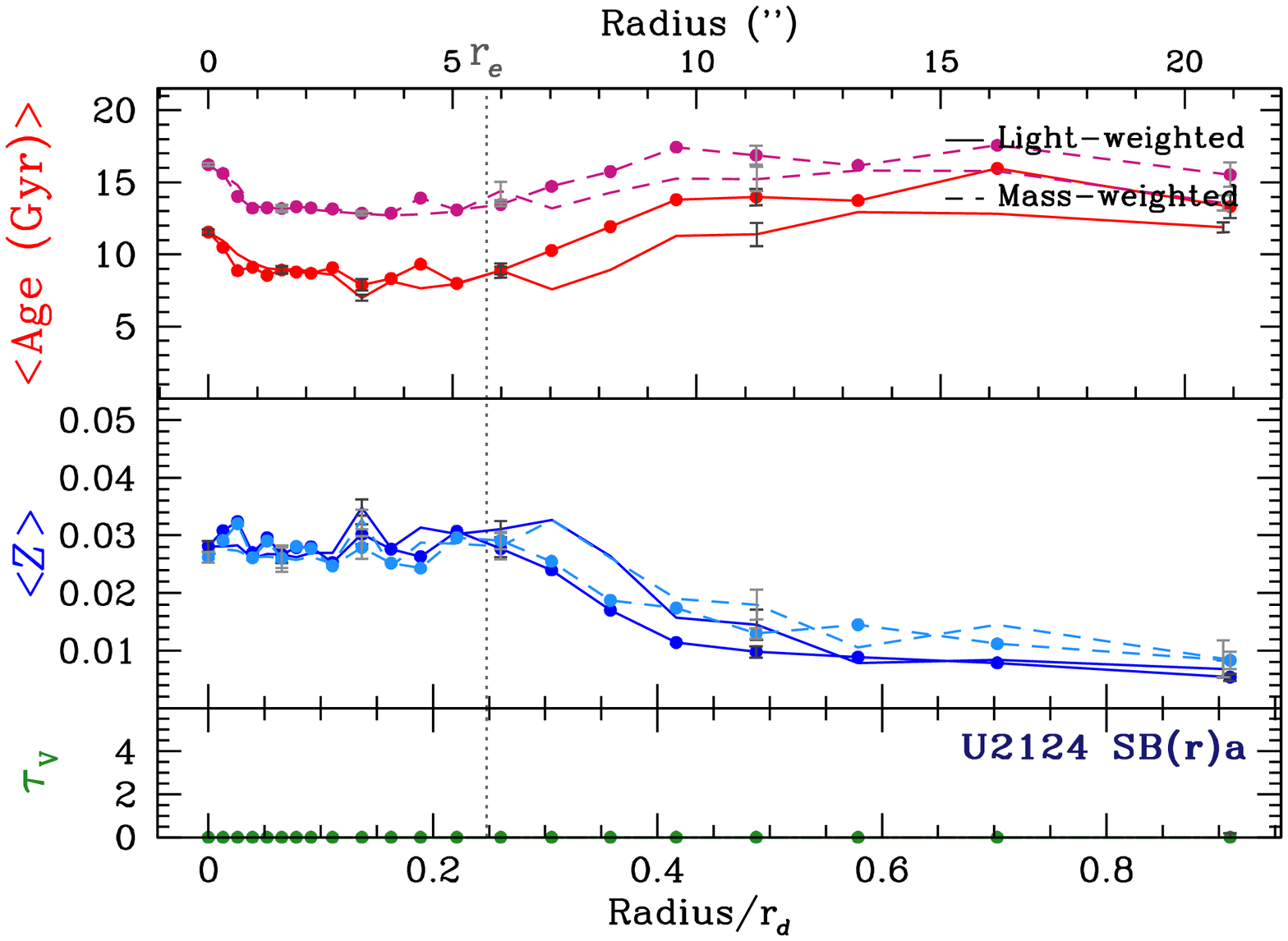}
\includegraphics[width=0.43\textwidth,bb=0 290 610 718]{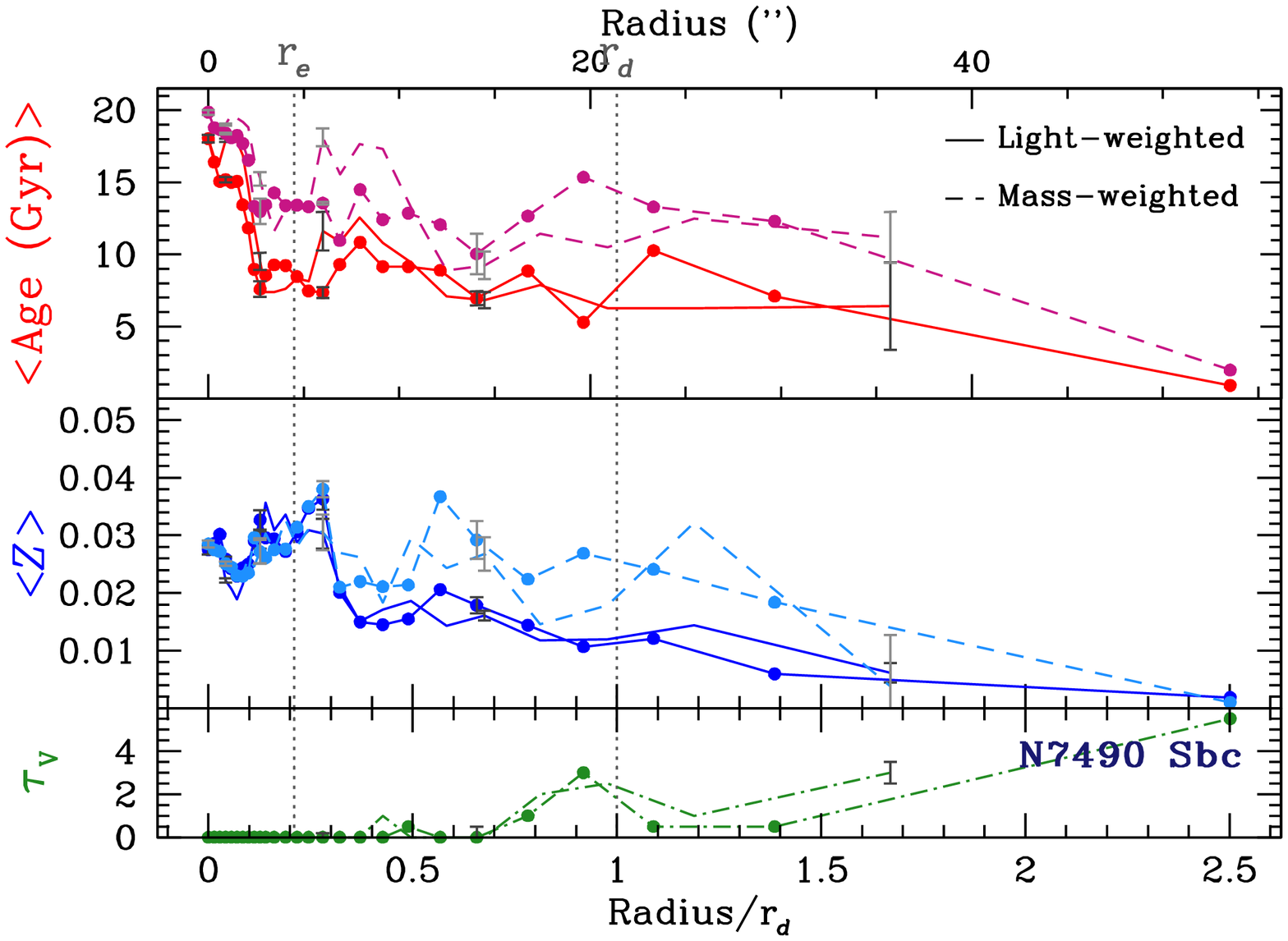}\hfill
\includegraphics[width=0.43\textwidth,bb=0 290 610 718]{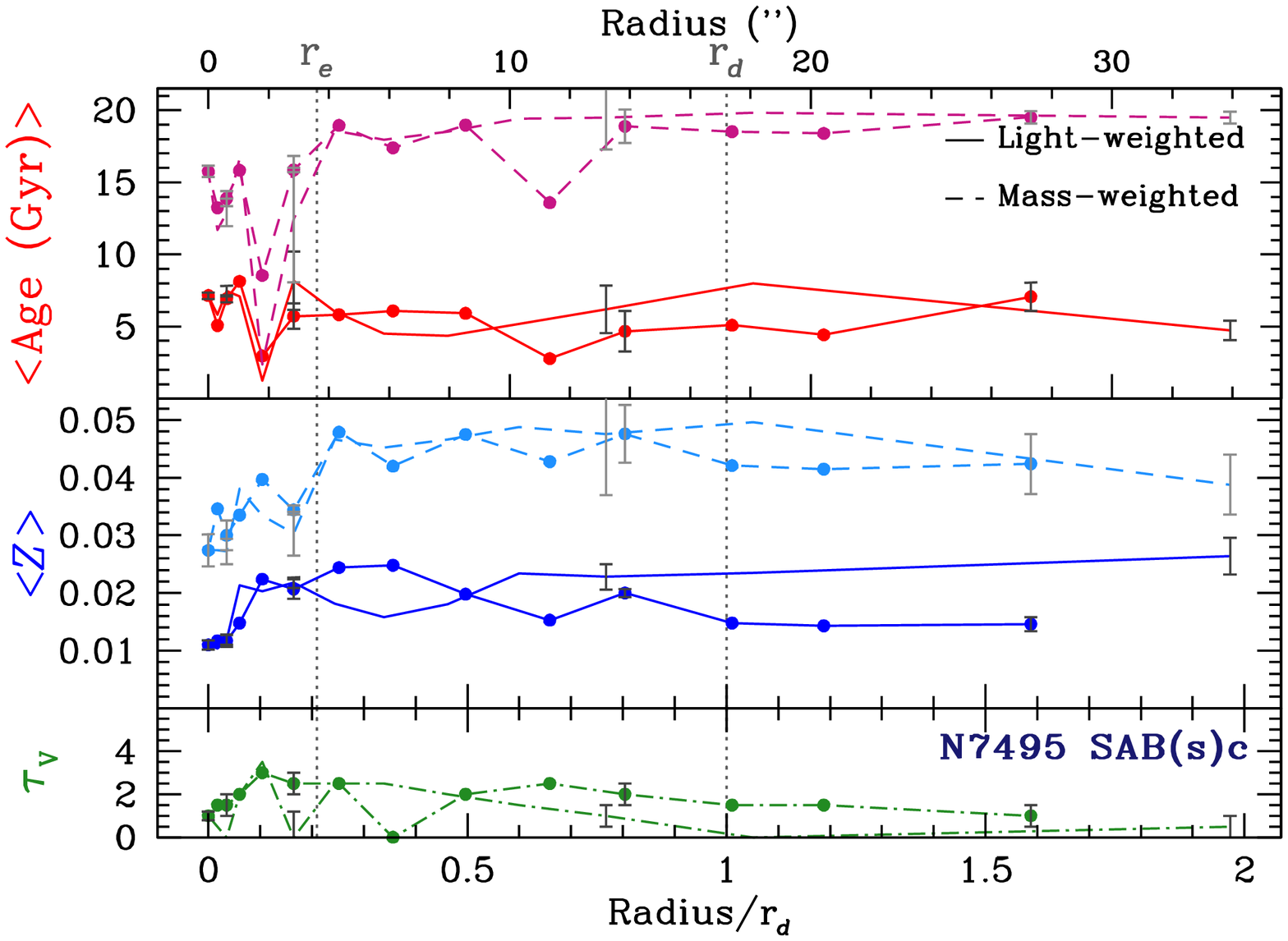}
\includegraphics[width=0.43\textwidth,bb=0 290 610 718]{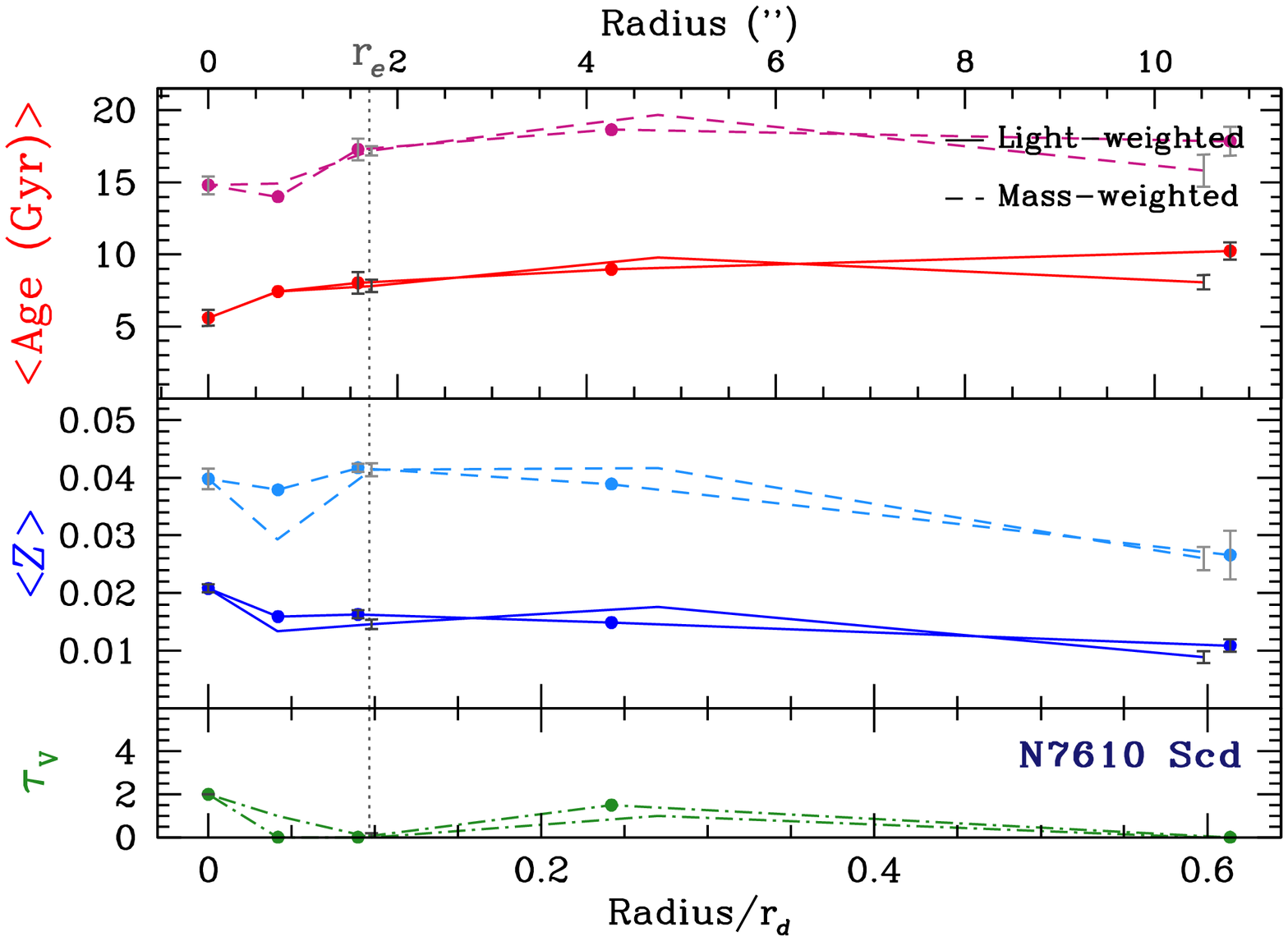}\hfill
\includegraphics[width=0.43\textwidth,bb=0 290 610 718]{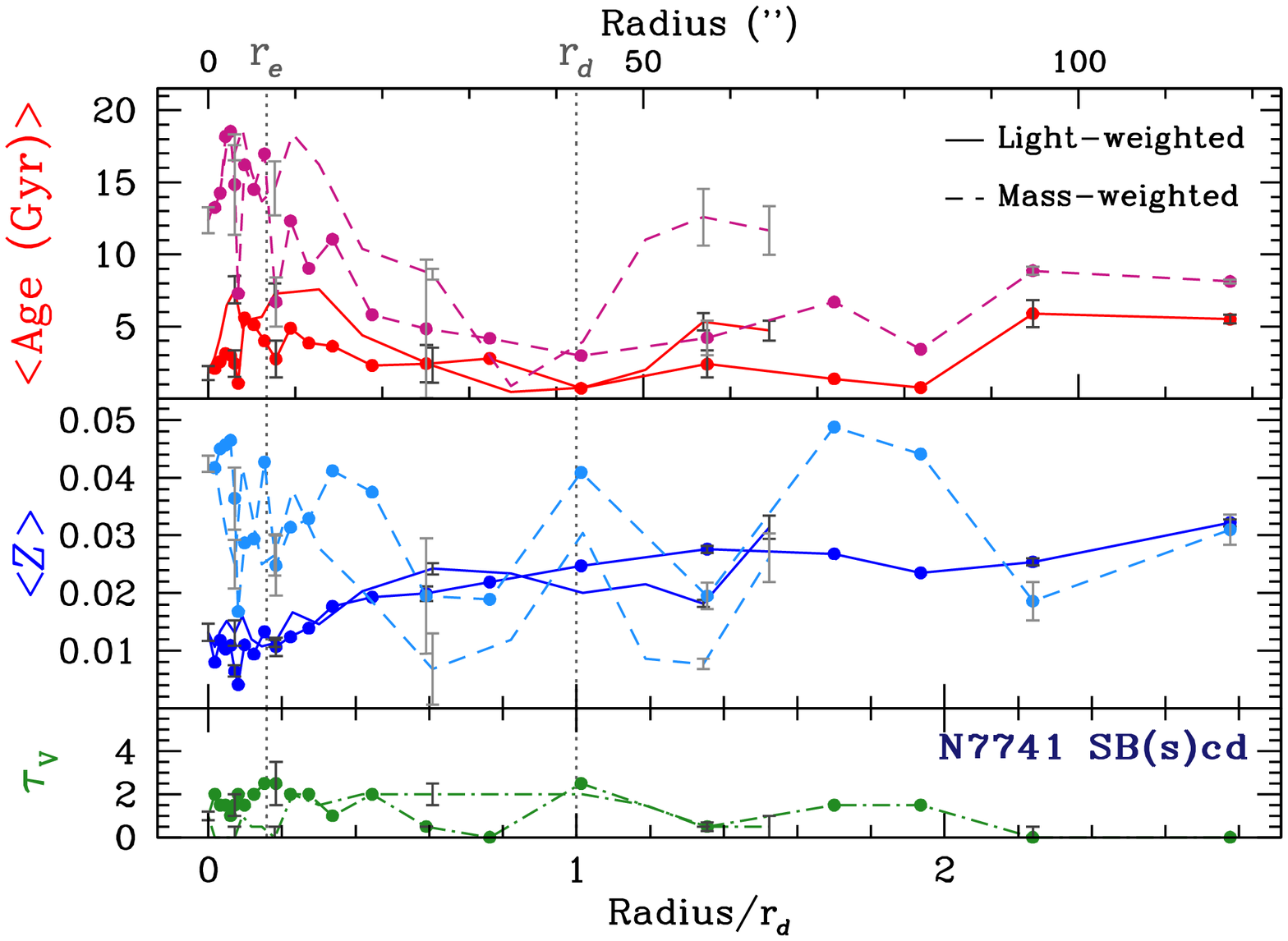}
\includegraphics[width=0.43\textwidth,bb=0 290 610 718]{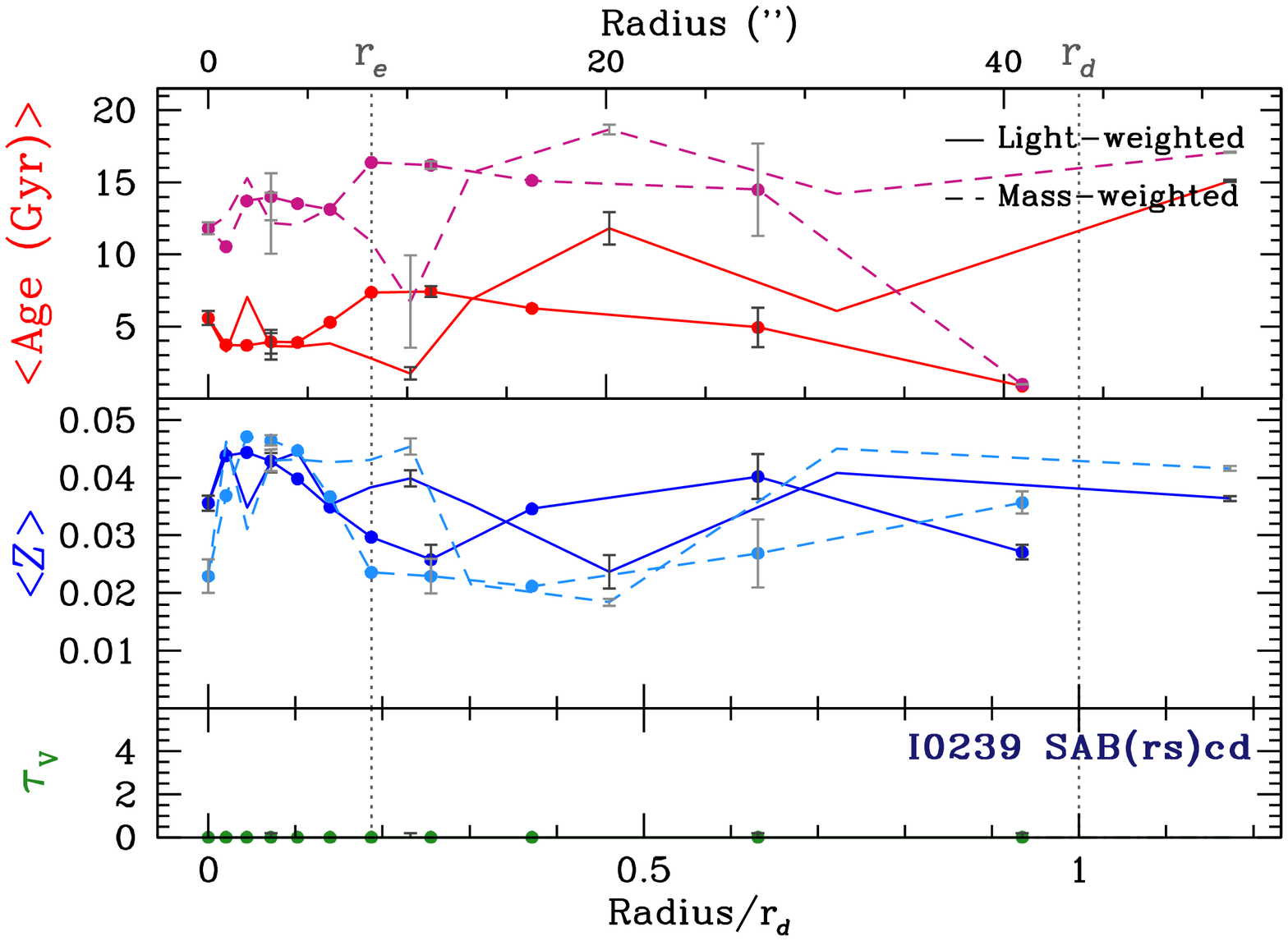}\hfill
\caption{Light and mass-weighted average age and metallicity gradients
         from our full population synthesis fits.  Plotted in each
         panel are the average age (top), average metallicity
         (middle), and \taueff$_V$ (bottom) as a function of radius.
         The solid lines indicate light-weighted quantities, while
         dashed lines are the corresponding mass-weighted values.  The
         variations for both sides of the slit are shown,
         distinguished with one side showing solid circles at each
         radial bin.  Perfect symmetry in the gradients is not
         expected due to the non-axisymmetric nature of spiral
         galaxies, but the consistency is quite remarkable. 
         Error bars are displayed (dark gray for light-weighted and
         light gray for mass-weighted quantities) at all radii for
         which MC simulations were done (see \S\ref{sec:errors}).
         Vertical dashed lines indicate the position of the effective
         radius and disk scale length (when reached).}
         \label{fig:grads}
\end{center}
\end{figure*}

Having refined our population synthesis procedure, we can now
investigate the measured age, metallicity, and dust extinction
gradients in our sample galaxies.  Fig.\@~\ref{fig:grads} presents the
radial profiles of the average age (top panels), average $Z$ (middle),
and \taueff$_V$ (bottom), for all our galaxies.  The solid lines
connect the radial bins for light-weighted quantities,
while the dashed lines are for the corresponding mass-weighted
quantities.  The vertical dotted lines indicate the position of the
bulge effective radius and disk scale length (when reached).  The
profiles for both sides of the slit about $r$\,=\,0 are shown, with
one side distinguished by solid circles at each bin.

A first glance of these gradients reveals the remarkable consistency
of the profiles on either side of the galaxy. A perfect match would not
be expected given that spiral galaxies are not axisymmetric systems,
but the similarities again provide confidence in the stability of our
fitting technique for assessing relative trends (as all radial bins
are modeled independently).

Any difference between the light- and mass-weighted quantities
indicates recent SF activity on top of an underlying older SP.
Indeed, most of the galaxies in our sample show evidence of recent SF
all the way to the center (the only exception being N7490 which, as
we will see below, also has the largest central velocity dispersion in
our sample).  Additionally, to the extent that we measure them, all
disks show evidence for some contribution from an underlying old
($\ga$\,10\,Gyr) population (see Table~\ref{tab:percent} for 
percentage contributions at $r_d$).

The dust content is generally small (\taueff$_V$\,$\la$\,3) and is
always associated with very young SSPs ($\sim$\,1\,Myr), as would be
expected since star forming regions are the sites of dust formation.  The
small implied dust extinction may seem at odds with the results of
Driver \etal\ (2007) who, based on inclination-dependent bulge and
disk luminosity functions for a sample of 10095 galaxies, infer
face-on $B$-band extinctions of up to 2.6\,mag for bulges and 1.1\,mag
for disks.  However, if the dust is largely concentrated in the plane
of the galaxy, as is the case for the models they use with the
scaleheight of the thickest dust component equal to $\sim$\,20\% of
the bulge effective radius, the overall effect for a face-on bulge is
that of near-complete attenuation of the stellar light behind the dust
lane, rather than a reddening from light passing through the dust.
Thus, even if such large attenuations exist in our bulges, because of
the face-on orientations, we are observing non-extincted light from
above the dust lane and thus would not expect our spectra to exhibit
strong signatures of dust reddening.

The gradients themselves show a wide variety of behaviors.  Within the
bulge-dominated region, the light-weighted age profiles decrease in
age with radius for N0173 \& N7490, increase for N0628 \& N7741, and
approximately flat for U2124, N7495, N7610, \& I0239, while the
light-weighted $Z$ profiles within the bulge decrease with radius for
N0173, N0628, \& I0239, increase for N7495 \& N7741, and are
approximately flat for U2124, N7490 \& N7610.  The disks, on the other
hand, almost always show mildly decreasing to flat profiles in both
age and metallicity.  This diversity in SP gradients, and the fact
that a trend in one is not necessarily associated with a trend in the
other, allow for a variety of formation scenarios.  This variety in
observed gradients is also in qualitative agreement with those based
on Lick-indices of Moorthy \& Holtzman (2006) for a sample of 38
spirals.  We further discuss the gradients for individual galaxies in
\S\ref{sec:indiv}, also in the context of their kinematic profiles
which are presented in the following section.

\subsection{Stellar Kinematics}\label{sec:kinem}

We now turn to an analysis of the kinematics of our spiral galaxies.
Given the intricacies involved in accurate measurement of absolute 
velocity dispersion and rotation profiles with long-slit observations,
we provide below a fairly detailed description of our procedure.

\subsubsection{Measurement of Stellar Velocity Dispersion \& Rotation Profiles}

For velocity dispersion, $\sigma$, and radial velocity, V$_{rot}$,
estimates we used the {\tt Movel} algorithm (updated from Gonz{\'a}lez 1993), 
first introduced in \S\ref{sec:templates}.  The population 
templates are now the composite stellar population fits obtained 
in \S\ref{sec:fits}.

In addition to a careful characterization of the resolution and
wavelength calibration of our GMOS data (\S\ref{sec:data}), in
\S\ref{sec:templates} we also explored in detail the absolute spectral
resolution, as a function of wavelength, as well as the wavelength
scale of the BC03 models.  For the derivation of
absolute $\sigma$s and rotation curves, the templates were first
corrected by the systematic error in their wavelength calibration
derived in \S\ref{sec:templates} (note that for the stellar population
fitting, it was the galaxy data that were distorted to the STELIB
wavelength scale).

The GMOS resolution is very well characterized by the convolution of a
top-hat slit resolution of 10.81\,$\pm$\,0.01\,\AA\ and a Gaussian
instrumental resolution of 0.8\,$\pm$\,0.02\,\AA.  Strictly speaking,
however, this resolution only applies to regions where the galaxy
effectively fills the 2\arcsec\ slit.  Towards the center of each
galaxy, where the galaxy structures become significantly smaller than
the slit width, an object can no longer be considered extended, and
the actual resolution is given by the convolution of the galaxy
surface brightness profile, the seeing FWHM, and the spectrograph
internal (after the slit) resolution (0.8\,\AA).  Using the observed
spectroscopic brightness profile of each galaxy and taking into
account the measured anamorphic amplification, we created 2D effective
resolution maps. In the {\tt Movel} procedure, each galaxy is first
convolved with the residual resolution of the templates
(FWHM\,=\,$\sqrt{3.4^2-0.8^2}$\,\AA\ and $\sigma$\,=\,11\,\kms) while
the template for each galaxy radial bin is convolved by its
corresponding row in the 2D-effective-resolution map of the galaxy. In
this fashion, {\tt Movel} can then extract absolute velocity
dispersions, since both galaxy and templates are now well matched at
all points along the slit.

The templates are generated for bins along the slit that have been
coadded to accumulate S/N/\AA\,$\geq$\,50.  To avoid artificially
broadening each radial bin, the bins must be coadded after
removing the rotation shape.  This is achieved by iteration, where we
first interpolate the template fits for each coadded bin and use them
in the unbinned data (continuous 0.07\arcsec\/pixel sampling along the
slit) to derive and remove the rotation curve.  We then reconstruct the
minimal S/N bins and derive the absolute $\sigma$s and the small
residual correction of the radial velocity at each coadded bin.

In Fig.\@~\ref{fig:kinem} we present the results from our {\tt Movel}
determination of the galaxy kinematic profiles.  For each galaxy, we
plot the velocity dispersion profile (top panel), rotation curve
(middle panel), and $\gamma$ profile (bottom panel).  Note the
different scales for each galaxy.  Recall that $\gamma$ is a measure
of the relative absorption strengths between galaxy and template, with
$\gamma$\,=\,1 representing a perfect match.  In addition to obvious
potential culprits inherent to the data reduction process (sky-line
subtraction, relative flux calibration, etc.), strong deviations from
$\gamma$\,=\,1 could indicate a region in parameter space that is not
represented in the models.  For example, a mismatch of abundance
ratios between the models (which only include solar-neighborhood
abundance patterns) and galaxies as described in \S\ref{sec:templates}
(see, \eg\ U2124), or at low age/$Z$ where the stellar libraries are
sparse (see, \eg\ N7741).  It could also indicate a spectrum whose
emission-line fill-in did not get fully masked out in our
``$\sigma$-clipping'' procedure described in \S\ref{sec:sigmaclip}.
For the most part, however, $\gamma$ is quite close to 1 indicating a
good match between galaxy and model fits.
\begin{figure*}
\begin{center}
\includegraphics[width=0.43\textwidth]{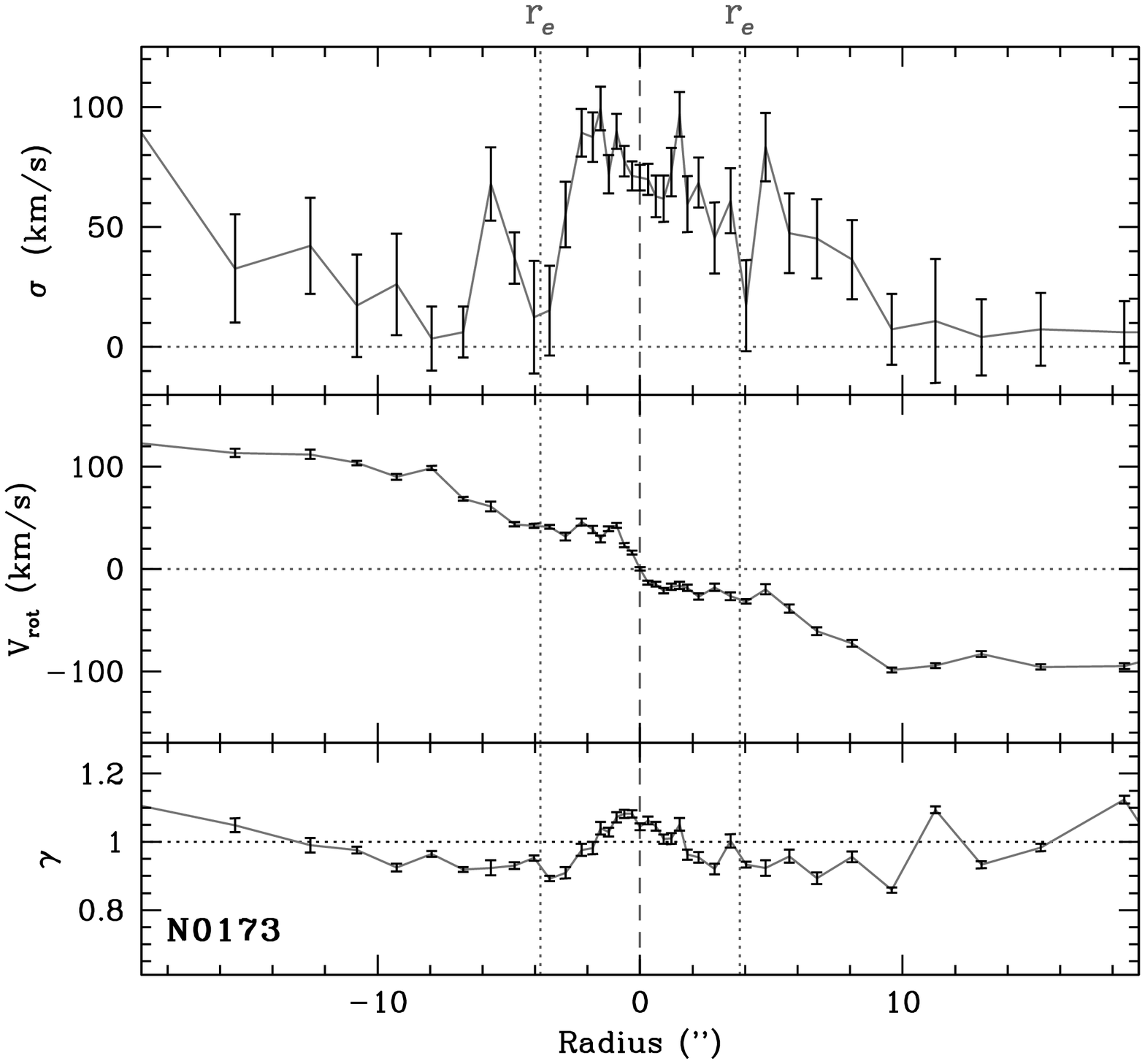}
\includegraphics[width=0.43\textwidth]{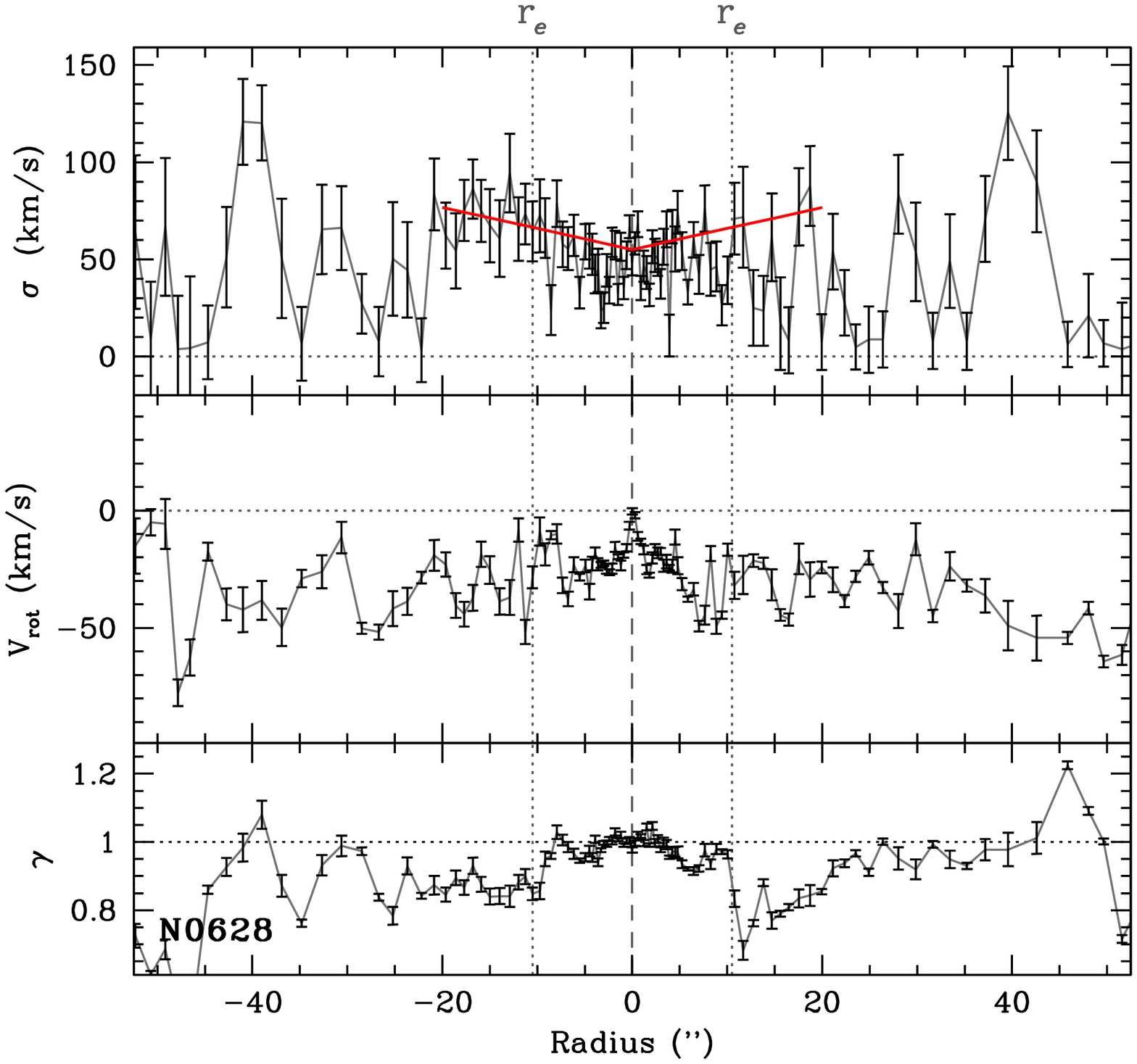}\\
\includegraphics[width=0.43\textwidth]{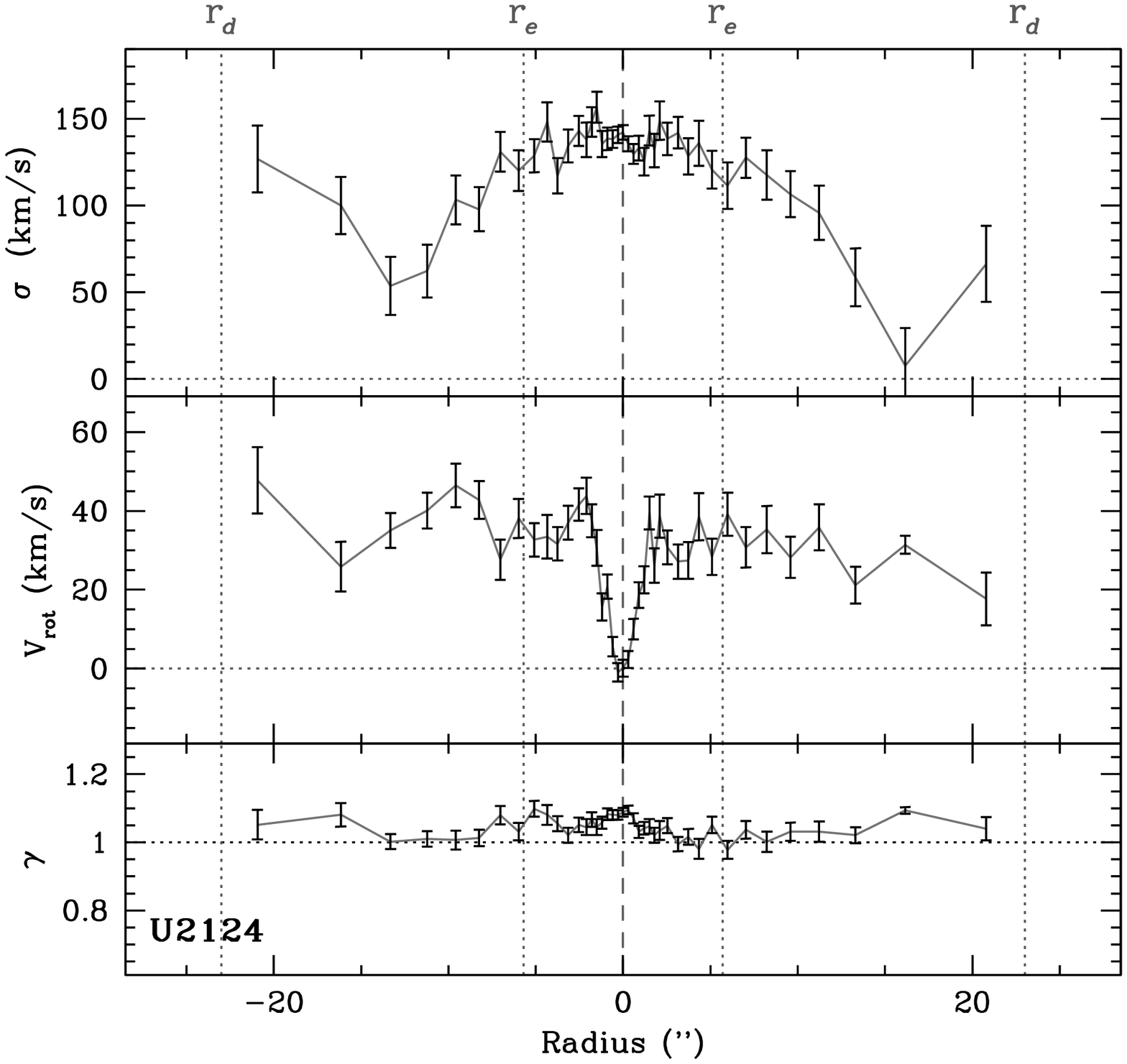}
\includegraphics[width=0.43\textwidth]{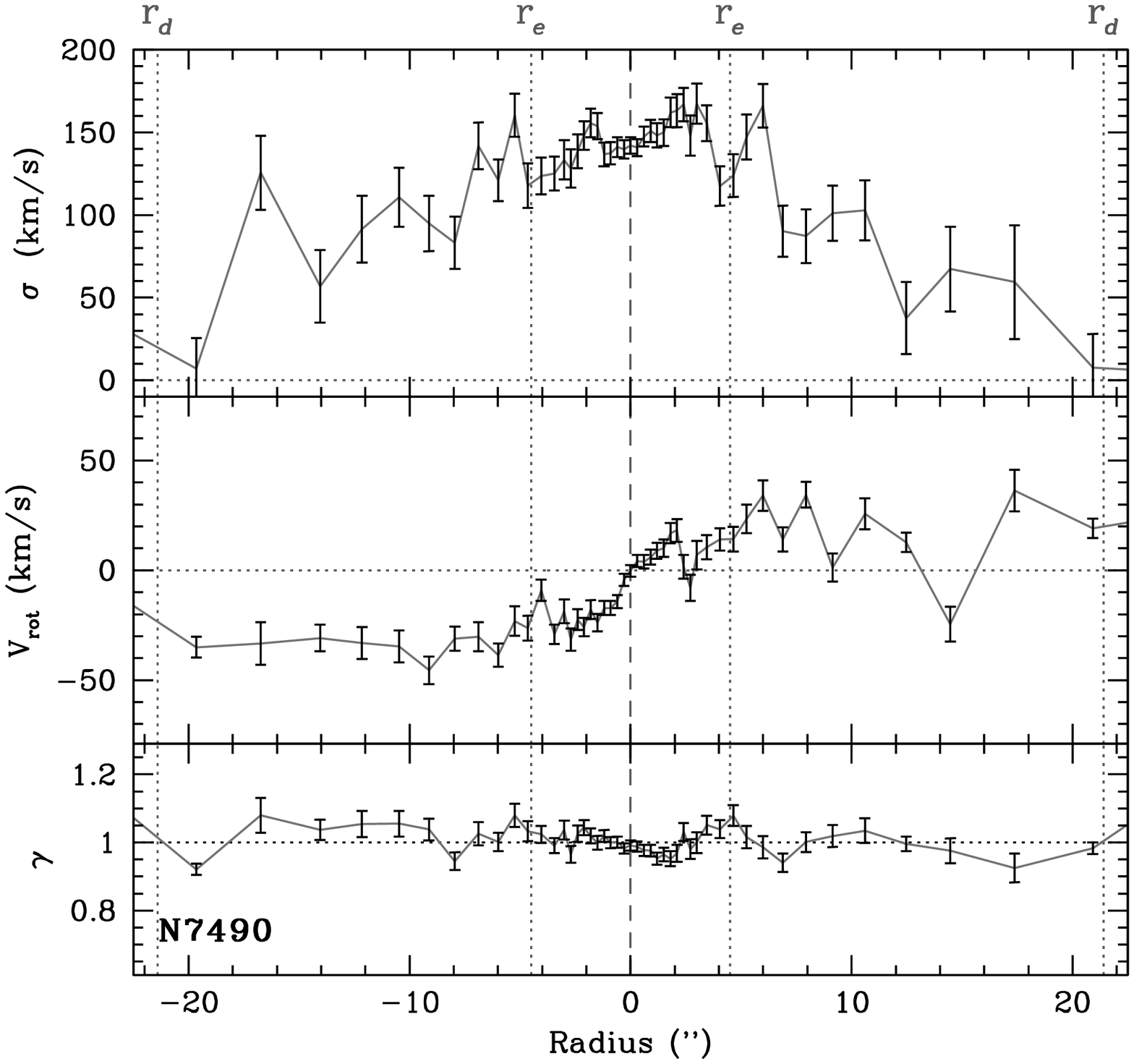}\\
\caption{Kinematic profiles for our eight galaxies.  Plotted in each panel
         are the velocity dispersion, $\sigma$, in \kms (top),
         rotational velocity, V$_{rot}$, in \kms (middle), and the
         $\gamma$ parameter (mean relative line intensity) of the
         kinematic fits (bottom) as a function of radius.  The radial
         scale extends to 5\,$r_{e}$ and note the different vertical
         scales for each galaxy.  The galaxy center is marked by the
         dashed line, while the bulge effective radius, $r_e$, and
         disk scale length, $r_d$, when reached, are marked by dotted
         lines.  For N0628, the red lines in the top panel indicate
         the velocity dispersion profile fit derived by Ganda \etal\
         (2007), showing good agreement.}
   \label{fig:kinem}
\end{center}
\end{figure*}
\begin{figure*}
\begin{center}
\includegraphics[width=0.43\textwidth]{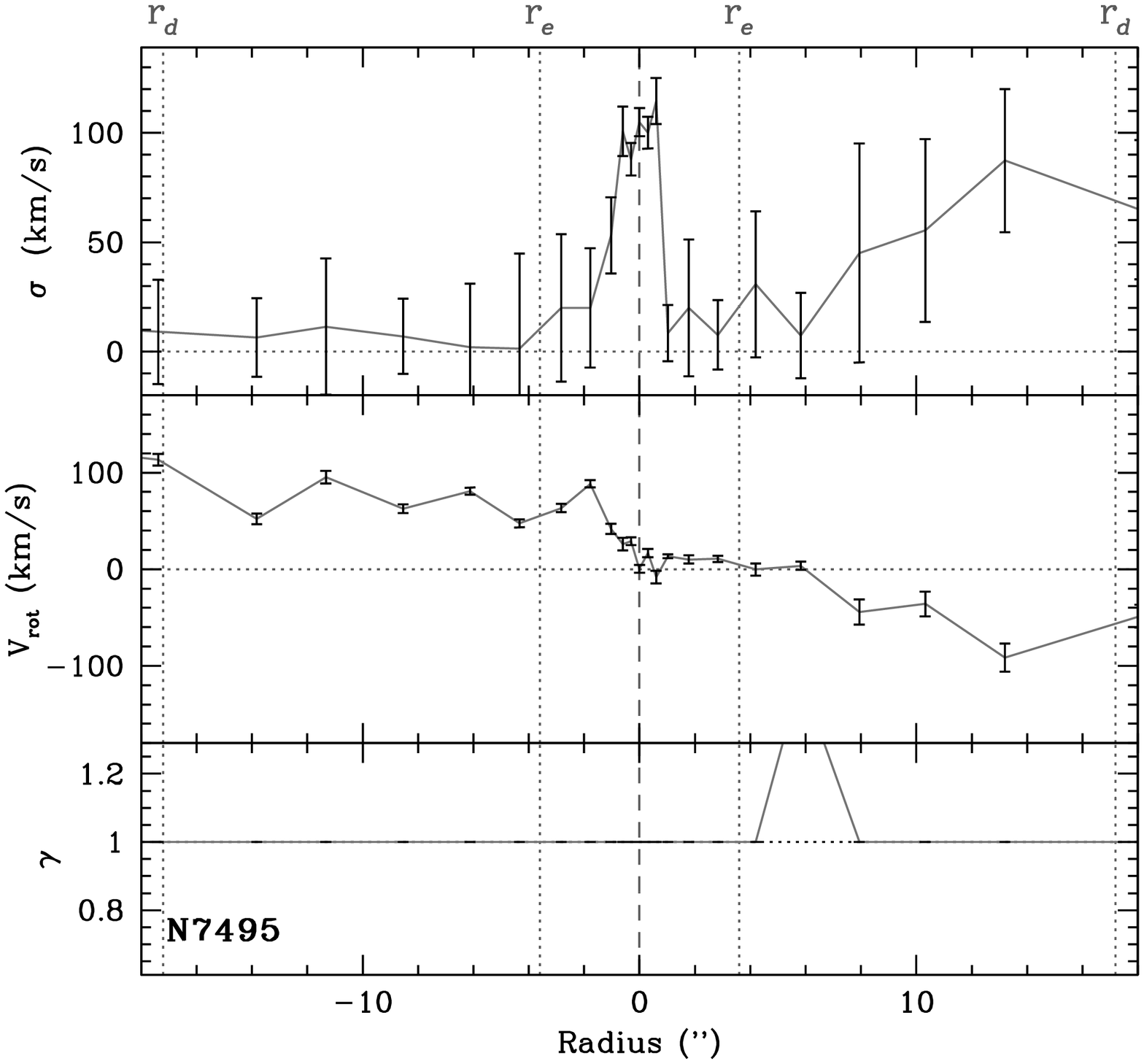}
\includegraphics[width=0.43\textwidth]{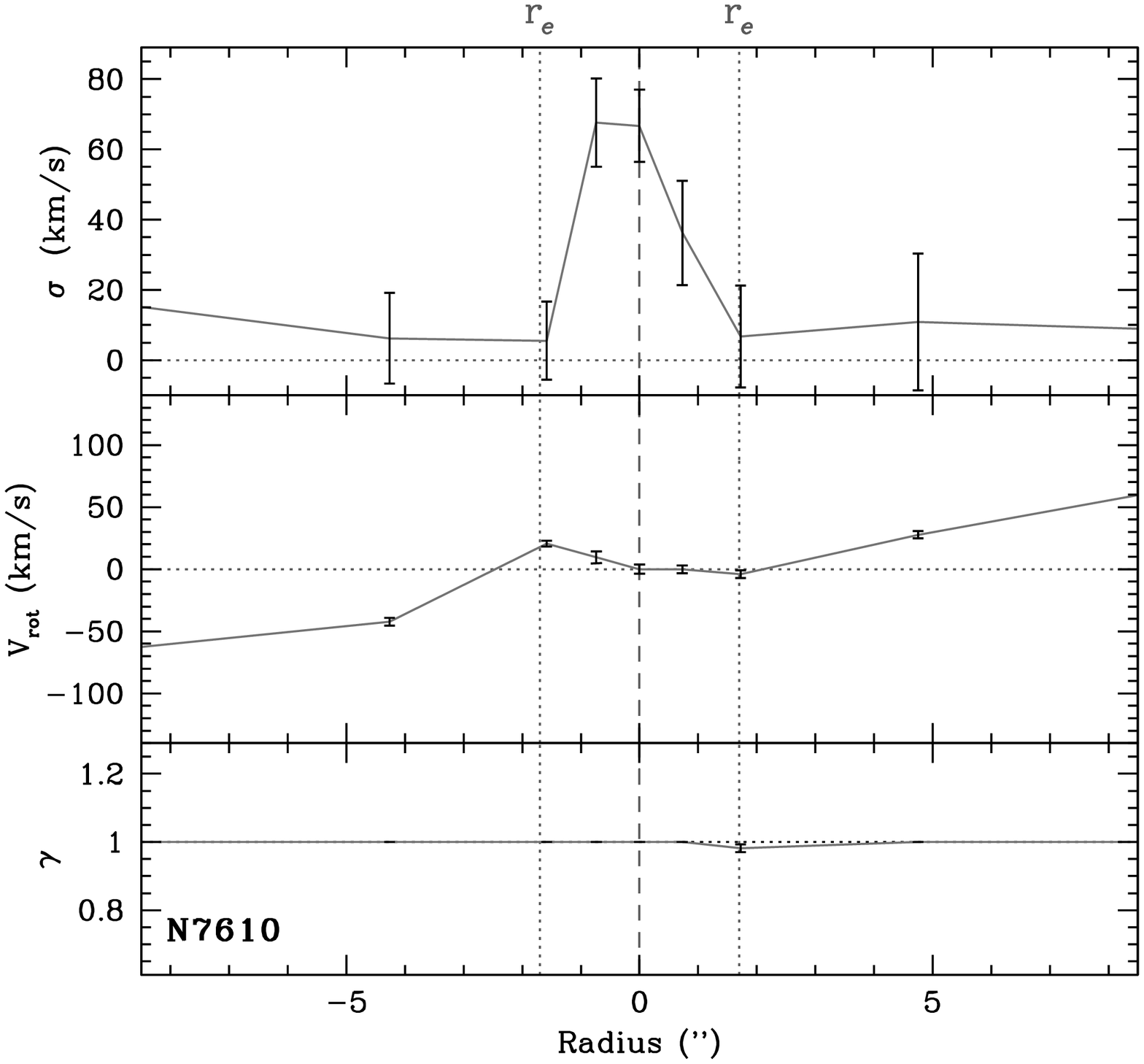}\\
\includegraphics[width=0.43\textwidth]{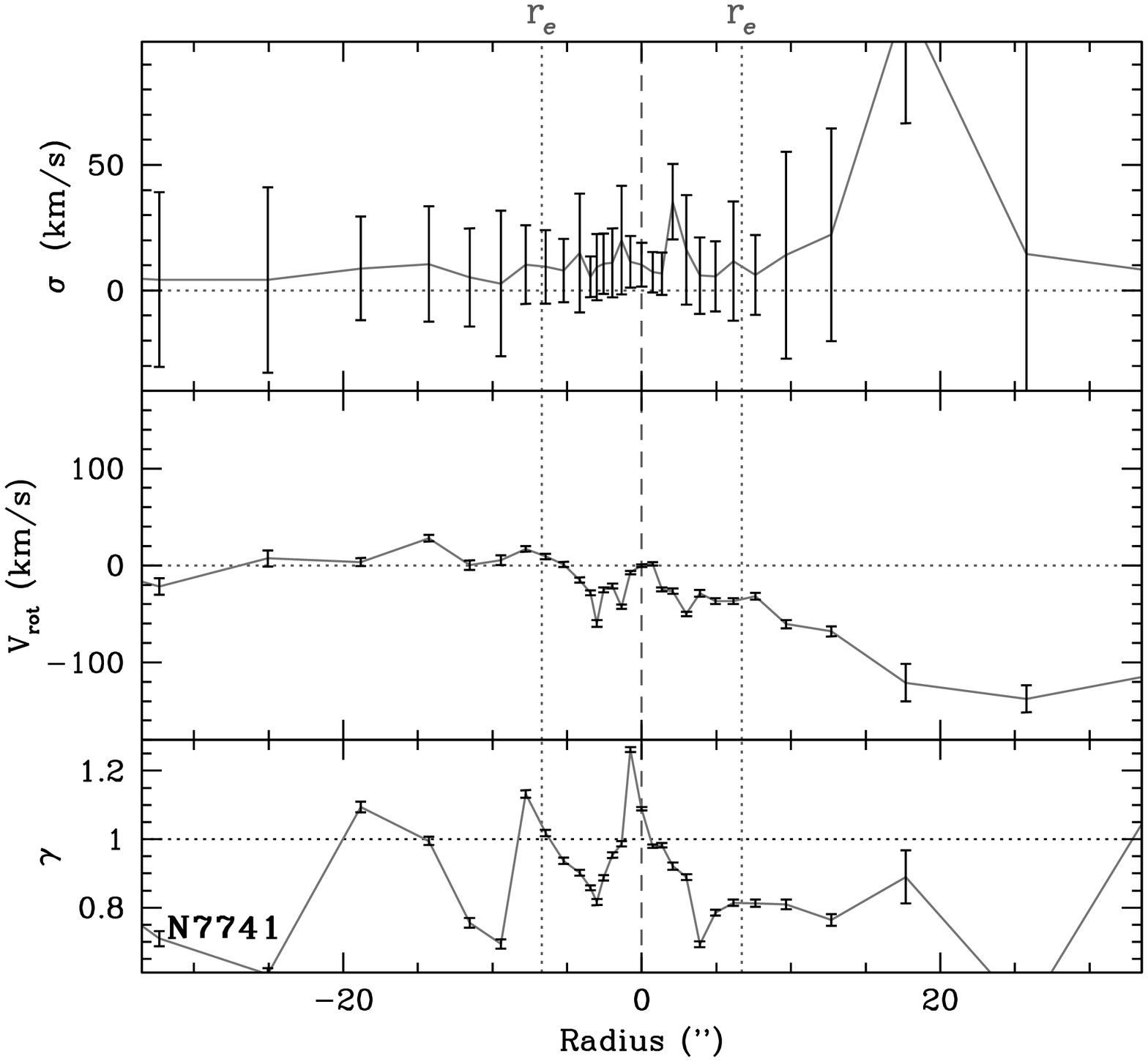}
\includegraphics[width=0.43\textwidth]{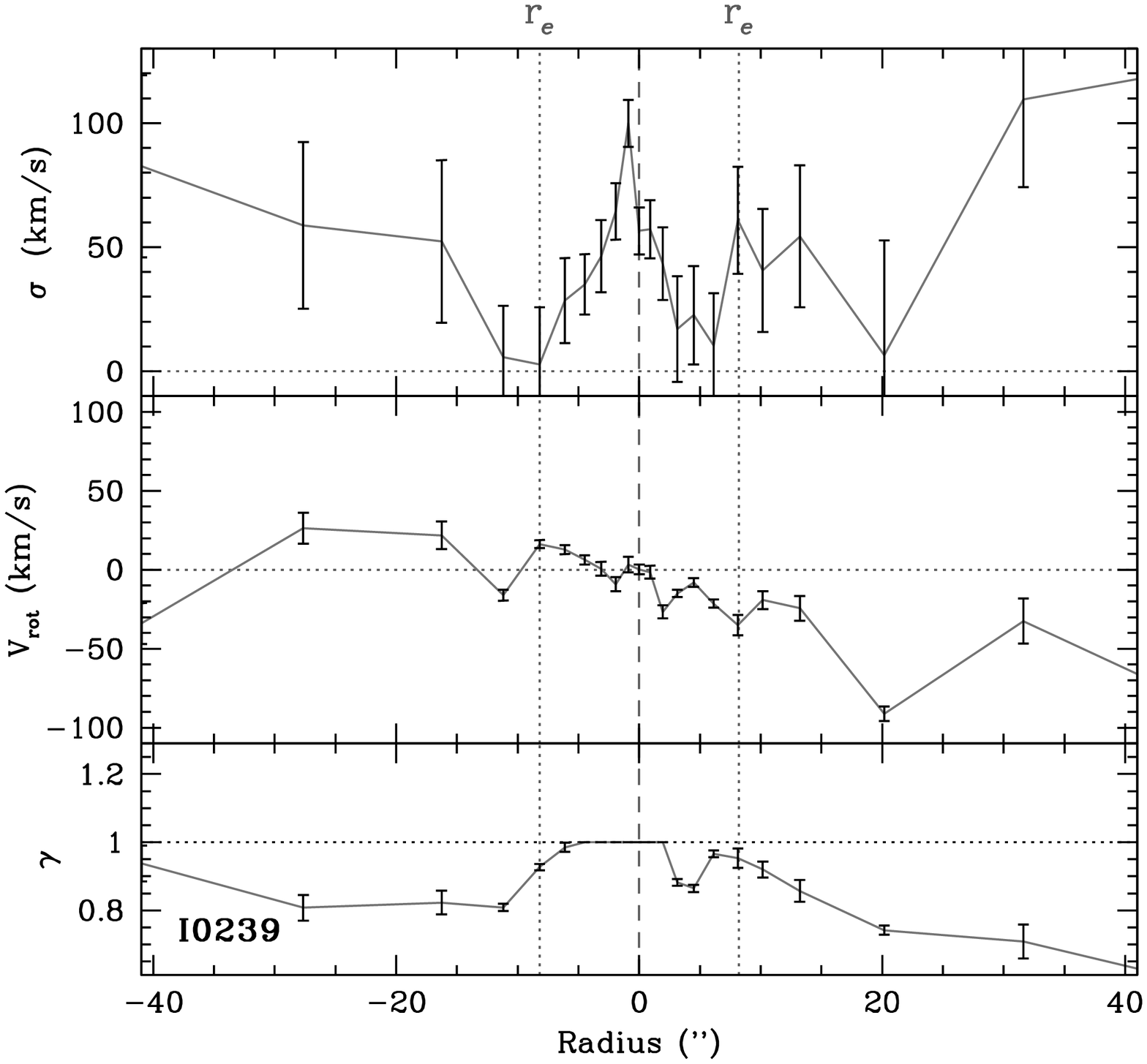}\\
    \contcaption{}
\end{center}
\end{figure*}

The kinematic profiles generally show a rapid decrease in velocity
dispersion outside the bulge effective radius and fairly well defined
rotation curves.  However, given the face-on nature of this sample,
the correction to absolute rotation speed (\ie\ $\sin(i)$) is quite
large and uncertain.  For N0628 and U2124, the odd rotation profiles
reveal a poor alignment of the slit on the nucleus of the galaxy.
Regardless, our $\sigma$ profile for N0628 is very well matched to
that of Ganda \etal\ (2007; hereafter Ganda07), indicated by the red
lines, which was derived from 2D SAURON data.

\begin{figure}
\begin{center}
\includegraphics[width=0.48\textwidth]{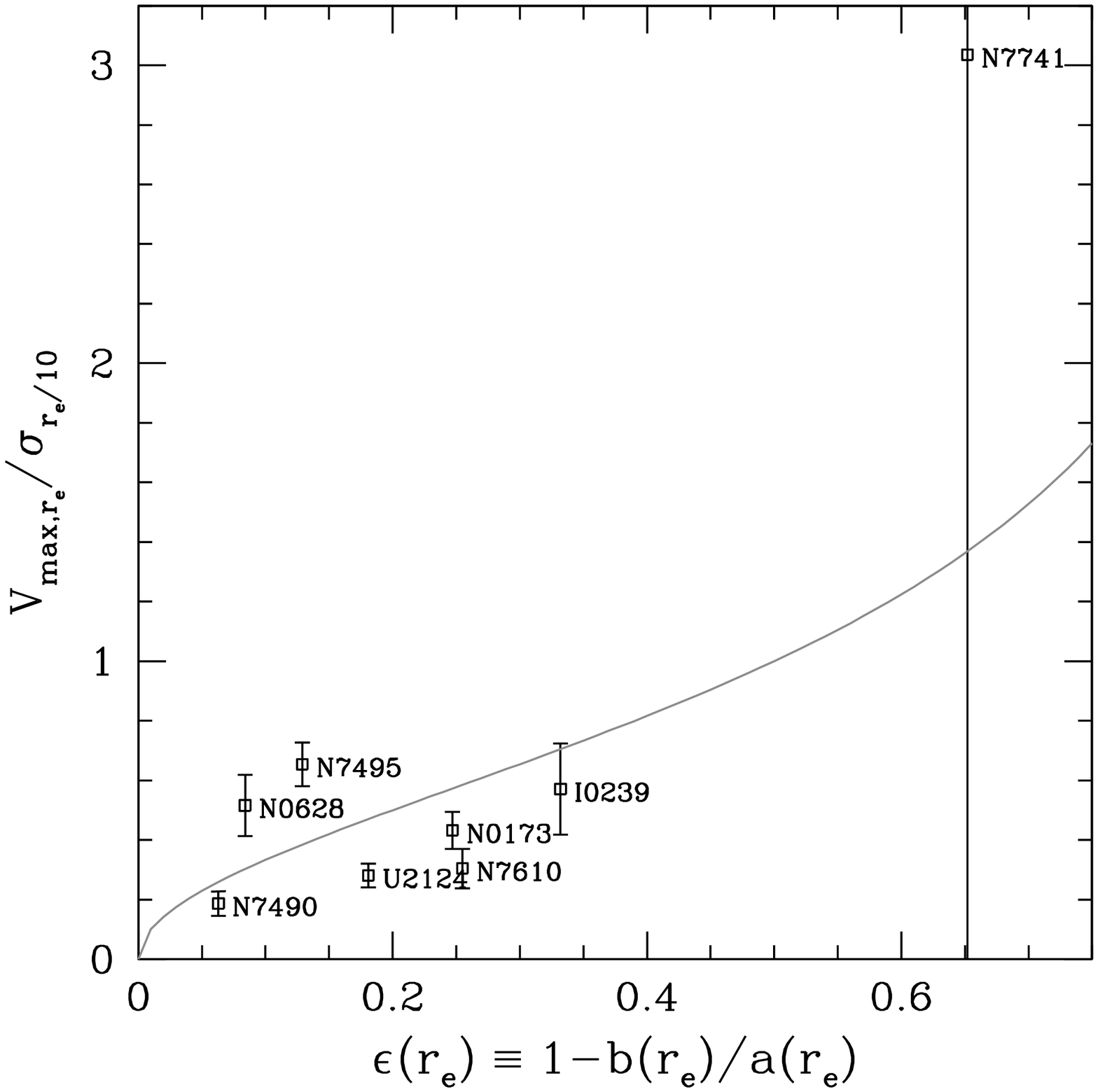}
\caption{Location of our bulges in the
         (V$_{max,r_e}/\sigma_0$, $\epsilon$) plane.  $\sigma_{re/10}$ is
         taken from Table~\ref{tab:kinem}.  V$_{max,r_e}$ and
         $\epsilon$\,$\equiv$\,$1-b/a$ were measured at the bulge effective
         radius from the kinematic profiles and isophotal SB fitting,
         respectively.  The solid curve (V$_{max}/\sigma_0$\,$\simeq$\,
         $[\epsilon/(1-\epsilon)]^{1/2}$, from Kormendy 1982) describes
         oblate-spheroidal systems with isotropic velocity dispersions
         flattened only by rotation (Binney \& Tremaine 1987).
         }
\label{fig:Vsigrat}
\end{center}
\end{figure}

The dominance of rotational versus random motion support of the bulges
of our spiral galaxies can be isolated and compared with pure
elliptical systems in the (V$_{max,r_e}/\sigma_0$, $\epsilon$) plane.
Here, V$_{max,r_e}$ is the maximum rotational velocity measured within
the effective radius of the bulge, $\sigma_0$ is the central velocity
dispersion, measured within an effective aperture of $r_e/10$
(and labeled as $\sigma_{r_{e}/10}$, see below).  The bulge ellipticity,
$\epsilon$, also measured at $r_e$, is derived from the isophotal SB
fitting (see Mac03 for details).  The location of our bulges in this
plane is shown in Fig.\@~\ref{fig:Vsigrat}.  The solid curve
represents the location of oblate spheroidal systems with isotropic
velocity dispersions that are flattened only by rotation (Binney \&
Tremaine 1987).  Three of our bulges lie above the oblate-line,
indicative of the disky behavior expected of bulges that have
undergone secular evolution, whereas a distinction between ``pseudo''
and ``classical'' for the other four bulges, which lie close to, but
below the oblate-line, is not as clear-cut.  Note that the large
uncertainties in N7741 result largely from the very low velocity
dispersion of this galaxy and its large ellipticity is due to its
dominant stellar bar.

The main kinematic parameter of interest for this work is the central
velocity dispersion.  This was measured as a light-weighted average
within a effective aperture of $r_e$/10 for direct comparison with the
elliptical galaxy sample of Thomas \etal\ (2005; hereafter Thomas05).
The average ages and metallicities within this central $r_e$/10
aperture were also computed and the values are listed in
Table~\ref{tab:kinem}.
\begin{table}
\begin{minipage}{0.48\textwidth}
\centering
\caption{Average Age, $Z$, and $\sigma$ within $r_{e}/10$} \label{tab:kinem}
\begin{tabular}{rrrrrrr}
\hline
\multicolumn{1}{c}{Name} &
\multicolumn{1}{c}{$\sigma_{r_{e/10}}$} &
\multicolumn{1}{c}{d$\sigma_{r_{e/10}}$} &
\multicolumn{1}{c}{\avgAl} &
\multicolumn{1}{c}{\avgAm} &
\multicolumn{1}{c}{\avgZl} &
\multicolumn{1}{c}{\avgZm} \\
\multicolumn{1}{c}{\scriptsize NGC} &
\multicolumn{2}{c}{(\kms)} &
\multicolumn{1}{c}{(Gyr)} &
\multicolumn{1}{c}{(Gyr)} &
\multicolumn{1}{c}{} &
\multicolumn{1}{c}{} \\
\multicolumn{1}{c}{(1)} & \multicolumn{1}{c}{(2)} & \multicolumn{1}{c}{(3)} & 
\multicolumn{1}{c}{(4)} &
\multicolumn{1}{c}{(5)} & \multicolumn{1}{c}{(6)} & \multicolumn{1}{c}{(7)}\\
\hline
 N0173 &   73.01 &   5.99 & 10.28 & 15.47 & 0.027 & 0.023 \\
 N0628 &   55.14 &   6.39 &  5.33 & 11.98 & 0.037 & 0.028 \\
 U2124 &  139.24 &   4.98 & 10.49 & 15.34 & 0.029 & 0.028 \\
 N7490 &  140.92 &   5.53 & 17.11 & 19.19 & 0.028 & 0.028 \\
 N7495 &   96.87 &   8.61 &  6.57 & 14.13 & 0.011 & 0.032 \\
 N7610 &   67.60 &  12.57 &  7.13 & 14.15 & 0.016 & 0.040 \\
 N7741 &   12.17 &  13.61 &  3.30 & 14.41 & 0.011 & 0.041 \\
 I0239 &   61.24 &  11.80 &  4.45 & 12.71 & 0.041 & 0.038 \\
\hline
\end{tabular}
\end{minipage}
{\scriptsize {\it Notes} --- Central galaxy parameters averaged within a 
radius of $r_e$/10\,$\times$\,2\arcsec\ (the slit-width).
Col.\@ (1): Galaxy ID.
Col.\@ (2): Velocity dispersion within $r_e$/10.
Col.\@ (3): Velocity dispersion error within $r_e$/10.
Col.\@ (4): Average $V$-band light-weighted age within $r_e$/10.
Col.\@ (5): Average $V$-band mass-weighted age within $r_e$/10.
Col.\@ (6): Average $V$-band light-weighted metallicity within $r_e$/10.
Col.\@ (7): Average $V$-band mass-weighted metallicity within $r_e$/10.
}
\end{table}

\subsection{Age and Metallicity vs. Central Velocity Dispersion}\label{sec:corr0}
\begin{figure}
\begin{center}
\includegraphics[width=0.48\textwidth]{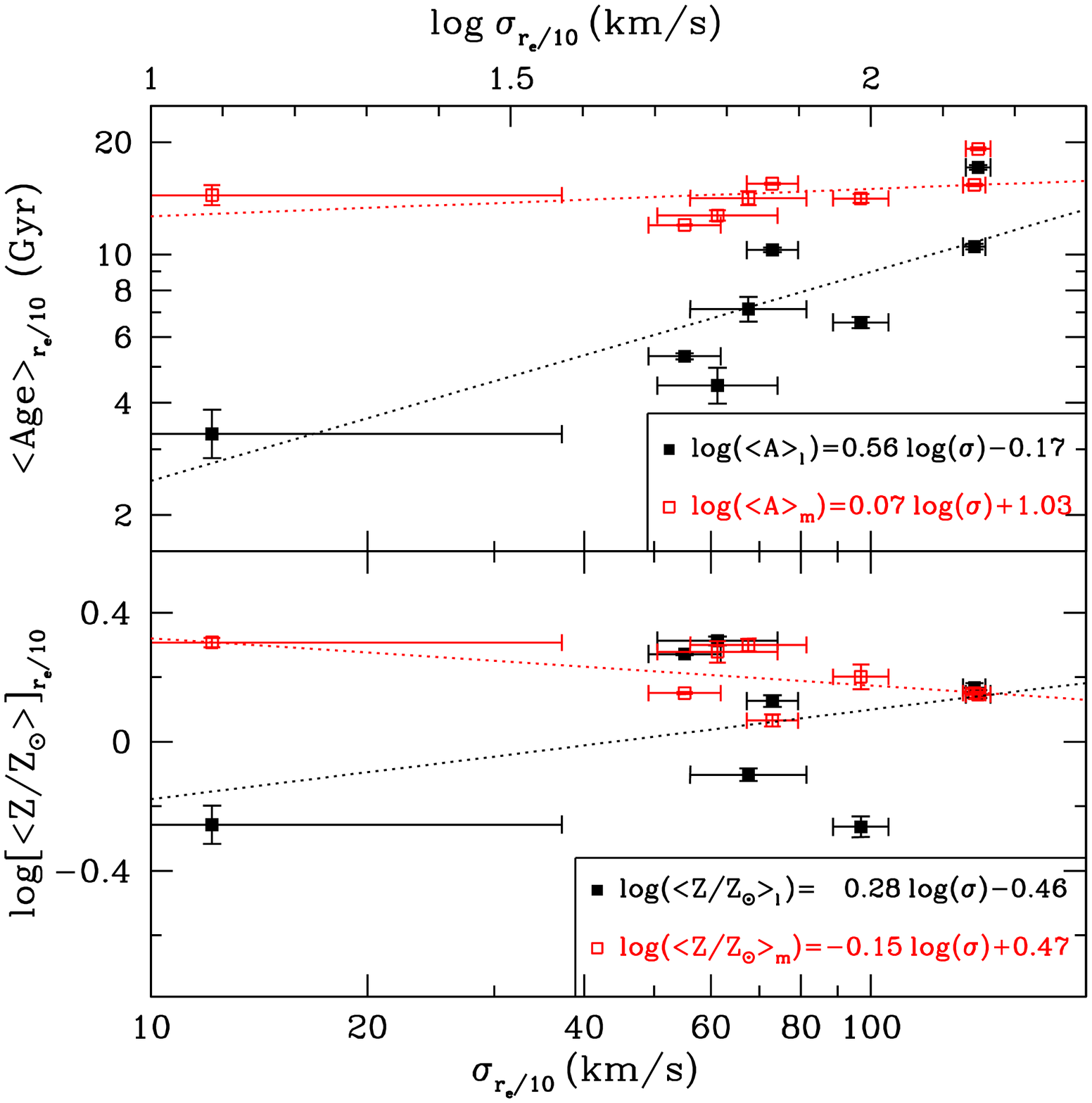}
\caption{Average age (top) and metallicity (bottom) as
         a function of the central velocity dispersion.  {\bf Black
         Solid Squares:} our data in a $r_e$/10 aperture (actually
         $r_e$/10\,$\times$\,2\arcsec\ due to the slit-width).  These
         are the light-weighted values from the popsynth fits.  {\bf
         Red Open Squares:} Same as black but for mass-weighted
         values.  The black and red dotted lines are simple linear
         regression fits to the light- and mass-weighted data
         respectively, and the fit parameters are indicated in the
         legends.}
\label{fig:AgeZsigma_lm}
\end{center}
\end{figure}

We can now assess any trends in the age and metallicity of our spiral
bulges as a function of velocity dispersion.  In
Fig.\@~\ref{fig:AgeZsigma_lm} we plot light- (solid black squares) and
mass- (red open squares) weighted average age (top panel) and $Z$
(bottom panel) for our eight bulges.  Simple linear least-squares fits to
both trends are shown in the legends.
The light-weighted values follow the expected trend of increasing
average age and metallicity with larger $\sigma_{r_{e}/10}$.  However,
when regarded from a mass-weighted point of view, both trends 
essentially disappear, indicating that the mass of all spiral bulges
is dominated by a population of very old and metal rich stars.

\subsection{Comparison With Other Studies}\label{sec:compare}
\begin{figure}
\begin{center}
\includegraphics[width=0.48\textwidth]{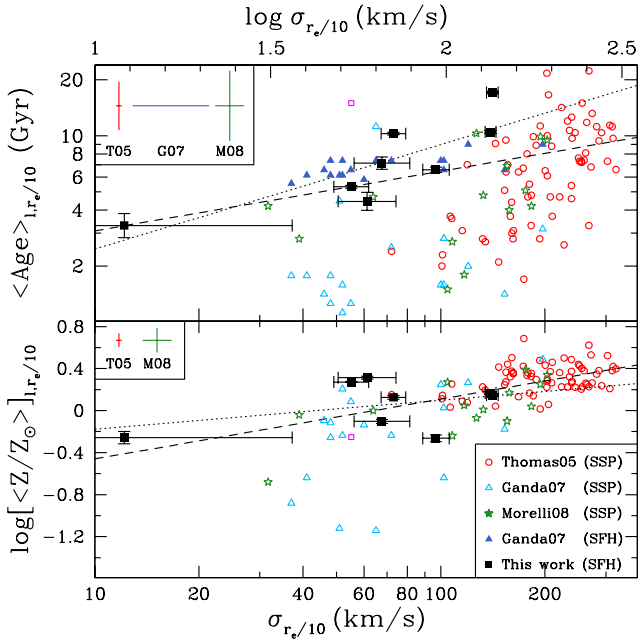}
\caption{Average light-weighted age (top) and metallicity (bottom) as
         a function of the central velocity dispersion.  {\bf Black
         Solid Squares:} our data in a $r_e$/10 aperture; these are
         the light-weighted values from the full population synthesis
         fits.  {\bf Red Open Circles:} The low-density environment
         ellipticals from Thomas \etal\ (2005) with ages measured from
         Lick indices and are light-weighted SSP values measured in
         apertures of $r_e$/10.  {\bf Light Blue Open Triangles:}
         Ganda \etal\ (2007) spirals (Sb\,--\,Sd) with ages and $Z$s
         in 1.5\arcsec\ apertures measured from Lick indices and are
         light-weighted SSP values. {\bf Dark Blue Solid Triangles:}
         Ganda \etal\ (2007) spirals where here they used the Lick
         indices and modeled them with an exponential SFH to find the
         best e-folding time, $\tau$ (see their Fig.\@~27), where we
         converted $\tau$ to \avgAl\ using Eq.\@ 11 in MacArthur
         \etal\ 2004 (they do not give the best fit $Z$s for this
         method).  {\bf Green Open Stars}: Morelli \etal\ (2008)
         cluster S0 and spirals (S0\,--\,Sbc) with ages and $Z$s
         measured from Lick indices and are light-weighted SSP values.
         {\bf Magenta Open Square:} Our results for N0628 {\it but}
         using the \avgAl\ and \avgZl\ measured at $\sim$\,1\,$r_e$
         (where the age peaks in our gradients).  This is to compare
         with the MW ``outlier'' in Thomas \& Davies (2006), Fig.\@~4.
         The deviation (from the central value; see square at same
         $\sigma$) is in the right direction in both age and $Z$, but
         not quite as strong in $Z$ to consider it a true ``outlier''.
         The black dashed lines are the best fit scaling relations for
         low-density environments derived in Thomas05 based on
         Monte-Carlo simulations (which assume linear correlations of
         the parameters age, metallicity, and $\alpha$/Fe ratio with
         log$\sigma$ and account for observational errors and
         intrinsic scatter in all 3 parameters).  The black dotted
         lines are the same as in Fig.\@~\ref{fig:AgeZsigma_lm} and
         are least-squares fits to our light-weighted values (black
         squares).  Error bars on all parameters for our data are
         shown on each point.  The legend in the top left corner of
         each panel shows the average errors on the samples from the
         literature, when provided (Ganda07 do not list errors on
         derived ages and {\it Z}s).}
\label{fig:AgeZsigma}
\end{center}
\end{figure}

Most studies of the SPs of spiral bulges and ellipticals to date have
focused on Lick-index measurements.  As outlined in the Appendix, such
measurements are subject to a number of limitations, particularly in
the case of star-forming systems.  The biggest limitation is due to
emission-line fill-in of the strongest age discriminators, the Balmer
absorption lines.  With high quality spectra, however, this limitation
can be largely overcome with a careful subtraction of the
emission-line contamination to the indices, as was done in the work of
Ganda07 based on SAURON data.  The method used is described in Sarzi
\etal\ (2006) and was also applied in the study of Morelli \etal\
(2008; hereafter Morelli08).  We thus consider the age and $Z$
estimates in the Ganda07 and Morelli08 samples to be appropriate for
direct comparison with the Lick-index-based estimates for early-type
galaxies in the sample of Thomas05. All of the above
studies provide SSP-equivalent values, and we note that these are
``light-weighted'' and not strictly ``average'' values in
the same sense as ours which implicitly measure the fractional
contribution of each SSP to the galaxy flux, normalized to the
$V$-band.  As such, small differences in the fractional light
contributions of different SPs can cause large scatter in the derived
parameters.  This effect is clearly seen in Fig.\@~\ref{fig:AgeZsigma}
where we plot the results from our study along with those from the
early-type sample of Thomas05 and the bulge samples of Ganda07 and
Morelli08.  See the figure caption for a description of the different 
data sets.  

A direct comparison of absolute ages is difficult due to differences
in the models used to derive them.  Of particular note are the extreme
ends of the age ranges used: the models of Thomas \etal\ (2003), that
are used in all three of our comparison studies, cover SSP of ages
1--15\,Gyr, where as the largest age SSP from the BC03 models is
20\,Gyr and we include those as well as 16\,Gyr SSPs in our modeling
(see Fig.\@~\ref{fig:templates}).  Thus, our age predictions will tend
to biased to larger ages whenever an SSP of age $\ga$\,13\,Gyr is
included.  At the young end, we include ages from BC03 down to
1.3\,Myr, thus our ages could be biased young whenever SSPs of age
$<$\,1\,Gyr are indicated.  This would have the overall effect of
increasing the slope of the age--$\sigma_0$ relation derived from our
methods.  The black dashed line in Fig.\@~\ref{fig:AgeZsigma} is the
best fit scaling relation for low-density environments derived in
Thomas05 based on Monte-Carlo simulations (which assume linear
correlations of the parameters age, metallicity, and $\alpha$/Fe ratio
with log\,$\sigma$ and account for observational errors and intrinsic
scatter in all 3 parameters). Indeed, if we compare the slope of the
Thomas05 fit with our own derived in Fig.\@~\ref{fig:AgeZsigma_lm} and
shown again as the black dotted line in Fig.\@~\ref{fig:AgeZsigma},
our slope is significantly steeper (we note, however, that our
statistics are very low).

Comparison of the age\,--\,$\sigma_0$ relation (upper panel in
Fig.\@~\ref{fig:AgeZsigma}), for each data set reveals a similar
general trend of increasing age with $\sigma_0$.  However, for the
early-type sample of Thomas05 (red open circles), the scatter in this
trend is much larger.  This can be easily explained in terms of the
above-mentioned difference in our quoted ages; ours are true
light-weighted {\it averages}, whereas those of Thomas05 are SSP-equivalent
values which are heavily weighted towards the most recent episode of
SF and largely insensitive to any underlying old population (even
though it may dominate the stellar mass).  The same applies for the
SSP-equivalent ages of the bulge samples of Morelli08 (green open
stars) and Ganda07 (light blue open triangles).  However, Ganda07 also
attempted to model their Lick-indices in the context of an exponential
SFH and derived the best-fit e-folding timescale, $\tau_{exp}$.  We
converted $\tau_{exp}$ into an average age using Eq.\@~11 in MacArthur
\etal\ (2004) and the results are shown in Fig.\@~\ref{fig:AgeZsigma}
as the solid blue triangles.  Not only do these show much smaller
scatter than their SSP-equivalent values, but they also follow a very
similar trend as our average ages.  Thus it seems that, when a
reasonable attempt to sample the SFH of a galaxy is made, the trend of
smaller ages with smaller $\sigma_0$ is much weaker and has
significantly less scatter.  We speculate that most of the scatter in
the Thomas05 and Morelli08 samples would disappear if they were
modeled in the context of a realistic SFH (as has been previously
suggest for pure ellipticals in the ``frosting'' models of, \eg, Serra
\& Trager 2007).

All data sets again follow the same, and expected, trend of increasing
metallicity with $\sigma_0$.  The scatter is quite small at high
velocity dispersion ($\sigma_0$\,$\ga$\,100\,\kms) but, for the
SSP-equivalent estimates becomes quite large at smaller $\sigma_0$,
again likely due to the fact that they are influenced by the stronger
recent SF.  On the other hand, our average values of $Z$ follow the
continuation of the Thomas05 slope and our derived slopes are almost
identical, despite the small overlap in the $\sigma_0$-range of our
respective samples.

The magenta square in both panels of Fig.\@~\ref{fig:AgeZsigma}
represents our results for N0628 {\it but} using the \avgAl and \avgZl
measured at $\sim$\,1\,$r_e$ (where the age peaks in our gradients,
see Fig.\@~\ref{fig:grads}).  The purpose here is to compare with the
MW ``outlier'' in Thomas \& Davies (2006; Fig.\@~4), as described in
\S\ref{sec:intro}.  The deviation (from the central value; see the
black square at the same $\sigma$) is in the right direction in age,
\ie\ if the age was determined at the bulge effective radius rather
than at the galaxy center, this galaxy would similarly stand out as an
outlier in the age\,--\,$\sigma_0$ relation.  We also see evidence of
super-solar [$\alpha$/Fe] at these radii in the \avgFe\,--\,Mg{\it b}
Lick-index plane.  The photometric (Zoccali \etal\ 2003) and
spectroscopic (Rich \& Origlia 2005; Zoccali \etal\ 2006) studies
finding old and $\alpha$-enhanced SPs of the MW bulge indeed sample a
larger physical radius, outside of the plane, which is different from
the ``central'' values measured here for external galaxies.  Thus, our
observation of a positive age gradient, along with the observations of
Peletier \etal\ (2007) of multicomponent (classical plus pseudo)
bulges, the position of the MW can be readily understood and is not a
major cause for concern.  We further discuss the reliability of this
result in \S\ref{sec:indiv}.

Finally, also of particular note here is that between all three
samples, a very large range in spheroid mass and type is covered; from
pure ellipticals, through S0 bulges, ``classical'' spiral bulges, and
to proto-typical spiral ``pseudobulges''.  However, in the context of
the SP parameters as a function of the central galaxy potential, when
a proper accounting of the spheroid's SFH is considered, there does
not seem to be a break in the trends going from pure ellipticals to
late-type bulges, nor is there a clear distinction between classical
and pseudobulges.

\section{Discussion}\label{sec:discuss}

Our study is the first to provide radially resolved spectra well into
galaxy disks, enabling a comparison of the bulge and inner disk
populations.  While the small size of our sample does not allow for a
detailed statistical study of galaxy parameters with inferred age \&
$Z$, we can still infer plausible SF mechanisms with a detailed 
look at each galaxy in the context of specific formation scenarios.
The currently favored scenarios for galaxy bulge formation 
can be summarized as follows:

\smallskip
\noindent{\bf ``CLASSICAL'' FORMATION}\\ For the context of the following
discussion, we consider ``classical'' formation to imply that the
bulges formed on a rapid timescale, as is thought to be the case for
pure ellipticals.  There are two classes that fall into this heading
as follows;

{\bf Monolithic collapse:} galaxies formed early through the
gravitational collapse of a single cloud of primordial gas (\eg\ Eggen,
Lyden-Bell, \& Sandage 1962; Larson 1974; Carlberg 1984; Thomas \etal\
1999).  The relevant predictions from this model include:\\
$\bullet$ spheroids are metal-rich in center and have a negative 
metallicity gradient\\
$\bullet$ spheroids are old and have shallow positive age gradients (centers 
slightly younger)\\
$\bullet$ stellar abundance ratios of spheroids are super-solar [$\alpha$/Fe] 
with positive gradients\\ 
$\bullet$ light profiles of spheroids have large \sersic\ $n$ values (\ie\ 
high concentration)\\
$\bullet$ disks accrete onto an already formed spheroid and are thus younger

{\bf Merging:} galaxies are gradually assembled in a hierarchical
process through multiple mergers of smaller subgalactic units.  This
process begins at early times and continues to the current epoch
(\eg\ White \& Rees 1978; Cole \etal\ 1994; Thomas \etal\ 1999;
De~Lucia \etal\ 2006).

Predictions for merger scenarios are clearly going to be much
more complicated and depend on many different factors including the
specific merger history of a given galaxy, the mass and gas content of
the infalling clumps at the time of merging, and the specific physics
describing the SF as a result of the merger and the overall shape of
the merger product.  In particular, predictions about SPs of galaxies,
particularly those of multiple-component (bulge, disk, bar), are still
in their infancy and must currently resort to semi-analytical recipes
to describe many of the complicated gastrophysical processes that
remain poorly understood and too difficult to model
explicitly.  However, certain global predictions based on
cosmologically-motivated merger trees are robust and include:\\
$\bullet$ a galaxy mass-metallicity relationship (more massive galaxies
have higher-$Z$)\\
$\bullet$ early mergers will have super-solar [$\alpha$/Fe], but more
recent merging leads to solar [$\alpha$/Fe] with flat profiles\\
$\bullet$ less massive spheroids have more extended SFHs leading to 
positive correlations of $\sigma_0$ with average age, $Z$, and [$\alpha$/Fe].

In the above two scenarios, the formed spheroid is considered a
``bulge'' if it happens to have formed a disk around it and/or if the disk
component of a progenitor galaxy survived the merger event.  Otherwise,
the spheroid is considered a pure elliptical galaxy.

\smallskip
\noindent{\bf SECULAR EVOLUTION}\\
In the secular scenario, galaxy bulges formed through a secular
redistribution of material from the disk (\eg\ Kormendy \& Kennicutt
2004, and references therein).

The predictions of SP parameters and their gradients in the context of
secular evolution are also complicated and remain in their infancy.  A
major contributor to the uncertainties involves the inclusion of gas
in the simulations leading to very different results for dissipational
versus dissipationless evolution.  Furthermore, processes that fall
under the guise of secular evolution can occur over a wide range of
timescales depending on bar presence/strength, spiral arm strength,
triaxiality of DM halo, etc.

The more tractable predictions include:\\
$\bullet$ dissipationless evolution rearranges
disk stars into a bulge component, thus both components would have the
same SP parameters \\
$\bullet$ dissipational evolution will result in fresh SF in
the center leading to bulges containing young stars.

\subsection{Notes On Individual Galaxies}\label{sec:indiv}
In the context of the above bulge formation scenarios and their
predictions for bulges properties, we now review each of our eight
galaxies individually.  In Table~\ref{tab:bulgeclass} we provide
a summary of the formation scenarios indicated by all of our
observations.

{\bf N0173:} The SP populations in the bulge region of this Sc galaxy
      reveal an old and metal-rich center with gradients to younger
      ages and smaller $Z$ within 1\,$r_e$, consistent with the
      ``classical'' monolithic picture of bulge formation.  The disk
      is somewhat younger with virtually no gradient in age, but a
      mild gradient to smaller $Z$, out to $\sim$\,2$r_d$.  The SFH in
      Fig.\@~\ref{fig:SFH} reveals roughly equal contributions in
      light (normalized to the $V$-band) of an old ($\sim$\,13\,Gyr)
      sub-solar ($Z$\,=\,0.004) population plus a young (1\,Gyr) metal
      rich ($Z$\,=\,0.05) population within $\sim$\,1\,$r_e$.  More
      recent metal-poor SF (possibly from accretion of unenriched gas)
      is evident in the outer region around 1\,$r_e$.  Throughout, the
      old population is entirely dominant in mass (right plot in
      Fig.\@~\ref{fig:SFH}).  There is evidence for slightly enhanced
      [$\alpha$/Fe] in the center (Fig.\@~\ref{fig:avgFeMgb}),
      pointing to short SF timescales.  This galaxy lies in the
      ``ambiguous'' region of the ($V_{max,r_e}/\sigma_0$, $\epsilon$)
      plane in Fig.\@~\ref{fig:Vsigrat}, but in general seems
      consistent with a classical monolithic bulge formation
      scenario.

{\bf N0628:} The prototypical grand-design Sc galaxy.  Here the inner
     bulge is quite young and metal rich, consistent with recent
     central SF, possibly due to secular funneling of gas to the
     central regions.  There is no sign of a bar component in this
     galaxy, but this does not rule out secular processes from, for
     example, a pre-existing bar or the strong spiral structure.
     The strong spiral dust lanes on the concave side of the arms are
     indicative of shocks from the pile up of gas when entering the
     spiral density wave (\eg\ Kormendy \& Kennicutt 2004).  The 
     gas loses energy at the shock front and sinks to the center resulting
     in central SF.  This process occurs on much longer timescales than 
     the action of a bar, but if the spiral structure is strong (high
     pitch angle), such a process will contribute to the building of
     a central component.

     As with N0173, we also see a light-dominant high-$Z$ 1\,Gyr
     component on top of an old stellar population.  Here, the
     young component has more weight by light and mass in the center,
     such that up to $\sim$\,40\% of the mass in the central regions
     was contributed by the SF episode that occurred 1\,Gyr ago.
     At increasing radii, we see stronger contribution from much
     younger and metal-poor SSPs.  This is indicative of accretion of
     unenriched material onto the disk.  However, the main
     contributor by mass out to $\sim$\,1$r_d$ is still from an
     old component.  One side of our slit is consistent with a disk
     age gradient toward younger ages with increasing radius,
     consistent with inside-out formation.  However, there are spikes
     to old ages and the other side of the slit is more consistently at
     old ages.  This is likely due to arm-interarm crossings, where
     the older regions coincide with an interarm region where only the
     underlying oldest SPs are present.

     Of particular note is the spike to older ages just beyond
     $\sim$\,1\,$r_e$ for NGC 628.  This is precisely the signature
     noted in \S\ref{sec:intro} required to reconcile the discrepant
     results for the MW bulge.  The fact that we observe this
     signature on both sides of the galaxy (whose spectra were fit
     independently) renders confidence that this signature is real.
     Additionally, using SAURON observations, Ganda07
     observed the same trend in age for NGC 628 using a fully
     independent approach.  In fact, this trend was first implied by
     the UV colour gradients presented in Cornett \etal\ (1994), from
     which they inferred that the SFH of this galaxy varies
     significantly as a function of radius.  Ganda \etal\ (2006)
     demonstrated that the stellar kinematics are characterized by
     slow projected rotation, and a mild central dip in the velocity
     dispersion profile, indicating a cold central region, which could
     be identified as an inner disk.  Our velocity dispersion profile
     agrees very well with theirs, as shown by the red lines in the 
     top panel of the top right plot in Fig.\@~\ref{fig:kinem}.

     As mentioned above, images reveal clear spiral dust lanes
     throughout the disk of this galaxy.  From the bottom panel in the
     top right plot in Fig.\@~\ref{fig:grads} we see that indeed, our
     fits indicate the presence of modest amounts of dust at various
     radii.

     In all, there seems to be a significant contribution from secular
     processes to the growth of this bulge.

{\bf U2124:} This is our earliest-type, strongly barred, spiral,
      SB(r)a; our slit was aligned along the bar.  Both the light- and
      mass-weighted age profiles are very similar; quite flat at old
      ages and super-solar $Z$ within 1\,$r_e$.  The flat profiles
      could be interpreted in the context of redistribution of
      material due to the bar.  However, the small weight of the
      younger SPs indicates that perhaps the gas supply is feeble such
      that only a small amount of SF is triggered by the bar action.
      There is no evidence for dust extinction in the spectra, which
      is also consistent with no recent SF and little gas content.
      There is also a clear gradient in $Z$ beyond 1\,$r_e$ which
      would be consistent with a bulge formation scenario similar to
      ellipticals (\ie\ early and short timescale of SF, also
      indicated by the super-solar [$\alpha$/Fe] in
      Fig.\@~\ref{fig:avgFeMgb}).  Unfortunately, our data do not extend
      past the bar radius ($\sim$\,21\arcsec), so we cannot examine
      the disk profile.

      The SFH indicates a dominant old population that is a mixture 
      of a 13\,Gyr $Z$\,=\,0.004 SSP and a $\sim$\,16\,Gyr solar/super-solar
      SSP.  On top of this, within $\sim$\,1\,$r_e$ is a metal-rich 1\,Gyr
      population extending all the way into the center.  This young
      population contributes $\sim$\,50\% in $V$-band light-weight, but
      only of order 5\% to the total stellar mass.
      
      In all, this bulge shows many classical formation features
      alongside pseudobulge-type structures, thus likely is composed
      of a mixture of both bulge types.

{\bf N7490:} This is a normal Sbc galaxy.  The spectra are quite red
      in the center with weak emission seen in \nii, and a general
      bluing to larger radii.  The age profile reveals a very old
      $\sim$solar $Z$ center with a strong gradient to younger ages
      out to $\sim$\,1\,$r_e$ and a weaker gradient in the disk to
      young ages out to $\sim$\,2\,$r_d$, but with a
      flat-to-increasing $Z$ within 1\,$r_e$ and decreasing in the
      disk.  These inner SP gradients are somewhat ambiguous in terms
      of their implication for the formation of the bulge, but the
      profiles are generally consistent with a merger scenario for the
      bulge with subsequent inside-out disk formation.

      The SFH reveals that the central $\sim$1/2\,$r_e$ is entirely
      dominated (in mass and light) by an old solar metallicity SSP,
      and this is the only galaxy in our sample that has almost no
      contribution from a younger SP in the very central region (to
      $\sim$\,1/2\,$r_e$). It also has the highest $\sigma_0$ of the
      sample.  This would also be consistent with formation by mergers
      at early times and/or the merger of gas-poor sub-components for
      the bulge component. This is further indicated by the
      super-solar abundance ratio in the center in
      Fig.\@~\ref{fig:avgFeMgb} as well as the strong features in the
      fit residuals in Fig.~\ref{fig:fits} similar to those shown in
      Fig.\@~\ref{fig:diffalphaFe}.  Beyond $\sim$1/2\,$r_e$, there is a
      slightly younger ($\sim$\,10\,Gyr) and more metal-poor
      population which is likely the accretion of a disk after the
      initial bulge collapse.  The SFH also reveals a SF episode
      1\,Gyr ago in a ring with a radius of about 1\,$r_e$, where we
      also see the ages dip to lower values and $Z$ to slightly
      higher. Such an episode of recent, enhanced, SF in a ring at
      1\,$r_e$ is a secular process.
      We also see from Fig.\@~\ref{fig:grads} that there is little to
      no dust in the central regions, but significant a ``spike'' at
      $\sim$\,1\,$r_d$ associated with the very young SSP contributing
      significantly to the light in this region.

      This bulge is most consistent with having formed
      ``classically'' via mergers at early times with subsequent disk
      accretion, with a minor, more recent, secular contribution in
      the form of a ring of SF.

{\bf N7495:} This SABc galaxy has strong emission throughout, being
      strongest at the very center (Fig.\@~\ref{fig:spec_rad}).  The
      current SF dominates the light, but is entirely insignificant by
      mass ($\sim$\,1--2\%), leaving an old and metal-rich population
      dominant by mass out to $\sim$\,1\,$r_d$.  There is a
      significant amount of dust out to $\sim$\,1\,$r_e$, again
      associated with the current SF.

      The age profile in the bulge is a bit erratic, jumping from old
      to young and back to old within 1\,$r_e$, while $Z$ shows a
      consistent rise in this region.  The disk profiles are both very
      flat and have significant contributions to the light of a very
      young \& metal-poor SP.  Significant amounts of dust throughout
      are evident, again in conjunction with the fresh SF.  In
      mass-weight, however, the disk is consistently very old and
      metal-rich.  This is difficult to interpret in terms of an
      inside-out scenario, but given the current level of SF throughout,
      this disk is still in formation.

      There is some evidence of recurrent episodes of SF in the very
      central regions; one contributing $\sim$30\% by mass 
      10\,Gyr ago and a less massive event ($\sim$1\%) at 0.4\,Gyr; 
      given the stochastic sampling of our SFHs, this could be 
      interpreted as a more extended SFH.

      The kinematics reveal a sharp decline in the velocity 
      dispersion from $\sim$\,100\,\kms\ at the center, to $\la$\,20 \kms\
      by 1\,$r_e$.  Its position in the $V/\sigma_0$ diagram
      (Fig.\@~\ref{fig:Vsigrat}) reveals a significant amount of 
      support against collapse from rotation, implying a disky 
      origin, whereas the bulge profile shape has our highest
      value of the \sersic\ $n$ shape parameter, pointing to a
      more classical bulge.

      Again, there seems to be contributions from both classical
      and secular components to the formation history of the bulge
      of this galaxy.

{\bf N7610:} This galaxy is our latest-type class, Scd.  There is
        significant SF in the center of this galaxy.  The bulge light
        profile is characterized by a \sersic\ $n$\,=\,0.8, highly
        suggestive of a disky origin.  The SP profiles are flat within
        the bulge region with a mild trend to younger \& more
        metal-rich inward (which is consistent with rapid E-like
        formation).  Again, the old population represents the bulk of
        the mass, but there is significant recent sub-solar SF (ages
        0.001--1\,Gyr).  Our data do not extend far enough to asses
        the disk profile, but they appear relatively flat out to
        several bulge radii.  Small levels of dust are seen coincident
        with the regions of strong current SF.

      The kinematics reveal a sharp decline in the velocity 
      dispersion from $\sim$\,70\,\kms\ at the center, to $\la$\,10 \kms\
      by 1\,$r_e$.  Its position in the $V/\sigma_0$ diagram in
      Fig.\@~\ref{fig:Vsigrat} is below the oblate line, but not
      far enough to place it unambiguously as occupying the E-type
      location of the plot.

      There is evidence for contributions from both classical and secular
      components to the formation history of the bulge of this galaxy.

{\bf N7741:} This is our only other strongly barred galaxy and our
             latest-type, SB(s)cd.  The light-weighted bulge is very
             young with sub-solar metallicity, but in mass-weight is
             old \& metal-rich.  The recent central SF indicated by
             the light-weighted values could be a result of
             in-funneling of fresh gas due to the bar.  Our slit is
             aligned perpendicular to the bar, thus no flattening of
             gradients would necessarily be expected.  The disk
	     profile is generally young and solar $Z$ with close to
	     flat profiles.

	     While there are clear secular processes contributing to the
	     central growth of this galaxy, the overwhelming dominance
	     by mass is that of a very old and metal rich whose gradients
	     are consistent with a classical picture.

{\bf I0239:} An SAB(rs)cd galaxy that also shows a significant
      population of 1\,Gyr metal-rich stars on top of an old
      metal-rich population.  In this case both contribute
      significantly to the stellar mass.  The age gradient in the
      bulge region is relatively flat out to $\sim$\,1\,$r_d$, while
      the $Z$ profile is decreasing.  Disk gradients are generally
      flat to mildly decreasing.  There is an indication of slightly
      super-solar [$\alpha$/Fe] (Fig.\@~\ref{fig:avgFeMgb}), indicative
      of rapid formation, while the \sersic\ $n$\,=\,0.9 light profile
      suggests a disky origin.
     
      Once again, this bulge displays a mix of classical and secular
      evolutionary processes.

\begin{table*}
\centering
\begin{minipage}{0.7\textwidth}
\caption{Summary of ``Secular/Pseudo'' (S) vs. ``Classical'' (C) Indicators 
for Each Galaxy} 
\label{tab:bulgeclass}
\begin{tabular}{rccccccccc}
\hline
\multicolumn{1}{c}{Name} &
\multicolumn{1}{c}{\sersic\ $n$} &
\multicolumn{1}{c}{Bar} &
\multicolumn{1}{c}{Young SP} &
\multicolumn{1}{c}{Old SP} &
\multicolumn{1}{c}{d\avgA/dr} &
\multicolumn{1}{c}{d\avgZ/dr} &
\multicolumn{1}{c}{$V/\sigma_0$} & 
\multicolumn{1}{c}{$\alpha$/Fe} &
\multicolumn{1}{c}{Overall Type} \\
\multicolumn{1}{c}{(1)} & \multicolumn{1}{c}{(2)} & \multicolumn{1}{c}{(3)} & 
\multicolumn{1}{c}{(4)} & \multicolumn{1}{c}{(5)} & 
\multicolumn{1}{c}{(6)} & \multicolumn{1}{c}{(7)} &
\multicolumn{1}{c}{(8)} & \multicolumn{1}{c}{(9)} &
\multicolumn{1}{c}{(10)} \\
\hline
 N0173 & S/C &    & S & C &  C  &  C  &  C  &  C  &  C  \\
 N0628 &  S  &    & S & C & S/C & S/C &  S  &  S  &  S  \\
 U2124 & S/C &  S & S & C & S/C & S/C &  C  &  C  &  C/S \\
 N7490 & S/C &    & S & C & C/S & S/C & S/C & S/C &  C/S \\
 N7495 &  C  &    & S & C & S/C & S/C &  S  &  S  &  S/C \\
 N7610 &  S  &    & S & C & C/S & C/S &  C  &  S  &  S/C \\
 N7741 &  S  &  S & S & C & S/C & S/C &  S  &  S  &  S \\
 I0239 &  S  &    & S & C & S/C & S/C & S/C & S/C &  S/C \\
\hline 
\end{tabular}
{\it Notes} --- 
Col.\@ (1): Galaxy ID.
Col.\@ (2): The dividing line in at $n$\,$\simeq$\,2. 
Col.\@ (3): If a bar is clearly present.
Col.\@ (4): Significant presence of recent SF (within the last $\sim$\,1--2\,Gyr).
Col.\@ (5): Significant presence of very old SP ($\ga$\,13\,Gyr).
Col.\@ (6): Indication from SP age gradient.
Col.\@ (7): Indication from SP $Z$ gradient.
Col.\@ (8): Indication from kinematics (support from rotation vs.\@ random
             motions).
Col.\@ (9): SF timescale implied by central abundance ratio (as inferred
            from the results of Fig.\@~\ref{fig:avgFeMgb}), whereby enhanced
	    $\alpha$/Fe implies a short formation timescale (as the Fe 
	elements produced by Type I supernovae have not had time to 
	enrich the gas in Fe before cessation of SF).
Col.\@ (10): Overall impression.
\end{minipage}
\end{table*}

\section{Summary}\label{sec:summary}
Using radially resolved long-slit spectra of eight star-forming spiral
galaxies, we have performed a detailed analysis of their stellar population
and kinematic profiles and provide interpretations based on currently
favored formation scenarios.  Central correlations in ages and metallicities
with velocity dispersion are compared to other studies of bulges and pure
elliptical galaxies.

The most pertinent observations are as follows:\\

\noindent{\bf Fitting Techniques:}
\begin{itemize}
\item With moderate spectral resolution, good $\lambda$ coverage, and
      high S/N/\AA\ ($\ge$50), measurement of light-weighted average ages \& 
      metallicities for star-forming galaxies is feasible.
\item Details are critical: calibration ($\lambda$ \& relative flux),
      resolution, velocity dispersion, and rotation must be treated
      self-consistently between the data and models.
\item Different fitting techniques weigh age, metallicity, and 
      abundance ratios differently:  Balmer emission limits age
      fitting from indices; age information is recovered (in the
      presence of emission) from full spectrum and continuum SED
      fitting (but compounds the caveats about dust extinction and 
      fluxing that are less important for indices).
\item SSPs are not a good match to late-type galaxies.  The degeneracies
      between age, metallicity, dust, etc., are extreme leading to 
      unstable fits.
\item Full population synthesis is the only method to provide reliable
      and consistent results, and is thus the method of choice for the
      establishment of light and/or mass-weighted average ages and
      metallicities of late-type galaxies.
\end{itemize}

\noindent{\bf Stellar Populations of Bulges and Inner Disks:}
\begin{itemize}
\item In a mass-weighted context, all bulges are predominantly 
composed of old and metal rich SPs ($\ga$\,80\% by mass). 
\item Some contribution to bulge growth by secular evolution is
clearly evident in most late-type bulges, with SPs of 0.001--1\,Gyr
contributing as much as 70\% to the optical light. The corresponding
contribution to the bulge mass, however, is generally small
($\sim$\,20\% by mass or less).
\item Spiral bulges display a wide variety of age and metallicity
gradients (from negative to positive) in the bulge region, allowing
for a range in formation mechanisms.  
\item The observation of positive age gradients within the effective
radius of some late-type bulges helps reconcile the long-standing
discrepancy of the secular-like kinematics and light profile shape
(including the presence of a bar) with ``classical''-like old and
$\alpha$-element enhanced SPs observed in the Milky Way bulge as being
due to SP sampling at different physical locations in the bulges.  
\item Spiral disks show mildly decreasing to flat profiles in
both age and metallicity, generally consistent with inside-out
formation.
\item Bulges follow the correlations of increasing light-weighted age
and $Z$ with central velocity dispersion as for elliptical galaxies
and early-type bulges found in other studies, but when a SFH more
complex and realistic than an SSP-equivalent is taken into account,
the trend is shallower and its scatter is much reduced.
\end{itemize}

The implication seems to be that bulge formation has been dominated by
processes that are common to all spheroids, whether or not they
currently reside in disks.  Monolithic collapse cannot be ruled out in
all cases, but merging must be invoked for most of our spiral bulges.
The process of formation occurs on shorter timescales for spheroids
with the highest central velocity dispersions, and the relative
contribution to the stellar mass budget in bulges via secular
processes or ``rejuvenated'' star formation is small, but generally
increases in weight with decreasing $\sigma_0$.

Ultimately, we desire a large enough sample to solidify the conclusions
inferred in this analysis and to assess any trends with galaxy 
parameters, but already these results provide important clues for
bulge formation scenarios and restrictions for future implementations
of galaxy formation models.

\section*{Acknowledgments}
We wish to thank Roberto Cid~Fernandes, St{\' e}phane Charlot,
Claudia Maraston, Richard Ellis, and Tommaso Treu for 
stimulating discussions.  Thanks also to the anonymous referee for
useful comments that led to valuable improvements to the paper.
We also owe a huge debt of gratitude to the Gemini staff, St{\'
e}phanie C{\^o}t{\' e}, Inger J{\o}rgensen, and Jean-Ren{\' e} Roy in
particular, for their instrumental contribution to the preparation and
execution of our queue-mode observations. LAM acknowledges financial
support from the National Science and Engineering Council of Canada
(NSERC).  SC acknowledges financial support through a Discover Grant
from the NSERC.  This research has made use of the NASA/IPAC
Extragalactic Database (NED) which is operated by the Jet Propulsion
Laboratory, California Institute of Technology, under contract with
the National Aeronautics and Space Administration.

\appendix
\section{1D Spectra}\label{sec:spectra}

In Fig.\@~\ref{fig:spec_rad} we present the fully calibrated and
velocity (redshift + rotation) subtracted spectra at a number of radii
for each galaxy.  The spectra are plotted on a logarithmic scale and
adjusted with an arbitrary constant to offset the spectra from each
other.  Only one side of the slit is shown, and not all radial bins are
plotted for the well sampled galaxies, but the $r$\,=\,0 and last two
bins are always plotted.  The spectra were coadded (radially) to a
minimum S/N/\AA\ $\ge$\,50 measured in the 5050--5450\,\AA\
wavelength interval (selected to avoid prominent emission lines such
as \oiiill, H$\alpha$\,$\lambda$6563,
\nii\,$\lambda\lambda$6583,6548, and the \oi\ sky line at
$\lambda$5577).  The minimum S/N/\AA\ of 50 was selected as a
compromise between accurate age and metallicity determinations, and
sufficient radial extent for gradient measurements.  The
light-weighted radius and the spectrum's corresponding S/N/\AA\ are
labeled at the right edge of each panel (note that the
$r$\,=\,0\arcsec\ bin is in reality closer to $r$\,=\,0\farcs04 due to
our binning of 0\farcs2/pixel and assuming a Gaussian light profile in
the galaxy center).  Many of the Lick indices (see
\S\ref{sec:AZindfits}) are marked as vertical shaded lines -- dark
gray delineates the central passband and light gray the
pseudo-continua.

A number of pertinent observations can be made from a visual examination of 
the spectra:
\begin{itemize}
\item Improper subtraction of prominent sky emission lines (falling in
and around the NaD and TiO indices) is evident at the largest radial
bins for all galaxies.

\item There is a significant overall bluing of the SED with radius
for most galaxies.  The two notable exceptions are N7741 \& I0239.
The former shows significant emission in \halpha, \hbeta, \& \hgammaA\
(the latter two appearing as fill-in to the underlying absorption
line), and at \nii\,$\lambda\lambda$6583,6548 and \oiiil\ in the
central spectrum, which could mask an underlying redder stellar
SED.  N7495 \& N7610 also show significant central emission and bluer
SEDs compared to those galaxies with less obvious central emission
(\eg\ N0628 \& U2124), but a significant (further) radial bluing is
still evident.

\item For some galaxies with significant emission in the H$\alpha$+\nii\
region, an emission spike superimposed on an underlying absorption can
be seen in the other Balmer lines (most noticeably in H$\beta$, but a
spike can be seen even in H$\gamma$ \& H$\delta$ in extreme cases,
\eg\ N7495 \& N7741).  Given that the Balmer lines are the most
sensitive age indicators in the Lick index system, it is likely that
index-based age estimates of the underlying stellar population will be
difficult to constrain in these galaxies.  In the full-spectrum fits,
these emission lines will need to be masked out as emission lines are
not included in the BC03 models.

The most likely source of the emission is due to normal, but intense,
current star formation.  We can rule out significant contributions
from active galactic nuclei (AGN) as the source of emission based on
the low \nii\,$\lambda$6583/\halpha\ and \oiii\,$\lambda$5007/\hbeta\
emission line ratios (see, \eg, Hao \etal\ 2005), but small
contributions in our most severe cases, N7495, N7610, \& especially
N7741, cannot be entirely ruled out (but these would be restricted to
Seyfert 2 or LINER AGN as the broad lines characteristic of Seyfert 1
spectra are not seen).

\item Clear radial variations in many Balmer and metal-line absorption
line strengths can be seen in many of our galaxies.  The interpretation
of these gradients in terms of age and metallicity require an
implementation of one of the fitting techniques described below in 
\S\ref{sec:AZfits}.
\end{itemize}

\begin{figure*}
  \begin{center}
   \includegraphics[width=0.99\textwidth,bb= 18 374 592 718]{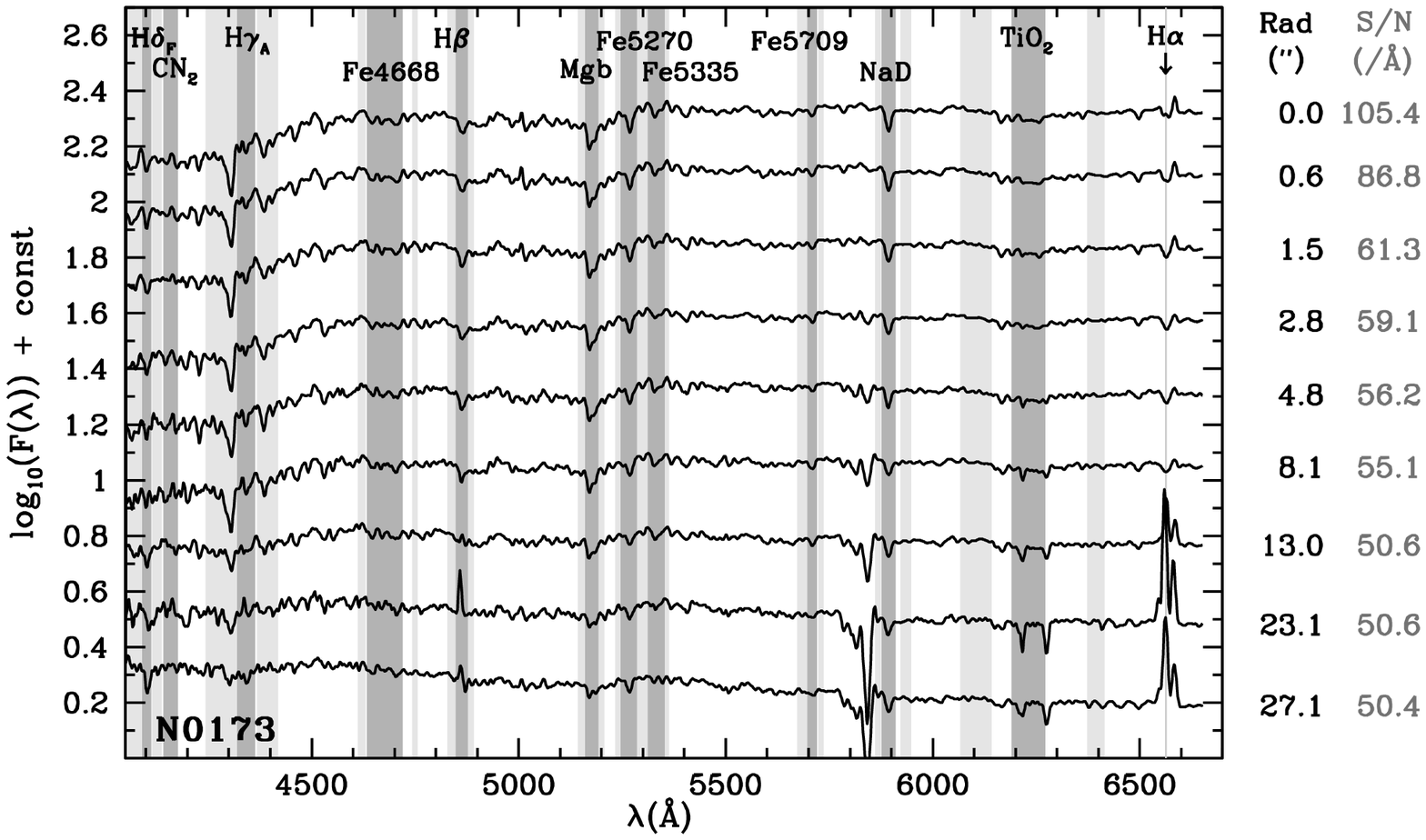}
   \includegraphics[width=0.99\textwidth,bb= 18 374 592 718]{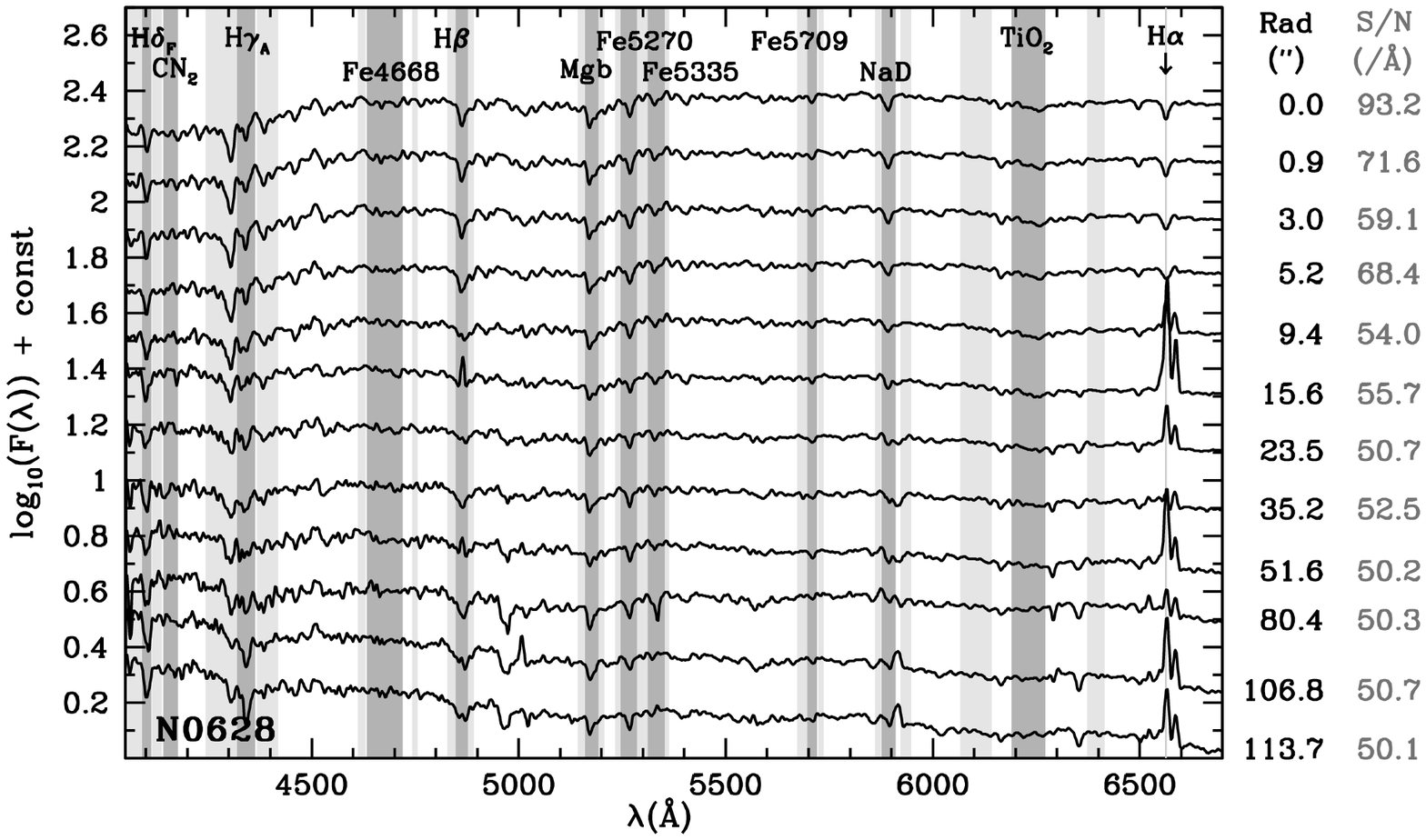}
    \caption{Spectra as a function of radius for all galaxies.  The
           spectra were coadded (radially) to a minimum S/N/\AA\ of
           50.  Not all radial bins are shown, but the first bin and
           last two bins are always plotted.  The light-weighted
           radius and the spectrum's corresponding S/N are labeled at
           the edge of the plot.  Many of the Lick indices are marked
           as vertical shading -- darker shading delineates the central
           passband and lighter shading the pseudo-continua.  The
           location of \halpha, often seen in emission, is also
           indicated.}
     \label{fig:spec_rad}
  \end{center}
\end{figure*}
\begin{figure*}
  \begin{center}
   \includegraphics[width=0.99\textwidth,bb= 18 374 592 718]{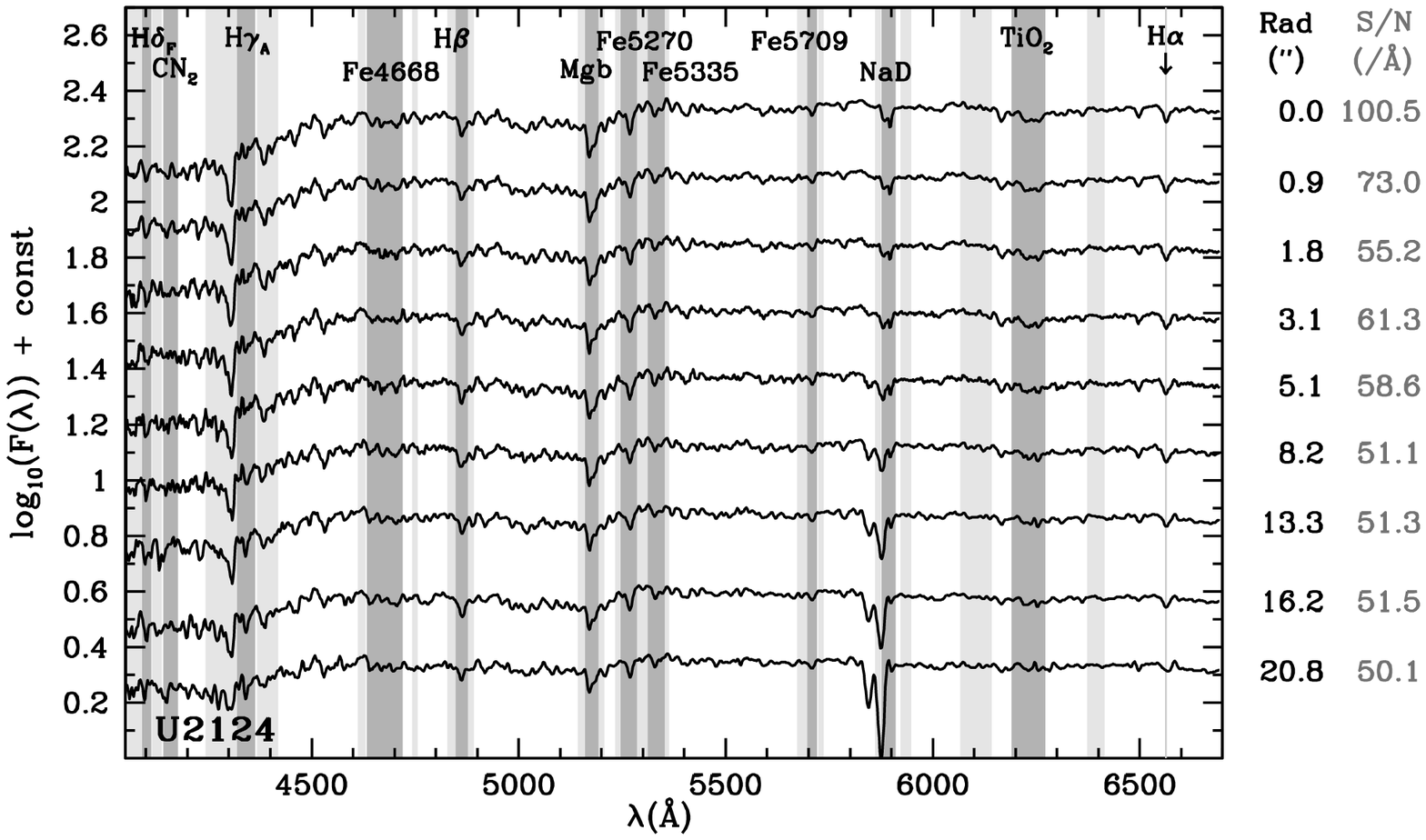}
   \includegraphics[width=0.99\textwidth,bb= 18 374 592 718]{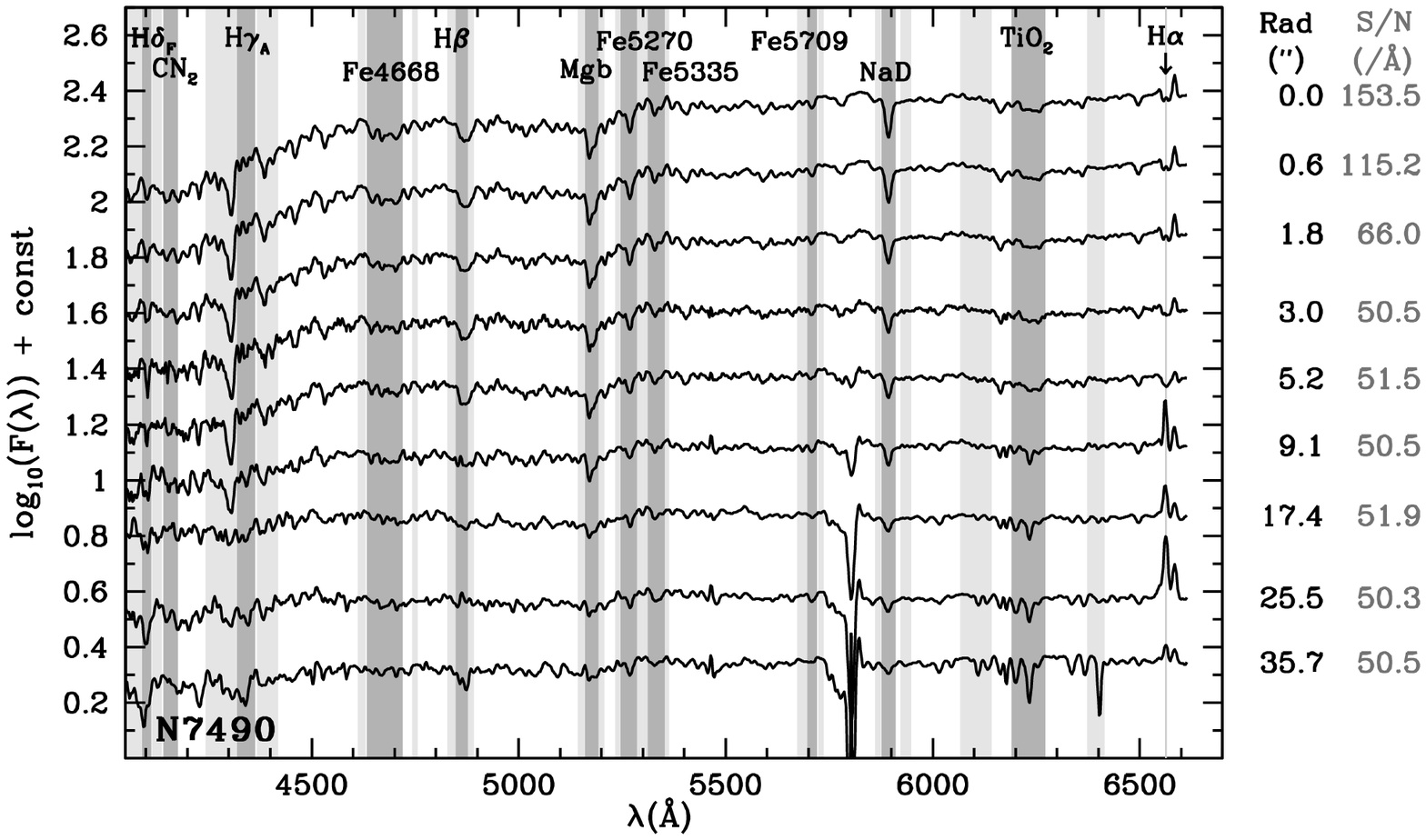}
    \contcaption{}
  \end{center}
\end{figure*}
\begin{figure*}
  \begin{center}
   \includegraphics[width=0.99\textwidth,bb= 18 374 592 718]{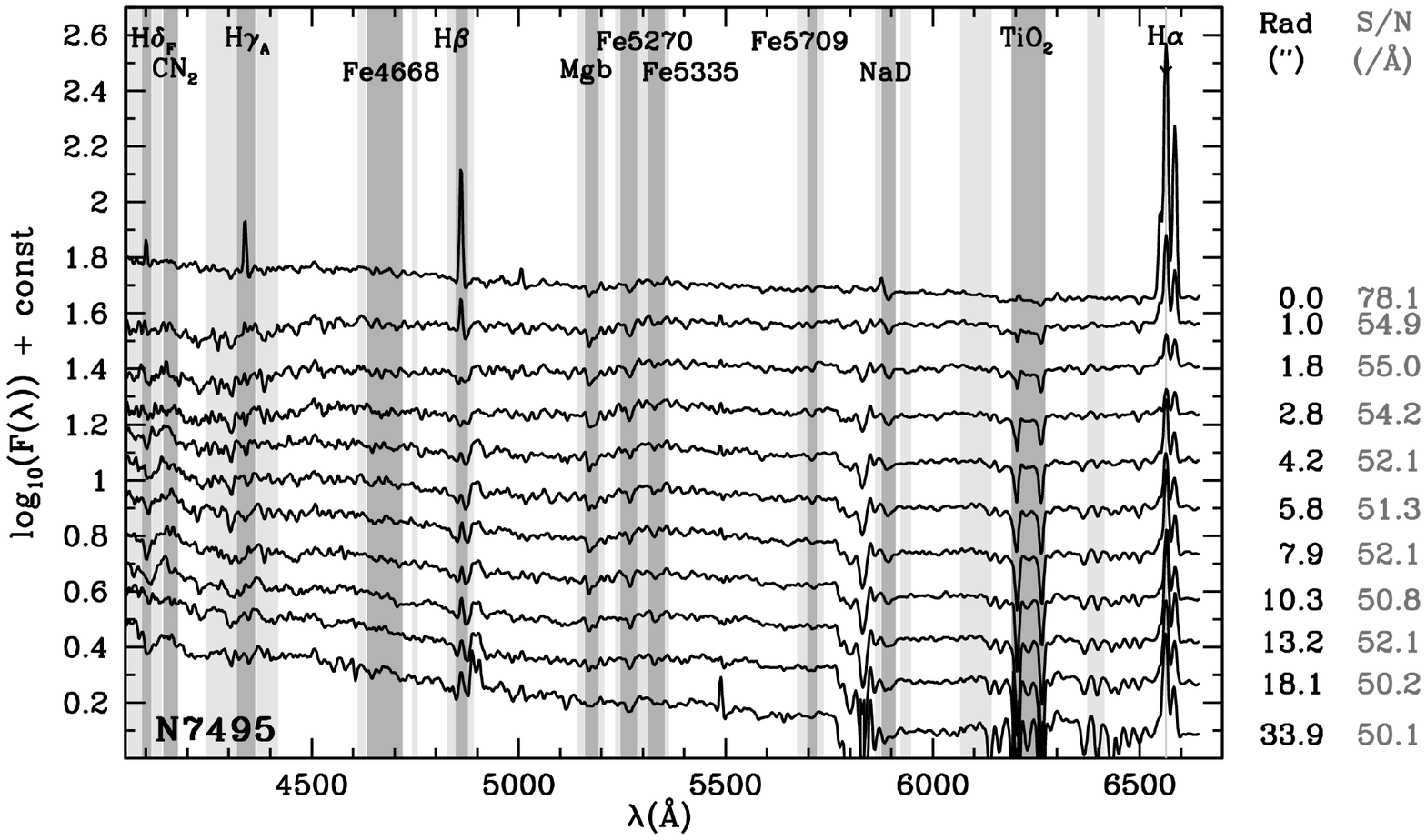}
   \includegraphics[width=0.99\textwidth,bb= 18 374 592 718]{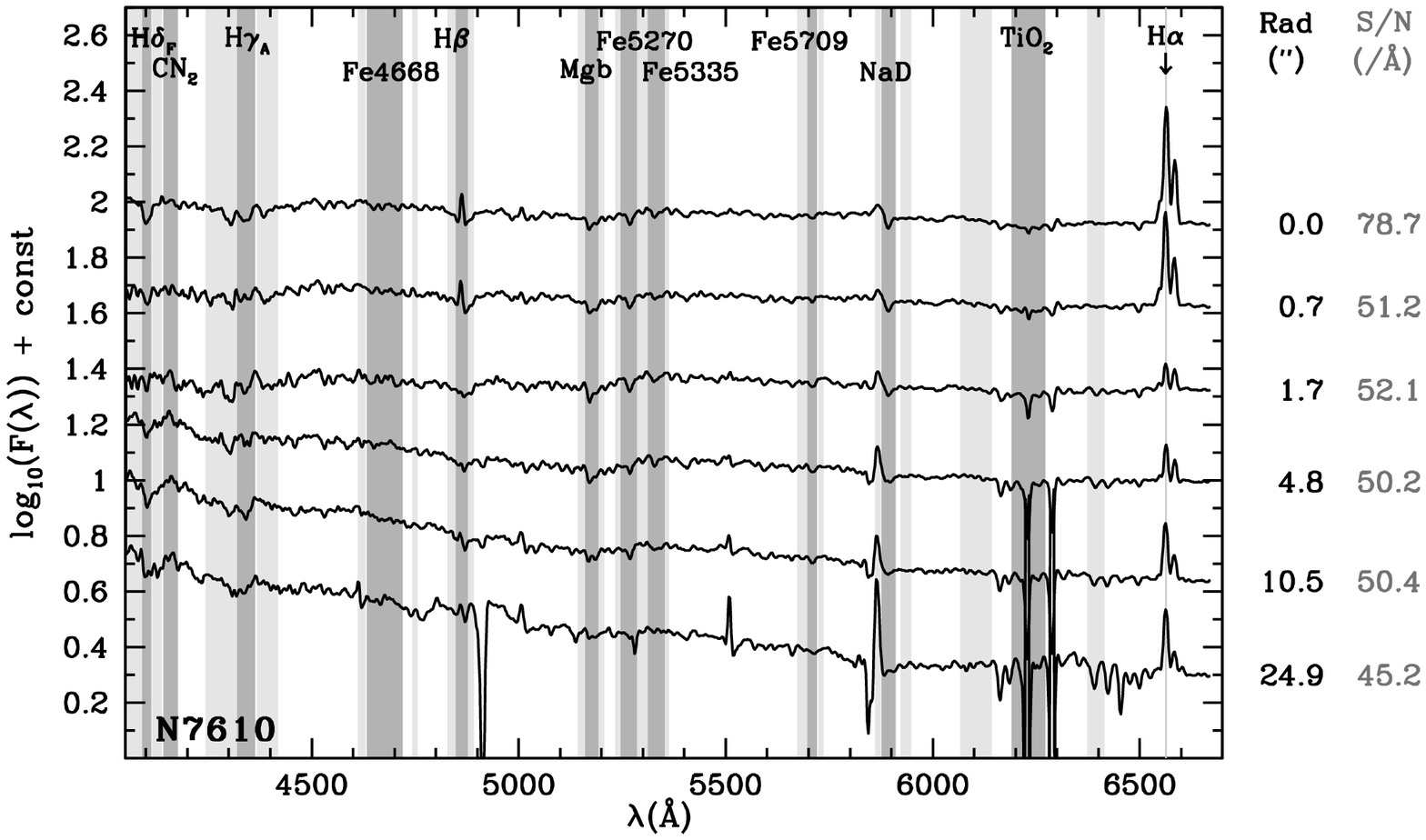}
    \contcaption{}
  \end{center}
\end{figure*}
\begin{figure*}
  \begin{center}
   \includegraphics[width=0.99\textwidth,bb= 18 374 592 718]{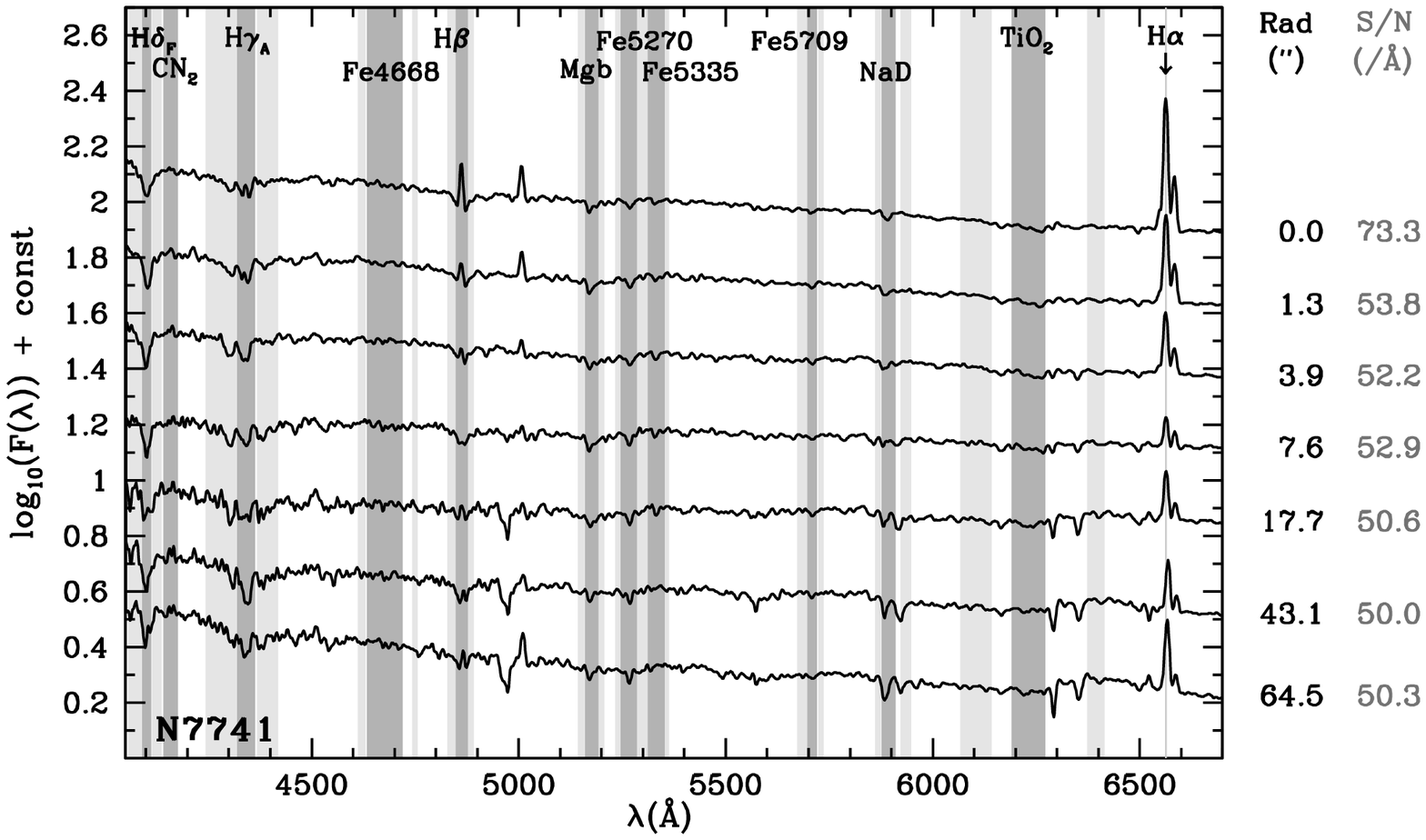}
   \includegraphics[width=0.99\textwidth,bb= 18 374 592 718]{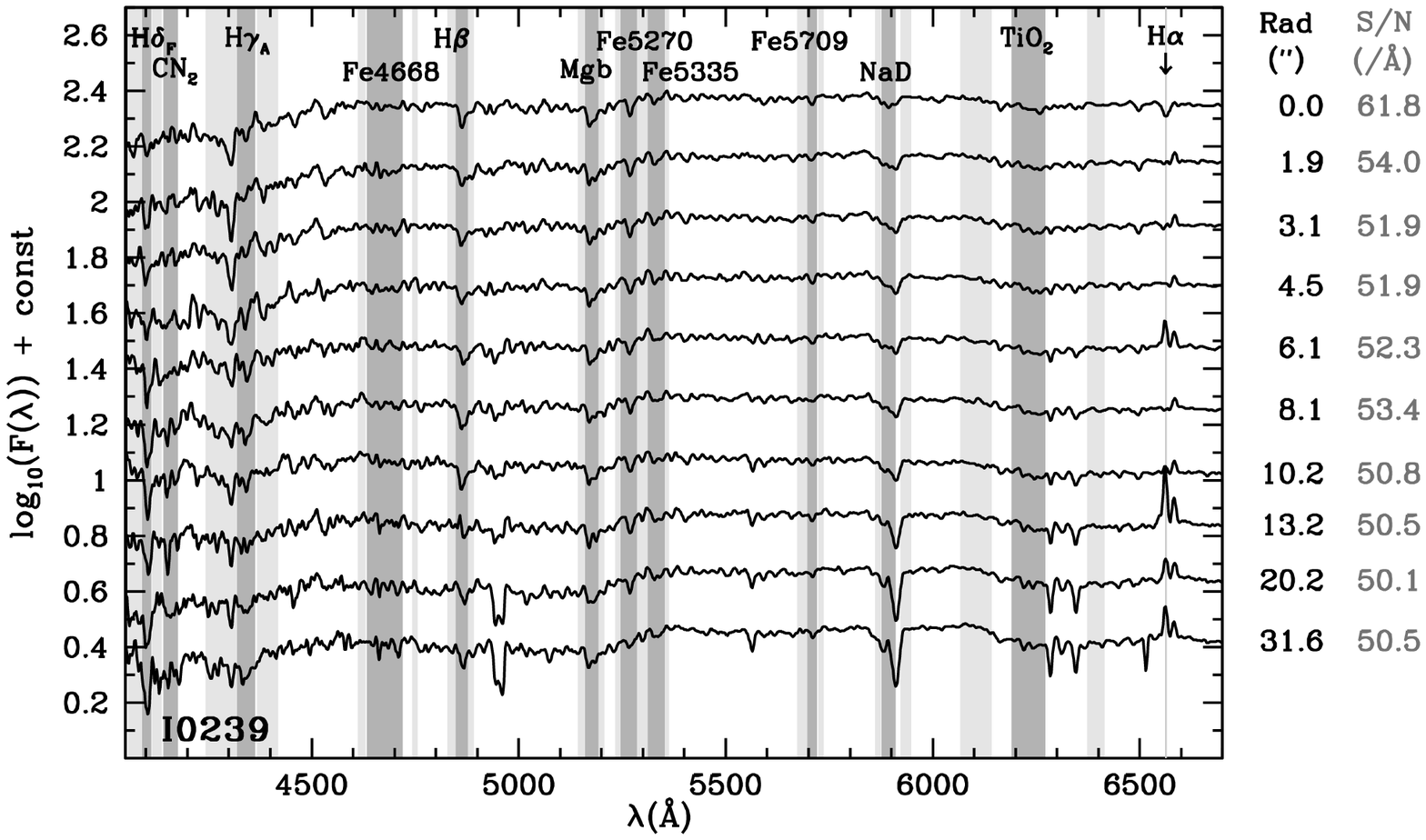}
    \contcaption{}
  \end{center}
\end{figure*}

\section{Age and Metallicity Fits}\label{sec:AZfits}

Several methods exist to assign an age
and metallicity to an observed integrated galaxy spectrum.  Each
technique has its own advantages and drawbacks depending on the
application.  One of the fundamental challenges any successful method
must overcome is the inherent degeneracy in stellar populations
between old age, high metallicity, and dust content, all leading to
redder SEDs.  Further difficulties can come from non-stellar sources
that are not included in the SP models, such as emission lines from
\hii\ regions, and contributions from a central AGN.  Below we
consider 3 methods for deriving the SP parameters from
our long-slit spectra of star-forming spiral galaxies: (i) the
Lick/IDS-index system; (ii) SSP fits to the entire spectrum; and (iii)
full synthesis fits to the entire spectrum.  The latter two also
include a prescription for dust extinction.  In the following, after
discussing the importance of understanding the resolution of both the
data and models when making comparisons, we describe the first two
methods individually and then compare their results with each other
and with those obtained from the full population synthesis fits in \S3.

\subsection{Resolution and Velocity Dispersion Effects}\label{sec:resolution}

In order to compare the data to the models, potential resolution
effects must be accounted for.  In the case where the models have
higher resolution than the data, this can be accomplished by degrading
the models to the resolution of the observations.  Alternatively, when
measuring absorption-line pseudo-equivalent widths, the indices could
be corrected for resolution effects using a calibration based on a
systematic broadening of templates of similar spectral type as the
data (\eg\ PS02).  The BC03 models used for this analysis have a
reported resolution FWHM$_{Gauss}$\,=\,3\,\AA\ that is constant with
$\lambda$.  However, as mentioned above in \S\ref{sec:BC03SSPs}, due
to an error in accounting for the relative velocities of the stars
before combining them into SSPs, the effective resolution is not only
a bit higher, but is best characterized by a constant FWHM with
$\lambda$ term plus a velocity (constant with $\Delta
\lambda/\lambda$) term.  The precise characterization will also be
slightly different for each model SSP, depending on the relative
velocity shifts and weights of the stars entering each SSP.  On
average, we find that the FWHM is closer to 4\,\AA, so it is still
much smaller than the 10.8\,\AA\ resolution of our data, which is
dominated by the slit width.  Thus, to account for resolution effects
due to the slit, we convolve all SSP models with a boxcar of width
10.8\,\AA.

Another broadening effect, independent of the instrumental setup,
that must be considered prior to any direct model/data comparison is
the galaxy velocity dispersion, $\sigma_{vd}$, along the
line-of-sight.  Galaxy disks are dynamically cold systems, flattened
and supported by rotation, and thus have small velocity dispersions
(of order \sm10--20\,\kms), whereas the spheroidal bulges of
spiral galaxies can have significant support from random motions.
Proctor \& Sansom (2002) measured velocity dispersions for late-type
spiral bulges in the range 50--200\,\kms\ for Hubble types
Sa\,--\,Sbc.  Because our galaxies are mostly of later type
(Sbc\,--\,Scd, with one Sa bulge), we can expect bulge velocity
dispersions closer to the small end of this range.

We have measured the velocity dispersion at each radial bin (see
\S\ref{sec:kinem}) for our sample and, in principle, a unique set of
models could be created for direct comparison for each radial bin of
each galaxy.  However, given that we are analyzing thousands of
individual spectra, constructing a model for each one becomes too
onerous.  As an alternative solution, we have opted to degrade the
resolution of all galaxy spectra to that with the highest dispersion,
so one set of models, properly convolved to this same velocity, can be
used for direct comparison with the data.  We thus account for
resolution by putting both the data and models on the same scale by
convolving all observed spectra to $\sigma$\,=\,168\,\kms, and all models
with a boxcar of width 10.8\,\AA\ (corresponding to the slit width)
and to $\sigma$\,=\,168\,\kms.

\subsection{Age and Metallicity Fits from Lick Indices}\label{sec:AZindfits}
The Lick/IDS system of spectral line indices was designed to calibrate
the strength of fundamental spectral features in stars and composite
systems (\eg, Gorgas \etal\ 1993). The indices measure the strength of
a particular spectral feature (either atomic and defined as an
equivalent width in \AA, or molecular and measured in magnitudes)
relative to a pseudo-continuum on each side of the feature.  The most
reliable indices have been calibrated as a function of stellar colour
(effective temperature), surface gravity, and metallicity (Gorgas
\etal\ 1993; Worthey \etal\ 1994) allowing for the construction of
semi-empirical population models (\eg\, Worthey 1994).  Each Lick
index is sensitive to the metallicity and age of stellar populations
to varying degrees. When compared with population models, diagnostic
plots of age versus metallicity-sensitive indices, such as \hbeta\
versus Mg{\it b} or Fe, help break the age-metallicity
degeneracy. However, measurements of many of the Lick indices are
quite sensitive to spectral resolution and, thus, to the velocity
dispersion of the system (Gonz{\' a}lez 1993; Trager \etal\ 1998;
Proctor \& Sansom 2002), and their use requires relatively high
signal-to-noise data (S/N/\AA\ $\ge$\,50; see Cardiel \etal\ 1998).
As mentioned above, we account for the former by convolving all data
and model spectra to the same resolution before measuring the indices,
and the latter by coadding the radial spectral bins to S/N/\AA\
$\ge$\,50.  In addition, certain indices, \hbeta\ in particular, can
suffer from nebular emission contamination, even in early-type
galaxies (Gonz{\' a}lez 1993; de~Zeeuw \etal\ 2002; Caldwell \etal\
2003).

To help overcome the problem of nebular emission fill-in of the
\hbeta\ feature, Worthey \& Ottaviani (1997, hereafter WO97) introduced
two pairs of indices that measure the higher order Balmer lines
\hgamma\ and \hdelta.  While their age-sensitivity is not as strong as
for \hbeta, the higher order Balmer lines are much less affected by
emission from ionized gases (\eg\, Osterbrock 1989; Osterbrock \&
Ferland 2006).  Thus, when combined with a metallicity-sensitive
index, the WO97 indices provide a more reliable age estimate for
star-forming galaxies.

Finally, another more subtle issue is the sensitivity of many of the
indices to non-solar(neighborhood) abundance ratios (\eg\ TMB03).  The
BC03 models used here do not account for variations in [$\alpha$/Fe]
which are expected for SFHs with different timescales.  Higher
[$\alpha$/Fe] is associated with very short SF timescales as there is
not enough time for the Fe produced in Type Ia supernovae to enrich
the star forming gas that is heavy with $\alpha$ elements produced in
the short-lived massive stars that end their lives as Type II
supernovae.  On the other hand, for more extended SF, there
is plenty of time for Fe to enrich the gas to solar levels of
[$\alpha$/Fe].  We attempt to address all of these issues in the
context of our spiral galaxy spectra in the following.

In all, there are currently 25 Lick/IDS indices defined and
calibrated\footnote{For up-to-date definitions of all 25 Lick/IDS
indices, see Guy Worthey's webpage at
http://astro.wsu.edu/worthey/html/index.table.html }.  Our spectra
cover 24 of the indices (missing only the bluest high-order Balmer
\hdeltaA\ index).  We measure all 24 and use various combinations
thereof in an attempt to assign reliable light-weighted age and
metallicities to each spectrum.

The obvious advantage of using the Lick/IDS-index system is that the
narrow wavelength span of each index measurement renders them less
sensitive to low-frequency effects such as overall flux calibration
errors and dust reddening (see MacArthur 2005).  One major drawback
when considering galaxies with current or recent SF is that the system
was inherently designed to study the old stellar populations
associated with elliptical galaxies and globular clusters.  

The age and metallicity fits based on Lick indices use same maximum
likelihood approach as in MacArthur (2005), but here the method has
been extended to include more than just two indices.  So, in the
figure of merit (Eq.\@~[A5] in MacArthur 2005), $N$ is now an arbitrary
number of indices (at least 2 and up to the 24 we measure here).

While the errors on the measured indices are independent, the model
tracks are not orthogonal.  This produces non-orthogonal errors in age
and metallicity, which can lead to spurious correlations if not
understood.  Additionally, the degree of non-orthogonality of the age
and metallicity tracks changes with position on the grid.  For this
reason we have used Monte Carlo methods to model the ``effective
ranges'' of the fitted ages and metallicities, taking into account a
normal distribution about the measured errors (see MacArthur \etal\
2004 for a more detailed description of this approach).

We have seen in \S\ref{sec:spectra} that many of our galaxies have
significant amounts of emission, most likely coming from star forming
\hii\ regions.  Many authors attempt to correct for emission by
fitting emission-free templates to their galaxy spectra.  The
templates can either be from linear combinations of stellar templates
(\eg\ Gonz{\' a}lez 1993), or from models (such as the BC03 models
used here).  Neither method is ideal.  The former requires a library
of stellar templates, ideally taken in the same observing run with
identical conditions, that match the galaxian spectra extremely well.
The latter imposes a model dependence since, if a model with a given
age is used to make the correction, the same model age will be
returned (you get out what you put in).  Additionally, a significant
amount of dust in the galaxy would cause a reddening of the spectrum
that must be included as another model-dependent free parameter in the
template fits (for both the empirical and model methods).

In a study of 40 elliptical galaxies, Gonz{\' a}lez (1993) created
individual templates from stellar spectra obtained with the same
observational set-up for each galaxy spectrum (at each radius) for the
purpose of measuring accurate velocity dispersions.  These templates
fit the observed spectra to within \sm1\%.  Division of the galaxy
spectrum by the best fit template enabled detection of faint levels of
emission (seen predominantly in \oiii\ and H$\beta$) in over 60\% of
his sample.  From the galaxy/template spectrum, he computed pseudo-EWs
for the \oiii\,$\lambda$\,5007 \& 4959 \AA\ and H$\beta$ emission and
found them to be strongly correlated as
EW(H$\beta$)\,=\,0.7\,EW(\oiii).  As such, the H$\beta$ index in
absorption could be corrected for emission line fill-in by adding the
correction 0.7\,EW(\oiii) to the measured index.  He also made a
correction to the Fe5015 index as the \oiii\,$\lambda$5007 line lies
within the limits of the central bandpass.  Gonz{\' a}lez (1993)
strongly emphasized the importance of having suitable templates for
reliable measurements of both velocity dispersion and emission
corrections.

The tight correlation between \oiii\ and H$\beta$ emission was tested
for later-type bulges (S0\,--\,Sb bulges) by PS02 using a method
similar to that of Gonz{\' a}lez (1993).  While they found the same
\oiii\,--\,H$\beta$ relation for their earliest-type bulges, the six
late-type bulges did not follow the Gonz{\' a}lez (1993) correlation
for ellipticals, with some bulges showing a stronger H$\beta$/\oiii\
ratio, while others showed clear \oiii\ emission, but none in
H$\beta$.  This is likely due in part to the stronger absorption in
the younger stellar populations, but could also be due to template
mismatch, or true physical differences in the emission line regions
for these galaxies.  Thus, no straight-forward correction seems to
apply to spiral bulges and, by extension, their disks as well.

The higher-order Balmer lines are much less affected by emission line
fill-in.  For a broad range of physical conditions expected in
galactic emission-line regions, the line-strength ratios of H$\delta$
\& H$\gamma$ relative to H$\beta$ are 0.25 and 0.5, respectively
(Osterbrock 1989), providing some relief from emission line fill-in.
Indeed, in many of our galaxies, the H$\beta$ index lies far off of
the model grids (from which ages $\gg$\,20\,Gyr would be inferred),
whereas the higher-order indices imply much younger ages
($\sim$\,5\,Gyr).  However, it is still possible that the higher-order
Balmer lines may suffer significant fill-in in strongly star forming
regions (and indeed we see this in a few of our spectra; see, \eg\
the central spectrum of N7495 in Fig.\@~\ref{fig:spec_rad}).

Given the above mentioned issues about emission-line corrections to
the measured Balmer-line indices, we have instead opted for scheme 
whereby the emission line affected indices are systematically eliminated
from the fit, and the age determinations are compared.  We prefer this 
method as it is less model-dependent and takes full advantage of the 
narrow baselines of the indices, while still using 
information from the entire observed SED.  

We also attempt to address the issue of abundance ratio variations.
While predictions for the effects of enhanced [$\alpha$/Fe] exist for
Lick indices (\eg\ Trager \etal\ 2000; TMB03; Thomas, Maraston, \&
Korn 2004; Schiavon 2007), these do not yet provide high-resolution
SSP spectra, thus we cannot properly account for resolution by
matching the models to the data, nor can we do full spectrum fits for
comparison.  To date, the only models that provide high-resolution
spectra for stars with non-solar abundance ratios are those of Coelho
\etal\ (2007).  Comparing the results obtained here with these new
models will be the focus of an upcoming paper, but for now, we
consider only the BC03 solar-scaled models for all fits.

Our fitting scheme for deriving light-weighted SSP ages and
metallicities from Lick indices considers 5 different cases as
follows:
\vskip 0.2in
\noindent{\bf Case A: Fit All Well-Measured Indices}

\begin{figure*}
\begin{center}
   \includegraphics[width=0.48\textwidth]{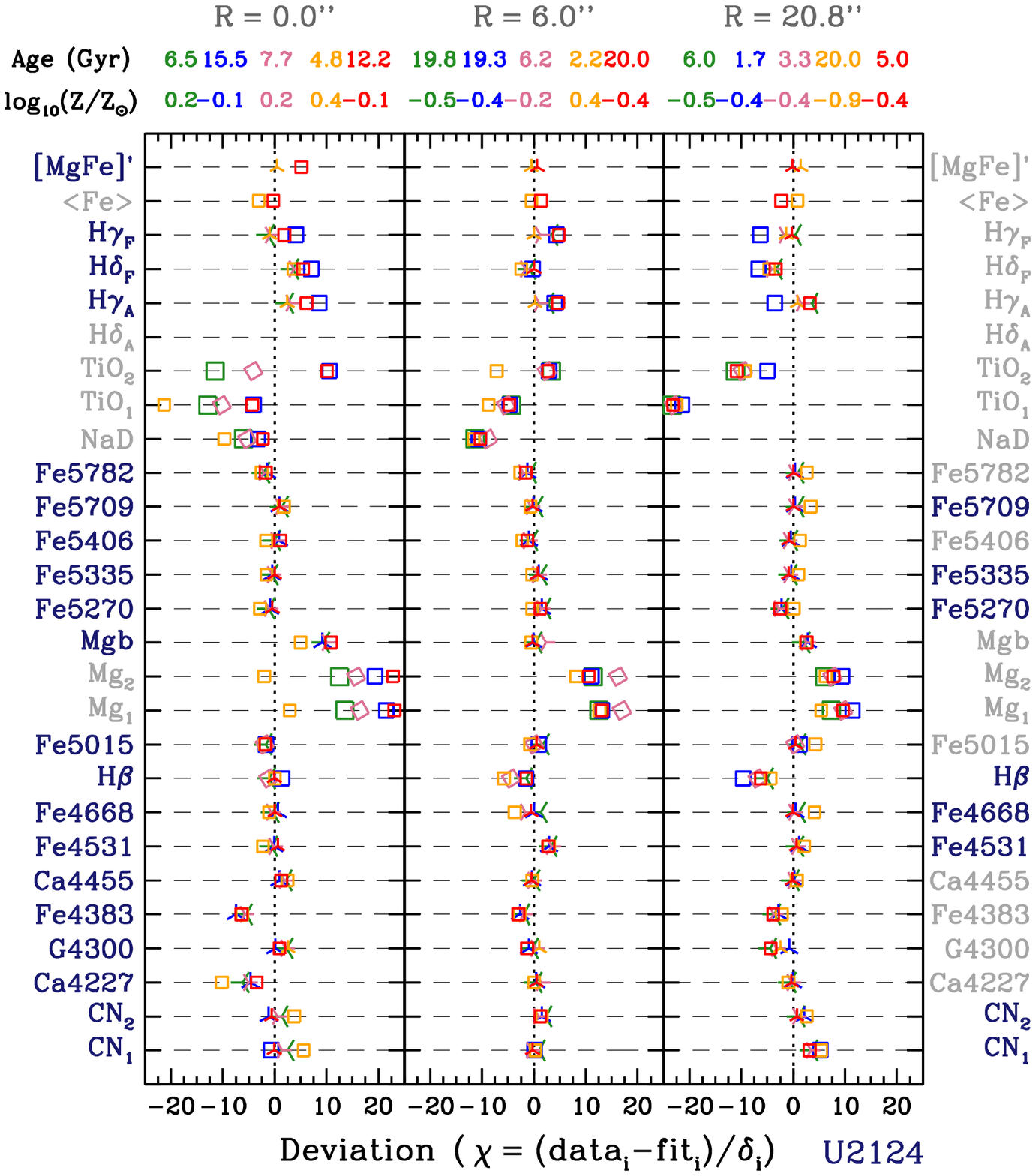}
   \includegraphics[width=0.48\textwidth]{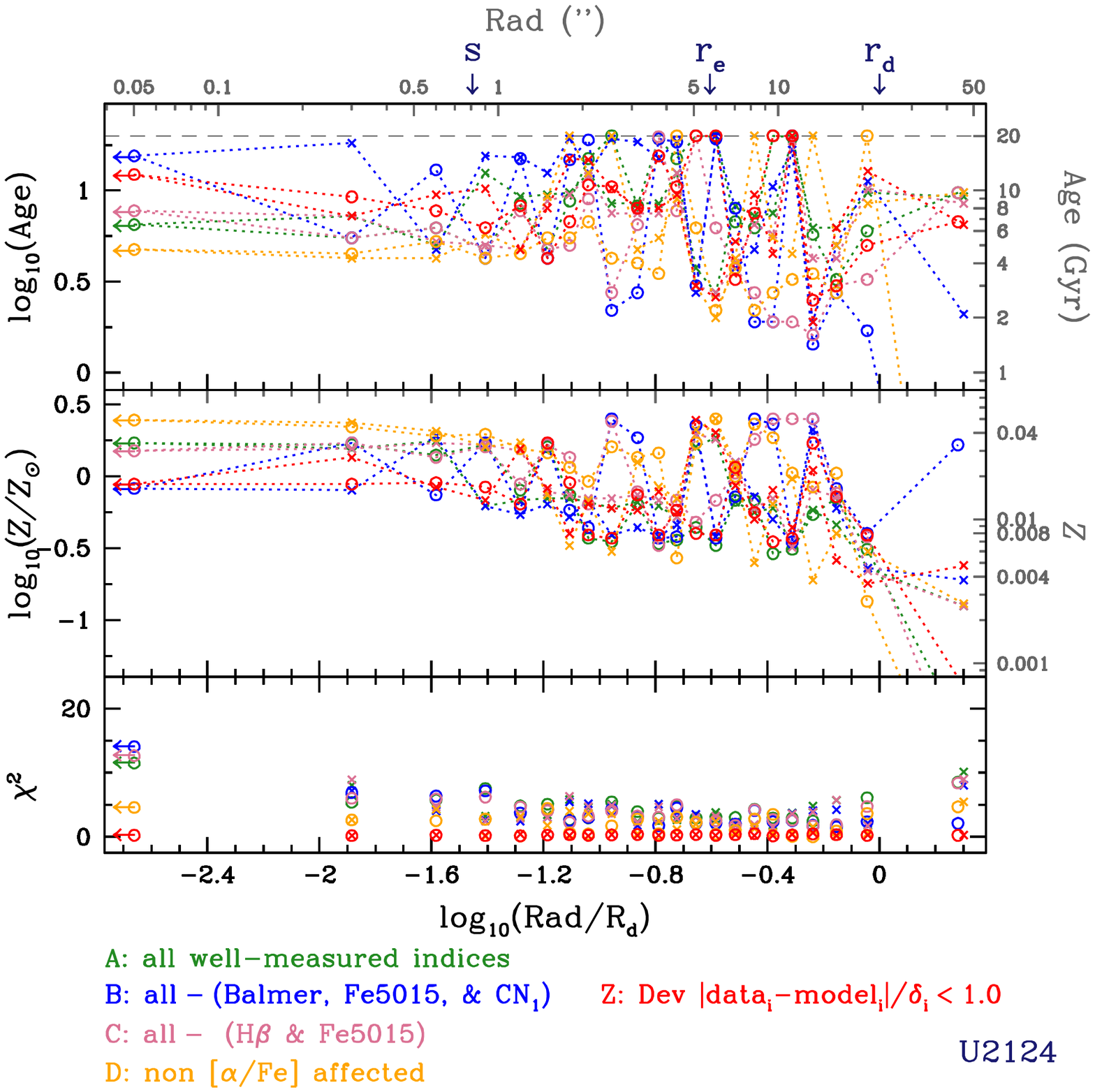}
    \caption{Results for age and metallicity fits from Lick indices
             for U2124.  See text for plot description.}
     \label{fig:AZindfits}
\end{center}
\end{figure*}

\begin{figure*}
\begin{center}
   \includegraphics[width=0.48\textwidth]{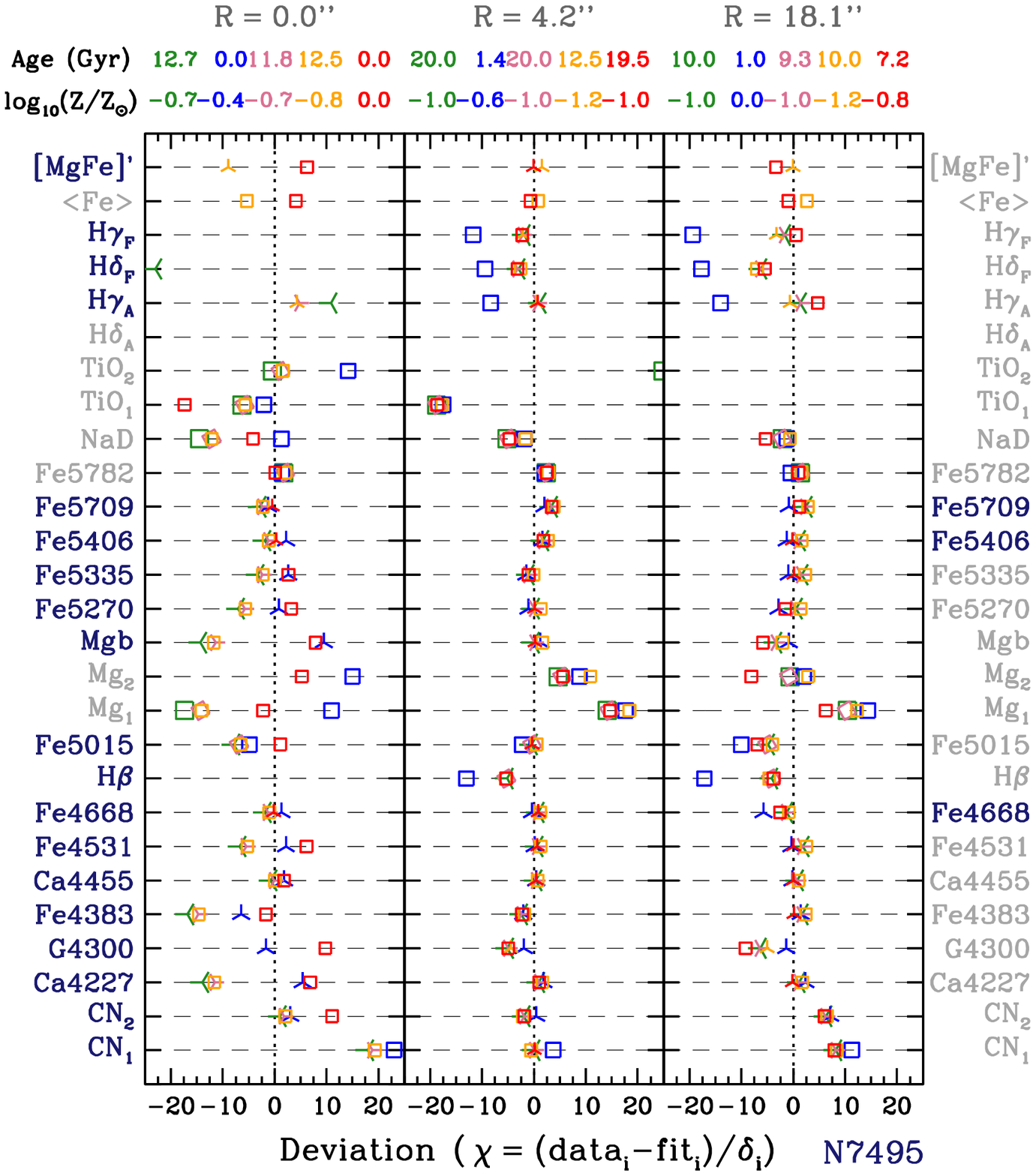}
   \includegraphics[width=0.48\textwidth]{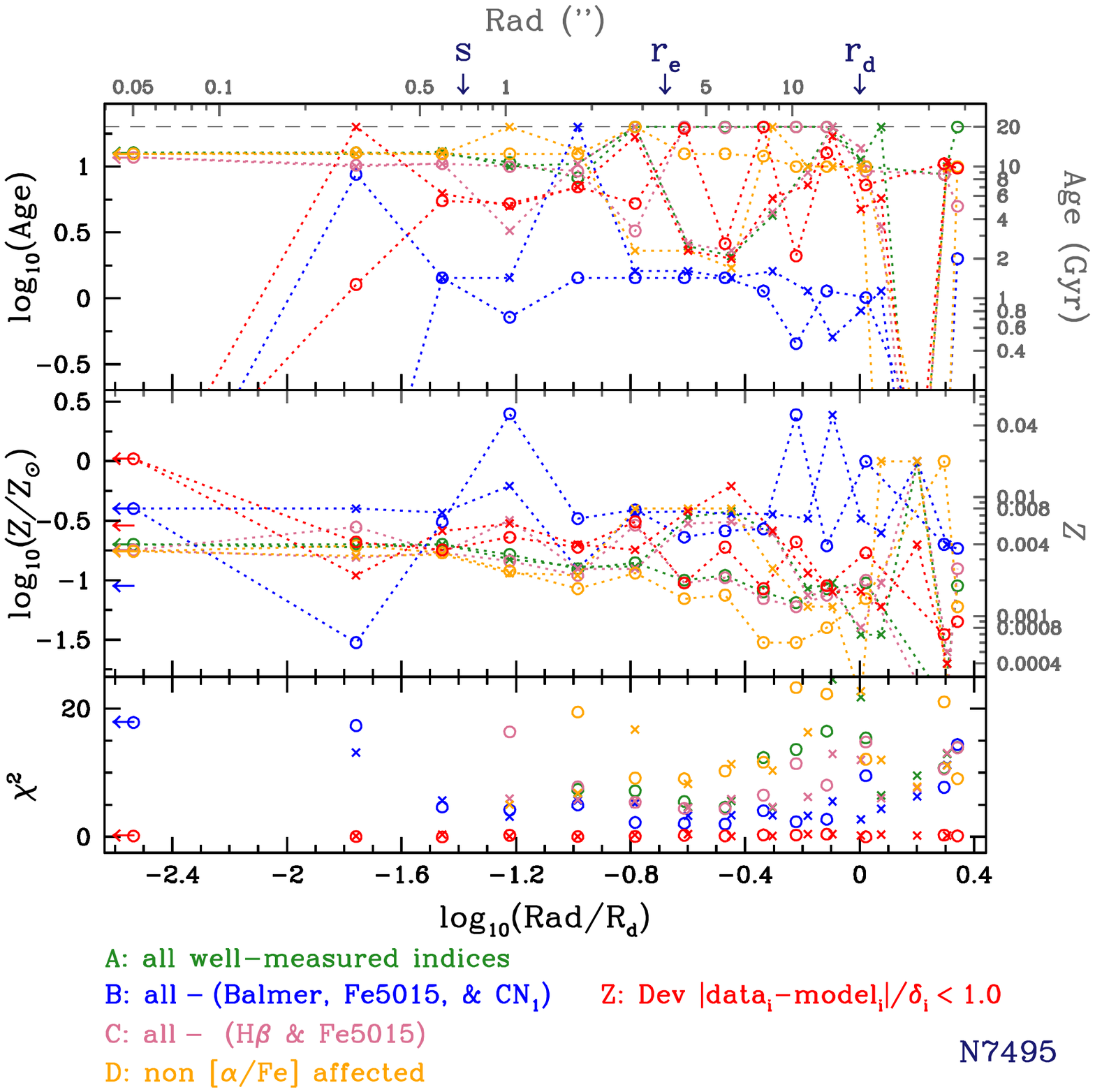}
    \caption{Same as Fig.\@~\ref{fig:AZindfits} except for N7495.}
     \label{fig:AZindfits8}
\end{center}
\end{figure*}

By ``well-measured'' we mean that the index (all 3 bands) is not severely
affected by systematic problems such as the wavelength gaps between
the 3 GMOS CCDs or strong sky lines (which are extremely
difficult to subtract accurately).  The indices that were
systematically eliminated from all fits are highlighted in gray in the
leftmost column in the deviation plots (left panels in
Figs.\@~\ref{fig:AZindfits}--\ref{fig:AZindfits8}, see below for plot
description).  Note that we never include the NaD index in any of the
fits as this index is affected by sky lines as well as poorly
understood absorption from the interstellar medium of the galaxy.  The
TiO indices were also excluded from all fits except for N0628 due to
sky line subtraction issues.

\vskip 0.2in
\noindent{\bf Case B: Case A $-$ (All Balmer-Line Indices, Fe5015, \& CN$_1$)}

\noindent Here we attempt to measure ages and metallicities by
eliminating all indices severely affected by emission.  This is
feasible because of the significant age dependence of G4300 and the
weak age dependence of the metal-line indices.  It is clear that all
Balmer-lines could suffer from emission line fill-in.  The Fe5015
index has the \oiii\,$\lambda$5007 line in its central bandpass thus
could also suffer fill-in.  While seldom mentioned in the
literature, we note that the CN$_1$ index could also be compromised by
a strong source of emission since it contains the
\hdelta\ line in its blue continuum (which forces the index to larger
values).  This is clearly the case for at least two of our galaxies,
N7495 \& N7741, where \hdelta\ is seen in emission (see their
respective spectra in Fig.\@~\ref{fig:spec_rad}).  As an example of
the strong influence that emission can have on CN$_1$, including
CN$_1$ in the fits, having removed all Balmer-line indices and Fe5015,
the age is forced to the maximum model age of 20\,Gyr to accommodate
the large CN$_1$.  When CN$_1$ is removed, the best-fit age drops to
1--2\,Gyr.  We also remark that, because of our resolution, if there
is a strong emission spike in \hgamma, its tails can bleed into the
red passband of the G4300 index, but G4300 was not eliminated from the
fits.

\vskip 0.2in
\noindent{\bf Case C: Case A $-$ (H$\beta$ \& Fe5015)}

\noindent Eliminate only the indices most severely affected by
emission.  This case should be appropriate for galaxies with only
small amounts of emission, which could be undetectable in the spectra,
because the higher-order Balmer-lines will be only weakly affected.
If the age estimates here are younger than those in Case A, 
emission is likely present, whether or not it is clear from
the spectra.

\vskip 0.2in
\noindent{\bf Case D: Fit Only H$\gamma_A$ H$\gamma_F$, [MgFe]$\arcmin$, 
\& G4300}

\noindent These indices are the least affected by abundance
ratios (TMB03; Thomas, Maraston, \& Korn 2004).  While only a few of
our galaxies showed evidence for significant [$\alpha$/Fe] enhancement
(N0173, U2124, \& N7490; see Fig.\@~\ref{fig:avgFeMgb}), many showed
CN$_2$ values far beyond of the model grids (N0628, N7490, I0239),
which this combination of indices also avoids.  Such CN enhancements
have been observed in globular clusters and elliptical galaxies (Henry
\& Worthey 1999), and are likely due to an enhancement in nitrogen
(rather than carbon).  However, the interpretation of
carbon-enhancement is challenging due to the many, and poorly
understood, production sites for carbon.

\vskip 0.2in
\noindent{\bf Case Z: $\sigma$ Clipping of High Deviators} 

\noindent Here we initiate the fit using all ``well-measured''
indices.  We then follow an iterative procedure removing at each step
indices that have large data$-$fit deviations: $|\chi_{i}|$\,=\,$
|{\mbox{data}}_{i} - {\mbox{fit}}_{i}|/\delta_{i}$\,$>$\,$\chi_{max}$,
where $\chi_{i}$ is the relative data/model fit deviation for index
$i$ in units of the error on the measured index,
$\delta_i$.  The threshold $\chi_{max}$ is set to 3.5 initially, and
reduced by 0.5 with each iteration until a minimum value of 1.  Thus,
all indices with $\chi_{i}$\,$\ge$\,1 get eliminated from the fit, and
$\geq$\,2 indices must remain for a successful fit.

\vskip 0.2in Results from the age and $Z$ fits for all five cases are
displayed in Figs.\@~\ref{fig:AZindfits}--\ref{fig:AZindfits8} for
U2124 and N7495, respectively.  In the left panels of each plot, we
show the deviation in units of error for each index in all four fits,
quantified as $\chi_{i}$\,=\,$({\mbox{data}}_{i} -
{\mbox{fit}}_{i})/\delta_{i}$.  Each case is denoted by a different
colour: Case A: green, Case B: blue, Case C: pink, Case D: orange, and
Case Z: red.  Three-pronged point-types mark the indices included in
the fit for each case, while open squares mark indices omitted from
the fit.  Point sizes and rotations were adjusted for visibility when
points are overlapping.  Thus, if a given colour is not visible for a
given index (except \avgFe, which is never explicitly fit, and
[MgFe]\arcmin\, for which we only show Case D results), it is
because the $\chi$ for that index is off scale. The dotted black
vertical line mark zero deviation.  The indices are labeled on the
y-axis (the dashed horizontal lines are there to guide the eye to the
index labels); those marked in gray are the poorly-measured indices
that are never included in the fits (note they are different for each
galaxy).  The index labels on the left indicate those excluded from
all fits, while the labels on the right indicate the indices that
ended up included (blue) in the Case Z fit for the $r$\,=\,0\arcsec\
spectrum.  For each galaxy we plot deviations at three different radii
(labeled in green above each panel): the central point [{\it left
panel}\,], the point closest to the bulge $r_e$ [{\it middle
panel}\,], and the point closest to the disk scale length $r_d$ [{\it
right panel}\,].  Below the radius label for each panel are the best
fit ages and metallicities for each case in their associated colours.

The right panels in
Figs.\@~\ref{fig:AZindfits}--\ref{fig:AZindfits8} show the age
gradients in log [{\it top panel}\,], metallicity gradients in log
[{\it middle panel}\,] and the $\chi^2$ figure of merit [{\it bottom
panel}\,], as a function of the logarithmic radius scaled to the disk
scale length (with labels in arcsec on the top panel) for all five fit
cases (the colour scheme is the same as in the deviation plots and is
labeled at the bottom of each plot).  The data points for each fit are
connected by dotted lines to guide the eye.  Labels on the right axes
show the linear values.  For each galaxy, the seeing FWHM, s, the bulge
$r_e$, and the disk $r_d$ are indicated with arrows at the top.  The
dashed lines in the age and metallicity plots denote the model limits,
so fits with these values should be treated with caution (recall that
we do not extrapolate beyond the model grids).  

Examining the fits for the different cases in
Figs.\@~\ref{fig:AZindfits}--\ref{fig:AZindfits8} we see that
the different cases can yield very different best-fit
parameters.  The reason for this in some cases is clear, for example,
when there is significant emission and the fits are different
depending on which of the Balmer-lines (if any) are included in the
fit (see, \eg, N7495 fits in Fig.\@~\ref{fig:AZindfits8}).  From
the deviation plots we also see that, in general, the indices affected
by prominent sky-lines are poorly fit; they were excluded from the
fits, but the deviation parameter was still computed.  For the
galaxies with very high central CN indices (\eg\ central region of
N7490), those indices were poorly fit in all cases.  

While no single case seems to be ideal for all situations, we take the
Case Z fits as our best age and metallicity estimates based on Lick
indices as, in most cases, these appear to be the most stable and are
the most desirable in terms of excluding highly deviant indices from
the fits.  Still, deriving SP parameters from Lick indices for star
forming galaxies is clearly not a secure method.  With our radially
resolved spectra, we can look at radial trends and identify the bins
that seem most deviant, but if we should consider only one bin per
galaxy, we would not be able to make such judgments.  Thus, let us now
consider SP fits to the full spectrum in hopes of finding a more
stable method for the measurement of age and metallicity in star
forming galaxies.

\subsection{Age and Metallicity Fits from Full Spectra Single SSP Fits}
\label{sec:SSPfits}

We now examine full-spectrum fits, considering here just the best-fit
single model SSP.  Again we use the BC03 models with all model and
galaxy spectra convolved to the same effective resolution.  While
fitting the full spectrum makes use of all the information in the
spectrum, there are some drawbacks as well.  Most notably, the wider
wavelength-baseline reinforces sensitivity to dust extinction and
relative fluxing errors.  However, in this case we can attempt to
model the dust content by incorporating a prescription for dust in the
models.  The fits performed here to single SSP templates are done in a
very similar manner to the full spectrum fits described in \S3, and we
refer the reader to that section for full details of the dust models
and the code.  Here, the number of templates in each fit is $M$\,=\,1,
and the best fit is determined by searching the \chisqr-space of the
fits to each of the 70 individual SSP templates shown in
Fig.\@~\ref{fig:templates}.  Additionally, for each SSP template, we
first compute the \chisqr\ of the best-fit dust-free model.  We then
consider the same SSP, but with increasing amounts of dust, as
described in \S\ref{sec:templates}.  At each dust level (increasing by
increments of \taueff$_V$\,=\,0.5 at each step), if the \chisqr\
decreases from the previous fit, the dust content is increased and the
SSP refit.  This procedure is repeated until we reach a minimum in the
\chisqr-space and only the lowest-\chisqr\ SSP+dust combination for each
template is kept for consideration in the final \chisqr\ grid search.
We imposed a maximum dust content of \taueff$_V$\,=\,8.5,
but this limit was never formally imposed, having always reached a
minimum \chisqr\ at a dust level lower than the limit.
\begin{figure*}
\begin{center}
\includegraphics[width=0.48\textwidth,bb=18 144 592 518]{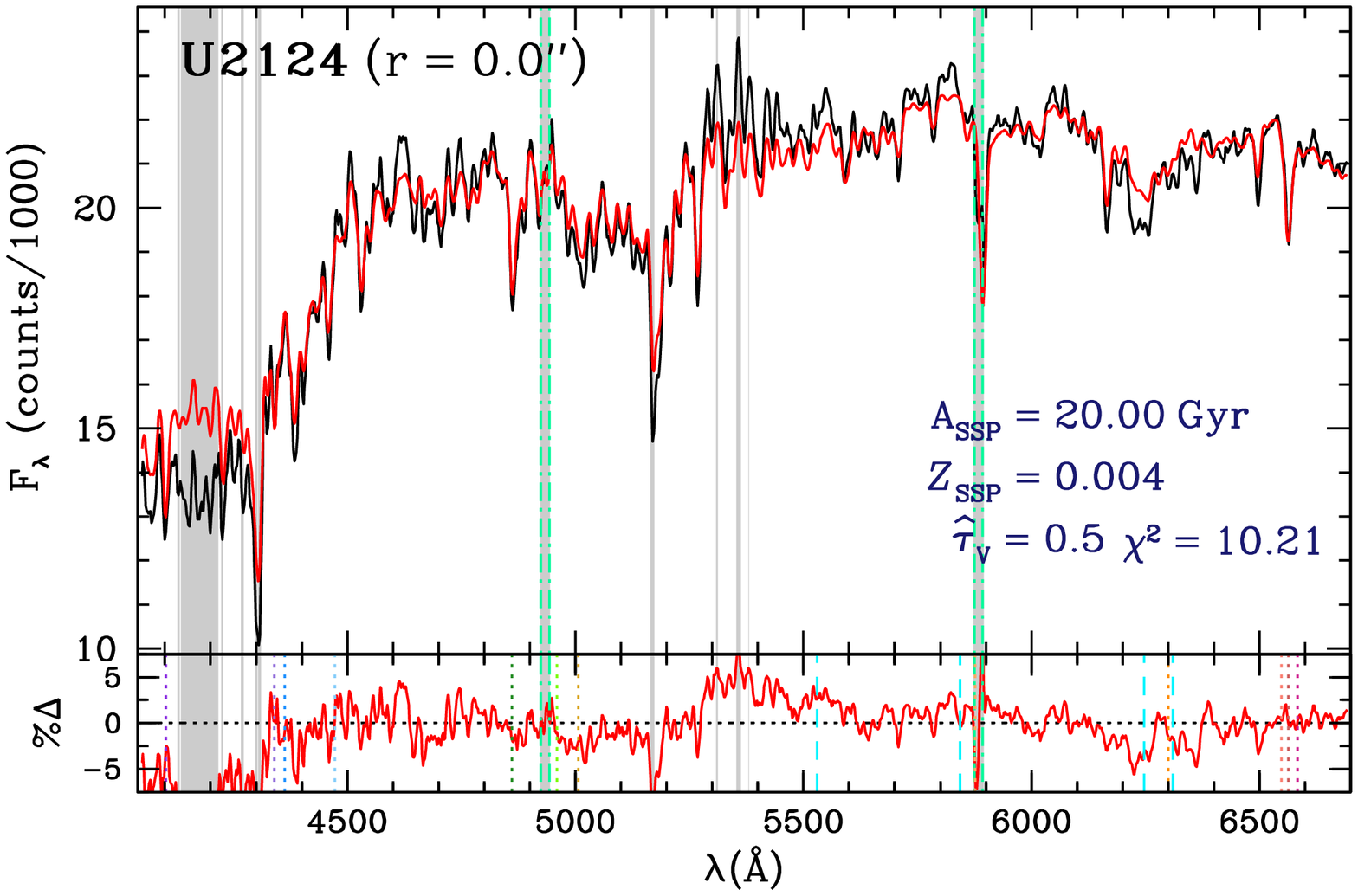}
\includegraphics[width=0.48\textwidth,bb=18 144 592 518]{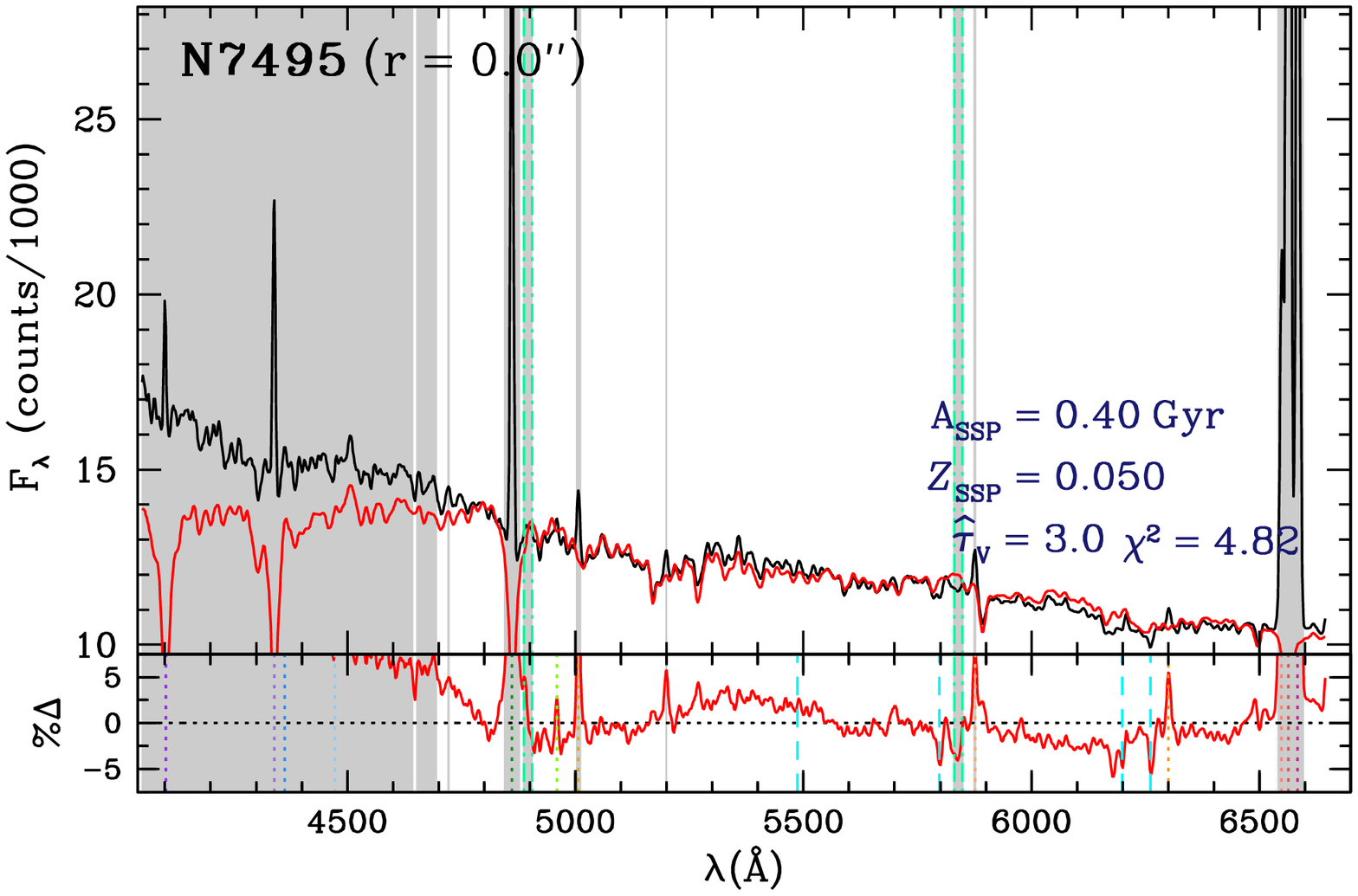}
    \caption{Examples of single SSP fits to central spectra of U2124
             (left) and the emission-line dominant N7495 (right).
             Colours, lines, shading, panels, and labels are as in
             Fig.\@~\ref{fig:fits}, but the ages and metallicities are
             now SSP-equivalent values.  To be compared with full
             synthesis fits in Fig.\@~\ref{fig:fits}.}
     \label{fig:SSPfits}
\end{center}
\end{figure*}
\begin{figure*}
\begin{center}
\includegraphics[width=0.48\textwidth]{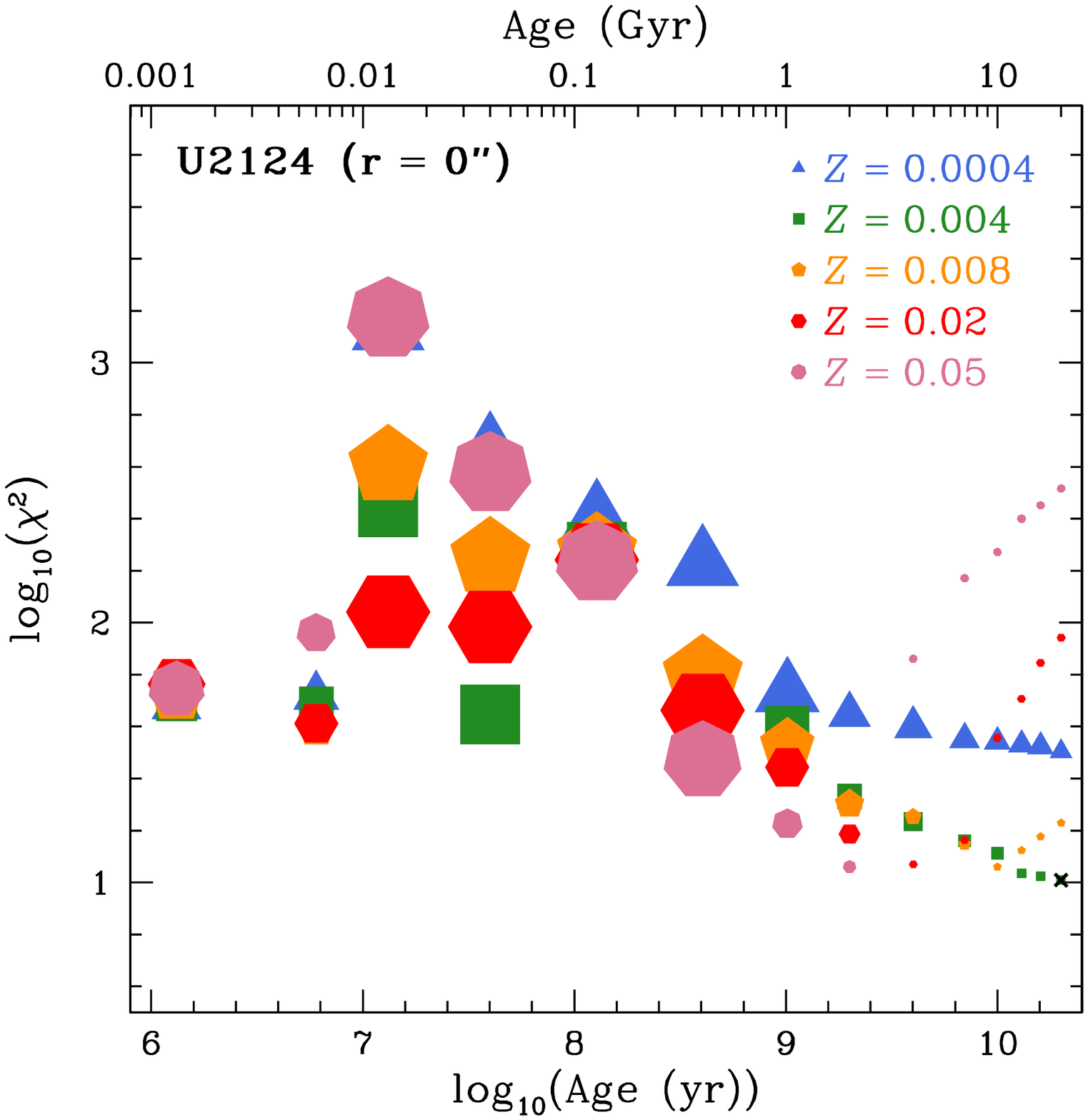}
\includegraphics[width=0.48\textwidth]{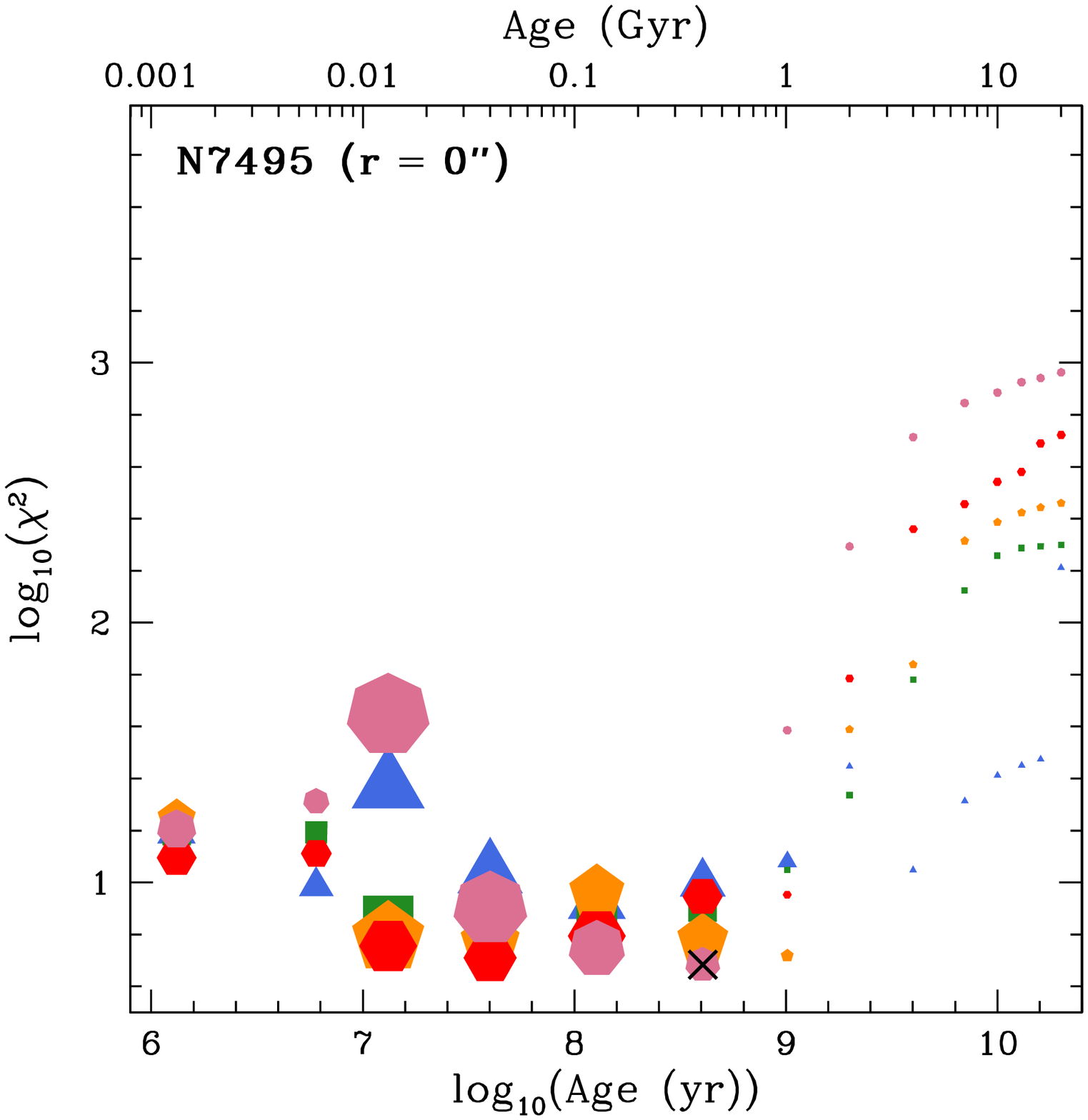}
    \caption{Examples of \chisqr\ distributions for single SSP fits to
     the central spectra of U2124 (left) and the emission-line
     dominant N7495 (right).  Colours and point types indicate
     metallicity as labeled in the upper right corner of the left
     panel.  Point size is proportional to the preferred \taueff$_V$
     of the fit to a given SSP.  The lowest \chisqr\ SSP is indicated
     by the black cross.}
     \label{fig:SSPchis}
\end{center}
\end{figure*}

In Fig.\@~\ref{fig:SSPfits} we show two examples of our SSP fits to
the central spectra of U2124 (left) and N7495 (right).  The central
spectrum of U2124 shows no evidence for emission when examined by-eye
(see Fig.\@~\ref{fig:spec_rad}), and the SED is generally well
represented by a very old (20\,Gyr) but low-metallicity
($Z$\,=\,0.004) SSP with a small amount of dust reddening
(\taueff$_V$\,=\,0.5).  The spectrum of N7495, on the other hand,
shows significant emission, even in the higher-order Balmer lines.
While these do get masked out by our $\sigma$-clipping procedure, 
this spectrum is clearly not well represented by a single SSP.
The entire spectrum blueward of $\sim$\,4700\,\AA\ is masked out in
the fit, there being no SSP that can simultaneously match the low- and
high-frequency characteristics of the spectrum.  

Further insight into the single SSP fits is revealed by examination of
the \chisqr\ distribution.  In Fig.\@~\ref{fig:SSPchis}, we show the
corresponding \chisqr\ distributions to the fits shown in
Fig.\@~\ref{fig:SSPfits}.  For U2124 (left), the lowest \chisqr\ is at
the 20\,Gyr $Z$\,=\,0.004 SSP (indicated by the black cross).
However, a clear minimum in the \chisqr\ distribution has not actually
been reached.  Since 20\,Gyr is the maximum age of the SSP models, we
cannot say whether the \chisqr\ would continue to decrease for even
older SSP ages\footnote{Note that we are not concerned here with model
ages that are older than that of the Universe.  Model ages are not
precisely calibrated and, as such, we are primarily concerned with
relative trends.}.  It is also clear that the \chisqr\ for the 2\,Gyr
$Z$\,=\,0.05 SSP is not all that different from the absolute minimum.
This is a clear example of the age/metallicity degeneracy in stellar
populations whereby an old/metal-poor SSP is as good a fit as a
young/metal-rich SSP.  Comparing the results to the full synthesis
fits (Fig.\@~\ref{fig:fits}), we see that they provide much better
overall matches to the SEDs.  Looking at the corresponding SFH for
$r$\,=\,0\arcsec\ (Fig.\@~\ref{fig:SFH}), we see that there is about
an equal contribution in terms of $V$-band light-weight from an
old/metal-poor SSP, and old/metal-rich SSP, and a young/metal-rich
SSP, indicating that this spectrum is not well-represented by a single
SSP.  On the other hand, the
\chisqr\ distribution for N7495 (Fig.\@~\ref{fig:SSPchis}, right)
reached a minimum within the SSP parameter space.  However, the
minimum is not well defined and a broad range of age/$Z$/\taueff\
combinations are essentially equally well fit.  Again, from the SFH
plot (Fig.\@~\ref{fig:SFH}, right), we see that the central spectrum
of N7495 has significant contributions from both very old and very
young SSPs, which cannot be distinguished in a single SSP fit.  Of
course, there are examples of single SSP fits where a clear and
unambiguous minimum in the age/$Z$/\taueff\ \chisqr-space exists, but
these are the exception for our spectra of star-forming spiral
galaxies.

\section{Comparison of Different Fitting Techniques}\label{sec:comparefits}
\begin{figure*}
\begin{center}
\includegraphics[width=0.9\textwidth,bb=18 310 592 718]{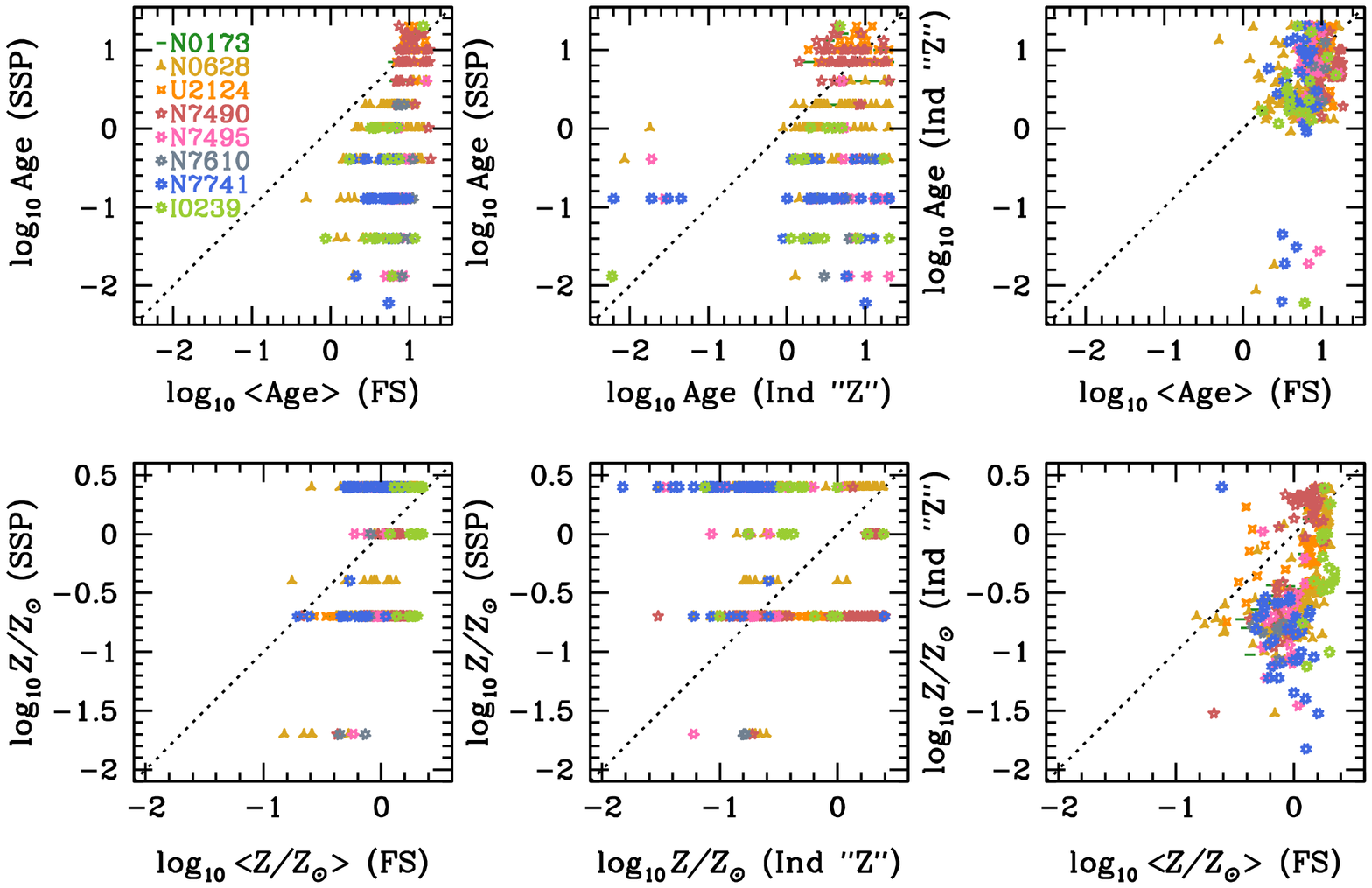}
\caption{Comparison of light-weighted ages and metallicities derived
         from the three different fitting techniques: single SSP vs.\@
         Full Synthesis (FS) [left panels], single SSP vs.\@ Case Z
         Lick index (Ind ``Z'') [middle panels], and Ind ``Z'' vs.\@
         FS [right panels].  Point types and colours are coded to a
         given galaxy, as labeled in the upper left panel.  The dotted
         line in each panel marks the one-to-one relation.}
     \label{fig:FSvsSSPvsInd}
\end{center}
\end{figure*}
\begin{figure*}
\begin{center}
\includegraphics[width=0.43\textwidth,bb=0 290 610 718]{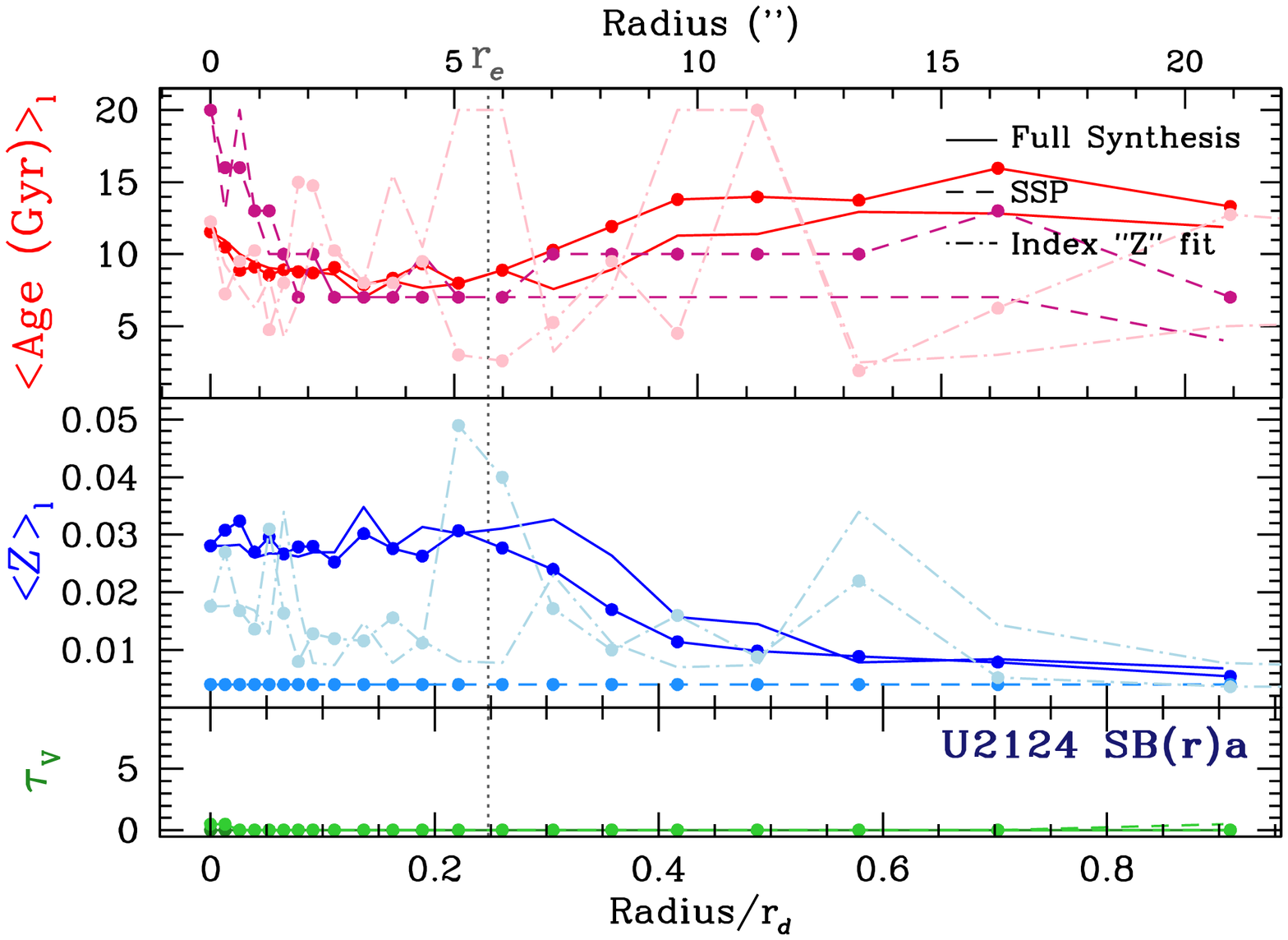}
\includegraphics[width=0.43\textwidth,bb=0 290 610 718]{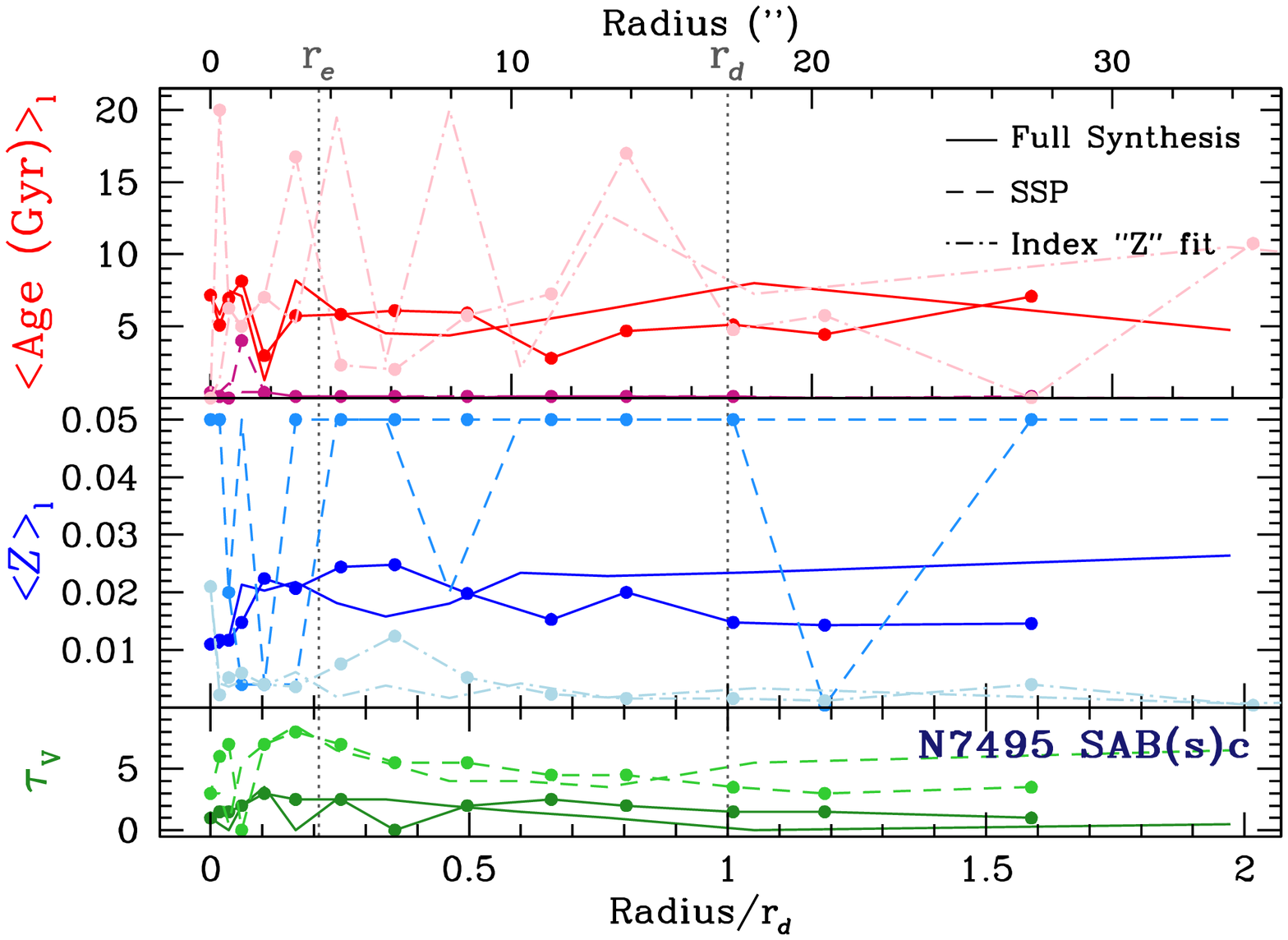}
    \caption{Comparison of the radial gradients from the three
             different fitting techniques.  Panels and symbols are as
             in Fig.\@~\ref{fig:grads}.  Different line types mark
             results from the 3 fitting techniques, as indicated in
             the top right of each panel.}
             \label{fig:gradsFSvsSSPvsInd}
\end{center}
\end{figure*}

It is already quite clear that the only truly representative fitting
technique to derive reliable relative SP ages and metallicities for
the integrated optical spectra of star forming spiral galaxies is the
full synthesis technique presented in \S\ref{sec:synthesis}.  However,
it is still useful to contrast the results from all 3 techniques to
examine any clear relative trends, thus allowing
the results from each method to be interpreted in the context of the
other techniques.  In Fig.\@~\ref{fig:FSvsSSPvsInd} we compare the
derived ages (top panels) and metallicities (bottom panels) between
the single SSP vs.\@ full synthesis (FS) fits (left panels), the
single SSP vs.\@ Case Z (Ind ``Z'') index fits (middle panels), and
the Ind ``Z'' vs.\@ FS fits (right panels).  All quantities are
light-weighted, and note that we label the FS parameters as average
quantities.  Note also that the single SSP ages and metallicities are
stochastically sampled in that they can only be exactly equal to one
of the 70 SSP templates considered, whereas the fits bases on Lick
indices are interpolated between the model grid space, and the FS fits
are weighted averages of all SSPs contributing to the fit.

There is a reasonable correlation between the SSP and FS ages (upper
left panel) but the slope is far from unity (dotted line).  This is
expected due to the difference between SSP-equivalent ages from those
derived from the average of a linear combination of SSPs.  The
difference and scatter become more severe the younger the age, and the
SSP ages reach much younger light-weighted values than those of the
FS fits.  This is once again due to the fact that the youngest stars
are bright and the single SSP fits are sensitive to the stars
contributing most dominantly in flux to the integrated spectrum,
whereas the FS fits are sensitive to any significant contribution from
an underlying old population.  On the other hand, there is no
correlation between the SSP vs.\@ Ind ``Z'' fits (top middle panel).
While these two fits both provide SSP-equivalent age estimates, their
primary source of age discrimination is quite different.  In the full
spectrum SSP fits, it comes largely from the overall shape of the SED.
For the index fits it comes from the age discriminating power
of the indices that do not get clipped from the fit, which may be
different for each fit, and are entirely insensitive to the overall
SED colour.  Finally, the Ind ``Z'' vs.\@ FS comparison (top right
panel) shows a weak correlation with large scatter.  The spread in the
age predictions is similar between these two methods, indicating that
neither is overly sensitive to just the most recent episode of SF, but
the slope of the correlation is not unity, with the FS fits tending to
older ages.  As expected, in all comparisons, the best match is found
for the oldest ages, corresponding to those spectra for which there is
no evidence of emission lines (\eg\ U2124 and N7490).

The situation for metallicity is quite similar, with a weak
and less than unity correlation between the SSP and FS fits 
(bottom left),  essentially no correlation for the SSP and 
Ind ``Z'' fits (bottom middle), and a stronger non-linear relation
for the Ind ``Z'' and FS fits (bottom right), with the FS fits 
tending to higher metallicities.  

Given the weak to null correlations between the derived SP parameters
between the three methods, it is also important to investigate how the
derived gradients would be affected.  In
Fig.\@~\ref{fig:gradsFSvsSSPvsInd} we show the radial variations in
age and $Z$ for the three different fits.  The dashed-dotted lines are
for the Ind ``Z'' fits and have erratic profiles that can jump
drastically from one radial bin to the next.  These jumps often land
at the SP model extremes, indicating that they are poorly determined.
The metallicity profiles seem to be much more stable that the age
profiles, and as we saw in Fig.\@~\ref{fig:FSvsSSPvsInd}, they are
biased to much lower values than the FS fits (solid lines).  The SSP
profiles (dashed lines), on the other hand, are much more stable from
one radial bin to the next.  For the relatively emission-line free
U2124 (left), the age profile has a similar overall form to that of the
FS fits, but because they are insensitive to SPs providing small
relative contributions to the SED flux, the SSP bulge looks much older
than is indicated by the average age from the FS fits.  Moreover, the
metallicity profile is constant at a sub-solar $Z$\,=\,0.004, whereas
both other fits indicate radial variation in $Z$, with higher
metallicity towards the center.  This is another manifestation of the
age/$Z$ degeneracy, particularly in the context of SEDs that are made
up of SPs of a range of ages and $Z$'s and are thus poorly represented
by a single SSP.

The emission-line dominated N7495 (right) shows a slightly different set
of pathologies.  The Ind ``Z'' age profile is again erratic, jumping from
very old to very young values at consecutive radial bins.  The single SSP
fit profile is flat, but saturated at the lowest SSP age, being only
sensitive to the significant amount of current SF and blind to any
underlying population.  The Ind ``Z'' metallicity profile again shows
a similar trend as the FS fits, but at systematically lower-$Z$.  The
single SSP fit profile here is erratic, jumping between high and
low-$Z$ extremes.

Finally, the radial profiles in the amount of dust extinction inferred
form the FS and SSP methods is shown in the bottom panels.  Both are
in agreement for U2124 which is consistent with very little or no dust
at all radii.  The profile shapes for N7495 are similar, but the SSP
derived dust levels are much higher.  The higher values of \taueff\ in
the single SSP fits could in part be absorbing the redness of the
underlying old SP that the single young SSP does not accommodate.

\smallskip
To summarize, there is no clear correlation between the three fitting
techniques for any spectrum with significant contributions to the
light from multiple SSPs.  The SSP-equivalent parameters derived from
either the index-based or single SSP fits reveal erratic behaviors in
the radial profiles, which are very sensitive to different sets of
characteristics of the spectrum.  Only the individual fits and radial
profiles from the full spectrum population synthesis fits yield
representative and stable results and are suitable for assessing
relative trends in age and metallicity.  Therefore, the analysis and
results presented in the main paper are based solely on the full
spectrum synthesis fits.

\label{lastpage}

\end{document}